\documentclass[a4paper,11pt]{article}
\usepackage{jheppub}

\usepackage[most]{tcolorbox}
\usepackage{blkarray, bigstrut} %
\usepackage[english]{babel}

\usepackage{amsmath}
\usepackage{microtype}

\usepackage{esint}
\usepackage{babel}
\usepackage{color}
\usepackage{enumitem}
\usepackage{subfigure}

\usepackage{float}
\usepackage[noabbrev]{cleveref}
\usepackage{bbold}
\usepackage{amsfonts}
\usepackage{amssymb}
\usepackage{amsthm}
\usepackage{enumitem}
\usepackage{shuffle}
\usepackage{comment}
\usepackage{cancel}
\usepackage{blkarray}
\usepackage{multirow}
\usepackage{subfigure}
\usepackage{longtable}
\usepackage{savesym}
\savesymbol{div}
\usepackage{physics}
\restoresymbol{TXF}{div}
\usepackage{tikz}
\usepackage{tikz-3dplot}
\usepackage{xcolor}
\usetikzlibrary{patterns, decorations.markings, arrows}
\usetikzlibrary{arrows.meta}
\usepackage{mathtools}
\usepackage[export]{adjustbox}
\usepackage{tcolorbox}
\usepackage{soul}

\usetikzlibrary{snakes}
\usetikzlibrary{calc}
\usetikzlibrary{decorations}

\usepackage[compat=1.1.0]{tikz-feynman}

\newcommand{\ep}{\epsilon}

\newcommand{\la}{\lambda}
\newcommand{\tl}{\widetilde{\lambda}}

\tikzset{
  rarrow/.style={
    decoration={,
      markings,
      mark=at position 0.65 with {\arrow{latex}}},
    postaction={decorate}
  },
  larrow/.style={
    decoration={,
      markings,
      mark=at position 0.55 with {\arrow{latex reversed}}
    },
    postaction={decorate}
  }
}

\newcommand{\ab}[1]{\langle #1 \rangle}
\newcommand{\Sb}[1]{[ #1 ]}

\def\eps{\epsilon}

\preprint{MPP-2026-2, USTC-ICTS/PCFT-26-09, \\ \rightline{LAPTH-005/26}}

\title{QCD Scattering Amplitudes and Prescriptive Unitarity}
\author[a]{Sérgio Carrôlo,}
\author[b]{Dmitry Chicherin,}
\author[a]{Johannes Henn,}
\author[a]{Qinglin Yang,}
\author[c,d,e]{Yang Zhang}

\affiliation[a]{Max-Planck-Institut für Physik, Werner-Heisenberg-Institut, Boltzmannstr.8, 85748 Garching, Germany}

\affiliation[b]{LAPTh, Université Savoie Mont Blanc, CNRS, B.P. 110, F-74941 Annecy-le-Vieux, France}

\affiliation[c]{Interdisciplinary Center for Theoretical Study, University of Science and Technology of China,
Hefei, Anhui 230026, China}

\affiliation[d]{Peng Huanwu Center for Fundamental Theory, Hefei, Anhui 230026, China}

\affiliation[e]{Center for High Energy Physics, Peking University, Beijing 100871, People’s Republic of China}

\emailAdd{scarrolo@mpp.mpg.de}
\emailAdd{chicherin@lapth.cnrs.fr}
\emailAdd{henn@mpp.mpg.de}
\emailAdd{qlyang@mpp.mpg.de}
\emailAdd{yzhphy@ustc.edu.cn}

\abstract{We present a systematic framework for the maximally-transcendental part of planar QCD scattering amplitudes and perform the first bootstrap computation of six-gluon MHV amplitudes in massless QCD at the symbol level. By analyzing the {\it maximal weight projection} of amplitudes at the integrand level, we relate their maximally-transcendental parts to {\it prescriptive unitarity} integrals. This reveals a novel analytic structure: the prefactors multiplying the functions of maximal transcendentality are identified with the four-dimensional leading singularities of the theory. As a consequence, these prefactors admit a complete classification and can be computed using {\it on-shell diagrams}, a formalism originally developed in $\mathcal{N}{=}4$ super Yang-Mills theory. As a concrete application, we determine the two-loop prefactors for planar MHV gluon amplitudes at arbitrary multiplicity. Combining these prefactors with recent advances in the planar two-loop six-point function space and explicit six-point prescriptive-unitarity input, we construct a complete symbol ansatz and uniquely fix the maximally-transcendental part of the two-loop six-gluon MHV QCD amplitudes by imposing physical constraints. The resulting symbols are expressible in a reduced 137-letter alphabet, suggesting that this alphabet is complete for two-loop six-point massless MHV scattering. We also discuss the implications for multi-collinear splitting and multi-soft functions.
}

\begin{document}

\maketitle

\section{Introduction}
Nowadays, our understanding of the fundamental interactions of elementary particles is deeply rooted in Quantum Field Theory (QFT), particularly described by Quantum Chromodynamics (QCD) within the Standard Model framework. Scattering amplitudes stand as one of the most crucial observables in perturbative QFT, serving as a vital bridge between theoretical predictions in high-energy particle physics and experimental observations at particle colliders such as the Large Hadron
Collider (LHC) (For reviews about modern amplitude study, see \cite{Elvang:2015rqa,Travaglini:2022uwo,Badger:2023eqz}.) The study and computation of scattering amplitudes therefore not only address phenomenological needs, but also deepen our understanding of the Standard Model and QCD, as well as the formal structure of QFT itself, ultimately advancing our grasp of the fundamental principles of nature.

The calculation of QCD amplitudes, particularly at loop level, remains a highly non-trivial task. At the integrand level, whether derived from Feynman rules and Feynman integrals or from unitarity-based methods \cite{Bern:1996je,Britto:2004nc,Ossola:2006us}, the result is invariably a complicated rational function. An even greater challenge lies in performing loop momentum integrations to obtain final results in terms of transcendental functions, which remains the primary bottleneck in analytic computations. Ultraviolet (UV) and infrared (IR) divergences in scattering amplitudes also make the choice of subtraction schemes a non-trivial issue. Currently, loop-level  calculations proceed through several steps: integral reduction via integration-by-parts (IBP) identities \cite{CHETYRKIN1981159}, evaluation of master integrals using the method of canonical differential equations \cite{Kotikov:1990kg,Henn:2013pwa,Henn:2014qga}, and reconstruction of kinematically-dependent series coefficients via numerical algorithms \cite{Peraro:2016wsq,Abreu:2017xsl,Peraro:2019svx}. For parton scattering processes, analytic results are currently available only for four-point amplitudes up to three loops \cite{Glover:2001af,Bern:2002tk,Bern:2003ck,Jin:2019nya,Ahmed:2019qtg,Caola:2021rqz,Caola:2021izf}, and five-point amplitudes up to two loops \cite{Badger:2013gxa,Badger:2015lda,Gehrmann:2015bfy,Abreu:2018zmy,Badger:2018enw,Badger:2019djh,Abreu:2019odu,Abreu:2021oya,Agarwal:2023suw,DeLaurentis:2023nss,DeLaurentis:2023izi}. Beyond five points, as the loop order increases and the number of independent kinematic variables grows, the computational complexity increases substantially. 

Recent years have witnessed remarkable advances in the study of scattering amplitudes in planar maximally supersymmetric Yang-Mills theory ($\mathcal{N}{=}4$ sYM theory \cite{Arkani-Hamed:2008owk}; for a recent review, see \cite{Arkani-Hamed:2022rwr}). As conformally-invariant quantities dual to null polygonal Wilson loops \cite{Alday:2007hr,Brandhuber:2007yx,Drummond:2008aq,Berkovits:2008ic,Caron-Huot:2010ryg,Mason:2010yk}, scattering amplitudes in $\mathcal{N}{=}4$ sYM possess elegant structures, enjoying exceptional physical and mathematical properties. Many pivotal modern methods in the study of scattering amplitudes either originated from, or were profoundly shaped by, studies  within this theory. These include on-shell methods and BCFW recursion relations \cite{Witten:2003nn,Britto:2004ap,Britto:2005fq,Mason:2009sa,Arkani-Hamed:2010zjl}, Grassmannian geometry and the classification of leading singularities \cite{Cachazo:2008vp,Bullimore:2009cb,Arkani-Hamed:2010pyv,Arkani-Hamed:2012zlh}, the Amplituhedron and Positive Geometry framework \cite{Arkani-Hamed:2013jha,Arkani-Hamed:2013kca} at integrand level, as well as symbol technology~\cite{Goncharov:2010jf, Duhr:2011zq}, cluster algebra structures for singularities \cite{Golden:2013xva,Golden:2014xqa,Caron-Huot:2020bkp,Drummond:2019cxm,He:2021eec} and the amplitude bootstrap \cite{Dixon:2011pw,Dixon:2014xca,Dixon:2014iba,Drummond:2014ffa,Dixon:2015iva,Caron-Huot:2016owq,Dixon:2016nkn,Drummond:2018caf,Caron-Huot:2019vjl,Caron-Huot:2019bsq,Caron-Huot:2020bkp,Dixon:2023kop,He:2025tyv} at integrated level.

Despite the remarkable success achieved within this formal theory, a significant gap remains when applying these advanced tools to experimentally relevant theories, and in particular to QCD scattering amplitudes. First, due to the constraints of dual conformal invariance (DCI) and Yangian invariance \cite{Drummond:2008vq,Drummond:2009fd}, scattering amplitudes in $ \mathcal{N} = 4$ sYM  are fully governed by the Bern-Dixon-Smirnov (BDS) ansatz at four and five points to all loops \cite{Bern:2005iz}. Non-trivial kinematics first appears at six points, where amplitudes depend on just three variables. In contrast, QCD scattering amplitudes, even in the planar limit, already exhibit intricate analytic structures at four and five points. Second, amplitudes in $ \mathcal{N} = 4$ sYM are uniform and expressed entirely in terms of maximal transcendental weight functions. QCD amplitudes, on the other hand, contain contributions of mixed transcendental weight as well as rational terms. Studying and computing these rational terms analytically remains highly challenging, even at one loop. So far, only limited analytic results and structural understanding have been obtained.

However, the 
{\it principle of maximal transcendentality} reveals striking connections between the observables of QCD and $ \mathcal{N} = 4$ sYM (for a recent review, see  \cite{Henn:2020omi}). It has been observed that the ``most complicated terms", i.e. the maximally-transcendental part, in
QCD twist-two anomalous dimensions are the same as in $\mathcal{N}{=}4$ sYM, due to the fact that gluons give the relevant dominant contribution in the BFKL equation \cite{Kotikov:2002ab,Kotikov:2004er}. Related heuristic observations of this principle also include duality between $H\to 3g$ processes and supersymmetric half-BPS form factors \cite{Brandhuber:2012vm,Jin:2019ile,Guo:2022pdw}, anomalous dimensions at higher loops \cite{Gehrmann:2011xn,Dixon:2017nat}, and a duality between Wilson loops with Lagrangian insertions and all-plus YM amplitudes \cite{Chicherin:2022bov,Chicherin:2022zxo}.
It is worth noting that, already at one loop, the maximally-transcendental part of the four-gluon QCD amplitude $({-}{+}{-}{+})$ exhibits a richer structure than its sYM counterpart. 
Nevertheless, the maximally-transcendental part of QCD amplitudes shares striking similarities with sYM amplitudes. In particular, the prefactors multiplying maximally-transcendental functions are conjectured to be closely tied to the leading singularities of the theory \cite{Henn:2021aco}, which are conformal invariant \cite{Henn:2019mvc}. This observation opens the possibility of applying on-shell methods and advanced tools originally developed in $\mathcal{N}{=}4$ sYM to the study of QCD amplitudes.

In this paper, we study the two-loop maximal-helicity-violating (MHV) six-gluon amplitude in planar massless QCD, a frontier problem in the analytic computation of QCD amplitudes. With particle momenta restricted to four dimensions, this scattering amplitude depends on seven independent scale-invariant kinematic variables. Important progress has been made recently on two-loop six-point massless integrals and their differential equations. All planar integral sectors have now been computed \cite{Henn:2025xrc,Abreu:2024fei}, and the corresponding planar massless hexagonal function space has been constructed. In principle, one could therefore attempt a direct computation by constructing the integrand via Feynman rules and reducing it to master integrals through IBP identities. However, 
such an analytic calculation would be prohibitively lengthy and cumbersome, while revealing very little structural insight into the amplitude itself. We therefore seek a modification of this procedure.

In this work, we adopt the {\it symbol bootstrap} method to approach this problem. The symbol bootstrap is a powerful method for analytically computing perturbative observables, by combining mathematical structures with physical constraints. The procedure begins by constructing a parametrized ansatz: the perturbative series expansion of the observable is expressed as a linear combination of complete symbol basis $f_{j}$, associated leading singularities $R_i$, with undetermined rational coefficients $c_{i,j}$, i.e.
\begin{equation}
    \mathcal{A}^{(L)}=\sum_{i,j}c_{i,j}R_if_j^{(L)}.
\end{equation}
Physical constraints are then systematically imposed to fix these parameters. These constraints include symmetry requirements, analytic behavior in various kinematic limits, and branch-cut consistency conditions. This approach has demonstrated remarkable computational power, enabling the analytic calculation of six- and seven-point scattering amplitudes \cite{Dixon:2011pw,Dixon:2014xca,Dixon:2014iba,Drummond:2014ffa,Dixon:2015iva,Caron-Huot:2016owq,Dixon:2016nkn,Drummond:2018caf,Caron-Huot:2019vjl,Caron-Huot:2019bsq,Caron-Huot:2020bkp,Dixon:2023kop,He:2025tyv}, as well as three- and four-point form factors of half-BPS operators in $\mathcal{N}{=}4$ sYM theory \cite{Dixon:2020bbt,Dixon:2021tdw,Dixon:2022rse,Dixon:2022xqh,Basso:2024hlx}. 
It has been applied for the first time to two-loop QCD amplitudes
in the Letter~\cite{Carrolo:2025agz}. The present paper refines the bootstrap procedure for QCD amplitudes developed there and generalizes it to more complicated helicity sectors.

To construct the bootstrap ansatz, two types of input are required. First, regarding the symbol basis $f_j^{(2)}$, bootstrap calculations for $\mathcal{N}{=}4$ sYM scattering amplitudes typically build such bases by forming integrable symbols from the anticipated singularity structure of the transcendental functions. In our case, since the planar hexagonal function space has been fully constructed,
it naturally constitutes a complete basis for our bootstrap procedure. Second, in this work we focus solely on the maximally-transcendental part of QCD amplitudes. All corresponding prefactors $R_i$ can be computed using on-shell methods. We demonstrate that these prefactors are completely classified by contours that satisfy triangle power counting, and by their associated maximal cut residues introduced in \cite{Bourjaily:2019gqu}. In that reference, such contours are argued to form a complete basis for MHV ultraviolet-finite amplitudes. 
In our context, since the maximally-transcendental part of the amplitude is always decoupled from UV-divergent Feynman integrals, these contours are sufficient to classify all relevant prefactors. This therefore provides the foundational ansatz for our bootstrap computations.

It is worth emphasizing that our bootstrap calculation establishes a refined connection between {\it prescriptive unitarity} \cite{Bourjaily:2013mma,Bourjaily:2017wjl,Bourjaily:2019gqu,Bourjaily:2019iqr} and the maximally-transcendental part of QCD amplitudes. Prescriptive unitarity refines the generalized unitarity approach by first classifying all independent $L$-loop contours and then constructing pure Feynman integrals that possess non-trivial, unit leading singularities only on specific contours. 
These integrals are called prescriptive unitarity integrals. The coefficients of the integrals in the construction of a scattering amplitude integrand are then determined by their corresponding leading singularities. In our approach, the two-loop prescriptive unitarity integrals of Ref.~\cite{Bourjaily:2019iqr} constitute a part of the transcendental function basis. 
The connection to prescriptive unitarity not only elucidates the analytic structure of the maximally-transcendental part of QCD amplitudes, but also significantly 
simplifies our bootstrap computations.

After constructing the ansatz, we proceed to determine the integrated scattering amplitude—bypassing the cumbersome IBP reduction—by imposing mathematical and physical consistency conditions such as spurious pole cancelation and (multi)collinear limits. 
We 
present the calculation for three independent helicity configurations of the two-loop six-gluon MHV amplitude, covering both pure Yang–Mills theory and QCD. The use of information from prescriptive unitarity integrals significantly reduces the number of undetermined coefficients in the ansatz, thereby enabling 
us to extend the method straightforwardly
to QCD amplitudes involving external quarks.
In the resulting symbols, we observe the reappearance of the 137-letter alphabet previously identified in related observables \cite{Carrolo:2025agz,Carrolo:2025pue,Chicherin:2025cua}.

Finally, we derive new results for the triple-collinear splitting and double-soft functions from our bootstrap analysis. 
In gauge theories, scattering amplitudes exhibit universal factorization properties when subsets of external particles become collinear or soft. In multi-collinear limits \cite{Campbell:1997hg,Catani:1998nv,DelDuca:1999iql,Kosower:2003bh,Catani:2003vu,Badger:2015cxa,Sborlini:2014hva} and multi-soft limits \cite{Kosower:2003bh,Volovich:2015yoa,Klose:2015xoa,Zhu:2020ftr,Czakon:2022dwk}, amplitudes factorize into lower-point amplitudes multiplied by universal process-independent splitting functions. These splitting functions encode the singular behavior associated with the emission of unresolved particles and depend only on the quantum numbers and kinematics of the collinear or soft partons. The universality of these factorization structures plays a central role in higher-order perturbative calculations, underpinning subtraction schemes for infrared divergences, and providing deep insight into the infrared structure of gauge theories. From our bootstrapped results, we extract two-loop triple-collinear splitting and double-soft gluonic functions at maximal weight. We find that these functions are closely related to their counterparts in $\mathcal{N}{=}4$ sYM theory, once again highlighting a connection between $\mathcal{N}{=}4$ sYM  and pure YM theory in the maximally transcendental sector.

The paper is organized as follows. In section~\ref{sec: Preliminaries} we review the basic ingredients and notation used throughout this work. In section~\ref{sec: maximal weight projection} we discuss the maximal weight projection and the corresponding decomposition of QCD amplitudes. Building on this general structure, in section \ref{sec: Leading singularities for MHV QCD amplitudes} we compute all linearly independent two-loop prefactors contributing to the maximally-transcendental part of MHV gluonic QCD amplitudes, arising from five types of maximal cuts and corresponding on-shell diagrams. In section~\ref{sec: Prescriptive unitarity integral} we discuss the relation between maximally-transcendental functions and prescriptive unitarity integrals. We list the prescriptive integrals required in this work and provide non-trivial checks of this correspondence using available perturbative data. In section~\ref{sec: Bootstrapping six-gluon QCD amplitudes} we present the bootstrap computation. We determine all independent MHV helicity sectors, namely $({-}{-}{+}{+}{+}{+})$, $({-}{+}{-}{+}{+}{+})$, and  $({-}{+}{+}{-}{+}{+})$, both for pure YM amplitudes and for the $N_f^k$ terms of six-gluon QCD amplitudes, and discuss the properties of the resulting symbols. In section~\ref{sec: behaviors}, we summarize essential features of multi-collinear/soft limits of QCD scattering amplitudes, which serve as necessary input for the bootstrap. We also extract new multi-collinear/soft splitting functions from our results. We conclude in section~\ref{sec: Discussion} and outline future directions. In Appendix \ref{sec: IBP reduction}, we provide more details on the computation of prescriptive unitarity integrals via IBP reduction. In Appendix~\ref{sec: review of on-shell diagrams}, we review on-shell diagrams and provide relevant references. Appendix~\ref{sec: Regge} contains a discussion of the multi-Regge behavior of the amplitudes.

\section{Preliminaries}
\label{sec: Preliminaries}

\subsection{Six-particle kinematics}

In this work, we calculate six-gluon planar massless scattering amplitudes in QCD. These amplitudes are defined on the kinematic space of six outgoing four-momenta satisfying momentum conservation and on-shell conditions, which read
\begin{equation}
    \sum_{i=1}^6p_i=0\,, \qquad p_i^2=0 \, , \qquad {i\in \{1,\dots,6\} }\,, \label{eq:mc_ll}
\end{equation}
and we denote $p_{6+i}:=p_i$.  However, instead of working with the constrained momenta, we employ variables that trivialize both of these conditions simultaneously. 

We start by making the massless condition of  $p_i$ manifest using the spinor-helicity variables $\lambda_i$ and $\widetilde{\lambda}_i$. These are introduced via the map
\begin{equation}
p_i^\mu \rightarrow p_i^{\alpha \dot{\beta}} \equiv (p_i^\mu \sigma_{\mu})^{\alpha\dot{\beta}}=\lambda_i^{\alpha}\widetilde{\lambda}_i^{\dot{\beta}} \, ,
\label{eq:spinHel}
\end{equation}
where the momentum is contracted with the four-vector of Pauli matrices $\sigma^\mu = (\mathbb{1}, \vec\sigma)$. Notice that the last equality in \eqref{eq:spinHel} follows from the determinant of $p^{\alpha \dot{\beta}}$ vanishing for a massless momentum. In these variables, Lorentz transformations act linearly. Thus, we can form invariants by contracting helicity spinors with the Levi-Civita tensor $\epsilon$  to yield
\begin{equation}
	\langle ab\rangle = \epsilon_{\alpha \beta} \lambda_a^\alpha \lambda_b^\beta \, , \qquad [ ab ] = \epsilon_{\dot{\alpha} \dot{\beta}} \widetilde{\lambda}_a^{\dot{\alpha}} \widetilde{\lambda}_b^{\dot{\beta}} \, .
\end{equation}
Helicity spinors are uniquely defined up to a little-group rescaling. This is particularly important because an amplitude involving a particle of helicity $h_a$ is homogeneous under the corresponding rescaling of $\lambda_a$ and $\widetilde{\lambda}_a$,
\begin{equation}
	A\left(t_a \lambda_a, \frac{1}{t_a}\widetilde{\lambda}_a ; h_a\right) = t^{-2h_a}A(\lambda_a, \widetilde{\lambda}_a; h_a) \, .
\end{equation} 
In this work, we focus exclusively on six-gluon scattering with two negative-helicity gluons. The corresponding tree-level amplitude $A^{(0)}$ is given by the well-known Parke-Taylor (PT) factor, for which we adopt the shorthand notation in the $n$-point case,
\begin{equation}\label{eq:PT}
    \text{PT}_{n,ij} := \frac{\langle ij\rangle^4 }{\langle12 \rangle\langle23\rangle \ldots
    \langle n1 \rangle} \, ,
\end{equation}
where $i,j$ label the negative-helicity gluons. 

Furthermore, since we consider planar amplitudes, we naturally choose the adjacent Mandelstam variables
as a basis of parity-even Lorentz invariants,
\begin{equation}\label{eq:sdef6}
    \{ s_{i,i{+}1}=(p_i+p_{i+1})^2 \ \}_{ i=1,\cdots,6}\, \cup\ \{ s_{i,i+1,i+2}=(p_i{+}p_{i{+}1}{+}p_{i{+}2})^2 \ \}_{i=1,2,3} \, .
\end{equation}
However, they are not linearly independent in four-dimensional kinematics, since the momenta of the scattered particles are constrained by a Gram-determinant condition. Because our parametrization of the kinematics is manifestly four-dimensional, this constraint is automatically satisfied.

In addition to solving the on-shell constraints, we can also trivialize momentum conservation. Given an ordering of scattered particles, we arrange their momenta to form a closed polygon. Focusing on the polygon vertices $x_i \in \mathbb{R}^4$ instead of the edges $p_i$,
\begin{equation}
\label{eq:mompoly}
	p_i = x_{i+1} - x_i \, , \qquad x_7 = x_1 \, ,
\end{equation}
ensures that momentum conservation is automatically satisfied. Combining eqs. \eqref{eq:spinHel} and \eqref{eq:mompoly}, we introduce momentum twistors $Z_i \in \mathbb{P}^3$,
\begin{equation}
	Z_i^I = \begin{pmatrix}
		\lambda_i^\alpha \\
		(x_i \lambda_i)^{\dot{\beta}}
	\end{pmatrix} \, , \qquad I\in\{1,\dots,4\} \, , \, {\alpha,\dot\beta} \in \{1,2\} \, .
\end{equation}
Thus, any collection of six momentum twistors $Z_{1,\dots,6}$ specifies six massless four-dimensional external momenta that satisfy momentum conservation. 

Both constraints in \eqref{eq:mc_ll} are invariant under conformal transformations, which take a particularly simple form in momentum twistors variables. Indeed, momentum twistors transform in the fundamental representation of $SL(4)$ under conformal transformations. The conformal invariants are four-brackets given by contractions of momentum twistors with the Levi-Civita tensor $\epsilon$ as follows,
\begin{equation}
	\langle a b c d \rangle = \epsilon_{IJKL} Z_a^I Z_b^J Z_c^K Z_d^L \, .
\end{equation}
In order to relate the momentum twistor four-brackets to the spinor-helicity two-brackets, we have to break conformal invariance. Thus, we introduce a point at infinity that corresponds to the infinity bi-twistor 
\begin{equation}
    I_{\infty}:=\left(\begin{matrix}
        0&0 & 1 &0 \\0&0 &0 & 1
    \end{matrix}\right) \,.
\end{equation}
Then, the spinor-helicity brackets are simply given by
\begin{equation}
	\langle i j I_\infty \rangle = \langle ij\rangle \, . 
\end{equation}
Finally, the Mandelstam variables are given by ratios of momentum twistor four-brackets, 
\begin{equation}
\label{eq:sdef63}
    s_{i,i{+}1}=\frac{\langle i{-}1\ i\ i{+}1\ i{+}2\rangle}{\langle i{-}1\,i\rangle\langle i{+}1\,i{+}2\rangle} \, , \quad  s_{i,i{+}1,i{+}2}=\frac{\langle i{-}1\,i\,i{+}2\,i{+}3\rangle}{\langle i{-}1\,i\rangle\langle i{+}2\,i{+}3\rangle} \, .
\end{equation}

\subsection{Iterated integrals and symbols}
\label{sec:iiandsymb}

All transcendental functions appearing in our calculation are Chen iterated integrals with dlog kernels \cite{Chen:1977oja}; multiple polylogarithms (MPLs) are among the functions of this type. The transcendental weight counts the number of iterated integrations.
The total differential of a weight-$w$ ${\rm d}\log$-iterated integral $\mathcal{F}^{(w)}$ satisfies the following recursive relation
\begin{equation}\label{eq:basictotald}
	\textrm{d} \mathcal{F}^{(w)} = \sum_i \mathcal{F}^{(w-1)}_i \textrm{d} \log \alpha_i \, ,
\end{equation}
where $\mathcal{F}_i^{(w{-}1)}$ are weight-$(w{-}1)$ ${\rm d}\log$-iterated integrals, and \emph{letters} $\alpha_i$ are algebraic functions in the kinematic variables of $\mathcal{F}^{(w)}$. Based on \eqref{eq:basictotald}, we introduce the {\it symbol} map ${\cal S}$ for ${\rm d}\log$-iterated integrals \cite{Goncharov:2010jf, Duhr:2011zq}. This map is linear and defined recursively in weight,
\begin{equation}
	\mathcal{S}(\mathcal{F}^{(w)})=\sum_i\mathcal{S}(\mathcal{F}_i^{(w{-}1)})\otimes \alpha_i \ , 
\end{equation}
and ${\cal S}(1) =1$. Thus, the symbol of $\mathcal{F}^{(w)}$ is a tensor product with $w$ entries,
\begin{equation} \label{eq:symbF}
    \mathcal{S}(\mathcal{F}^{(w)})=\sum_k c_k \,  
    \alpha_1^{k}\otimes\cdots\otimes \alpha_w^{k} \ ,
\end{equation}
where the coefficients $c_k$ are rational numbers. More generally, if the function ${\cal F}^{(w)}$ is a homogeneous linear combination of weight-$w$ $d\log$-iterated integrals, then the coefficients $c_k$ in eq.~\eqref{eq:symbF} are the so-called \emph{leading singularities}, and they are typically rational functions of external kinematics. 
If coefficients $c_k$ are rational numbers, then the corresponding function $\mathcal{F}^{(w)}$ is called \emph{pure}. Furthermore, the full collection of $\alpha_i^k$ forms the \emph{symbol alphabet} of $\mathcal{F}^{(w)}$. The zero loci of the letters are branch-cut singularities of $\mathcal{F}^{(w)}$ and its analytic continuation.

In the method of differential equations \cite{Henn:2013pwa}, for a fixed number of scattered particles and fixed loop order, one constructs a basis of dimensionally regularized \emph{master integrals}. In the \emph{uniform transcendental} (UT) basis, the master integrals satisfy canonical differential equations
\begin{equation}
    {\rm d}\mathbf{I}=\epsilon\ {\rm d}A\cdot\mathbf{I}\,, \label{eq:cDE}
\end{equation}
with the dimensional-regularization parameter $\epsilon$ factored out. The matrix ${\rm d}A$ depends  on the external kinematic variables only in the following way,
\begin{align}
    {\rm d}A=\sum_{i}A_i\ {\rm d}\log W_i \,,
\end{align}
where $A_i$ are matrices whose entries are rational numbers, and $W_i$ are symbol letters (also denoted as $\alpha_i$ above). 

With the symbol alphabet known prior to performing loop integrations, one can attempt to compute the amplitude, as well as other observables, using the \emph{symbol bootstrap} strategy. This calculation usually starts with constructing the function space from the symbol letters and imposing the integrability conditions. 
Given the function space and the collection of leading singularities, one then builds an ansatz for the observable containing a finite number of free parameters. To fix these parameters, one typically uses the known behaviour of the observable in singular kinematic regions. After imposing these physical conditions, all parameters can ideally be determined, yielding a unique result. Following this approach, outstanding progress has been made in bootstrapping scattering amplitudes of planar $\mathcal{N}{=}4$ sYM theory \cite{Dixon:2011pw,Dixon:2014xca,Dixon:2014iba,Drummond:2014ffa,Dixon:2015iva,Caron-Huot:2016owq,Dixon:2016nkn,Drummond:2018caf,Caron-Huot:2019vjl,Caron-Huot:2019bsq,Caron-Huot:2020bkp,Dixon:2023kop,He:2025tyv}, both at the symbol and function level. However, 
despite these remarkable advances for $\mathcal{N}{=}4$ sYM, these methods remain largely unexplored for physically realistic theories, such as the theory of the strong interactions.

\subsection{Color decomposition and planar limit}

For the sake of generality, we consider a gauge theory with an $SU(N_c)$ gauge group. The generators are denoted by $T^a$ and the gauge bosons are referred to as gluons. The scattering amplitudes contain both color, associated with the gauge group, and kinematic structures. At tree-level, we disentangle color and kinematics via the standard color decomposition \cite{Bern:1994zx}, which takes the following form for $n$-point gluon scattering
\begin{equation}
    A_n^{(0)} = g^{n-2} \sum_{\text{perm }\sigma} \Tr (T^{\sigma(a_1 \dots} T^{a_n)}) A_n^{(0)}( \sigma(1 \dots n)) \, , 
\end{equation}
where the summation is over $(n-1)!$ permutations $\sigma$ of $n$ elements modulo cyclic shifts and $A_n^{(0)}$ are called the color-ordered partial amplitudes. The latter are given by the PT factors \eqref{eq:PT} in the MHV helicity case. 

The tree-level amplitudes involve only single-trace terms. Starting from one-loop, multi-trace color structures contribute. However, they are suppressed in the large $N_c$ approximation. For example, a one-loop gluonic amplitude in the notation of \cite{Bern:1994zx}
\begin{equation}
    A^{(1)}_n = g^{n} \left[  \sum_{ \sigma} N_c \Tr (T^{\sigma(a_1 \dots} T^{a_n)}) A_{n;1}^{(1)} + \sum_c \sum_{\sigma} \Tr(T^{\sigma[1\dots}T^{(c-1)]})\Tr(T^{\sigma[c\dots}T^{n]}) A_{n;c}^{(1)}\right] \, .
\end{equation}
In the previous expression, in the large $N_c$ limit at fixed 't Hooft coupling $\lambda = g^2N_c$, the double-trace contributions are suppressed relative to the single-trace contributions. Only Feynman integrals of planar topology contribute in this limit, motivating the term planar limit. Similarly, in the planar limit at $L$-loop order, only single-trace contributions survive,
\begin{align}
A^{(L)}_n = g^{n-2+2L} \left[  \sum_{ \sigma} N^L_c \Tr (T^{\sigma(a_1 \dots} T^{a_n)}) A_{n;1}^{(L)} + \text{subleading}\right] . \label{eq:Aleadcolor} 
\end{align}
The corresponding partial amplitudes $A^{(L)}_{n;1}$ are called the leading-color partial amplitudes.

For gluon scattering in QCD, it is necessary to include contributions with quarks running in the loops. The quarks are fermions in the fundamental representation of $SU(N_c)$, and there are $N_f$ quark flavors. To extract the planar quark contributions, we take both $N_f$ and $N_c$ to be large, with the ratio $N_f/N_c$ held fixed and finite,
\begin{align}
\text{planar limit :} \qquad N_c \rightarrow \infty \,,\quad \lambda=g^2N_c = \text{fixed} \,,\quad
N_f/N_c = \text{fixed} \,. \label{eq:planar}
\end{align}
For example, in the planar limit, the two-loop partial amplitudes receive contributions from both gluons and quarks running in the loops in the form of the following expansion
\begin{equation}
    A_{n;1}^{(2)} =  A_{n;1}^{[1]} + \left(\frac{N_f}{N_c}\right) A_{n;1}^{[\frac12]} + \left(\frac{N_f}{N_c}\right)^2 A_{n;1}^{[\frac12,\frac12]} \,  . \label{eq:A2dec}
\end{equation}
All partial amplitudes in the previous equation are understood to be at two-loop order. $A_{n;1}^{[1]}$ is the contribution from gluons circulating in both loops, which corresponds to the pure YM theory. $A_{n;1}^{[\frac12]}$ denotes the contribution from diagrams involving a single quark loop and a single gluon loop, and $A_{n;1}^{[\frac12,\frac12]}$ corresponds to quarks running in both loops. Note that there is a subtlety in the on-shell diagrams that one has to take into account in the planar limit. Namely, not all on-shell diagrams with a quark loop contribute to the $N_f/N_c$-term. We will comment on this below when discussing relevant leading singularities in Section~\ref{sec:QCDls}.

\subsection{IR subtraction, UV renormalisation and hard function}
\label{sec:IR}

Two-loop color-ordered amplitudes contain IR poles up to the fourth order in the dimensional regulator $\epsilon$, where $D=4-2\epsilon$. Since we assign transcendental weight $-1$ to $\epsilon$, the maximally-transcendental part of the two-loop $n$-point amplitude is given schematically by the following sum
\begin{equation}
    A_n^{(2)} = \sum_{\ell=0}^{2} \frac{1}{\epsilon^{2\ell}} f_n^{(4-2\ell)} + \mathcal{O}(\epsilon) \, ,
\end{equation}
where $f_n^{(w)}$ is a function (which is not pure) of transcendental weight $w$. 

The infrared structure of gauge theory amplitudes is well-understood \cite{Sterman:2002qn,Becher:2009cu,Magnea:1990zb,Gardi:2009qi}. Their IR poles exponentiate, and this exponentiation is governed by the cusp \cite{Korchemsky:1987wg,Korchemskaya:1992je} and collinear \cite{Dixon:2008gr} anomalous dimensions. Consequently, the IR divergences of the amplitudes have a universal form and can be systematically subtracted. This generic treatment of IR divergences considerably simplifies in our calculation set up, because we  
\begin{itemize}
    \item consider the planar limit,
    \item keep only the maximally-transcendental terms,
    \item and omit transcendental constants.
\end{itemize}
Thus, we can rely on the dipole formula to implement the IR subtraction of the amplitudes~\cite{Catani:1998bh}. Moreover, we can ignore the collinear anomalous dimension and retain only the leading order expression for the cusp anomalous dimension. Thus, at one-loop order, we choose the IR subtraction function 
\begin{equation}
    I_n^{(1)} =- \frac{1}{\epsilon^2}\sum_{i=1}^{n}\left(-\frac{s_{i,i+1}}{\mu^2}\right)^{-\epsilon} \,. \label{eq:I61}
\end{equation}
Working at the symbol level and
retaining only the maximally-transcendental contributions, the two-loop IR subtraction 
can be simplified to
 (cf. \cite{Aybat:2006mz})
\begin{equation}
    I_n^{(2)} \longrightarrow -\frac{1}{2}\left(I^{(1)}_n\right)^2 \, . \label{eq:I62}
\end{equation}
In addition to IR divergences, loop integrations also generate UV divergences. 
These are typically eliminated by expressing the bare coupling in terms of the renormalized coupling. Diagrammatically, UV divergences of scattering amplitudes are always contributed by bubble-counting integrals, which always leads to weight-drop after integration due to the double-pole at $\ell\to\infty$. Since we focus on the leading transcendentality part, UV poles can be safely ignored in the present analysis. Accordingly, $g$ denotes the renormalized coupling.

Using this subtraction scheme,
we then define the finite parts,  so-called {\it hard functions}, of the amplitudes as
\begin{equation}\label{eq:Hdef}
    H_n^{(L)} = \lim_{\ep \to 0} \left( A_n^{(L)} - \sum_{\ell = 1}^{L} A_n^{(L-\ell)}I_n^{(\ell)} \right).
\end{equation}
Usually, loop integrations in dimensional regularization introduce explicit dependence on the renormalization mass scale $\mu^2$ in the form of powers of $\log(\mu^2)$. With our choice of the IR subtraction scheme, the maximally-transcendental parts of hard functions $H_n^{(L)}$ contain no explicit dependence of this mass scale, and thus they are  
well-defined in four dimensions.

Finally, the two-loop QCD hard functions allow for the following decomposition in the planar limit \eqref{eq:planar},
\begin{equation}\label{eq:hardde}
    H_{n}^{(2)} =  H_{n}^{[1],(2)} + \left(\frac{N_f}{N_c}\right) H_{n}^{[\frac12],(2)} + \left(\frac{N_f}{N_c}\right)^2 H_{n}^{[\frac12,\frac12],(2)} \,,
\end{equation}
which is analogous to \eqref{eq:A2dec}.
In this work, we consider gluonic amplitudes in QCD. The term $H_{n}^{[1]}$ in \eqref{eq:hardde} corresponds to the pure YM theory, while the second $H_{n}^{[\frac12]}$ and third $H_{n}^{[\frac12,\frac12]}$ terms originate from Feynman diagrams with quark loops. Since in the following sections we mainly discuss the details of pure YM amplitudes, if we omit the $[k]$ superscript in an expression, it should be understood that we are referring to the results for the pure YM hard function, $H_n^{[1],(L)}$ .

We must also specify the helicity of the scattered particles.
Our primary interest in the present work is six-gluon scattering amplitudes in the MHV helicity configuration, which involve two negative helicity gluons. There are 15 distinct ways to assign negative helicity to a pair of gluons among the six. Fortunately, the corresponding helicity amplitudes are not all independent. There are three inequivalent MHV helicity configurations,
\begin{align}
1^- 2^- 3^+ 4^+ 5^+ 6^+ \,,\qquad   
1^- 2^+ 3^- 4^+ 5^+ 6^+ \,,\qquad  
1^- 2^+ 3^+ 4^- 5^+ 6^+  \,, \label{eq:config}
\end{align}
and the remaining MHV configurations are cyclic permutations thereof. Indeed, recall that we consider the planar limit and the leading color structures in \eqref{eq:Aleadcolor} are invariant under dihedral permutations of the scattered particles. The dihedral transformations $\mathbb{Z}_6 \times \mathbb{Z}_2$ are generated by the cyclic shift $i \to i+1$ (assuming $7 \equiv 1$) and reflection $i \to 7-i$. Modding out these dihedral symmetries therefore leaves three distinct configurations \eqref{eq:config}. Furthermore, the helicity configurations in \eqref{eq:config} retain residual dihedral symmetries, which in turn imply dihedral symmetry relations for the six-gluon hard functions, 
\begin{align}
& H(1^- 2^- 3^+ 4^+ 5^+ 6^+ )  =  H(2^- 1^- 6^+ 5^+ 4^+ 3^+ ) \,, \notag \\
& H(1^- 2^+ 3^- 4^+ 5^+ 6^+ )  =  H(3^- 2^+ 1^- 6^+ 5^+ 4^+ ) \,, \notag\\
& H(1^- 2^+ 3^+ 4^- 5^+ 6^+ )  =  H(4^- 3^+ 2^+ 1^- 6^+ 5^+ ) =  H(4^- 5^+ 6^+ 1^- 2^+ 3^+ ) \, \label{eq:Hsym}
\end{align}
where $H$ stands for $L$-loop planar $H_6^{(L)}$, as well as individual terms in its $N_f$-decomposition, see \eqref{eq:hardde}. We tacitly understand the hard function to refer to the symbol projection of its maximally-transcendental part, so that $H^{(L)}_6$ is a weight-$2L$ symbol.

\begin{table}[t]
\centering
\begin{tabular}{|l|c|c|c|c|}
\hline
Transcendental weight  & 1 & 2 & 3 & 4 \\ \hline
\hline
\# Two-loop six-point symbols  & 9 & 59  & 266  & 639 \\ \hline
\# Two-loop five-point one-mass symbols   & 9 & 59 & 263 & 594\\ \hline
\# Genuine two-loop six-point symbols  & 0 & 0 & 3 & 45 \\ \hline \hline
\# One-loop six-point symbols & 9 & 26 & 32 & 32 \\ \hline
\end{tabular}
\caption{Dimensions of symbol basis for planar six-point massless Feynman integrals.
}
\label{tab:symbol_weights}
\end{table}

\subsection{Basis of planar six-point two-loop functions}
\label{sec:basis-symbols}

In this subsection, we summarize the key features of the planar two-loop six-particle Feynman integral function space, which serves as our basis for symbol bootstrap calculation.

A symbol bootstrap calculation typically begins with a given symbol alphabet. It then proceeds by constructing integrable symbols of a given transcendental weight dictated by the loop order. The size of the resulting symbol ansatz is determined by the number of integrable symbols, as briefly outlined in subsection~\ref{sec:iiandsymb}. In this approach, computational challenges grow rapidly with both the size of the symbol alphabet and the loop order. In the present work, by contrast, we exploit that the amplitudes must lie within the function space of the planar two-loop six-particle Feynman integrals.
A similar strategy has been employed in Refs.~\cite{Guo:2021bym,Guo:2022qgv} for bootstrapping two-loop four-point form factors, and more recently in six-point two-loop \cite{Carrolo:2025pue,Carrolo:2025agz} and five-point three-loop  \cite{Chicherin:2025jej} bootstrap computations.

In a series of recent works \cite{Henn:2021cyv,Henn:2022ydo,Henn:2024ngj,Henn:2025xrc,Abreu:2024fei}, the planar two-loop six-particle Feynman integrals have been computed analytically, and the corresponding space of transcendental functions has been determined. These Feynman integrals satisfy the canonical DE given by \eqref{eq:cDE}, which are solved in terms of Chen's iterated integrals.
In the first three rows of Table~\ref{tab:symbol_weights}, we present the counting of the corresponding symbols up to weight four \cite{Henn:2025xrc}. Note that two-loop five-point Feynman integrals with one massive leg belong to the same planar two-loop six-point massless integral families. As a result, their symbols span a linear subspace of the six-point symbol space. The dimension of the corresponding quotient space therefore counts the number of genuine six-point two-loop symbols. 

At maximal transcendentality, the finite term in the $\eps$-expansion of the two-loop amplitude belongs to the ${ 639}$-dimensional space of weight-four symbols counted in the ``two-loop six-point symbols" row of Table~\ref{tab:symbol_weights}. However, the two-loop hard functions require a larger symbol space. Specifically, we also have to include the weight-four symbols arising from the subtraction terms $I_6^{(1)}A_6^{(1)}$ and $I_6^{(2)}$. For this purpose, we need to include an extended symbol basis. The latter consists of the weight-four symbol space of the one-loop master integrals,  together with cyclically symmetric sums of powers of logarithms $\log(s_{i,i{+}1})$ and their products with one-loop symbols of weights one through three. The result of counting one-loop symbols (from Ref.~\cite{Henn:2025xrc}) is listed in the ``one-loop six-point symbols" row of Table~\ref{tab:symbol_weights}. Extending the basis of the two-loop weight-four symbols in this manner, we obtain a total of ${712}$ independent weight-four symbols, which form the basis for our six-gluon hard function ansätze.

When counting the independent symbols in Table~\ref{tab:symbol_weights}, we 
imposed that the external kinematics is four-dimensional. We identify $167$ symbol letters appearing in the ${712}$-dimensional space required for the two-loop hard functions. In the notation of Ref.~\cite{Henn:2025xrc}, these symbol letters form a subset of $\{W_1, \ldots W_{289}\}$, which constitutes the complete planar two-loop six-point alphabet in $D$-dimensional kinematics.
Among the $167$ letters, $99$ letters are parity even, while $68$ are parity odd. Of these, $154$ letters arise from two-loop five-point Feynman integrals with one massive leg, while $13$ letters are genuine six-point letters, 
\begin{equation}\label{eq:letter6}
\begin{aligned}
&W_{100}=-s_{23} s_{34} s_{56}+s_{23} s_{345} s_{56}-s_{12} s_{45} s_{61}-s_{34} s_{61} s_{123}+s_{12} s_{45} s_{234}+s_{34} s_{123} s_{234} \\
&\phantom{aaaaaaaaaaaaaaaaaaaaaaaaaaaaaaaaaaaaaaa}+s_{61} s_{123} s_{345}-s_{123} s_{234} s_{345}\,, \\
 &W_{100+i}=\mathcal \tau^i (W_{100}), \quad i=1,\ldots,5 \,, \\
&W_{138}=\Delta_6=
\langle12\rangle [23]\langle34\rangle [45]\langle 56\rangle [61]-[12]\langle 23\rangle [34]\langle45\rangle [56]\langle61\rangle \,, \\
&W_{242}=\frac{s_{12} \left(s_{234}{-}s_{45}{-}s_{61}\right){+}s_{23} \left(s_{34}{+}s_{56}{-}s_{345}\right){+}s_{123} \left({-}s_{34}{+}s_{61}{-}s_{234}{+}s_{345}\right){-}\epsilon(1,2,3,5)}{s_{12} \left(s_{234}{-}s_{45}{-}s_{61}\right){+}s_{23} \left(s_{34}{+}s_{56}{-}s_{345}\right){+}s_{123} \left({-}s_{34}{+}s_{61}{-}s_{234}{+}s_{345}\right){+}\epsilon(1,2,3,5)} \,,
\\
 &W_{242+i}=\mathcal \tau^i (W_{242}), \quad i=1,\ldots,5 \,,
\end{aligned}
\end{equation}
where $\tau$ denotes the cyclic-shift generator, $\tau(p_i) = p_{i+1}$, and the parity-odd Lorentz-invariants are defined through the antisymmetric Levi-Civita tensor as
\begin{equation}
    \label{eq:epsijkl1}             \epsilon(i,j,k,l):=4i\varepsilon_{\mu\nu\rho\sigma}p_i^\mu p_j^\nu p_k^\rho p_l^\sigma \, , \qquad {i,j,k,l \in \{1,\dots,6\}} \, .
\end{equation}
At we will see at the end of our bootstrap calculation, the two-loop hard functions in fact involve a slightly smaller subset of 137 symbol letters.

\newpage

\section{Maximal weight projection}
\label{sec: maximal weight projection}

We begin this section by reviewing the basics of the \textit{maximal weight projection}
for QCD amplitudes and their hard functions \cite{Henn:2021aco}, which is a primary focus of this work. We show that the maximally-transcendental part of a hard function naturally admits a decomposition into two contributions: an IR-finite term that is unaffected by IR subtraction, and an IR-subtracted term. This decomposition holds universally for the maximally-transcendental contributions to QCD hard functions of arbitrary helicity and multiplicity. The prefactors appearing in both terms of this decomposition are four-dimensional leading singularities, which are computed by taking maximal residues along closed contours in loop-momentum space.  As proposed in Ref.~\cite{Bourjaily:2017wjl}, the basis integrals used in generalized unitarity calculations can be chosen more optimally such that their maximal residues are diagonalized on a set of linearly independent maximal-cut contours.  This approach was called prescriptive unitarity by the authors. 
In Refs.~\cite{Bourjaily:2019iqr,Bourjaily:2019gqu}, the maximal cut contours for all basis integrals contributing to the IR-finite part and unaffected by IR subtractions were classified within this framework. In the present paper, combining insights from prescriptive unitarity and the maximal weight projection of Ref.~\cite{Henn:2021aco} plays a central role.

\subsection{Maximally-transcendental part of hard functions}

In this work, we focus exclusively on the maximally-transcendental part of the QCD hard functions. To understand their general structure, let us recall how the maximal weight projection is implemented at the level of individual Feynman integrals, which may be both UV- and IR-divergent. It is understood that these Feynman integrals evaluate to Chen iterated integrals, for which the notion of transcendental weight is well defined. We adopt the algorithm introduced in Ref.~\cite{Henn:2021aco}, which builds on ideas pioneered in Ref.~\cite{Arkani-Hamed:2010pyv} on four-dimensional local integrals that evaluate, upon loop integration, to functions of maximal transcendentality. One starts from the four-dimensional $L$-loop integrand of a given Feynman integral,
\begin{align} \label{eq:I4L}
\prod_{i=1}^L{\rm d}^4\ell_i\ \mathcal{I}(\ell_i,p_e)\ ,  
\end{align} 
and then chooses a convenient parameterization of the loop momenta $\ell_i$ and identifies all propagator poles. Next, one constructs a basis of $L$-loop ${\rm d}\log$ integrands with propagators appearing in the original integrand \eqref{eq:I4L}. In this parametrization, the basis elements take the form
\begin{equation}
    \{\mathcal{I}_j\}_{j=1,\cdots,N},\ \ \mathcal{I}_j=\sum_i b_i\bigwedge_{k=1}^{4L}{\rm d}\log(\alpha_{i,j,k}) \, ,
\end{equation}
where coefficients $b_j$ are rational numbers and $\alpha_{i,j,k}$ are poles of the integrand. By construction, these integrands possess only simple poles. Upon performing loop integrations, they evaluate to pure functions of weight $2L$, which is the maximal transcendental weight attainable at $L$ loops\footnote{Equivalently, the integrated basis elements can be represented by $2L$-fold ${\rm d}\log$ iterated integrals.}.

The original integrand may contain double poles (and higher) in loop momenta, as is typical for QCD loop integrands. Consequently, it cannot be expanded solely in terms of the ${\rm d}\log$ basis, which we schematically write  as \begin{equation}\label{eq:generalpro}
   \prod_{i=1}^{L}{\rm d}^4\ell_i\ \mathcal{I}(\ell_i,p_e)=\sum^{N}_{j=1}R_j\,\mathcal{I}_j\ {+}\ \text{double pole terms} \, .
\end{equation}
The terms containing double poles give rise, upon loop integration, to contributions of subleading transcendental weight. Since we are interested only in the maximally-transcendental part, these terms can therefore be safely neglected. As a result, the maximal weight projection is implemented directly at the integrand level, without performing any loop integrations, as 
\begin{equation}
    \mathcal{P}_{\text{MT}}\left[\prod_{i=1}^{L}{\rm d}^4\ell_i\ \mathcal{I}(\ell_i,p_e)\right]=\sum_{j}R_j\,\mathcal{I}_j\, .
\end{equation}
The remaining tasks are to classify the rational prefactors $R_j$ and to perform the loop integrations of the basis integrands $\mathcal{I}_j$. In what follows, we accomplish both goals without requiring explicit knowledge of the full QCD loop integrand.

For loop integrands of QCD amplitudes, the maximal weight projection admits a further refinement. UV divergences originate from double poles at infinity and are therefore projected out by $\mathcal{P}_{\text{MT}}$. By contrast, IR divergences contribute to the maximally-transcendental part. 
The infrared structure of QCD amplitudes is well understood and was briefly summarized in subsection~\ref{sec:IR}. In particular, the IR poles of an $L$-loop amplitude factorize into products of lower-loop amplitudes and universal IR-subtraction operators. Consequently, the maximal weight projection of an $L$-loop QCD amplitude integrand $\mathcal{I}(\ell_i,p_e)$   naturally decomposes into an IR-divergent part and an IR-finite part, as follows,
\begin{tcolorbox}
  \begin{equation}\label{eq:main}
    \mathcal{P}_{\text{MT}}\left[\prod_{i=1}^L{\rm d}^4\ell_i\ \mathcal{I}(\ell_i,p_e)\right]=\underbrace{\sum_{\ell=0}^{L{-}1}\sum_{j=1}^{k_\ell}R^{(\ell)}_{j}\ \mathcal{I}_{j,\ell}}_{\text{IR-subtraction needed}}{+}\underbrace{\sum_{k=1}^{k_L}R_{k}^{(L)}\ \mathcal{I}_k}_{\text{free from IR-divergence}},
\end{equation}  
\end{tcolorbox}
\noindent where $R_j^{(\ell)}$, with $\ell=0,\cdots,L{-}1$ and $j=1,\cdots,k_\ell$, denote the $k_{\ell}$ independent rational prefactors associated with the $\ell$-loop amplitudes, whereas $R_k^{(L)}$ are genuinely $L$-loop prefactors that appear for the first time in the maximally-transcendental part of the $L$-loop amplitude. This distinction among the prefactors determines whether the accompanying loop integrals are IR divergent, owing to the universal structure of IR divergences in QCD amplitudes. In particular, the loop integrals accompanying the genuinely $L$-loop prefactors $R_k^{(L)}$ are unaffected by IR subtraction and are finite and well defined in four dimensions.

As explained in \cite{Henn:2021aco}, since the basis elements $\mathcal{I}_j$ are ${\rm d}\log$ forms, the expansion coefficients $R_j$ can be efficiently computed as maximal residues of the integrand. However, a purely four-dimensional analysis may be insufficient for the integrand $\mathcal{I}_{j,\ell}$ whose loop integrations are divergent in four dimensions and therefore require regularization. In dimensional regularization with $D=4{-}2\epsilon$, the $\ell_i^{[-2\ep]}$-components of the loop momenta must be taken into account, and {\it evanescent terms} may in general be missed by a strictly four-dimensional maximal-residue calculation. By contrast, the IR finite terms in \eqref{eq:main} are completely characterized by four-dimensional $L$-loop maximal residues, {\it i.e.} by $L$-loop maximal-cut contours. The corresponding prefactors $R_k^{(L)}$ can therefore be computed using standard four-dimensional generalized unitarity methods. In what follows, we first discuss the IR-finite part of the maximal weight projection and then turn to the IR-divergent part.

\begin{figure}[t]
    \centering
    \begin{tikzpicture}
        \draw[line width=1pt,black] (0,0)--(1.5,0)--(1.5,1.5)--(0,1.5)--(0,0);
          \draw[line width=1pt,black] (0,0)--(-0.5,-0.5);
        \draw[line width=1pt,black] (0,2)--(0,1.5)--(-0.5,1.5);
         \draw[line width=1pt,black] (1.5,-0.5)--(1.5,0)--(2,0);
          \draw[line width=1pt,black] (1.5,1.5)--(2,2);
        \path[fill=gray] (0,0) circle[radius=0.11];
        \path[fill=black] (1.5,0) circle[radius=0.11];
        \path[fill=gray] (1.5,1.5) circle[radius=0.11];
        \path[fill=black] (0,1.5) circle[radius=0.11];
        \path[fill=gray] (-0.3,1.6) circle[radius=0.03];
        \path[fill=gray] (-0.2,1.7) circle[radius=0.03];
        \path[fill=gray] (-0.1,1.8) circle[radius=0.03];
         \path[fill=gray] (1.6,-0.3) circle[radius=0.03];
        \path[fill=gray] (1.7,-0.2) circle[radius=0.03];
        \path[fill=gray] (1.8,-0.1) circle[radius=0.03];
        \node[anchor=north east] at (-0.5,-0.5) {\small{$k$}};
         \node[anchor=south west] at (2,2) {\small{$j$}};
         \node[anchor=east] at (-0.5,1.5) {\small{$(k{+}1)$}};
          \node[anchor=south] at (0,2) {\small{$(j{-}1)$}};
        \node[anchor=west] at (2,0) {\small{$(j{+}1)$}};
          \node[anchor=north] at (1.5,-0.5) {\small{$(k{-}1)$}};
          \node[anchor=north] at (0.75,0) {\small{$\ell_d$}};
    \node[anchor=west] at (1.5,0.75) {\small{$\ell_a$}};
    \node[anchor=south] at (0.75,1.5) {\small{$\ell_b$}};
   \node[anchor=east] at (0,0.75) {\small{$\ell_c$}};
    \end{tikzpicture}
    \caption{One-loop MHV contour.}
    \label{fig:1Lcut}
\end{figure}
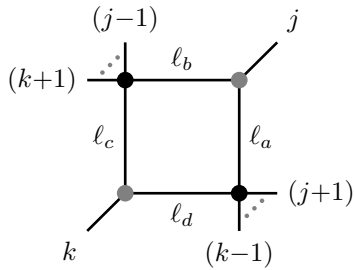

\subsection{IR-finite contributions}
\label{sec:IRfinTerms}

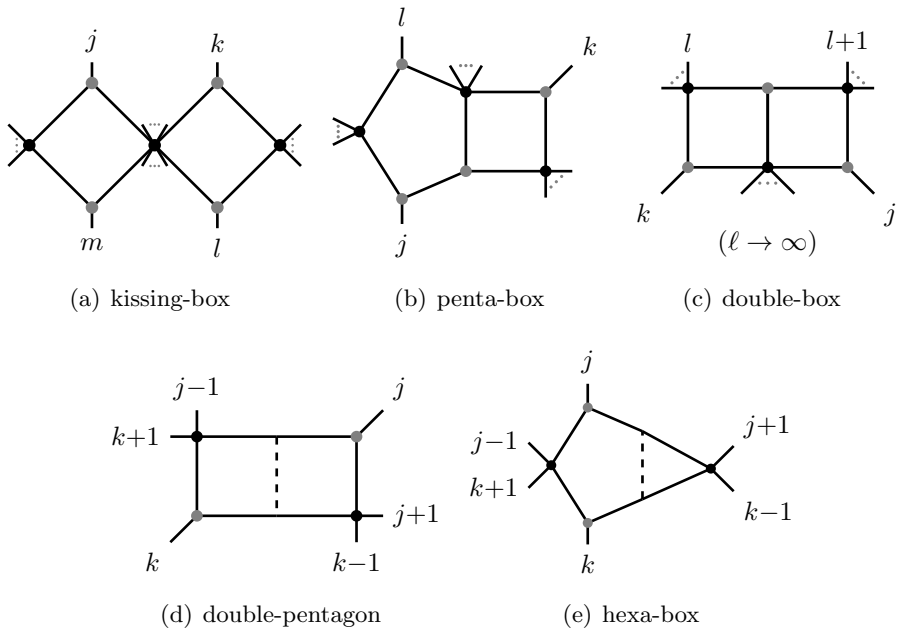
\begin{figure}[t]
\centering
\subfigure[kissing-box]{
\begin{tikzpicture}[scale=0.55]
     \draw[line width=1pt,black] (0,0)--(1.5,1.5)--(3,0)--(1.5,-1.5)--(0,0); 
        \draw[line width=1pt,black] (3,0)--(4.5,1.5)--(6,0)--(4.5,-1.5)--(3,0); 
        \draw[line width=1pt,black] (-0.5,-0.5)--(0,0)--(-0.5,0.5);
         \draw[line width=1pt,black] (6.5,-0.5)--(6,0)--(6.5,0.5);
         \draw[line width=1pt,black] (2.7,0.5)--(3,0)--(3.3,0.5);
         \draw[line width=1pt,black] (2.7,-0.5)--(3,0)--(3.3,-0.5);
     \draw[line width=1pt,black] (1.5,1.5)--(1.5,2);
     \draw[line width=1pt,black] (1.5,-1.5)--(1.5,-2);
     \draw[line width=1pt,black] (4.5,1.5)--(4.5,2);
     \draw[line width=1pt,black] (4.5,-1.5)--(4.5,-2);
    \path[fill=gray] (1.5,1.5) circle[radius=0.15];
    \path[fill=gray] (1.5,-1.5)circle[radius=0.15];
    \path[fill=gray] (4.5,1.5) circle[radius=0.15];
    \path[fill=gray] (4.5,-1.5)circle[radius=0.15];
     \path[fill=black] (0,0) circle[radius=0.15];
    \path[fill=black] (3,0)circle[radius=0.15];
    \path[fill=black] (6,0) circle[radius=0.15];
    \path[fill=gray] (-0.3,0) circle[radius=0.03];
    \path[fill=gray] (-0.3,0.15) circle[radius=0.03];
    \path[fill=gray] (-0.3,-0.15)circle[radius=0.03];
        \path[fill=gray] (6.3,0) circle[radius=0.03];
        \path[fill=gray] (6.3,0.15) circle[radius=0.03];
        \path[fill=gray] (6.3,-0.15)circle[radius=0.03];
    \path[fill=gray] (3,0.5) circle[radius=0.03];
   \path[fill=gray] (2.9,0.5) circle[radius=0.03];
   \path[fill=gray] (3.1,0.5) circle[radius=0.03];
       \path[fill=gray] (3,-0.5) circle[radius=0.03];
   \path[fill=gray] (2.9,-0.5) circle[radius=0.03];
   \path[fill=gray] (3.1,-0.5) circle[radius=0.03];
\node[anchor=north] at (1.5,-2) {\small{$m$}};
\node[anchor=north] at (4.5,-2) {\small{$l$}};
\node[anchor=south] at (1.5,2) {\small{$j$}};
\node[anchor=south] at (4.5,2) {\small{$k$}};
\end{tikzpicture}}\quad
\subfigure[penta-box]{\begin{tikzpicture}[scale=0.7]
    \draw[line width=1pt,black] (0,0)--(1.5,0)--(1.5,1.5)--(0,1.5)--(0,0);
    \draw[line width=1pt,black] (-0.3,2)--(0,1.5)--(0.3,2);
    \draw[line width=1pt,black] (2,0)--(1.5,0)--(1.5,-0.5);
    \draw[line width=1pt,black] (-2.5,1)--(-2,0.75)--(-2.5,0.5);
     \draw[line width=1pt,black] (-1.2,-0.525)--(-1.2,-1);
      \draw[line width=1pt,black] (-1.2,2.025)--(-1.2,2.55);
\draw[line width=1pt,black] (1.5,1.5)--(2,2);
    \draw[line width=1pt,black] (0,0)--(-1.2,-0.525)--(-2,0.75)--(-1.2,2.025)--(0,1.5);
    \path[fill=gray] (-1.2,2.025) circle[radius=0.11];
     \path[fill=gray] (-1.2,-0.525) circle[radius=0.11];
    \path[fill=black] (0,1.5) circle[radius=0.11];
    \path[fill=black] (1.5,0) circle[radius=0.11];
    \path[fill=black] (-2,0.75) circle[radius=0.11];
    \path[fill=gray] (1.5,1.5) circle[radius=0.11];
    \path[fill=gray] (0,0) circle[radius=0.11];
    \node[anchor=south west] at (2,2) {\small{$k$}};
    \node[anchor=south] at (-1.2,2.55){\small{$l$}};
     \node[anchor=north] at (-1.2,-1){\small{$j$}};
      \path[fill=gray] (0,2) circle[radius=0.03];
   \path[fill=gray] (-0.1,2) circle[radius=0.03];
   \path[fill=gray] (0.1,2) circle[radius=0.03];
   \path[fill=gray] (1.6,-0.3) circle[radius=0.03];
        \path[fill=gray] (1.7,-0.2) circle[radius=0.03];
        \path[fill=gray] (1.8,-0.1) circle[radius=0.03];
    \path[fill=gray] (-2.4,0.65) circle[radius=0.03];
    \path[fill=gray] (-2.4,0.75) circle[radius=0.03];
    \path[fill=gray] (-2.4,0.85) circle[radius=0.03];
\end{tikzpicture}}
\subfigure[double-box]{\label{fig:doubleboxcut}
\begin{tikzpicture}[scale=0.7]
        \draw[line width=1pt,black] (0,0)--(1.5,0)--(1.5,1.5)--(0,1.5)--(0,0);
         \draw[line width=1pt,black] (1.5,0)--(3,0)--(3,1.5)--(1.5,1.5)--(1.5,0);
          \draw[line width=1pt,black] (0,0)--(-0.5,-0.5);
            \draw[line width=1pt,black] (3,0)--(3.5,-0.5);
        \draw[line width=1pt,black] (0,2)--(0,1.5)--(-0.5,1.5);
        \draw[line width=1pt,black] (1,-0.5)--(1.5,0)--(2,-0.5);
         \draw[line width=1pt,black] (3,2)--(3,1.5)--(3.5,1.5);
        \path[fill=gray] (3,0) circle[radius=0.11];
          \path[fill=black] (3,1.5) circle[radius=0.11];
        \path[fill=gray] (0,0) circle[radius=0.11];
        \path[fill=black] (1.5,0) circle[radius=0.11];
        \path[fill=gray] (1.5,1.5) circle[radius=0.11];
        \path[fill=black] (0,1.5) circle[radius=0.11];
        \path[fill=gray] (-0.3,1.6) circle[radius=0.03];
        \path[fill=gray] (-0.2,1.7) circle[radius=0.03];
        \path[fill=gray] (-0.1,1.8) circle[radius=0.03];
\path[fill=gray] (3.3,1.6) circle[radius=0.03];
        \path[fill=gray] (3.2,1.7) circle[radius=0.03];
        \path[fill=gray] (3.1,1.8) circle[radius=0.03];
\path[fill=gray] (1.35,-0.3) circle[radius=0.03];
        \path[fill=gray] (1.5,-0.3) circle[radius=0.03];
        \path[fill=gray] (1.65,-0.3) circle[radius=0.03];
        \node[anchor=north east] at (-0.5,-0.5) {\small{$k$}};
          \node[anchor=south] at (0,2) {\small{$l$}};
 \node[anchor=south] at (3,2) {\small{$l{+}1$}};
\node[anchor=north west] at (3.5,-0.5) {\small{$j$}};
\node[anchor=north] at (1.5,-1) {\small{$(\ell\to\infty)$}};
\end{tikzpicture}}
\subfigure[double-pentagon]{\label{fig:doublepentagoncut}  \begin{tikzpicture}[scale=0.7]
        \draw[line width=1pt,black] (1.5,1.5)--(0,1.5)--(0,0)--(1.5,0);
        \draw[line width=1pt,black,dashed] (1.5,1.5)--(1.5,0);
         \draw[line width=1pt,black] (1.5,0)--(3,0)--(3,1.5)--(1.5,1.5);
          \draw[line width=1pt,black] (0,0)--(-0.5,-0.5);
            \draw[line width=1pt,black] (3,-0.5)--(3,0)--(3.5,0);
        \draw[line width=1pt,black] (0,2)--(0,1.5)--(-0.5,1.5);
         \draw[line width=1pt,black] (3.5,2)--(3,1.5);
        \path[fill=black] (3,0) circle[radius=0.11];
          \path[fill=gray] (3,1.5) circle[radius=0.11];
        \path[fill=gray] (0,0) circle[radius=0.11];
        \path[fill=black] (0,1.5) circle[radius=0.11];
        \node[anchor=north east] at (-0.5,-0.5) {\small{$k$}};
          \node[anchor=south] at (0,2) {\small{$j{-}1$}};
             \node[anchor=east] at (-0.5,1.5) {\small{$k{+}1$}};
 \node[anchor=south west] at (3.5,2) {\small{$j$}};
\node[anchor=north] at (3,-0.5) {\small{$k{-}1$}};
\node[anchor=west] at (3.5,0) {\small{$j{+}1$}};
\end{tikzpicture}}
\subfigure[hexa-box]{\label{fig:hexaboxcut}\begin{tikzpicture}[scale=0.6]
    \draw[line width=1pt,black] (0,0)--(1.5,0.675)--(0,1.5);
    \draw[line width=1pt,black,dashed] (0,0)--(0,1.5);
    \draw[line width=1pt,black] (2,0.175)--(1.5,0.675);
    \draw[line width=1pt,black] (-2.5,1.25)--(-2,0.75)--(-2.5,0.25);
     \draw[line width=1pt,black] (-1.2,-0.525)--(-1.2,-1);
      \draw[line width=1pt,black] (-1.2,2.025)--(-1.2,2.55);   
\draw[line width=1pt,black] (1.5,0.675)--(2,1.175);
    \draw[line width=1pt,black] (0,0)--(-1.2,-0.525)--(-2,0.75)--(-1.2,2.025)--(0,1.5);
    \path[fill=gray] (-1.2,2.025) circle[radius=0.11];
     \path[fill=gray] (-1.2,-0.525) circle[radius=0.11];
    \path[fill=black] (1.5,0.675) circle[radius=0.11];
    \path[fill=black] (-2,0.75) circle[radius=0.11];
    \node[anchor=south west] at (2,1.125) {\small{$j{+}1$}};
    \node[anchor=south] at (-1.2,2.55){\small{$j$}};
     \node[anchor=north] at (-1.2,-1){\small{$k$}};
    \node[anchor=north west] at (2,0.175){\small{$k{-}1$}};
        \node[anchor=east] at (-2.5,1.25){\small{$j{-}1$}};
            \node[anchor=east] at (-2.5,0.25){\small{$k{+}1$}};
\end{tikzpicture}}
\caption{Five types of two-loop contours for MHV amplitudes.}
\label{fig:2Lcuts}
\end{figure}

To classify the genuinely two-loop leading singularities $R_k^{(2)}$ in \eqref{eq:main}, we need to identify a complete basis of two-loop maximal-cut integration contours that contribute to the IR-finite part, and evaluate the maximal residues along them. According to the general framework of prescriptive unitarity in \cite{Bourjaily:2017wjl,Bourjaily:2019iqr}, the IR-finite part receives contributions only from those contours whose maximal residue does not localize the loop momenta to integration regions that generate IR divergences in the amplitudes, {\it i.e.} collinear regions where $\ell_i\propto p_j$ for certain loop momentum $\ell_i$ and external momentum $p_j$ \cite{Arkani-Hamed:2010pyv,Bourjaily:2011hi,Henn:2019rmi}. For two-loop MHV gauge-theory amplitudes, such contours have been already fully classified in \cite{Bourjaily:2019gqu}. We collect all independent maximal-cut contours for MHV amplitudes in pure YM theory at one and two loops in Figures \ref{fig:1Lcut} and \ref{fig:2Lcuts}, respectively. In these figures, every edge of a diagram represents placing the corresponding propagator on-shell, while the black and gray vertices impose proportionality relations among the helicity spinors of the momenta flowing into the vertex. A solution to the cut conditions imposed by the on-shell propagators typically has several branches. The chirality of each branch is determined by the coloring of the vertices in the bipartite graph. 

At one loop, the only maximal cut with a non-vanishing residue contributing to the IR-finite part is shown in Fig.~\ref{fig:1Lcut}. Let us use this example to illustrate the notation for contour diagrams employed in the figures. 
The two-mass-easy box maximal-cut imposes the following conditions on the loop momenta:
\begin{equation}
\ell_a^2=\ell_b^2=\ell_c^2=\ell_d^2=0.
\end{equation} 
Solving these four-dimensional equations, we find two distinct maximal cut solutions, denoted by $\ell_a^{(1)}$ and $\ell_a^{(2)}$, for the loop momentum $\ell_a$,
\begin{equation}\label{eq:twobranchsol}
    \left(\ell^{(1)}_a \right)^{\alpha \dot{\beta}}=\frac{\lambda_j^{\alpha}\ (p_{j{+}1,\cdots,k{-}1}\cdot\lambda_k)^{\dot{\beta}}}{\langle jk\rangle},\ \ \left(\ell_a^{(2)}\right)^{\alpha \dot{\beta}}=\frac{(\tilde{\lambda}_k\cdot p_{j{+}1,\cdots,k{-}1})^{\alpha}\tilde{\lambda}^{\dot{\beta}}_j}{[jk]} \, ,
\end{equation}
where we have written all spinor-helicity indices explicitly. We see that in the first solution branch $\lambda_{\ell_a}\propto\lambda_j$, whereas in the second branch $\tilde{\lambda}_{\ell_a}\propto\tilde{\lambda}_j$. These two solutions are also commonly referred to as the MHV and $\overline{\text{MHV}}$ branches, respectively. For the MHV amplitude integrand, only the first solution $\ell_a^{(1)}$ yields a non-trivial leading singularity \cite{Arkani-Hamed:2010pyv,Bourjaily:2013mma}. In Fig.~\ref{fig:1Lcut}, we use gray blobs in these contours to denote the proportionality relations $\lambda_j \propto \lambda_{\ell_a} \propto \lambda_{\ell_b}$ and $\lambda_k \propto \lambda_{\ell_c} \propto \lambda_{\ell_d}$ among the holomorphic helicity spinors attached to the corresponding three-vertices. Neither of the two maximal-cut solutions accesses the collinear region $\ell_a\propto p_j$ or $\ell_c\propto p_k$, indicating that these cuts do not probe soft or collinear regions.

It is worth noting that the contour diagrams shown in Figs.~\ref{fig:1Lcut} and \ref{fig:2Lcuts} closely resemble the on-shell diagrams \cite{Arkani-Hamed:2010zjl}, which are  discussed in section~\ref{sec: Leading singularities for MHV QCD amplitudes}. In fact, on-shell diagrams are in one-to-one correspondence with chiral solutions of maximal cuts. In this work, we deliberately distinguish between these two diagrammatic notations in order to clearly indicate whether we are referring to the on-shell propagator conditions associated with a given contour or to the leading singularities computed by that contour.

At two loops, there are five types of maximal cuts with non-vanishing residues, shown in Fig.~\ref{fig:2Lcuts}. When considering QCD amplitudes, one must include the double-box cuts shown in Fig.~\ref{fig:doubleboxcut}, which are absent for amplitude integrands in $\mathcal{N}{=}4$ sYM (cf. Ref.~\cite{Bourjaily:2019gqu}.) This difference is closely related to the fact that $\mathcal{N}{=}4$ sYM amplitude integrands have vanishing residues at poles at infinity. For QCD amplitudes, by contrast, the double-box cut contours contribute precisely because of non-vanishing residues at poles at infinity. The cuts shown in Figs.~\ref{fig:doublepentagoncut} and~\ref{fig:hexaboxcut} involve residues taken in a soft region, which set the momentum flowing through the dashed propagator to zero and thereby identify the loop momenta as $\ell_1=\ell_2$ in the corresponding diagrams\footnote{The cuts in Figs.~\ref{fig:doublepentagoncut} and~\ref{fig:hexaboxcut} are special planar cases of IR-finite nonplanar cuts, which explains their naming.  See Tables III and IV in Ref.~\cite{Bourjaily:2019gqu} for further details.}. Consequently, two-loop cuts in Figs.~\ref{fig:doublepentagoncut} and~\ref{fig:hexaboxcut} are equivalent to the one-loop two-mass-easy box cut shown in Fig.~\ref{fig:1Lcut}. In section~\ref{sec: Leading singularities for MHV QCD amplitudes}, we present a systematic calculation of the two-loop leading singularities in QCD based on these contours, their associated maximal residues, and the corresponding on-shell diagrams. In section~\ref{sec: Prescriptive unitarity integral}, we further show that their accompanying integrands $\mathcal{I}_k$ belong to a class of two-loop Feynman integrals known as prescriptive unitarity integrals. These integrals are IR finite, possess unit leading singularities, and evaluate to pure functions of uniform transcendentality. This structure enables their efficient computation via IBP reduction techniques.

\subsection{IR-divergent contributions}

Having addressed the IR-finite part of the amplitude in \eqref{eq:main}, we now turn to its IR-divergent contribution. From a four-dimensional perspective, the corresponding prefactors $R_{j,\ell}$ originate from lower-loop leading singularities. For two-loop MHV amplitudes, they are given by one-loop leading singularities together with the tree-level Parke-Taylor factors \eqref{eq:PT}. Since this contribution is IR divergent, it must be treated within dimensional regularization, and the corresponding integrands are therefore defined in $D=4{-}2\epsilon$ dimensions. Consequently, a naive four-dimensional computation of the associated leading singularities is 
insufficient in general, and the hard function may receive contributions from evanescent terms. In other words, certain prefactors can be missed by the four-dimensional unitarity. Nevertheless, as discussed in \cite{Henn:2021aco}, with an appropriate choice of IR-subtraction scheme, one may expect these terms to be artifacts of dimensional regularization that 
cancel out in the hard function, leaving only four-dimensional information.

In what follows, we adopt a bootstrap strategy to determine the IR-subtracted part, while reserving a tailored basis of prescriptive unitarity integrals for the IR-finite contribution. For each helicity sector, we begin with the tree-level $\text{PT}_{n,1i}$ factors and the corresponding one-loop prefactors $R_j^{(1)}$, construct a weight-four symbol ansatz for the accompanying pure transcendental functions, and fix all unknown coefficients by imposing physical constraints. We find that this strategy consistently leads to a unique solution satisfying all known physical requirements and exhibiting the expected analytic structure.

\newpage

\section{Prefactors for MHV QCD amplitudes}
\label{sec: Leading singularities for MHV QCD amplitudes}

Having reviewed the classification of integration contours for IR-finite parts of one-loop and two-loop MHV amplitudes in the previous section, we now turn to the computation of the rational prefactors $R_k^{(\ell)}$ in \eqref{eq:main}. As discussed earlier, these prefactors are closely related to the maximal residues of the integrand and to the leading singularities. The one-loop and two-loop rational prefactors are determined by the maximal cuts shown in Fig.~\ref{fig:1Lcut} and Fig.~\ref{fig:2Lcuts}, respectively. Our calculation employs the formalism of {\it on-shell diagrams} and {\it on-shell functions} \cite{Arkani-Hamed:2012zlh}. This formalism was pioneered in $\mathcal{N}{=}4$ sYM theory, but it can also be adopted to pure YM and QCD amplitudes.

We begin with a brief review of the basic aspects of on-shell diagrams needed for our discussion. Then we work out one-loop examples as a warm-up for two-loop computations. As discussed in \cite{Arkani-Hamed:2012zlh,Bourjaily:2021ujs}, on-shell diagrams giving rise to leading singularities beyond those of $\mathcal{N}{=}4$ sYM theory exhibit distinctive graphical features. These graphical rules greatly simplify the task of identifying all distinct prefactors of QCD amplitude. Finally, we consider the two-loop MHV QCD amplitudes, presenting a complete set of on-shell diagrams that determine all distinct amplitude prefactors and computing them explicitly. These results provide the foundational input for constructing the ansatz for hard functions in subsequent bootstrap calculations. A more detailed review of on-shell diagrams is presented in Appendix~\ref{sec: review of on-shell diagrams}.

\subsection{Non-singlet on-shell diagrams and one-loop prefactors}

The unitarity of scattering amplitudes implies that the loop integrand factorizes
into simpler and lower-loop objects when certain propagators are put on shell. Putting the maximal allowed number of propagators on shell, thereby freezing all $4L$ degrees of freedom of the loop momenta, defines a maximal cut of the integrand. In this situation, all external and internal momenta are on shell, with the internal momenta localized on the maximal-cut solutions. The resulting factorized building blocks are represented as vertices in on-shell diagrams and correspond to tree-level amplitudes of the theory. 

In pure YM theory, on-shell diagrams for MHV amplitudes involve two types of vertices: three-point $\overline{\text{MHV}}$ YM amplitudes $\bar{\mathcal{A}}_3$ and $n$-point MHV YM amplitudes $\mathcal{A}_n$. They are given by the PT factor, cf. eq. \eqref{eq:PT},
\begin{equation}
\begin{aligned}
    &\vcenter{\hbox{\begin{tikzpicture}[decoration={,markings,mark=at position 0.7 with {\arrow{latex}}},scale=0.8]
        \draw[thick] (0,0) circle (0.2);
        \coordinate (p1) at ($({cos(0)},{sin(0)})$);
        \coordinate (p2) at ($({cos(120)},{sin(120)})$);
        \coordinate (p3) at ($({cos(-120)},{sin(-120)})$);
        \coordinate (p1a) at ($0.2*({cos(0)},{sin(0)})$);
        \coordinate (p2a) at ($0.2*({cos(120)},{sin(120)})$);
        \coordinate (p3a) at ($0.2*({cos(-120)},{sin(-120)})$);
        \node (p1t) at ($1.3*(p1)$){$b^+$};
        \node (p2t) at ($1.3*(p2)$){$a^+$};
        \node (p3t) at ($1.3*(p3)$){$c^-$};
        \draw[very thick, postaction={decorate}] (p1a.center) to (p1.center);
        \draw[very thick, postaction={decorate}] (p2a.center) to (p2.center);
        \draw[very thick, postaction={decorate}] (p3.center) to (p3a.center);
    \end{tikzpicture}}}:\ \bar{\mathcal{A}}_3(a^+,b^+,c^-)=\frac{[ab]^4}{[ab][bc][ca]}\delta^{2\times2}(P),\\
    &\vcenter{\hbox{\begin{tikzpicture}[scale=0.8]
        \coordinate (p1) at (160:1);
        \coordinate (p2) at (45:1);
        \coordinate (p3) at (-45:1);        
        \coordinate (p4) at (-160:1);

        \node (p1t) at ($1.3*(p1)$){$1^+$};
        \node (p2t) at ($1.3*(p2)$){$i^-$};
        \node (p3t) at ($1.3*(p3)$){$j^-$};
        \node (p34) at ($1.3*(p4)$){$n^+$};

        \draw[fill,thick] (0,0) circle (0.2);
        \draw[very thick, larrow] (p1) to (0,0);
        \draw[very thick, rarrow] (p2) to (0,0);
        \draw[very thick, rarrow] (p3) to (0,0);
        \draw[very thick, larrow] (p4) to (0,0);
        
        	\node[rotate=10] at (0,0.6) {$\dots$};
        	\node[rotate=-10] at (0,-0.6) {$\dots$};
        	\node[rotate=90] at (0.7,0) {$\dots$};

    \end{tikzpicture}}}:\ \mathcal{A}_n(1^+,\cdots, i^-,\cdots,j^-,\cdots,n^+)=\text{PT}_{n,ij}\delta^{2\times2}(P).
\end{aligned}
\end{equation}
Here, the arrows indicate the helicity of the on-shell particles, and $\delta^{2\times 2}(P)$ imposes momentum conservation. The resulting on-shell diagrams are bipartite, i.e. built from black and white vertices sewn together by cut propagators.
The three-point MHV and $\overline{\text{MHV}}$ vertices, ${\cal A}_3(a,b,c)$ and $\bar{{\cal A}}_3(a,b,c)$, play distinct roles in this formalism. They are related by parity conjugation, and complex kinematics is assumed.
These vertices impose proportionality constraints on the momenta: for a three-point $\overline{\text{MHV}}$ vertex one has $\lambda_a\propto\lambda_b\propto\lambda_c$ , while for a three-point MHV vertex 
$\tilde{\lambda}_a\propto\tilde{\lambda}_b\propto\tilde{\lambda}_c$.

As a simple example, by calculating the box on-shell diagram we reproduce the four-point PT factor as follows,
\begin{equation}
	\begin{aligned}
		\vcenter{\hbox{\begin{tikzpicture}[scale=0.8]
			
        		\coordinate (p1) at (-0.6,0.6);
        		\coordinate (p2) at (0.6,0.6);
        		\coordinate (p3) at (0.6,-0.6);
        		\coordinate (p4) at (-0.6,-0.6);

        		\draw[very thick, rarrow] (p1) to (p2);
			\draw[very thick, rarrow] (p2) to (p3);
			\draw[very thick, larrow] (p3) to (p4);
			\draw[very thick, rarrow] (p4) to (p1);
			
			\draw[very thick, rarrow] ($(p1) + (135:1)$) to (p1);
			\draw[very thick, larrow]  ($(p2) + (45:1)$) to (p2);
			\draw[very thick, larrow] ($(p3)+(-45:1)$) to (p3);
			\draw[very thick, rarrow] ($(p4) + (-135:1)$) to (p4);
			\draw[fill,thick] (p1) circle (0.2);
        		\draw[fill=white, thick] (p2) circle (0.2);
        		\draw[fill, thick] (p3) circle (0.2);
        		\draw[fill=white, thick] (p4) circle (0.2);
        		\node at ($(p1) + (135:1.3)$) {$2$};
        		\node at ($(p2) + (45:1.3)$) {$3$};
        		\node at ($(p3) + (-45:1.3)$) {$4$};
        		\node at ($(p4) + (-135:1.3)$) {$1$};
        		\node at ($(p1)!0.5!(p4) + (-0.35,0)$) {$\ell_1$};
        		\node at ($(p1)!0.5!(p2) + (0,0.35)$) {$\ell_2$};
        		\node at ($(p2)!0.5!(p3) + (0.35,0)$) {$\ell_3$};
        		\node at ($(p3)!0.5!(p4) + (0,-0.35)$) {$\ell_4$};
        
    \end{tikzpicture}}}
    &=\bar{\mathcal{A}}_3(\ell_1^+,\ell_4^+,1^-)\bar{\mathcal{A}}_3(3^+,\ell_3^+,\ell_2^-)\mathcal{A}_3(2^-,\ell_1^-,\ell_2^+)\mathcal{A}_3(\ell_3^-,\ell_4^-,4^+)\\
   &=\frac{\langle12\rangle^4}{\langle12\rangle\langle23\rangle\langle34\rangle\langle41\rangle}\delta^{2\times2}(P)
	\end{aligned}
    \label{eq:onshell1}
\end{equation}
Here, we substitute the maximal-cut solution for the loop momenta flowing through the on-shell propagators,
\begin{equation}\label{eq:sol1}
    \ell_1=\frac{\langle23\rangle}{\langle13\rangle}\lambda_1\tilde{\lambda}_2,\quad \ell_2=\frac{\langle12\rangle}{\langle13\rangle}\lambda_3\tilde{\lambda}_2,\quad \ell_3=\frac{\langle14\rangle}{\langle13\rangle}\lambda_3\tilde{\lambda}_4,\quad \ell_4=\frac{\langle34\rangle}{\langle13\rangle}\lambda_1\tilde{\lambda}_4.
\end{equation}
We conclude that $\text{PT}_{4,12}$, see \eqref{eq:PT}, is a one-loop leading singularity of the four-point pure YM amplitude in the helicity configuration $(1^-2^-3^+4^+)$. The three-point vertices in a bipartite on-shell diagram impose proportionality relations on the momenta. In the present example,
\begin{align}
\la_1  \propto \la_{\ell_1} \propto \la_{\ell_4} \,,\quad 
\tilde\la_2  \propto \tilde\la_{\ell_1} \propto \tilde\la_{\ell_2} \,,\quad 
\la_3  \propto \la_{\ell_2} \propto \la_{\ell_3} \,,\quad 
\tilde\la_4  \propto \tilde\la_{\ell_3} \propto \tilde\la_{\ell_4} \,.
\end{align}
As a result, a bipartite graph also specifies the chirality of the corresponding maximal cut solution. This is the basic logic of Ref.~\cite{Arkani-Hamed:2012zlh} for classifying the leading singularities of gauge theory amplitudes at a given loop order. More details of this method are provided in Appendix~\ref{sec: review of on-shell diagrams}.

Several points are worth noting here. First, when focusing on a specific helicity sector and an on-shell diagram of fixed topology, the helicities of the external legs constrain the helicity assignments of the internal lines. Diagrammatically, this restricts the allowed bipartite colorings of the on-shell diagram. For example, a box on-shell diagram in the MHV helicity sector can never contain a next-to-MHV (NMHV) corner. This is because the required helicity distribution would necessarily force another corner to be single-minus, which gives a vanishing residue at one loop. From a combinatorial perspective, this implies that any admissible maximal cut contributing to MHV amplitudes necessarily exhibits MHV helicity counting. At one loop, any box MHV on-shell diagram necessarily consists of two MHV and two $\overline{\text{MHV}}$ tree-level amplitudes.
\begin{figure}[t]
\centering	
\begin{tikzpicture}[]
			
        		\coordinate (p2) at (-0.6,0.6);
        		\coordinate (p3) at (0.6,0.6);
        		\coordinate (p4) at (0.6,-0.6);
        		\coordinate (p1) at (-0.6,-0.6);
        		\coordinate (p1aux) at ($(p1) + (-135:1)$);
        		\coordinate (p3aux) at ($(p3)+(45:1)$);
        		\coordinate (p2aux1) at ($(p2) + (180:1)$);
        		\coordinate (p2aux2) at ($(p2) + (135:1)$);
        		\coordinate (p2aux3) at ($(p2) + (90:1)$);
        		
        		\coordinate (p4aux1) at ($(p4) + (0:1)$);
        		\coordinate (p4aux2) at ($(p4) + (-45:1)$);
        		\coordinate (p4aux3) at ($(p4) + (-90:1)$);
        		
        		\draw[very thick,] (p1) to (p2);
			\draw[very thick, ] (p2) to (p3);
			\draw[very thick, ] (p3) to (p4);
			\draw[very thick, ] (p4) to (p1);
			
			\draw[very thick, larrow] (p1aux) to (p1);
			
			\draw[very thick, ]  (p2aux1) to (p2);
			\draw[very thick, ]  (p2aux3) to (p2);

			\draw[very thick, larrow] (p3aux) to (p3);
			
			\draw[very thick, ]  (p4aux1) to (p4);
			\draw[very thick, ]  (p4aux3) to (p4);

			\draw[fill=white,thick] (p1) circle (0.2);
        		\draw[fill, thick] (p2) circle (0.2);
        		\draw[fill=white, thick] (p3) circle (0.2);
        		\draw[fill, thick] (p4) circle (0.2);
        		\node at ($(p1aux) + (-135:0.3)$) {$k$};
        		\node at ($(p2aux2) + (135:0.3)$) {$1^-$};
        		\node at ($(p3aux) + (45:0.3)$) {$j$};
        		\node at ($(p4aux2) + (-45:0.3)$) {$i^-$};
        		\node[rotate=45] at ($(p2)!0.6!(p2aux2)$) {$\dots$};
        		\node[rotate=45] at ($(p4)!0.6!(p4aux2)$) {$\dots$};
    \end{tikzpicture}
    \caption{Non-singlet two-mass-easy box on-shell diagrams contributing genuinely one-loop prefactors.}
    \label{fig:1Lonshell}
\end{figure}
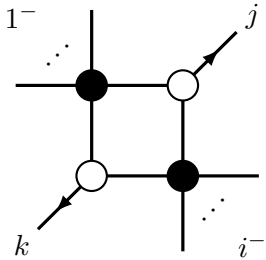

Second, it is crucial to keep track of the helicity of the internal states, which are indicated by arrows on the on-shell propagators. When computing the leading singularity associated with a given maximal cut, one must sum over all possible helicity assignments flowing through the on-shell diagram. 
As a result, it becomes clear how one-loop leading singularities beyond PT factors arise from on-shell diagrams. For the diagram in eq.~\eqref{eq:onshell1}, the helicity assignment of the internal lines is fixed uniquely by the helicity configuration of the external gluons and by the chirality of the three-point vertices. 
However, ``{\it non-singlet}" on-shell diagrams contribute for certain other helicity configurations of the external gluons \cite{Arkani-Hamed:2012zlh,Bourjaily:2021ujs,Brown:2022wqr},
\begin{equation}
\begin{aligned}
		\vcenter{\hbox{\begin{tikzpicture}[scale=0.8]
			
        		\coordinate (p1) at (-0.6,0.6);
        		\coordinate (p2) at (0.6,0.6);
        		\coordinate (p3) at (0.6,-0.6);
        		\coordinate (p4) at (-0.6,-0.6);

        		\draw[very thick] (p1) to (p2);
			\draw[very thick] (p2) to (p3);
			\draw[very thick] (p3) to (p4);
			\draw[very thick] (p4) to (p1);
			
			\draw[very thick, larrow] ($(p1) + (135:1)$) to (p1);
			\draw[very thick, rarrow]  ($(p2) + (45:1)$) to (p2);
			\draw[very thick, larrow] ($(p3)+(-45:1)$) to (p3);
			\draw[very thick, rarrow] ($(p4) + (-135:1)$) to (p4);
			\draw[fill=white,thick] (p1) circle (0.2);
        		\draw[fill, thick] (p2) circle (0.2);
        		\draw[fill=white, thick] (p3) circle (0.2);
        		\draw[fill, thick] (p4) circle (0.2);
        		\node at ($(p1) + (135:1.35)$) {$2^+$};
        		\node at ($(p2) + (45:1.35)$) {$3^-$};
        		\node at ($(p3) + (-45:1.35)$) {$4^+$};
        		\node at ($(p4) + (-135:1.35)$) {$1^-$};
        
    	\end{tikzpicture}}}
    	&=
    	\vcenter{\hbox{\begin{tikzpicture}[scale=0.8]
			
        		\coordinate (p1) at (-0.6,0.6);
        		\coordinate (p2) at (0.6,0.6);
        		\coordinate (p3) at (0.6,-0.6);
        		\coordinate (p4) at (-0.6,-0.6);

        		\draw[very thick, rarrow] (p1) to (p2);
			\draw[very thick, rarrow] (p2) to (p3);
			\draw[very thick, rarrow] (p3) to (p4);
			\draw[very thick, rarrow] (p4) to (p1);
			
			\draw[very thick, larrow] ($(p1) + (135:1)$) to (p1);
			\draw[very thick, rarrow]  ($(p2) + (45:1)$) to (p2);
			\draw[very thick, larrow] ($(p3)+(-45:1)$) to (p3);
			\draw[very thick, rarrow] ($(p4) + (-135:1)$) to (p4);
			\draw[fill=white,thick] (p1) circle (0.2);
        		\draw[fill, thick] (p2) circle (0.2);
        		\draw[fill=white, thick] (p3) circle (0.2);
        		\draw[fill, thick] (p4) circle (0.2);
        		\node at ($(p1) + (135:1.35)$) {$2^+$};
        		\node at ($(p2) + (45:1.35)$) {$3^-$};
        		\node at ($(p3) + (-45:1.35)$) {$4^+$};
        		\node at ($(p4) + (-135:1.35)$) {$1^-$};
        
    	\end{tikzpicture}}}+
    	\vcenter{\hbox{\begin{tikzpicture}[scale=0.8]
			
        		\coordinate (p1) at (-0.6,0.6);
        		\coordinate (p2) at (0.6,0.6);
        		\coordinate (p3) at (0.6,-0.6);
        		\coordinate (p4) at (-0.6,-0.6);

        		\draw[very thick, larrow] (p1) to (p2);
			\draw[very thick, larrow] (p2) to (p3);
			\draw[very thick, larrow] (p3) to (p4);
			\draw[very thick, larrow] (p4) to (p1);
			
			\draw[very thick, larrow] ($(p1) + (135:1)$) to (p1);
			\draw[very thick, rarrow]  ($(p2) + (45:1)$) to (p2);
			\draw[very thick, larrow] ($(p3)+(-45:1)$) to (p3);
			\draw[very thick, rarrow] ($(p4) + (-135:1)$) to (p4);
			\draw[fill=white,thick] (p1) circle (0.2);
        		\draw[fill, thick] (p2) circle (0.2);
        		\draw[fill=white, thick] (p3) circle (0.2);
        		\draw[fill, thick] (p4) circle (0.2);
        		\node at ($(p1) + (135:1.35)$) {$2^+$};
        		\node at ($(p2) + (45:1.35)$) {$3^-$};
        		\node at ($(p3) + (-45:1.35)$) {$4^+$};
        		\node at ($(p4) + (-135:1.35)$) {$1^-$};
        
    	\end{tikzpicture}}}
    	\end{aligned}
\end{equation}
In this example, a gluon on-shell state circulates inside the loop. There are two possible helicity assignments, corresponding to the two possible 
orientations of the loop. In ${\cal N} =4$ sYM theory, however, the entire vector on-shell supermultiplet circulates in the loop of these on-shell diagrams, rather than only the gluon. This leads to a discrepancy between the corresponding expressions for the leading singularities in pure YM and ${\cal N} =4$ sYM theories.

Sewing the three-point amplitudes evaluated on the maximal cut solutions, we obtain the following leading singularity,
\begin{equation} \label{eq:LS1L-+-+}
      \vcenter{\hbox{\begin{tikzpicture}[scale=0.8]
			
        		\coordinate (p1) at (-0.6,0.6);
        		\coordinate (p2) at (0.6,0.6);
        		\coordinate (p3) at (0.6,-0.6);
        		\coordinate (p4) at (-0.6,-0.6);

        		\draw[very thick] (p1) to (p2);
			\draw[very thick] (p2) to (p3);
			\draw[very thick] (p3) to (p4);
			\draw[very thick] (p4) to (p1);
			
			\draw[very thick, larrow] ($(p1) + (135:1)$) to (p1);
			\draw[very thick, rarrow]  ($(p2) + (45:1)$) to (p2);
			\draw[very thick, larrow] ($(p3)+(-45:1)$) to (p3);
			\draw[very thick, rarrow] ($(p4) + (-135:1)$) to (p4);
			\draw[fill=white,thick] (p1) circle (0.2);
        		\draw[fill, thick] (p2) circle (0.2);
        		\draw[fill=white, thick] (p3) circle (0.2);
        		\draw[fill, thick] (p4) circle (0.2);
        		\node at ($(p1) + (135:1.35)$) {$2^+$};
        		\node at ($(p2) + (45:1.35)$) {$3^-$};
        		\node at ($(p3) + (-45:1.35)$) {$4^+$};
        		\node at ($(p4) + (-135:1.35)$) {$1^-$};
        
    	\end{tikzpicture}}}=\text{PT}_{4,13}\times\left[\left(\frac{\langle12\rangle\langle34\rangle}{\langle13\rangle\langle24\rangle}\right)^4{+}\left(\frac{\langle14\rangle\langle23\rangle}{\langle13\rangle\langle24\rangle}\right)^4\right].
\end{equation}
In the ${\cal N} =4$ sYM case, this on-shell diagram evaluates to the supersymmetric PT factor, in agreement with the general statement that MHV leading singularities at any loop order in this theory are PT factors \cite{Arkani-Hamed:2008owk,Arkani-Hamed:2012zlh}. In contrast, in pure YM theory the leading singularity \eqref{eq:LS1L-+-+} appears as a rational prefactor (together with the PT factor $\text{PT}_{4,13}$, owing to IR subtractions) in the maximally-transcendental part of the one-loop hard function for the helicity configuration $(1^-2^+3^-4^+)$.

The previous expression is readily extended to the $n$-point one-loop MHV case. In this setting, all leading singularities beyond the PT factors arise from the two-mass-easy box type on-shell diagrams in Fig.~\ref{fig:1Lonshell}. 
The two negative helicity legs, $1^-$ and $i^-$, are placed in the separate MHV corners, and two possible orientations of the gluon circulating in the loop give rise to the following leading singularity, 
\begin{tcolorbox}
\vspace{-2ex}
\begin{equation}\label{eq:L1LS}
    R_{1i,jk,n}^{(1)}=\text{PT}_{n,1i}\times\left[\left(\frac{\langle 1j\rangle\langle ik\rangle}{\langle jk\rangle\langle 1i\rangle}\right)^4+\left(\frac{\langle 1k\rangle\langle ji\rangle}{\langle jk\rangle\langle 1i\rangle}\right)^4\right],\ \ \  1< j<i<k\leq n.
\end{equation}
\end{tcolorbox}
\noindent This leading singularity contributes only for helicity configurations satisfying $i-1\geq2$, i.e. legs $1^-$ and $i^-$ are nonadjacent. Consequently, the maximally-transcendental parts of the one-loop $n$-point MHV split-helicity hard functions $H_{--+\cdots+}^{(1)}$ are pure functions, normalized by the PT factor. This observation is in agreement with the explicit expressions for the one-loop pure YM hard functions available in the literature.

In the following sections, we will see that MHV leading singularities beyond the PT factors always arise from non-singlet on-shell diagrams. This observation provides a guiding principle for classifying the prefactors in the hard function ansatz.

As a final comment, note that the on-shell diagrams are built from conformally-invariant tree-level amplitudes, which are sewn together in a conformally invariant manner. As a result, the corresponding leading singularities are conformally invariant and are therefore annihilated by the conformal boost generator \cite{Witten:2003nn,Henn:2019mvc},
\begin{align}\label{eq:Wittenconformaloperater}
k_{\alpha \dot \alpha} = \sum_{i=1}^{n} \frac{\partial}{\partial \lambda_{i}^{\alpha}} \frac{\partial}{\partial \tilde\lambda_{i}^{\dot \alpha}} \,.
\end{align}
In fact, this annihilation is trivial for MHV leading singularities considered in this work, since the latter involve only angular spinor-helicity brackets $\langle\la_a \la_b\rangle$. One can see it explicitly for the one-loop leading singularity given in eq.~\eqref{eq:L1LS}, as well as for all two-loop MHV leading singularities, which we compute explicitly below. Let us also note that the conformal invariance of NMHV (and higher-helicity) leading singularities, which involve both $\la$ and $\tilde\la$ helicity spinors, is nontrivial.

\subsection{Two-loop MHV prefactors for pure YM amplitudes}

\begin{figure}[t]
    \centering
	\subfigure[Kissing-box]{\begin{tikzpicture}[scale=0.7]
			
			\coordinate (d45) at (45:1);
			\coordinate (dm45) at (-45:1);
			\coordinate (d135) at (135:1);
			\coordinate (dm135) at (-135:1);
			
	  		\coordinate (p3) at (0,0);
	  		\coordinate (p2) at (135:1.2);
	  		\coordinate (p1) at (180:1.70);
	  		\coordinate (p7) at (-135:1.2);
	  		
	  		\coordinate (p4) at (45:1.2);
	  		\coordinate (p5) at (0:1.70);
	  		\coordinate (p6) at (-45:1.2);
	  		
	  		\coordinate (p1aux1) at ($(p1)+(dm135)$);
	  		\coordinate (p1aux2) at ($(p1)+(-1,0)$);
	  		\coordinate (p1aux3) at ($(p1)+(d135)$);

	  		\coordinate (p3aux1) at ($(p3)+(70:1)$);
	  		\coordinate (p3aux2) at ($(p3)+(90:1)$);
	  		\coordinate (p3aux3) at ($(p3)+(110:1)$);

	  		\coordinate (p3aux4) at ($(p3)+(-70:1)$);
	  		\coordinate (p3aux5) at ($(p3)+(-90:1)$);
	  		\coordinate (p3aux6) at ($(p3)+(-110:1)$);

	  		\coordinate (p5aux1) at ($(p5)+(d45)$);
	  		\coordinate (p5aux2) at ($(p5)+(1,0)$);
	  		\coordinate (p5aux3) at ($(p5)+(dm45)$);

	  		\coordinate (p2aux) at ($(p2)+(0,1)$);
	  		\coordinate (p4aux) at ($(p4)+(0,1)$);
	  		\coordinate (p7aux) at ($(p7)+(0,-1)$);
	  		\coordinate (p6aux) at ($(p6)+(0,-1)$);

	  		\draw[very thick, ] (p1) to (p2);
	  		\draw[very thick,  ] (p2) to (p3);
	  		\draw[very thick, ] (p3) to (p7);
	  		\draw[very thick, ] (p7) to (p1);
	  		\draw[very thick, ] (p3) to (p4);
	  		\draw[very thick, ] (p4) to (p5);
	  		\draw[very thick, ] (p5) to (p6);
	  		\draw[very thick, ] (p6) to (p3);
	  		
	  		\draw[very thick, ] (p1aux1) to (p1);
	  		\draw[very thick, ] (p1aux3) to (p1);
	  		
	  		\draw[very thick, ] (p3aux1) to (p3);
	  		\draw[very thick, ] (p3aux3) to (p3);
	  		\draw[very thick, ] (p3aux4) to (p3);
	  		\draw[very thick, ] (p3aux6) to (p3);
	  		
	  		\draw[very thick, ] (p5aux1) to (p5);
	  		\draw[very thick, ] (p5aux3) to (p5);
	  		
	  		\draw[very thick, ] (p2aux) to (p2);
	  		\draw[very thick, ] (p4aux) to (p4);
	  		\draw[very thick, ] (p7aux) to (p7);
	  		\draw[very thick, ] (p6aux) to (p6);

        		\draw[fill, thick] (p1) circle (0.2);
        		\draw[fill=white,thick] (p2) circle (0.2);
            \draw[fill, thick] (p3) circle (0.2);
 		    \draw[fill=white, thick] (p6) circle (0.2);
        		\draw[fill=white, thick] (p7) circle (0.2);
        		\draw[fill, thick] (p5) circle (0.2);
        		\draw[fill=white, thick] (p4) circle (0.2);

        		\node at ($(p1aux2)+(180:0.2)$) {$1^-$};
%
%
%
        		\node at ($(p5aux2)+(0:0.3)$) {$i^{-}$};
%
        		\node at ($(p4aux)+(0,0.35)$) {$k$};
        		\node at ($(p2aux)+(0,0.35)$) {$j$};
        		\node at ($(p7aux)+(0,-0.35)$) {$m$};
        		\node at ($(p6aux)+(0,-0.35)$) {$l$};

        		\node[rotate=90] at ($(p1)!0.5!(p1aux2)$) {$\dots$};
        		\node[] at ($(p3aux2)$) {...};
        		\node[rotate = 90] at ($(p5)!0.5!(p5aux2)$) {$\dots$};
        		\node[rotate = 0] at ($(p3aux5)$) {...};

    \end{tikzpicture}}
	\subfigure[Penta-box]{
    \begin{tikzpicture}[scale=0.7]
			
			\coordinate (d45) at (45:1);
			\coordinate (dm45) at (-45:1);
			\coordinate (d135) at (135:1);
			\coordinate (dm135) at (-135:1);
			
	  		\coordinate (p1) at (180:1);
	  		\coordinate (p2) at (108:1);
	  		\coordinate (p3) at (36:1);
	  		\coordinate (p6) at (-36:1);
	  		\coordinate (p7) at (-108:1);
	  		
	  		\coordinate (p4) at ($(p3)+(1.2,0)$);
	  		\coordinate (p5) at ($(p6)+(1.2,0)$);
	  		
	  		\coordinate (p1aux1) at ($(p1)+(dm135)$);
	  		\coordinate (p1aux2) at ($(p1)+(-1,0)$);
	  		\coordinate (p1aux3) at ($(p1)+(d135)$);

	  		\coordinate (p3aux1) at ($(p3)+(60:1)$);
	  		\coordinate (p3aux2) at ($(p3)+(90:1)$);
	  		\coordinate (p3aux3) at ($(p3)+(120:1)$);

	  		\coordinate (p5aux1) at ($(p5)+(1,0)$);
	  		\coordinate (p5aux2) at ($(p5)+(dm45)$);
	  		\coordinate (p5aux3) at ($(p5)+(0,-1)$);

	  		\coordinate (p2aux1) at ($(p2)+(108:1)$);
	  		\coordinate (p4aux1) at ($(p4)+(45:1)$);
	  		\coordinate (p7aux1) at ($(p7)+(-108:1)$);

	  		\draw[very thick, ] (p1) to (p2);
	  		\draw[very thick,  ] (p2) to (p3);
	  		\draw[very thick, ] (p3) to (p6);
	  		\draw[very thick, ] (p6) to (p7);
	  		\draw[very thick, ] (p7) to (p1);
	  		\draw[very thick, ] (p5) to (p6);
	  		\draw[very thick, ] (p3) to (p4);
	  		\draw[very thick, ] (p4) to (p5);
	  		
	  		\draw[very thick, ] (p1aux1) to (p1);
	  		\draw[very thick, ] (p1aux3) to (p1);
	  		
	  		\draw[very thick, ] (p3aux1) to (p3);
	  		\draw[very thick, ] (p3aux3) to (p3);
	  		
	  		\draw[very thick, ] (p5aux1) to (p5);
	  		\draw[very thick, ] (p5aux3) to (p5);
	  		
	  		\draw[very thick, ] (p2aux1) to (p2);
	  		\draw[very thick, ] (p4aux1) to (p4);
	  		\draw[very thick, ] (p7aux1) to (p7);

        		\draw[fill, thick] (p1) circle (0.2);
        		\draw[fill=white,thick] (p2) circle (0.2);
            \draw[fill, thick] (p3) circle (0.2);
 		    \draw[fill=white, thick] (p6) circle (0.2);
        		\draw[fill=white, thick] (p7) circle (0.2);
        		\draw[fill, thick] (p5) circle (0.2);
        		\draw[fill=white, thick] (p4) circle (0.2);

        		\node at ($(p1aux2)+(180:0.2)$) {$1^-$};
%
%
        		\node at ($(p5aux2)+(-45:0.3)$) {$i^{-}$};
%
        		\node at ($(p4aux1)+(45:0.35)$) {$k$};
        		\node at ($(p2aux1)+(108:0.35)$) {$l$};
        		\node at ($(p7aux1)+(-108:0.35)$) {$j$};
        		
        		\node[rotate=90] at ($(p1)!0.5!(p1aux2)$) {$\dots$};
        		\node[] at ($(p3)!0.8!(p3aux2)$) {...};
        		\node[rotate = 45] at ($(p5)!0.5!(p5aux2)$) {$\dots$};
        
    \end{tikzpicture}}
	\subfigure[Double-box]{
\begin{tikzpicture}[scale=0.7]
			
			\coordinate (d45) at (45:1);
			\coordinate (dm45) at (-45:1);
			\coordinate (d135) at (135:1);
			\coordinate (dm135) at (-135:1);
			
	  		\coordinate (p1) at (-1.2,0.6);
	  		\coordinate (p2) at (0,0.6);
	  		\coordinate (p3) at (1.2,0.6);
	  		\coordinate (p4) at (1.2,-0.6);
	  		\coordinate (p5) at (0,-0.6);
	  		\coordinate (p6) at (-1.2,-0.6);
	  		
	  		\coordinate (p1aux1) at ($(p1)+(-1,0)$);
	  		\coordinate (p1aux2) at ($(p1)+(d135)$);
	  		\coordinate (p1aux3) at ($(p1)+(0,1)$);
	  		\coordinate (p3aux1) at ($(p3)+(1,0)$);
	  		\coordinate (p3aux2) at ($(p3)+(d45)$);
	  		\coordinate (p3aux3) at ($(p3)+(0,1)$);
	  		\coordinate (p5aux2) at ($(p5)+(0,-1)$);
	  		\coordinate (p5aux1) at ($(p5)+(-60:1)$);
	  		\coordinate (p5aux3) at ($(p5)+(-120:1)$);
	  		
	  		\coordinate (p2aux) at ($(p2)+(0,1)$);
	  		\coordinate (p4aux) at ($(p4)+(dm45)$);
	  		\coordinate (p6aux) at ($(p6)+(dm135)$);
	  		
	  		\draw[very thick, ] (p1) to (p2);
	  		\draw[very thick,  ] (p2) to (p3);
	  		\draw[very thick, ] (p3) to (p4);
	  		\draw[very thick, ] (p4) to (p5);
	  		\draw[very thick, ] (p5) to (p6);
	  		\draw[very thick, ] (p6) to (p1);
	  		\draw[very thick, ] (p2) to (p5);
	  		
        		\draw[very thick] (p1aux1) to (p1);
        		\draw[very thick] (p1aux3) to (p1);
        		
        		\draw[very thick] (p3aux3) to (p3);
        		\draw[very thick] (p3aux1) to (p3);

        		\draw[very thick] (p5aux1) to (p5);
        		\draw[very thick] (p5aux3) to (p5);
        		
        		\draw[very thick, ] (p4aux) to (p4);
        		\draw[very thick,] (p6aux) to (p6);

        		\draw[fill, thick] (p1) circle (0.2);
        		\draw[fill=white,thick] (p2) circle (0.2);
            \draw[fill, thick] (p3) circle (0.2);
 		    \draw[fill=white, thick] (p4) circle (0.2);
        		\draw[fill, thick] (p5) circle (0.2);
        		\draw[fill=white, thick] (p6) circle (0.2);

        		
        		\node at ($(p1aux2)+(135:0.2)$) {$1^-$};
       		\node at ($(p1aux3)+(90:0.4)$) {$l$};
       		\node at ($(p3aux3)+(90:0.4)$) {$l+1$};
        		\node at ($(p3aux2)+(45:0.3)$) {$i^{-}$};
%
%
        		\node at ($(p4aux)+(-45:0.35)$) {$j$};
        		\node at ($(p6aux)+(-135:0.35)$) {$k$};
        		
        		\node[rotate=45] at ($(p1)!0.5!(p1aux2)$) {$\dots$};
        		\node[rotate=-45] at ($(p3)!0.5!(p3aux2)$) {$\dots$};
        		\node[] at ($(p5)!0.8!(p5aux2)$) {...};
        
    \end{tikzpicture}}
    \caption{Non-singlet on-shell diagrams contributing to genuinely two-loops prefactors.}
    \label{fig:2Lonshell}
\end{figure}
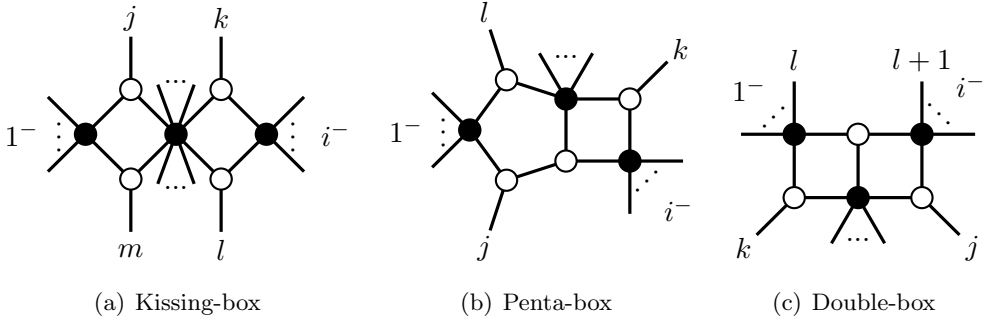

We next analyze the two-loop MHV leading singularities. As discussed in section~\ref{sec: maximal weight projection}, when considering the maximal weight projection of the MHV amplitudes, the relevant contours for computing two-loop four-dimensional maximal cuts are the kissing-box, double-box, and penta-box cut, shown in Fig.~\ref{fig:2Lcuts}. Thus, it suffices to study 
these on-shell diagrams. By enumerating all possible helicity assignments on the external legs and internal lines, one finds that genuinely new two-loop leading singularities, beyond both PT factors and one-loop leading singularities, arise from on-shell diagrams containing exactly two oriented cycles of helicity flow. We also find that the pair of negative-helicity legs must be chosen as shown in Fig.~\ref{fig:2Lonshell}. Note that, since the massless positive-helicity corners are present in these on-shell diagrams, the kissing-box cut contributes only starting from six points and for $i-1\geq3$, while the penta-box cut contributes only starting from five points and for $i-1\geq2$. By contrast, the double-box cut contributes starting from four points and in all MHV sectors.

We further compute all genuinely new two-loop leading singularities for pure YM MHV amplitudes at arbitrary multiplicity, which are represented by on-shell diagrams in Fig.~\ref{fig:2Lonshell}. The kissing-box on-shell diagrams are readily evaluated by sewing together seven tree-level amplitudes that constitute them. For each facet of the kissing-box on-shell diagram, there are two possible helicity assignments, which form oriented cycles of the helicity flow, 
\begin{equation}
	\begin{aligned}
		\vcenter{\hbox{\begin{tikzpicture}[scale=0.7]
			
			\coordinate (d45) at (45:1);
			\coordinate (dm45) at (-45:1);
			\coordinate (d135) at (135:1);
			\coordinate (dm135) at (-135:1);
			
	  		\coordinate (p3) at (0,0);
	  		\coordinate (p2) at (135:1.2);
	  		\coordinate (p1) at (180:1.70);
	  		\coordinate (p7) at (-135:1.2);
	  		
	  		\coordinate (p4) at (45:1.2);
	  		\coordinate (p5) at (0:1.70);
	  		\coordinate (p6) at (-45:1.2);
	  		
	  		\coordinate (p1aux1) at ($(p1)+(dm135)$);
	  		\coordinate (p1aux2) at ($(p1)+(-1,0)$);
	  		\coordinate (p1aux3) at ($(p1)+(d135)$);

	  		\coordinate (p3aux1) at ($(p3)+(70:1)$);
	  		\coordinate (p3aux2) at ($(p3)+(90:1)$);
	  		\coordinate (p3aux3) at ($(p3)+(110:1)$);

	  		\coordinate (p3aux4) at ($(p3)+(-70:1)$);
	  		\coordinate (p3aux5) at ($(p3)+(-90:1)$);
	  		\coordinate (p3aux6) at ($(p3)+(-110:1)$);

	  		\coordinate (p5aux1) at ($(p5)+(d45)$);
	  		\coordinate (p5aux2) at ($(p5)+(1,0)$);
	  		\coordinate (p5aux3) at ($(p5)+(dm45)$);

	  		\coordinate (p2aux) at ($(p2)+(0,1)$);
	  		\coordinate (p4aux) at ($(p4)+(0,1)$);
	  		\coordinate (p7aux) at ($(p7)+(0,-1)$);
	  		\coordinate (p6aux) at ($(p6)+(0,-1)$);

	  		\draw[very thick, rarrow ] (p1) to (p2);
	  		\draw[very thick,  rarrow] (p2) to (p3);
	  		\draw[very thick, rarrow] (p3) to (p7);
	  		\draw[very thick, rarrow] (p7) to (p1);
	  		\draw[very thick, rarrow] (p3) to (p4);
	  		\draw[very thick, rarrow] (p4) to (p5);
	  		\draw[very thick, rarrow] (p5) to (p6);
	  		\draw[very thick, rarrow] (p6) to (p3);
	  		
	  		\draw[very thick, ] (p1aux1) to (p1);
	  		\draw[very thick, ] (p1aux3) to (p1);
	  		
	  		\draw[very thick, ] (p3aux1) to (p3);
	  		\draw[very thick, ] (p3aux3) to (p3);
	  		\draw[very thick, ] (p3aux4) to (p3);
	  		\draw[very thick, ] (p3aux6) to (p3);
	  		
	  		\draw[very thick, ] (p5aux1) to (p5);
	  		\draw[very thick, ] (p5aux3) to (p5);
	  		
	  		\draw[very thick,larrow ] (p2aux) to (p2);
	  		\draw[very thick,larrow ] (p4aux) to (p4);
	  		\draw[very thick,larrow ] (p7aux) to (p7);
	  		\draw[very thick,larrow ] (p6aux) to (p6);

        		\draw[fill, thick] (p1) circle (0.2);
        		\draw[fill=white,thick] (p2) circle (0.2);
            \draw[fill, thick] (p3) circle (0.2);
 		    \draw[fill=white, thick] (p6) circle (0.2);
        		\draw[fill=white, thick] (p7) circle (0.2);
        		\draw[fill, thick] (p5) circle (0.2);
        		\draw[fill=white, thick] (p4) circle (0.2);

        		\node at ($(p1aux2)+(180:0.2)$) {$1^-$};
%
%
%
        		\node at ($(p5aux2)+(0:0.3)$) {$i^{-}$};
%
        		\node at ($(p4aux)+(0,0.35)$) {$k$};
        		\node at ($(p2aux)+(0,0.35)$) {$j$};
        		\node at ($(p7aux)+(0,-0.35)$) {$m$};
        		\node at ($(p6aux)+(0,-0.35)$) {$l$};

        		\node[rotate=90] at ($(p1)!0.5!(p1aux2)$) {$\dots$};
        		\node[] at ($(p3aux2)$) {...};
        		\node[rotate = 90] at ($(p5)!0.5!(p5aux2)$) {$\dots$};
        		\node[rotate = 0] at ($(p3aux5)$) {...};

    \end{tikzpicture}}} \quad + \quad 
    \vcenter{\hbox{\begin{tikzpicture}[scale=0.7]
			
			\coordinate (d45) at (45:1);
			\coordinate (dm45) at (-45:1);
			\coordinate (d135) at (135:1);
			\coordinate (dm135) at (-135:1);
			
	  		\coordinate (p3) at (0,0);
	  		\coordinate (p2) at (135:1.2);
	  		\coordinate (p1) at (180:1.70);
	  		\coordinate (p7) at (-135:1.2);
	  		
	  		\coordinate (p4) at (45:1.2);
	  		\coordinate (p5) at (0:1.70);
	  		\coordinate (p6) at (-45:1.2);
	  		
	  		\coordinate (p1aux1) at ($(p1)+(dm135)$);
	  		\coordinate (p1aux2) at ($(p1)+(-1,0)$);
	  		\coordinate (p1aux3) at ($(p1)+(d135)$);

	  		\coordinate (p3aux1) at ($(p3)+(70:1)$);
	  		\coordinate (p3aux2) at ($(p3)+(90:1)$);
	  		\coordinate (p3aux3) at ($(p3)+(110:1)$);

	  		\coordinate (p3aux4) at ($(p3)+(-70:1)$);
	  		\coordinate (p3aux5) at ($(p3)+(-90:1)$);
	  		\coordinate (p3aux6) at ($(p3)+(-110:1)$);

	  		\coordinate (p5aux1) at ($(p5)+(d45)$);
	  		\coordinate (p5aux2) at ($(p5)+(1,0)$);
	  		\coordinate (p5aux3) at ($(p5)+(dm45)$);

	  		\coordinate (p2aux) at ($(p2)+(0,1)$);
	  		\coordinate (p4aux) at ($(p4)+(0,1)$);
	  		\coordinate (p7aux) at ($(p7)+(0,-1)$);
	  		\coordinate (p6aux) at ($(p6)+(0,-1)$);

	  		\draw[very thick,rarrow ] (p1) to (p2);
	  		\draw[very thick, rarrow ] (p2) to (p3);
	  		\draw[very thick, rarrow] (p3) to (p7);
	  		\draw[very thick, rarrow] (p7) to (p1);
	  		\draw[very thick, larrow] (p3) to (p4);
	  		\draw[very thick,larrow ] (p4) to (p5);
	  		\draw[very thick, larrow] (p5) to (p6);
	  		\draw[very thick, larrow] (p6) to (p3);
	  		
	  		\draw[very thick, ] (p1aux1) to (p1);
	  		\draw[very thick, ] (p1aux3) to (p1);
	  		
	  		\draw[very thick, ] (p3aux1) to (p3);
	  		\draw[very thick, ] (p3aux3) to (p3);
	  		\draw[very thick, ] (p3aux4) to (p3);
	  		\draw[very thick, ] (p3aux6) to (p3);
	  		
	  		\draw[very thick, ] (p5aux1) to (p5);
	  		\draw[very thick, ] (p5aux3) to (p5);
	  		
	  		\draw[very thick,larrow ] (p2aux) to (p2);
	  		\draw[very thick,larrow ] (p4aux) to (p4);
	  		\draw[very thick,larrow ] (p7aux) to (p7);
	  		\draw[very thick,larrow ] (p6aux) to (p6);

        		\draw[fill, thick] (p1) circle (0.2);
        		\draw[fill=white,thick] (p2) circle (0.2);
            \draw[fill, thick] (p3) circle (0.2);
 		    \draw[fill=white, thick] (p6) circle (0.2);
        		\draw[fill=white, thick] (p7) circle (0.2);
        		\draw[fill, thick] (p5) circle (0.2);
        		\draw[fill=white, thick] (p4) circle (0.2);

        		\node at ($(p1aux2)+(180:0.2)$) {$1^-$};
%
%
%
        		\node at ($(p5aux2)+(0:0.3)$) {$i^{-}$};
%
        		\node at ($(p4aux)+(0,0.35)$) {$k$};
        		\node at ($(p2aux)+(0,0.35)$) {$j$};
        		\node at ($(p7aux)+(0,-0.35)$) {$m$};
        		\node at ($(p6aux)+(0,-0.35)$) {$l$};

        		\node[rotate=90] at ($(p1)!0.5!(p1aux2)$) {$\dots$};
        		\node[] at ($(p3aux2)$) {...};
        		\node[rotate = 90] at ($(p5)!0.5!(p5aux2)$) {$\dots$};
        		\node[rotate = 0] at ($(p3aux5)$) {...};

    \end{tikzpicture}}} \\
    \vcenter{\hbox{\begin{tikzpicture}[scale=0.7]
			
			\coordinate (d45) at (45:1);
			\coordinate (dm45) at (-45:1);
			\coordinate (d135) at (135:1);
			\coordinate (dm135) at (-135:1);
			
	  		\coordinate (p3) at (0,0);
	  		\coordinate (p2) at (135:1.2);
	  		\coordinate (p1) at (180:1.70);
	  		\coordinate (p7) at (-135:1.2);
	  		
	  		\coordinate (p4) at (45:1.2);
	  		\coordinate (p5) at (0:1.70);
	  		\coordinate (p6) at (-45:1.2);
	  		
	  		\coordinate (p1aux1) at ($(p1)+(dm135)$);
	  		\coordinate (p1aux2) at ($(p1)+(-1,0)$);
	  		\coordinate (p1aux3) at ($(p1)+(d135)$);

	  		\coordinate (p3aux1) at ($(p3)+(70:1)$);
	  		\coordinate (p3aux2) at ($(p3)+(90:1)$);
	  		\coordinate (p3aux3) at ($(p3)+(110:1)$);

	  		\coordinate (p3aux4) at ($(p3)+(-70:1)$);
	  		\coordinate (p3aux5) at ($(p3)+(-90:1)$);
	  		\coordinate (p3aux6) at ($(p3)+(-110:1)$);

	  		\coordinate (p5aux1) at ($(p5)+(d45)$);
	  		\coordinate (p5aux2) at ($(p5)+(1,0)$);
	  		\coordinate (p5aux3) at ($(p5)+(dm45)$);

	  		\coordinate (p2aux) at ($(p2)+(0,1)$);
	  		\coordinate (p4aux) at ($(p4)+(0,1)$);
	  		\coordinate (p7aux) at ($(p7)+(0,-1)$);
	  		\coordinate (p6aux) at ($(p6)+(0,-1)$);

	  		\draw[very thick,larrow ] (p1) to (p2);
	  		\draw[very thick, larrow ] (p2) to (p3);
	  		\draw[very thick, larrow] (p3) to (p7);
	  		\draw[very thick, larrow] (p7) to (p1);
	  		\draw[very thick, rarrow] (p3) to (p4);
	  		\draw[very thick,rarrow ] (p4) to (p5);
	  		\draw[very thick, rarrow] (p5) to (p6);
	  		\draw[very thick, rarrow] (p6) to (p3);
	  		
	  		\draw[very thick, ] (p1aux1) to (p1);
	  		\draw[very thick, ] (p1aux3) to (p1);
	  		
	  		\draw[very thick, ] (p3aux1) to (p3);
	  		\draw[very thick, ] (p3aux3) to (p3);
	  		\draw[very thick, ] (p3aux4) to (p3);
	  		\draw[very thick, ] (p3aux6) to (p3);
	  		
	  		\draw[very thick, ] (p5aux1) to (p5);
	  		\draw[very thick, ] (p5aux3) to (p5);
	  		
	  		\draw[very thick,larrow ] (p2aux) to (p2);
	  		\draw[very thick,larrow ] (p4aux) to (p4);
	  		\draw[very thick,larrow ] (p7aux) to (p7);
	  		\draw[very thick,larrow ] (p6aux) to (p6);

        		\draw[fill, thick] (p1) circle (0.2);
        		\draw[fill=white,thick] (p2) circle (0.2);
            \draw[fill, thick] (p3) circle (0.2);
 		    \draw[fill=white, thick] (p6) circle (0.2);
        		\draw[fill=white, thick] (p7) circle (0.2);
        		\draw[fill, thick] (p5) circle (0.2);
        		\draw[fill=white, thick] (p4) circle (0.2);

        		\node at ($(p1aux2)+(180:0.2)$) {$1^-$};
%
%
%
        		\node at ($(p5aux2)+(0:0.3)$) {$i^{-}$};
%
        		\node at ($(p4aux)+(0,0.35)$) {$k$};
        		\node at ($(p2aux)+(0,0.35)$) {$j$};
        		\node at ($(p7aux)+(0,-0.35)$) {$m$};
        		\node at ($(p6aux)+(0,-0.35)$) {$l$};

        		\node[rotate=90] at ($(p1)!0.5!(p1aux2)$) {$\dots$};
        		\node[] at ($(p3aux2)$) {...};
        		\node[rotate = 90] at ($(p5)!0.5!(p5aux2)$) {$\dots$};
        		\node[rotate = 0] at ($(p3aux5)$) {...};

    \end{tikzpicture}}} \quad + \quad 
    \vcenter{\hbox{\begin{tikzpicture}[scale=0.7]
			
			\coordinate (d45) at (45:1);
			\coordinate (dm45) at (-45:1);
			\coordinate (d135) at (135:1);
			\coordinate (dm135) at (-135:1);
			
	  		\coordinate (p3) at (0,0);
	  		\coordinate (p2) at (135:1.2);
	  		\coordinate (p1) at (180:1.70);
	  		\coordinate (p7) at (-135:1.2);
	  		
	  		\coordinate (p4) at (45:1.2);
	  		\coordinate (p5) at (0:1.70);
	  		\coordinate (p6) at (-45:1.2);
	  		
	  		\coordinate (p1aux1) at ($(p1)+(dm135)$);
	  		\coordinate (p1aux2) at ($(p1)+(-1,0)$);
	  		\coordinate (p1aux3) at ($(p1)+(d135)$);

	  		\coordinate (p3aux1) at ($(p3)+(70:1)$);
	  		\coordinate (p3aux2) at ($(p3)+(90:1)$);
	  		\coordinate (p3aux3) at ($(p3)+(110:1)$);

	  		\coordinate (p3aux4) at ($(p3)+(-70:1)$);
	  		\coordinate (p3aux5) at ($(p3)+(-90:1)$);
	  		\coordinate (p3aux6) at ($(p3)+(-110:1)$);

	  		\coordinate (p5aux1) at ($(p5)+(d45)$);
	  		\coordinate (p5aux2) at ($(p5)+(1,0)$);
	  		\coordinate (p5aux3) at ($(p5)+(dm45)$);

	  		\coordinate (p2aux) at ($(p2)+(0,1)$);
	  		\coordinate (p4aux) at ($(p4)+(0,1)$);
	  		\coordinate (p7aux) at ($(p7)+(0,-1)$);
	  		\coordinate (p6aux) at ($(p6)+(0,-1)$);

	  		\draw[very thick,larrow ] (p1) to (p2);
	  		\draw[very thick, larrow ] (p2) to (p3);
	  		\draw[very thick, larrow] (p3) to (p7);
	  		\draw[very thick, larrow] (p7) to (p1);
	  		\draw[very thick, larrow] (p3) to (p4);
	  		\draw[very thick, larrow ] (p4) to (p5);
	  		\draw[very thick, larrow] (p5) to (p6);
	  		\draw[very thick, larrow] (p6) to (p3);
	  		
	  		\draw[very thick, ] (p1aux1) to (p1);
	  		\draw[very thick, ] (p1aux3) to (p1);
	  		
	  		\draw[very thick, ] (p3aux1) to (p3);
	  		\draw[very thick, ] (p3aux3) to (p3);
	  		\draw[very thick, ] (p3aux4) to (p3);
	  		\draw[very thick, ] (p3aux6) to (p3);
	  		
	  		\draw[very thick, ] (p5aux1) to (p5);
	  		\draw[very thick, ] (p5aux3) to (p5);
	  		
	  		\draw[very thick,larrow ] (p2aux) to (p2);
	  		\draw[very thick,larrow ] (p4aux) to (p4);
	  		\draw[very thick,larrow ] (p7aux) to (p7);
	  		\draw[very thick,larrow ] (p6aux) to (p6);

        		\draw[fill, thick] (p1) circle (0.2);
        		\draw[fill=white,thick] (p2) circle (0.2);
            \draw[fill, thick] (p3) circle (0.2);
 		    \draw[fill=white, thick] (p6) circle (0.2);
        		\draw[fill=white, thick] (p7) circle (0.2);
        		\draw[fill, thick] (p5) circle (0.2);
        		\draw[fill=white, thick] (p4) circle (0.2);

        		\node at ($(p1aux2)+(180:0.2)$) {$1^-$};
%
%
%
        		\node at ($(p5aux2)+(0:0.3)$) {$i^{-}$};
%
        		\node at ($(p4aux)+(0,0.35)$) {$k$};
        		\node at ($(p2aux)+(0,0.35)$) {$j$};
        		\node at ($(p7aux)+(0,-0.35)$) {$m$};
        		\node at ($(p6aux)+(0,-0.35)$) {$l$};

        		\node[rotate=90] at ($(p1)!0.5!(p1aux2)$) {$\dots$};
        		\node[] at ($(p3aux2)$) {...};
        		\node[rotate = 90] at ($(p5)!0.5!(p5aux2)$) {$\dots$};
        		\node[rotate = 0] at ($(p3aux5)$) {...};

    \end{tikzpicture}}}
	\end{aligned}
    \label{eq:fig_kb}
\end{equation} 
Substituting the maximal cut solutions for the eight on-shell propagators (which coincide with those of the one-loop two-mass-easy box) into the MHV and $\overline{\text{MHV}}$ vertices of the on-shell diagrams, we obtain the following expression for the sum of four on-shell diagrams in \eqref{eq:fig_kb},
\begin{tcolorbox}
\begin{equation}\label{eq:Rkb}
\begin{aligned}
\resizebox{0.95\width}{!}{$R^{(2),\text{kb}}_{1i,jklm,n}:=\text{PT}_{n,1i}\times\left[\left(\frac{\langle1j\rangle\langle ki\rangle\langle ml\rangle}{\langle1i\rangle\langle jm\rangle\langle kl\rangle}\right)^4
    {+}\left(\frac{\langle1j\rangle\langle il\rangle\langle km\rangle}{\langle1i\rangle\langle jm\rangle\langle kl\rangle}\right)^4
    {+}\left(\frac{\langle1m\rangle\langle ki\rangle\langle jl\rangle}{\langle1i\rangle\langle jm\rangle\langle kl\rangle}\right)^4
    {+}\left(\frac{\langle1m\rangle\langle il\rangle\langle jk\rangle}{\langle1i\rangle\langle jm\rangle\langle kl\rangle}\right)^4\right]$,}\\
    1<j<k<i<l<m\leq n\ .\phantom{aaaaaaaaa}
\end{aligned}
\end{equation}
\end{tcolorbox}

The calculation for the penta-box cut proceeds analogously.
Putting eight propagators on shell completely freezes the loop momenta.
In spinor-helicity notations, these eight on-shell internal states carry the momenta (see \eqref{eq:Ipb} for the definition of the loop momenta)
\begin{equation} \label{eq:l_pb_sol}
\begin{aligned}
    &\ell_a=\frac{\lambda_{j}(P_{l{+}1,\cdots,j{-}1}\cdot\lambda_l)}{\langle jk\rangle},\ \ell_b=\frac{\lambda_{j}(P_{j{+}1,\cdots,l{-}1}\cdot\lambda_l)}{\langle jk\rangle},\ \ell_c=\frac{\lambda_{l}(P_{j{+}1,\cdots,l{-}1}\cdot\lambda_j)}{\langle jk\rangle},\\
    &\ell_d=\frac{\lambda_{l}(P_{l{+}1,\cdots,j{-}1}\cdot\lambda_j)}{\langle jk\rangle}, \ell_e=\frac{\lambda_k (P_{l{+}1,\cdots,k{-}1}{+}\ell_d)\cdot\lambda_a}{\langle ak\rangle}, \ \ell_f=\frac{\lambda_k (P_{k{+}1,\cdots,j{-}1}\cdot\lambda_a)}{\langle ak\rangle}, \\
    &\ell_g=\frac{\lambda_a (P_{k{+}1,\cdots,j{-}1}\cdot\ell_k)}{\langle ak\rangle},\ \ell_h=\frac{\lambda_a (P_{l{+}1,\cdots,k{-}1}{+}\ell_d)\cdot\lambda_k}{\langle ak\rangle},
\end{aligned}
\end{equation}
where $\lambda_a:=\lambda_{\ell_a}$ and $P_{i,\cdots,j}:=p_i{+}p_{i{+1}}{+}\cdots{+}p_j$.
By sewing the vertices of the on-shell diagram, we identify three possible cyclically oriented helicity assignments: either both facets of the graph carry an oriented cycle, or only one of the two facets carries an oriented cycle while the other oriented cycle flows through the larger loop, as depicted in eq.~\eqref{eq:pb_onshell}.

\begin{equation}
\vcenter{\hbox{\begin{tikzpicture}[scale=0.7]
			
			\coordinate (d45) at (45:1);
			\coordinate (dm45) at (-45:1);
			\coordinate (d135) at (135:1);
			\coordinate (dm135) at (-135:1);
			
	  		\coordinate (p1) at (180:1);
	  		\coordinate (p2) at (108:1);
	  		\coordinate (p3) at (36:1);
	  		\coordinate (p6) at (-36:1);
	  		\coordinate (p7) at (-108:1);
	  		
	  		\coordinate (p4) at ($(p3)+(1.2,0)$);
	  		\coordinate (p5) at ($(p6)+(1.2,0)$);
	  		
	  		\coordinate (p1aux1) at ($(p1)+(dm135)$);
	  		\coordinate (p1aux2) at ($(p1)+(-1,0)$);
	  		\coordinate (p1aux3) at ($(p1)+(d135)$);

	  		\coordinate (p3aux1) at ($(p3)+(60:1)$);
	  		\coordinate (p3aux2) at ($(p3)+(90:1)$);
	  		\coordinate (p3aux3) at ($(p3)+(120:1)$);

	  		\coordinate (p5aux1) at ($(p5)+(1,0)$);
	  		\coordinate (p5aux2) at ($(p5)+(dm45)$);
	  		\coordinate (p5aux3) at ($(p5)+(0,-1)$);

	  		\coordinate (p2aux1) at ($(p2)+(108:1)$);
	  		\coordinate (p4aux1) at ($(p4)+(45:1)$);
	  		\coordinate (p7aux1) at ($(p7)+(-108:1)$);

	  		\draw[very thick, rarrow ] (p1) to (p2);
	  		\draw[very thick, rarrow ] (p2) to (p3);
	  		\draw[very thick, rarrow ] (p3) to (p6);
	  		\draw[very thick,rarrow ] (p6) to (p7);
	  		\draw[very thick,rarrow ] (p7) to (p1);
	  		\draw[very thick,larrow ] (p5) to (p6);
	  		\draw[very thick,larrow ] (p3) to (p4);
	  		\draw[very thick,larrow ] (p4) to (p5);
	  		
	  		\draw[very thick, ] (p1aux1) to (p1);
	  		\draw[very thick, ] (p1aux3) to (p1);
	  		
	  		\draw[very thick, ] (p3aux1) to (p3);
	  		\draw[very thick, ] (p3aux3) to (p3);
	  		
	  		\draw[very thick, ] (p5aux1) to (p5);
	  		\draw[very thick, ] (p5aux3) to (p5);
	  		
	  		\draw[very thick, larrow ] (p2aux1) to (p2);
	  		\draw[very thick, larrow] (p4aux1) to (p4);
	  		\draw[very thick, larrow] (p7aux1) to (p7);

        		\draw[fill, thick] (p1) circle (0.2);
        		\draw[fill=white,thick] (p2) circle (0.2);
            \draw[fill, thick] (p3) circle (0.2);
 		    \draw[fill=white, thick] (p6) circle (0.2);
        		\draw[fill=white, thick] (p7) circle (0.2);
        		\draw[fill, thick] (p5) circle (0.2);
        		\draw[fill=white, thick] (p4) circle (0.2);

        		\node at ($(p1aux2)+(180:0.2)$) {$1^-$};
%
%
        		\node at ($(p5aux2)+(-45:0.3)$) {$i^{-}$};
%
        		\node at ($(p4aux1)+(45:0.35)$) {$k$};
        		\node at ($(p2aux1)+(108:0.35)$) {$l$};
        		\node at ($(p7aux1)+(-108:0.35)$) {$j$};
        		
        		\node[rotate=90] at ($(p1)!0.5!(p1aux2)$) {$\dots$};
        		\node[] at ($(p3)!0.8!(p3aux2)$) {...};
        		\node[rotate = 45] at ($(p5)!0.5!(p5aux2)$) {$\dots$};
        
    \end{tikzpicture}}}  + \,
    \vcenter{\hbox{\begin{tikzpicture}[scale=0.7]
			
			\coordinate (d45) at (45:1);
			\coordinate (dm45) at (-45:1);
			\coordinate (d135) at (135:1);
			\coordinate (dm135) at (-135:1);
			
	  		\coordinate (p1) at (180:1);
	  		\coordinate (p2) at (108:1);
	  		\coordinate (p3) at (36:1);
	  		\coordinate (p6) at (-36:1);
	  		\coordinate (p7) at (-108:1);
	  		
	  		\coordinate (p4) at ($(p3)+(1.2,0)$);
	  		\coordinate (p5) at ($(p6)+(1.2,0)$);
	  		
	  		\coordinate (p1aux1) at ($(p1)+(dm135)$);
	  		\coordinate (p1aux2) at ($(p1)+(-1,0)$);
	  		\coordinate (p1aux3) at ($(p1)+(d135)$);

	  		\coordinate (p3aux1) at ($(p3)+(60:1)$);
	  		\coordinate (p3aux2) at ($(p3)+(90:1)$);
	  		\coordinate (p3aux3) at ($(p3)+(120:1)$);

	  		\coordinate (p5aux1) at ($(p5)+(1,0)$);
	  		\coordinate (p5aux2) at ($(p5)+(dm45)$);
	  		\coordinate (p5aux3) at ($(p5)+(0,-1)$);

	  		\coordinate (p2aux1) at ($(p2)+(108:1)$);
	  		\coordinate (p4aux1) at ($(p4)+(45:1)$);
	  		\coordinate (p7aux1) at ($(p7)+(-108:1)$);

	  		\draw[very thick, rarrow ] (p1) to (p2);
	  		\draw[very thick, rarrow ] (p2) to (p3);
	  		\draw[very thick, larrow ] (p3) to (p6);
	  		\draw[very thick,rarrow ] (p6) to (p7);
	  		\draw[very thick,rarrow ] (p7) to (p1);
	  		\draw[very thick,rarrow ] (p5) to (p6);
	  		\draw[very thick,rarrow ] (p3) to (p4);
	  		\draw[very thick, rarrow] (p4) to (p5);
	  		
	  		\draw[very thick, ] (p1aux1) to (p1);
	  		\draw[very thick, ] (p1aux3) to (p1);
	  		
	  		\draw[very thick, ] (p3aux1) to (p3);
	  		\draw[very thick, ] (p3aux3) to (p3);
	  		
	  		\draw[very thick, ] (p5aux1) to (p5);
	  		\draw[very thick, ] (p5aux3) to (p5);
	  		
	  		\draw[very thick, larrow] (p2aux1) to (p2);
	  		\draw[very thick, larrow] (p4aux1) to (p4);
	  		\draw[very thick, larrow] (p7aux1) to (p7);

        		\draw[fill, thick] (p1) circle (0.2);
        		\draw[fill=white,thick] (p2) circle (0.2);
            \draw[fill, thick] (p3) circle (0.2);
 		    \draw[fill=white, thick] (p6) circle (0.2);
        		\draw[fill=white, thick] (p7) circle (0.2);
        		\draw[fill, thick] (p5) circle (0.2);
        		\draw[fill=white, thick] (p4) circle (0.2);

        		\node at ($(p1aux2)+(180:0.2)$) {$1^-$};
%
%
        		\node at ($(p5aux2)+(-45:0.3)$) {$i^{-}$};
%
        		\node at ($(p4aux1)+(45:0.35)$) {$k$};
        		\node at ($(p2aux1)+(108:0.35)$) {$l$};
        		\node at ($(p7aux1)+(-108:0.35)$) {$j$};
        		
        		\node[rotate=90] at ($(p1)!0.5!(p1aux2)$) {$\dots$};
        		\node[] at ($(p3)!0.8!(p3aux2)$) {...};
        		\node[rotate = 45] at ($(p5)!0.5!(p5aux2)$) {$\dots$};
        
    \end{tikzpicture}}}  + \, 
    \vcenter{\hbox{\begin{tikzpicture}[scale=0.7]
			
			\coordinate (d45) at (45:1);
			\coordinate (dm45) at (-45:1);
			\coordinate (d135) at (135:1);
			\coordinate (dm135) at (-135:1);
			
	  		\coordinate (p1) at (180:1);
	  		\coordinate (p2) at (108:1);
	  		\coordinate (p3) at (36:1);
	  		\coordinate (p6) at (-36:1);
	  		\coordinate (p7) at (-108:1);
	  		
	  		\coordinate (p4) at ($(p3)+(1.2,0)$);
	  		\coordinate (p5) at ($(p6)+(1.2,0)$);
	  		
	  		\coordinate (p1aux1) at ($(p1)+(dm135)$);
	  		\coordinate (p1aux2) at ($(p1)+(-1,0)$);
	  		\coordinate (p1aux3) at ($(p1)+(d135)$);

	  		\coordinate (p3aux1) at ($(p3)+(60:1)$);
	  		\coordinate (p3aux2) at ($(p3)+(90:1)$);
	  		\coordinate (p3aux3) at ($(p3)+(120:1)$);

	  		\coordinate (p5aux1) at ($(p5)+(1,0)$);
	  		\coordinate (p5aux2) at ($(p5)+(dm45)$);
	  		\coordinate (p5aux3) at ($(p5)+(0,-1)$);

	  		\coordinate (p2aux1) at ($(p2)+(108:1)$);
	  		\coordinate (p4aux1) at ($(p4)+(45:1)$);
	  		\coordinate (p7aux1) at ($(p7)+(-108:1)$);

	  		\draw[very thick, larrow] (p1) to (p2);
	  		\draw[very thick,larrow  ] (p2) to (p3);
	  		\draw[very thick, larrow] (p3) to (p6);
	  		\draw[very thick, larrow] (p6) to (p7);
	  		\draw[very thick, larrow] (p7) to (p1);
	  		\draw[very thick, larrow] (p5) to (p6);
	  		\draw[very thick, larrow] (p3) to (p4);
	  		\draw[very thick, larrow] (p4) to (p5);
	  		
	  		\draw[very thick, ] (p1aux1) to (p1);
	  		\draw[very thick, ] (p1aux3) to (p1);
	  		
	  		\draw[very thick, ] (p3aux1) to (p3);
	  		\draw[very thick, ] (p3aux3) to (p3);
	  		
	  		\draw[very thick, ] (p5aux1) to (p5);
	  		\draw[very thick, ] (p5aux3) to (p5);
	  		
	  		\draw[very thick, larrow] (p2aux1) to (p2);
	  		\draw[very thick, larrow] (p4aux1) to (p4);
	  		\draw[very thick, larrow] (p7aux1) to (p7);

        		\draw[fill, thick] (p1) circle (0.2);
        		\draw[fill=white,thick] (p2) circle (0.2);
            \draw[fill, thick] (p3) circle (0.2);
 		    \draw[fill=white, thick] (p6) circle (0.2);
        		\draw[fill=white, thick] (p7) circle (0.2);
        		\draw[fill, thick] (p5) circle (0.2);
        		\draw[fill=white, thick] (p4) circle (0.2);

        		\node at ($(p1aux2)+(180:0.2)$) {$1^-$};
%
%
        		\node at ($(p5aux2)+(-45:0.3)$) {$i^{-}$};
%
        		\node at ($(p4aux1)+(45:0.35)$) {$k$};
        		\node at ($(p2aux1)+(108:0.35)$) {$l$};
        		\node at ($(p7aux1)+(-108:0.35)$) {$j$};
        		
        		\node[rotate=90] at ($(p1)!0.5!(p1aux2)$) {$\dots$};
        		\node[] at ($(p3)!0.8!(p3aux2)$) {...};
        		\node[rotate = 45] at ($(p5)!0.5!(p5aux2)$) {$\dots$};
        
    \end{tikzpicture}}}
\label{eq:pb_onshell}
\end{equation}
Using the maximal cut solutions \eqref{eq:l_pb_sol} to
evaluate the three on-shell diagrams, we finally obtain,
\begin{tcolorbox}
\vspace{-2ex}
\begin{equation}\label{eq:Rpb} 
\begin{aligned}
R_{1i,jkl,n}^{(2),\text{pb}}:=\text{PT}_{n,1i}\times\left[\left(\frac{\langle 1j\rangle\langle ij\rangle\langle kl\rangle}{\langle 1i\rangle\langle jl\rangle\langle jk\rangle}\right)^4{+}\left(\frac{\langle1l\rangle\langle ij\rangle}{\langle 1i\rangle\langle jl\rangle}\right)^4{+}\left(\frac{\langle1j\rangle\langle ik\rangle}{\langle1i\rangle\langle jk\rangle}\right)^4\right].\ \\
1<j<i<k<l\leq n\phantom{a}\cup\phantom{a} 1<l<k<i<j\leq n\phantom{a}
\end{aligned}
\end{equation}
\end{tcolorbox}
\noindent We would like to draw the reader’s attention to the fact that this penta-box diagram is not invariant under reflection. As a result, clockwise and counterclockwise orderings of the external legs give rise to the two allowed ranges of $j,k,l$ in the expression above. However, once the planarity ordering condition on the external legs is imposed, the arrangement of the three legs $j,k,l$ effectively determines the positions of all other external legs. Hence, in what follows, we simply use the ordered triple of indices $jkl$ as subscripts to specify the positions of external legs.

Notably, the sum of three terms inside the square brackets is symmetric under the exchange $1\leftrightarrow i$ and $l\leftrightarrow k$. Therefore, this leading singularity is invariant under reflection symmetry, and it is always given by the sum of two distinct penta-box maximal cuts, i.e.
\begin{align}
R_{1i,jkl,n}^{(2),\text{pb}}=R_{i1,jlk,n}^{(2),\text{pb}}  \,. \label{eq:Rpbsym}  
\end{align}

The double-box on-shell diagrams are more subtle to compute, since the on-shell constraints imposed by their seven propagators do not completely localize the loop momenta. The leading singularities arising from this type of on-shell diagram are called {\it composite}. They originate from poles of the integrand at $\ell\to\infty$ and are 
extracted by choosing integration contours that evaluate the maximal residues in a loop-by-loop manner. We identify three distinct helicity assignments for the internal lines of the double-box that contain exactly two oriented cycles of helicity flow: 
\begin{equation}
	\label{eq:dbonshell}\vcenter{\hbox{\begin{tikzpicture}[scale=0.7]
			
			\coordinate (d45) at (45:1);
			\coordinate (dm45) at (-45:1);
			\coordinate (d135) at (135:1);
			\coordinate (dm135) at (-135:1);
			
	  		\coordinate (p1) at (-1.2,0.6);
	  		\coordinate (p2) at (0,0.6);
	  		\coordinate (p3) at (1.2,0.6);
	  		\coordinate (p4) at (1.2,-0.6);
	  		\coordinate (p5) at (0,-0.6);
	  		\coordinate (p6) at (-1.2,-0.6);
	  		
	  		\coordinate (p1aux1) at ($(p1)+(-1,0)$);
	  		\coordinate (p1aux2) at ($(p1)+(d135)$);
	  		\coordinate (p1aux3) at ($(p1)+(0,1)$);
	  		\coordinate (p3aux3) at ($(p3)+(1,0)$);
	  		\coordinate (p3aux2) at ($(p3)+(d45)$);
	  		\coordinate (p3aux1) at ($(p3)+(0,1)$);
	  		\coordinate (p5aux2) at ($(p5)+(0,-1)$);
	  		\coordinate (p5aux1) at ($(p5)+(-60:1)$);
	  		\coordinate (p5aux3) at ($(p5)+(-120:1)$);
	  		
	  		\coordinate (p2aux) at ($(p2)+(0,1)$);
	  		\coordinate (p4aux) at ($(p4)+(dm45)$);
	  		\coordinate (p6aux) at ($(p6)+(dm135)$);
	  		
	  		\draw[very thick, larrow,purple] (p1) to (p2);
	  		\draw[very thick, larrow] (p2) to (p5);
	  		\draw[very thick, larrow] (p5) to (p6);
	  		\draw[very thick, larrow] (p6) to (p1);
	  		
	  		\draw[very thick,  rarrow] (p2) to (p3);
	  		\draw[very thick, rarrow] (p3) to (p4);
	  		\draw[very thick, rarrow] (p4) to (p5);

        		\draw[very thick] (p1aux1) to (p1);
        		\draw[very thick] (p1aux3) to (p1);
        		
        		\draw[very thick] (p3aux3) to (p3);
        		\draw[very thick] (p3aux1) to (p3);

        		\draw[very thick] (p5aux1) to (p5);
        		\draw[very thick] (p5aux3) to (p5);
        		
        		\draw[very thick,larrow ] (p4aux) to (p4);
        		\draw[very thick,larrow] (p6aux) to (p6);

        		\draw[fill, thick] (p1) circle (0.2);
        		\draw[fill=white,thick] (p2) circle (0.2);
            \draw[fill, thick] (p3) circle (0.2);
 		    \draw[fill=white, thick] (p4) circle (0.2);
        		\draw[fill, thick] (p5) circle (0.2);
        		\draw[fill=white, thick] (p6) circle (0.2);

        		
        		\node at ($(p1aux2)+(135:0.2)$) {$1^-$};
       		\node at ($(p1aux3)+(90:0.4)$) {$l$};
       		\node at ($(p3aux1)+(90:0.4)$) {$l+1$};
        		\node at ($(p3aux2)+(45:0.3)$) {$i^{-}$};
%
%
        		\node at ($(p4aux)+(-45:0.35)$) {$j$};
        		\node at ($(p6aux)+(-135:0.35)$) {$k$};
        		
        		\node[rotate=45] at ($(p1)!0.5!(p1aux2)$) {$\dots$};
        		\node[rotate=-45] at ($(p3)!0.5!(p3aux2)$) {$\dots$};
        		\node[] at ($(p5)!0.8!(p5aux2)$) {...};
        
    \end{tikzpicture}}}
	+\,
	\vcenter{\hbox{\begin{tikzpicture}[scale=0.7]
			
			\coordinate (d45) at (45:1);
			\coordinate (dm45) at (-45:1);
			\coordinate (d135) at (135:1);
			\coordinate (dm135) at (-135:1);
			
	  		\coordinate (p1) at (-1.2,0.6);
	  		\coordinate (p2) at (0,0.6);
	  		\coordinate (p3) at (1.2,0.6);
	  		\coordinate (p4) at (1.2,-0.6);
	  		\coordinate (p5) at (0,-0.6);
	  		\coordinate (p6) at (-1.2,-0.6);
	  		
	  		\coordinate (p1aux1) at ($(p1)+(-1,0)$);
	  		\coordinate (p1aux2) at ($(p1)+(d135)$);
	  		\coordinate (p1aux3) at ($(p1)+(0,1)$);
	  		\coordinate (p3aux3) at ($(p3)+(1,0)$);
	  		\coordinate (p3aux2) at ($(p3)+(d45)$);
	  		\coordinate (p3aux1) at ($(p3)+(0,1)$);
	  		\coordinate (p5aux2) at ($(p5)+(0,-1)$);
	  		\coordinate (p5aux1) at ($(p5)+(-60:1)$);
	  		\coordinate (p5aux3) at ($(p5)+(-120:1)$);
	  		
	  		\coordinate (p2aux) at ($(p2)+(0,1)$);
	  		\coordinate (p4aux) at ($(p4)+(dm45)$);
	  		\coordinate (p6aux) at ($(p6)+(dm135)$);
	  		
	  		\draw[very thick, larrow] (p1) to (p2);
	  		\draw[very thick, rarrow] (p2) to (p5);
	  		\draw[very thick, larrow] (p5) to (p6);
	  		\draw[very thick, larrow] (p6) to (p1);
	  		
	  		\draw[very thick,  larrow] (p2) to (p3);
	  		\draw[very thick, larrow] (p3) to (p4);
	  		\draw[very thick, larrow] (p4) to (p5);
	  		
        		\draw[very thick] (p1aux1) to (p1);
        		\draw[very thick] (p1aux3) to (p1);
        		
        		\draw[very thick] (p3aux3) to (p3);
        		\draw[very thick] (p3aux1) to (p3);

        		\draw[very thick] (p5aux1) to (p5);
        		\draw[very thick] (p5aux3) to (p5);
        		
        		\draw[very thick,larrow ] (p4aux) to (p4);
        		\draw[very thick,larrow] (p6aux) to (p6);

        		\draw[fill, thick] (p1) circle (0.2);
        		\draw[fill=white,thick] (p2) circle (0.2);
            \draw[fill, thick] (p3) circle (0.2);
 		    \draw[fill=white, thick] (p4) circle (0.2);
        		\draw[fill, thick] (p5) circle (0.2);
        		\draw[fill=white, thick] (p6) circle (0.2);

        		
        		\node at ($(p1aux2)+(135:0.2)$) {$1^-$};
       		\node at ($(p1aux3)+(90:0.4)$) {$l$};
       		\node at ($(p3aux1)+(90:0.4)$) {$l+1$};
        		\node at ($(p3aux2)+(45:0.3)$) {$i^{-}$};
%
%
        		\node at ($(p4aux)+(-45:0.35)$) {$j$};
        		\node at ($(p6aux)+(-135:0.35)$) {$k$};
        		
        		\node[rotate=45] at ($(p1)!0.5!(p1aux2)$) {$\dots$};
        		\node[rotate=-45] at ($(p3)!0.5!(p3aux2)$) {$\dots$};
        		\node[] at ($(p5)!0.8!(p5aux2)$) {...};
        
    \end{tikzpicture}}}
	+\,
	\vcenter{\hbox{\begin{tikzpicture}[scale=0.7]
			
			\coordinate (d45) at (45:1);
			\coordinate (dm45) at (-45:1);
			\coordinate (d135) at (135:1);
			\coordinate (dm135) at (-135:1);
			
	  		\coordinate (p1) at (-1.2,0.6);
	  		\coordinate (p2) at (0,0.6);
	  		\coordinate (p3) at (1.2,0.6);
	  		\coordinate (p4) at (1.2,-0.6);
	  		\coordinate (p5) at (0,-0.6);
	  		\coordinate (p6) at (-1.2,-0.6);
	  		
	  		\coordinate (p1aux1) at ($(p1)+(-1,0)$);
	  		\coordinate (p1aux2) at ($(p1)+(d135)$);
	  		\coordinate (p1aux3) at ($(p1)+(0,1)$);
	  		\coordinate (p3aux3) at ($(p3)+(1,0)$);
	  		\coordinate (p3aux2) at ($(p3)+(d45)$);
	  		\coordinate (p3aux1) at ($(p3)+(0,1)$);
	  		\coordinate (p5aux2) at ($(p5)+(0,-1)$);
	  		\coordinate (p5aux1) at ($(p5)+(-60:1)$);
	  		\coordinate (p5aux3) at ($(p5)+(-120:1)$);
	  		
	  		\coordinate (p2aux) at ($(p2)+(0,1)$);
	  		\coordinate (p4aux) at ($(p4)+(dm45)$);
	  		\coordinate (p6aux) at ($(p6)+(dm135)$);
	  		
	  		\draw[very thick, rarrow] (p1) to (p2);
	  		\draw[very thick, rarrow] (p2) to (p5);
	  		\draw[very thick, rarrow] (p5) to (p6);
	  		\draw[very thick, rarrow] (p6) to (p1);
	  		
	  		\draw[very thick,  rarrow] (p2) to (p3);
	  		\draw[very thick, rarrow] (p3) to (p4);
	  		\draw[very thick, rarrow] (p4) to (p5);
	  		
        		\draw[very thick] (p1aux1) to (p1);
        		\draw[very thick] (p1aux3) to (p1);
        		
        		\draw[very thick] (p3aux3) to (p3);
        		\draw[very thick] (p3aux1) to (p3);

        		\draw[very thick] (p5aux1) to (p5);
        		\draw[very thick] (p5aux3) to (p5);
        		
        		\draw[very thick, larrow] (p4aux) to (p4);
        		\draw[very thick,larrow] (p6aux) to (p6);

        		\draw[fill, thick] (p1) circle (0.2);
        		\draw[fill=white,thick] (p2) circle (0.2);
            \draw[fill, thick] (p3) circle (0.2);
 		    \draw[fill=white, thick] (p4) circle (0.2);
        		\draw[fill, thick] (p5) circle (0.2);
        		\draw[fill=white, thick] (p6) circle (0.2);

        		
        		\node at ($(p1aux2)+(135:0.2)$) {$1^-$};
       		\node at ($(p1aux3)+(90:0.4)$) {$l$};
       		\node at ($(p3aux1)+(90:0.4)$) {$l+1$};
        		\node at ($(p3aux2)+(45:0.3)$) {$i^{-}$};
%
%
        		\node at ($(p4aux)+(-45:0.35)$) {$j$};
        		\node at ($(p6aux)+(-135:0.35)$) {$k$};
        		
        		\node[rotate=45] at ($(p1)!0.5!(p1aux2)$) {$\dots$};
        		\node[rotate=-45] at ($(p3)!0.5!(p3aux2)$) {$\dots$};
        		\node[] at ($(p5)!0.8!(p5aux2)$) {...};
        
    \end{tikzpicture}}}.
\end{equation}
Solving the cut conditions for seven on-shell propagators, we obtain the following one-form expression for the sum of the three on-shell diagrams,
\begin{tcolorbox}
\vspace{-2ex}
\begin{align}
    \Omega^{(2) \text{db}}_{1i,jkl,n}= &\text{PT}_{n,1i}\times \frac{\langle l \, l+1 \rangle \langle k \, \ell(z) \rangle^2 \, {\rm d}z }{\langle 1 i \rangle^4 \langle l \, \ell(z) \rangle \langle l+1 \, \ell(z) \rangle }\phantom{aaaaaaaaaaaaaaa} 1\leq l<i<j<k\leq n,  \notag \\
    & \times\biggl[ \langle k j \rangle^4 \left( \frac{ \langle 1 \ell(z) \rangle \langle i \ell(z) \rangle }{\langle j \ell(z) \rangle \langle k \, \ell(z) \rangle  } \right)^4   
    +\langle i j \rangle^4 \left( \frac{ \langle 1 \ell(z) \rangle}{\langle j \ell(z) \rangle} \right)^4
    + \langle 1 k \rangle^4 \left( \frac{\langle i \ell(z) \rangle}{\langle k \, \ell(z) \rangle} \right)^4 \biggr],\label{eq:LSdb}    
\end{align}
\end{tcolorbox}
\noindent where the variable $z$ parametrizes the loop momentum 
\begin{equation}
(\ell(z))^{\alpha\dot{\alpha}} = \left( z  |k \rangle - \frac{P|k]}{\langle k |P| k]} \right)^{\alpha} (\langle k| P)^{\dot{\alpha}},
\end{equation}
flowing through the upper horizontal edge of left box subdiagram (colored in the first diagram)  and $P := p_{k+1} + \ldots + p_{l} $ is the total momentum flowing in the upper-left corner. The genuinely two-loop composite leading singularity is then obtained by taking the residue of this form at $z=\infty$ . The general expression of this residue is too lengthy to present here. We denote the resulting invariant by $R_{1i,jkl,n}^{(2),\text{db}}$ in what follows. At six points, their explicit results can be found in the ancillary files.

In summary, all genuinely two-loop prefactors contributing to the maximally-transcendental part of pure YM MHV hard functions at arbitrary multiplicity fall into three classes: $R^{(2),\text{kb}}$, $R^{(2),\text{pb}}$, and $R^{(2),\text{db}}$. Together with the one-loop prefactors $R^{(1)}_{1i,jk,n}$ \eqref{eq:L1LS} and tree-level $\text{PT}_{n,1i}$ \eqref{eq:PT}, they form a complete basis of prefactors for our computation of pure YM MHV hard functions. All these prefactors involve only angular spinor-helicity brackets and  are therefore manifestly conformally invariant, being annihilated by the conformal boost generator \eqref{eq:Wittenconformaloperater}. In the cases of four- and five-gluon scattering, we have verified that our results are in complete agreement with the prefactors appearing in the corresponding one-loop and two-loop hard functions available in the literature, referring to hard function results in \cite{Jin:2019nya,Abreu:2019odu}.
Moreover, the spinor-helicity representations of the rational prefactors computed in this section provide a substantial simplification compared to the corresponding lower-point expressions written in terms of Mandelstam invariants.

\begin{table}[t]
    \centering
\begin{tabular}{|c|c|c|c|}
\hline
 & $({-}{-}{+}{+}{+}{+})$ & $({-}{+}{-}{+}{+}{+})$ & $({-}{+}{+}{-}{+}{+})$ \\ \hline
$L=1$ & 1 & 4 & 5 \\ \hline
$L=2$ & 7 & 13 & 16 \\ \hline
\end{tabular}
\caption{Counting of independent prefactors for the maximally-transcendental part of the six-point $L$-loop pure YM hard functions in the MHV helicity sectors.}\label{tab:1}
\end{table}

Our previous computations of the prefactors applies to arbitrary multiplicity $n$. We now specialize to the six-point case. By enumerating all possible  distributions of external legs in the on-shell diagrams, we obtain the counting of independent prefactors for the six-point MHV sectors, summarized in Table~\ref{tab:1}. At six points, there are three inequivalent helicity sectors, see~\eqref{eq:config}. There are no genuinely one-loop leading singularities in the sector $({-}{-}{+}{+}{+}{+})$. By contrast, the two remaining sectors contain three and four such leading singularities, respectively, arising from the two-mass-easy box cuts. For the  $({-}{-}{+}{+}{+}{+})$ sector, only double-box on-shell diagrams contribute at two loops, yielding six genuinely two-loop leading singularities,
\begin{equation}\label{eq:Rdb12}
    R_{12,ij1,6}^{(2),{\rm db}}=\text{PT}_{6,12}\times\left[-2 + 
 12  \frac{\langle1i\rangle\langle 2j\rangle}{\langle12\rangle\langle ij\rangle}  -  30 \left(\frac{\langle1i\rangle\langle 2j\rangle}{\langle12\rangle\langle ij\rangle}\right)^2 + 20 \left(\frac{\langle1i\rangle\langle 2j\rangle}{\langle12\rangle\langle ij\rangle}\right)^3 \right],
\end{equation}
with $2<i<j\leq6$. They featured in the two-loop bootstrap of Ref.~\cite{Carrolo:2025agz} \footnote{Note that \eqref{eq:Rdb12} differ from the leading singularities $R_{i,j}$ defined in \cite{Carrolo:2025agz} by a PT term.}. It is worth mentioning that this expression vanishes when the cross ratio $\frac{\ab{1i}\ab{2j}}{\ab{12}\ab{ij}}$ is one. This is equivalent to having $\lambda_1 \sim \lambda_j$ or $\lambda_2 \sim \lambda_i$, which corresponds to a non-planar collinear configuration. It turns out that other double-box prefactors also enjoy this property, see appendix \ref{sec: review of on-shell diagrams}. We leave further exploration of these properties to future work.

\begin{figure}[t]
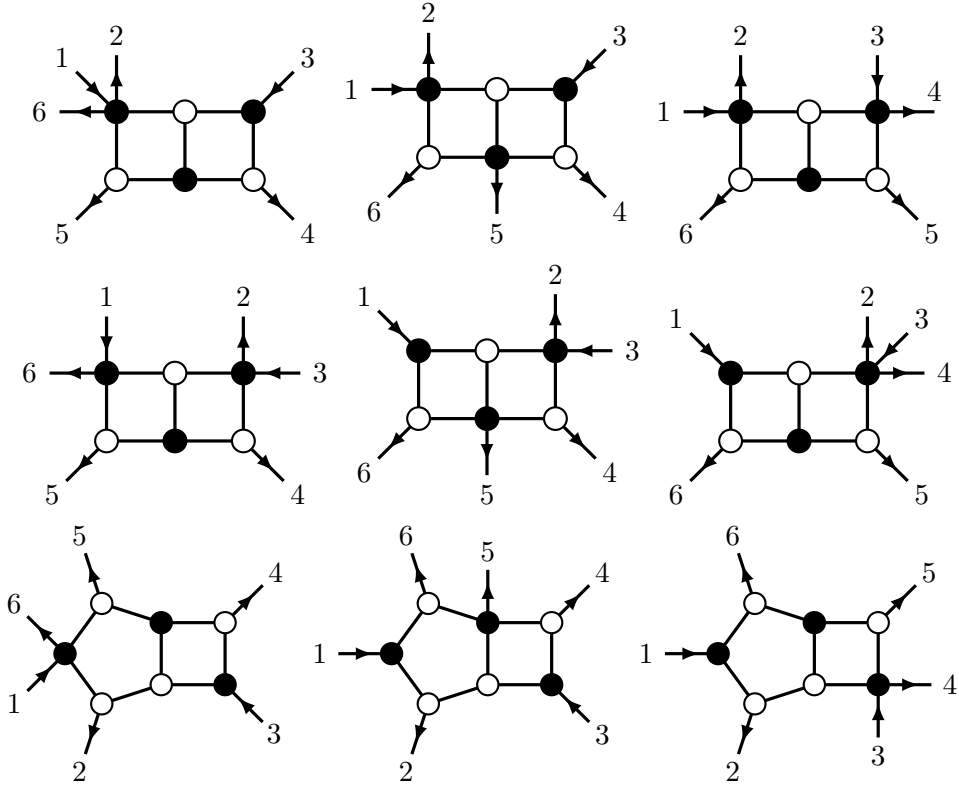

\centering	

    \caption{Two-loop on-shell diagrams for the MHV sector $({-}{+}{-}{+}{+}{+})$.}
    \label{fig:onshell22}
\end{figure}

\begin{figure}[h!]
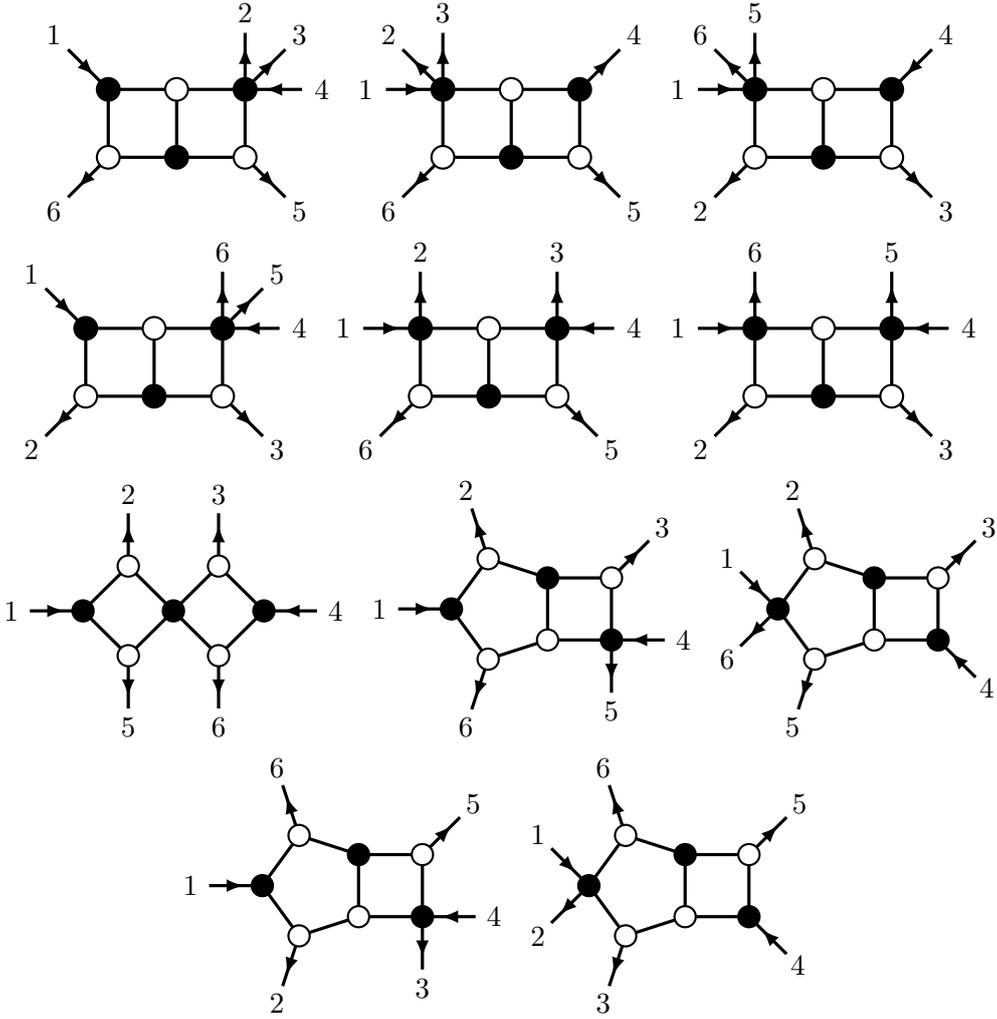

\centering	
%
    \caption{Two-loop on-shell diagrams for the MHV sector $({-}{+}{+}{-}{+}{+})$.}
    \label{fig:onshell33}
\end{figure}

For the remaining two helicity sectors, more types of two-loop on-shell diagrams contribute. For the sector $({-}{+}{-}{+}{+}{+})$, there is one $\text{PT}$-factor, three two-mass-easy box prefactors, six double-box prefactors, and three penta-box prefactors. By contrast, for the  $({-}{+}{+}{-}{+}{+})$ sector, there is one $\text{PT}$-factor, four two-mass-easy box prefactors, six double-box prefactors, one kissing-box prefactors, and four penta-box prefactors.\footnote{Note that for each penta-box on-shell diagrams, its reflection generates the same leading singularity. We only count and show independent ones in Fig.~\ref{fig:onshell22} and Fig.~\ref{fig:onshell33}.}
The corresponding on-shell diagrams are shown in Fig.~\ref{fig:onshell22} and Fig.~\ref{fig:onshell33}.

Note that IR-divergent MHV amplitudes may involve a larger set of prefactors as compared to the IR-finite hard functions. For example, the maximally-transcendental part of the one-loop five-point pure YM amplitude in the split-helicity configuration $(--+++)$ contains an additional prefactor at order $\mathcal{O}(\epsilon)$, see Ref.~\cite{Abreu:2019odu},
\begin{equation}
    R^{(1)}_\star=\text{PT}_{6,12}\times\left( 1{-}\frac{4s_{15}s_{23}s_{34}s_{45}}{\text{tr}_5^2}{+} \frac{2s_{15}^2s_{23}^2s_{34}^2s_{45}^2}{\text{tr}_5^4}\right) \, ,
\end{equation}
where $\tr_5 := 4 \textup{i} \ep_{1234}$.
This amplitude prefactor is not conformally-invariant and therefore cannot be captured by four-dimensional on-shell diagrams. Detecting it would instead require a $D$-dimensional leading singularity analysis, see \cite{Bern:1995db,Bern:1996je,Badger:2008cm,Ellis:2011cr}. Furthermore, the corresponding two-loop five-point amplitude also involves $R^{(1)}_\star$ at order $\mathcal{O}(\epsilon^0)$. However, upon implementing the two-loop IR subtraction~\cite{Catani:1998bh}, one finds that $R^{(1)}_\star$ cancels out from the four-dimensional hard function. Its maximally-transcendental part involves only the conformally-invariant prefactors computed from on-shell diagrams.

\subsection{Two-loop MHV prefactors for QCD amplitudes}
\label{sec:QCDls}

The above computation of leading singularities straightforwardly extends to two-loop MHV amplitudes in QCD, requiring only minor modifications. The sole adjustment is the inclusion of tree-level amplitudes with both gluonic and fermionic legs as vertices in the on-shell diagrams. For instance, the basic building blocks involving a single quark-antiquark pair are given by the following tree-level amplitudes,
\begin{equation}\label{eq:PTfermion}
	\begin{aligned}
		&\vcenter{\hbox{\begin{tikzpicture}[scale=0.8]
        \draw[thick] (0,0) circle (0.2);
        \coordinate (p1) at ($({cos(0)},{sin(0)})$);
        \coordinate (p2) at ($({cos(120)},{sin(120)})$);
        \coordinate (p3) at ($({cos(-120)},{sin(-120)})$);
        \coordinate (p1a) at ($0.2*({cos(0)},{sin(0)})$);
        \coordinate (p2a) at ($0.2*({cos(120)},{sin(120)})$);
        \coordinate (p3a) at ($0.2*({cos(-120)},{sin(-120)})$);
        \node (p1t) at ($1.35*(p1)$){$b^+$};
        \node (p2t) at ($1.35*(p2)$){$c^{-}_q$};
        \node (p3t) at ($1.35*(p3)$){$a^{+}_q$};
        \draw[very thick, rarrow] (p1a.center) to (p1.center);
        \draw[blue, very thick, rarrow] (p2.center) to (p2a.center);
        \draw[blue,very thick, rarrow] (p3a.center) to (p3.center);
    \end{tikzpicture}}} : \bar{\mathcal{A}}_3(a^+_g,b^+_q,c^-_q)=\frac{[ab]^3[ac]}{[ab][bc][ca]}\delta^{2\times2}(P)\\
     &\vcenter{\hbox{\begin{tikzpicture}[rotate=180,scale=0.8]

        \coordinate (p1) at (135:1);
        \coordinate (p2) at (45:1);
        \coordinate (p3) at (-45:1);
        \coordinate (p4) at (-135:1);
        	
        \draw[blue,very thick, rarrow] (p1) to (0,0);
        \draw[ very thick, larrow] (p2) to (0,0);
        \draw[very thick, larrow] (p3) to (0,0);
        	\draw[blue,very thick, larrow] (p4) to (0,0);
        	
        	\draw[fill,thick] (0,0) circle (0.2);

        	\node[rotate=90] at ($(0,0)!0.7!(1,0)$) {$\dots$};
        	
        \node (p1t) at ($1.35*(p1)$){$2^{-}_q$};
        \node (p2t) at ($1.35*(p2)$){$3^+$};
        \node (p3t) at ($1.35*(p3)$){$n^+$};
		\node (p3t) at ($1.35*(p4)$){$1^{+}_q$};
		\node[rotate=90] at ($(0,0)!0.7!(1,0)$) {$\dots$};
		\node at (1.3,0) {$i^-$};
    \end{tikzpicture}}}: \mathcal{A}_n(1_q^+,2_q^-,3_g^+,\cdots,i_g^-,\cdots,n_g^+)=\frac{\langle2i\rangle^3\langle1i\rangle}{\langle12\rangle\cdots\langle n1\rangle}\delta^{2\times2}(P)
	\end{aligned}
\end{equation}
We represent quarks by blue lines, while thick black lines continue to denote gluons, and we also refer to the second expression as PT factors for tree-level QCD amplitudes with a pair of fermionic external legs. For the $N_f^2$ contribution of helicity sector $({-}{+}{+}{-}{+}{+})$, we also need MHV tree-level amplitudes with two quark-antiquark pairs, which take the form

\begin{equation}
	\vcenter{\hbox{\begin{tikzpicture}[scale=0.8]
        \coordinate (p1) at (150:1);
        \coordinate (p2) at (90:1);
        \coordinate (p3) at (30:1);
        	\coordinate (p4) at (-30:1);
        	\coordinate (p5) at (-90:1);
        	\coordinate (p6) at (-150:1);
        	
        \node (p1t) at ($1.35*(p1)$){$n$};
        \node (p2t) at ($1.35*(p2)$){$1^+_q$};
        \node (p3t) at ($1.35*(p3)$){$2^-_q$};
        \node (p4t) at ($1.35*(p4)$){$3^+_Q$};
		\node (p5t) at ($1.35*(p5)$) {$4^-_Q$};
		\node (p6t) at ($1.35*(p6)$) {$5$};

        \draw[very thick, larrow] (p1) to (0,0);
        \draw[very thick, blue,larrow] (p2) to (0,0);
        \draw[very thick, blue,rarrow] (p3) to (0,0);
        \draw[very thick, blue,larrow] (p4) to (0,0);
        \draw[very thick, blue,rarrow] (p5) to (0,0);
        \draw[very thick, larrow] (p6) to (0,0);
        
        	\draw[fill,thick] (0,0) circle (0.2);
        	\node[rotate=90] at (-0.7,0) {$\dots$};

    \end{tikzpicture}}}: \mathcal{A}_n(1_q^+,2_q^-,3_Q^+,4_Q^-,\cdots,n_g^+)=\frac{\langle24\rangle^3\langle13\rangle}{\langle12\rangle\cdots\langle n1\rangle}\delta^{2\times2}(P)
\end{equation}
For simplicity, we do not distinguish between different quark flavors in our diagrammatic notations.
In what follows, we discuss leading singularities arising from $N_f$- and $N_f^2$-contributions, which distinguish QCD from pure YM.

Tree-level gluon amplitudes are independent of quark content, while $N_f$-terms of gluonic amplitudes first arise at one loop. At one loop, the corresponding prefactors originate from non-singlet on-shell diagrams containing a closed cycle of quark lines,
\begin{equation}
	\vcenter{\hbox{\begin{tikzpicture}[scale=0.8]
			
        		\coordinate (p2) at (-0.6,0.6);
        		\coordinate (p3) at (0.6,0.6);
        		\coordinate (p4) at (0.6,-0.6);
        		\coordinate (p1) at (-0.6,-0.6);
        		\coordinate (p1aux) at ($(p1) + (-135:1)$);
        		\coordinate (p3aux) at ($(p3)+(45:1)$);
        		\coordinate (p2aux1) at ($(p2) + (180:1)$);
        		\coordinate (p2aux2) at ($(p2) + (135:1)$);
        		\coordinate (p2aux3) at ($(p2) + (90:1)$);
        		
        		\coordinate (p4aux1) at ($(p4) + (0:1)$);
        		\coordinate (p4aux2) at ($(p4) + (-45:1)$);
        		\coordinate (p4aux3) at ($(p4) + (-90:1)$);
        		
        		\draw[very thick, blue] (p1) to (p2);
			\draw[very thick, blue] (p2) to (p3);
			\draw[very thick, blue] (p3) to (p4);
			\draw[very thick, blue] (p4) to (p1);
			
			\draw[very thick, larrow] (p1aux) to (p1);
			
			\draw[very thick, ]  (p2aux1) to (p2);
			\draw[very thick, ]  (p2aux3) to (p2);

			\draw[very thick, larrow] (p3aux) to (p3);
			
			\draw[very thick, ]  (p4aux1) to (p4);
			\draw[very thick, ]  (p4aux3) to (p4);

			\draw[fill,thick] (p1) circle (0.2);
        		\draw[fill=white, thick] (p2) circle (0.2);
        		\draw[fill, thick] (p3) circle (0.2);
        		\draw[fill=white, thick] (p4) circle (0.2);
        		\node at ($(p1aux) + (-135:0.3)$) {$k$};
        		\node at ($(p2aux2) + (135:0.3)$) {$1^-$};
        		\node at ($(p3aux) + (45:0.3)$) {$j$};
        		\node at ($(p4aux2) + (-45:0.3)$) {$i^-$};
        		\node[rotate=45] at ($(p2)!0.6!(p2aux2)$) {$\dots$};
        		\node[rotate=45] at ($(p4)!0.6!(p4aux2)$) {$\dots$};
    \end{tikzpicture}}} = 
    \vcenter{\hbox{\begin{tikzpicture}[scale=0.8]
			
        		\coordinate (p2) at (-0.6,0.6);
        		\coordinate (p3) at (0.6,0.6);
        		\coordinate (p4) at (0.6,-0.6);
        		\coordinate (p1) at (-0.6,-0.6);
        		\coordinate (p1aux) at ($(p1) + (-135:1)$);
        		\coordinate (p3aux) at ($(p3)+(45:1)$);
        		\coordinate (p2aux1) at ($(p2) + (180:1)$);
        		\coordinate (p2aux2) at ($(p2) + (135:1)$);
        		\coordinate (p2aux3) at ($(p2) + (90:1)$);
        		
        		\coordinate (p4aux1) at ($(p4) + (0:1)$);
        		\coordinate (p4aux2) at ($(p4) + (-45:1)$);
        		\coordinate (p4aux3) at ($(p4) + (-90:1)$);
        		
        		\draw[very thick, blue, rarrow] (p1) to (p2);
			\draw[very thick, blue, rarrow] (p2) to (p3);
			\draw[very thick, blue, rarrow] (p3) to (p4);
			\draw[very thick, blue, rarrow] (p4) to (p1);
			
			\draw[very thick, larrow] (p1aux) to (p1);
			
			\draw[very thick, ]  (p2aux1) to (p2);
			\draw[very thick, ]  (p2aux3) to (p2);

			\draw[very thick, larrow] (p3aux) to (p3);
			
			\draw[very thick, ]  (p4aux1) to (p4);
			\draw[very thick, ]  (p4aux3) to (p4);

			\draw[fill,thick] (p1) circle (0.2);
        		\draw[fill=white, thick] (p2) circle (0.2);
        		\draw[fill, thick] (p3) circle (0.2);
        		\draw[fill=white, thick] (p4) circle (0.2);
        		\node at ($(p1aux) + (-135:0.3)$) {$k$};
        		\node at ($(p2aux2) + (135:0.3)$) {$1^-$};
        		\node at ($(p3aux) + (45:0.3)$) {$j$};
        		\node at ($(p4aux2) + (-45:0.3)$) {$i^-$};
        		\node[rotate=45] at ($(p2)!0.6!(p2aux2)$) {$\dots$};
        		\node[rotate=45] at ($(p4)!0.6!(p4aux2)$) {$\dots$};
    \end{tikzpicture}}} + 
    \vcenter{\hbox{\begin{tikzpicture}[scale=0.8]
			
        		\coordinate (p2) at (-0.6,0.6);
        		\coordinate (p3) at (0.6,0.6);
        		\coordinate (p4) at (0.6,-0.6);
        		\coordinate (p1) at (-0.6,-0.6);
        		\coordinate (p1aux) at ($(p1) + (-135:1)$);
        		\coordinate (p3aux) at ($(p3)+(45:1)$);
        		\coordinate (p2aux1) at ($(p2) + (180:1)$);
        		\coordinate (p2aux2) at ($(p2) + (135:1)$);
        		\coordinate (p2aux3) at ($(p2) + (90:1)$);
        		
        		\coordinate (p4aux1) at ($(p4) + (0:1)$);
        		\coordinate (p4aux2) at ($(p4) + (-45:1)$);
        		\coordinate (p4aux3) at ($(p4) + (-90:1)$);
        		
        		\draw[very thick, blue, larrow] (p1) to (p2);
			\draw[very thick, blue, larrow] (p2) to (p3);
			\draw[very thick, blue, larrow] (p3) to (p4);
			\draw[very thick, blue, larrow] (p4) to (p1);
			
			\draw[very thick, larrow] (p1aux) to (p1);
			
			\draw[very thick, ]  (p2aux1) to (p2);
			\draw[very thick, ]  (p2aux3) to (p2);

			\draw[very thick, larrow] (p3aux) to (p3);
			
			\draw[very thick, ]  (p4aux1) to (p4);
			\draw[very thick, ]  (p4aux3) to (p4);

			\draw[fill,thick] (p1) circle (0.2);
        		\draw[fill=white, thick] (p2) circle (0.2);
        		\draw[fill, thick] (p3) circle (0.2);
        		\draw[fill=white, thick] (p4) circle (0.2);
        		\node at ($(p1aux) + (-135:0.3)$) {$k$};
        		\node at ($(p2aux2) + (135:0.3)$) {$1^-$};
        		\node at ($(p3aux) + (45:0.3)$) {$j$};
        		\node at ($(p4aux2) + (-45:0.3)$) {$i^-$};
        		\node[rotate=45] at ($(p2)!0.6!(p2aux2)$) {$\dots$};
        		\node[rotate=45] at ($(p4)!0.6!(p4aux2)$) {$\dots$};
    \end{tikzpicture}}}.
    \label{eq:box-quark-loop}
\end{equation}
The internal helicity assignment necessarily forms a closed cycle as a consequence of flavor conservation. Accordingly, PT factors are no longer allowed prefactors for $N_f$-terms. In the helicity sector $(--+\cdots)$, no non-singlet on-shell diagram exists. As a result, the $N_f$-term of the corresponding one-loop hard function vanishes, $H_{--++\ldots+}^{[\frac12],(1)}=0$, in agreement with the five-point expressions in Ref.~\cite{Abreu:2019odu}. By contrast, the leading singularities in the sector $(1^-2^+\cdots i^-\cdots n^+)$ are obtained by evaluating the on-shell diagrams in eq.~\eqref{eq:box-quark-loop},
\begin{equation}\label{eq:1LSjk}   S_{1i,jk,n}^{(1)}=\text{PT}_{n,1i}\times\left(\frac{\langle 1j\rangle^3\langle ik\rangle^3\langle 1k\rangle\langle ji\rangle}{\langle 1i\rangle^4\langle jk\rangle^4}{+}\frac{\langle 1k\rangle^3\langle ji\rangle^3\langle 1j\rangle\langle ik\rangle}{\langle 1i\rangle^4\langle jk\rangle^4}\right),
\end{equation}
which contributes in the $N_f$-term of the hard function starting at one loop for $i-1>1$.

Before proceeding to the two-loop analysis in QCD, we clarify a subtle point concerning on-shell diagrams involving quark lines.
Since we work in the planar limit defined by $N_c$, $N_f\to\infty$ and fixed ratio $N_f/N_c$, see \eqref{eq:planar}, special care is required in tracking the color and flavor structure of the on-shell diagrams. To this end, we review the double-line notation to determine which contributions survive in the planar limit and which are subleading.

In QCD, there are $N_f$ quarks transforming in the fundamental representation of $SU(N_c)$, while gluons transform in the adjoint representation. Since all $N_f$ quarks interact identically with the gluons, the theory enjoys an additional global $SU(N_f)$ flavor symmetry. Under this symmetry, quarks transform in the fundamental representation, whereas the gluons are singlets, as they are neutral under flavor transformations. Subtleties of the planar limit arise when, in addition to taking $N_c \rightarrow \infty$, one also considers $N_f \sim N_c$. In this case, it is no longer sufficient to count only the number of loops in a diagram, as the scaling depends on whether the loops are formed by quarks or by gluons. 

To make this explicit, we employ 't Hooft's double-line notation. In this framework, all color structures are expressed in terms of (anti-)fundamental representations, and lines denote index contractions dictated by the Feynman rules. Moreover, since two symmetry groups are involved---$SU(N_c)$ and $SU(N_f)$---we use black and red lines to denote indices in the corresponding (anti-)fundamental representations. With this conventions, the gluon and convention propagators are represented as follows
\begin{equation}
	\vcenter{\hbox{
	\begin{tikzpicture}
		\begin{feynman}
			\node at (-0.7,0) {$a$};
			\node at (0.7,0) {$b$};
			\coordinate (p1) at (-0.5,0);
			 \coordinate (p2) at (0.5,0);
			\diagram{
			 (p1) --[gluon] (p2);
			 };
		\end{feynman}
	\end{tikzpicture}}} \sim \delta_{a,b}  \rightarrow \delta^{i}_k\delta_{j}^l=
	\vcenter{\hbox{\begin{tikzpicture}
		\draw[very thick, larrow] (-0.5,0.2) to (0.5,0.2);
		\draw[very thick, rarrow] (-0.5,-0.2) to (0.5,-0.2);
		\node at (-0.7,0.2) {$i$};
		\node at (-0.7,-0.2) {$j$};
		\node at (0.7,0.2) {$k$};
		\node at (0.7,-0.2) {$l$};
	\end{tikzpicture}}}
	\, , \quad
	\vcenter{\hbox{\begin{tikzpicture}
		\begin{feynman}
			\node at (-0.85,0) {$i$,\textcolor{red}{$m$}};
			\node at (0.85,0) {$k$,\textcolor{red}{$n$}};
			\coordinate (p1) at (-0.5,0);
			\coordinate (p2) at (0.5,0);
			\diagram{
			 	(p1) --[very thick,fermion] (p2);
			 };
		\end{feynman}	
	\end{tikzpicture}}} \sim \delta^i_l \textcolor{red}{\delta_n^m} \rightarrow
	\vcenter{\hbox{\begin{tikzpicture}
		\draw[very thick, larrow] (-0.5,0.2) to (0.5,0.2);
		\draw[red,very thick, rarrow] (-0.5,-0.2) to (0.5,-0.2);
		\node at (-0.7,0.2) {$i$};
		\node at (-0.7,-0.2) {\textcolor{red}{$m$}};
		\node at (0.7,0.2) {$j$};
		\node at (0.7,-0.2) {\textcolor{red}{$n$}};
	\end{tikzpicture}}} \, ,
\end{equation}
where $i,j,k,l$ label indices in the (anti-)fundamental representation of $SU(N_c)$, whereas $m,n$ label indices in the (anti-)fundamental of $SU(N_f)$. Following the same graphical convention, in which index contractions are represented by oriented lines connecting the corresponding indices, the three-point gluon vertex is depicted as
\begin{equation}
	\vcenter{\hbox{\begin{tikzpicture}
		\begin{feynman}
			\node at (120:1.3) {$a$};
			\node at (0:1.3) {$b$};
			\node at (-120:1.3) {$c$};
			\coordinate (p12) at (120:1);
			\coordinate (p23) at (0:1);
			\coordinate (p31) at (-120:1);
			\coordinate (O) at (0,0);
			\diagram{
				(p12) --[gluon] (O);
				(p23) --[gluon] (O);
				(p31) --[gluon] (O);
			};
		\end{feynman}
	\end{tikzpicture}}} \sim f_{a,b,c} \sim \Tr{T_a T_b T_c} = (T_a)^i_j(T_b)^j_k(T_c)^k_l \rightarrow 
		\vcenter{\hbox{\begin{tikzpicture}
		\coordinate (p12) at (120:1);
		\coordinate (p23) at (0:1);
		\coordinate (p31) at (-120:1);
		\coordinate (O) at (0,0);
		
		\coordinate (r21) at ($(p12)+(30:0.25)$);
		\coordinate (r12) at ($(p12)-(30:0.25)$);
		
		\coordinate (r22) at ($(p23)+(90:0.25)$);
		\coordinate (r31) at ($(p23)-(90:0.25)$);
		
		\coordinate (r32) at ($(p31)+(-30:0.25)$);
		\coordinate (r11) at ($(p31)-(-30:0.25)$);
		
		\coordinate (r1c) at (180:0.29);
		\coordinate (r2c) at (60:0.29);
		\coordinate (r3c) at (-60:0.29);
		
		\draw[very thick,rarrow] (r11) to (r1c.center);
		\draw[very thick, rarrow] (r1c.center) to (r12);
		
		\draw[very thick,rarrow] (r21) to (r2c.center);
		\draw[very thick, rarrow] (r2c.center) to (r22);
		
		\draw[very thick,rarrow] (r31) to (r3c.center);
		\draw[very thick, rarrow] (r3c.center) to (r32);

		\node at ($(r11)+0.3*(p31)$) {$j$};
		\node at ($(r12)+0.3*(p12)$) {$j$};
		
		\node at ($(r21)+0.3*(p12)$) {$i$};
		\node at ($(r22)+0.3*(p23)$) {$i$};
		
		\node at ($(r31)+0.3*(p23)$) {$k$};
		\node at ($(r32)+0.3*(p31)$) {$k$};
	\end{tikzpicture}}}  \, ,
\end{equation}
and the three-point quark-gluon interaction vertex as
\begin{equation}
	 \vcenter{\hbox{\begin{tikzpicture}
		\begin{feynman}
			\node at (120:1.3) {$i,$\textcolor{red}{$m$}};
			\node at (0:1.3) {$a$};
			\node at (-120:1.3) {$j,$\textcolor{red}{$n$}};
			\coordinate (p12) at (120:1);
			\coordinate (p23) at (0:1);
			\coordinate (p31) at (-120:1);
			\coordinate (O) at (0,0);
			\diagram{
				(p12) --[thick, fermion] (O);
				(p23) --[gluon] (O);
				(p31) --[thick, anti fermion] (O);
			};
		\end{feynman}
	\end{tikzpicture}}} \sim (T_a)^i_j \textcolor{red}{\delta^m_n} \rightarrow 
		\vcenter{\hbox{\begin{tikzpicture}
		\coordinate (p12) at (120:1);
		\coordinate (p23) at (0:1);
		\coordinate (p31) at (-120:1);
		\coordinate (O) at (0,0);
		
		\coordinate (r21) at ($(p12)+(30:0.25)$);
		\coordinate (r12) at ($(p12)-(30:0.25)$);
		
		\coordinate (r22) at ($(p23)+(90:0.25)$);
		\coordinate (r31) at ($(p23)-(90:0.25)$);
		
		\coordinate (r32) at ($(p31)+(-30:0.25)$);
		\coordinate (r11) at ($(p31)-(-30:0.25)$);
		
		\coordinate (r1c) at (180:0.29);
		\coordinate (r2c) at (60:0.29);
		\coordinate (r3c) at (-60:0.29);
		
		\draw[very thick,rarrow,red] (r11) to (r1c.center);
		\draw[very thick, rarrow,red] (r1c.center) to (r12);
		
		\draw[very thick,rarrow] (r21) to (r2c.center);
		\draw[very thick, rarrow] (r2c.center) to (r22);
		
		\draw[very thick,rarrow] (r31) to (r3c.center);
		\draw[very thick, rarrow] (r3c.center) to (r32);

		\node at ($(r11)+0.3*(p31)$) {\textcolor{red}{$m$}};
		\node at ($(r12)+0.3*(p12)$) {\textcolor{red}{$n$}};
		
		\node at ($(r21)+0.3*(p12)$) {$i$};
		\node at ($(r22)+0.3*(p23)$) {$i$};
		
		\node at ($(r31)+0.3*(p23)$) {$j$};
		\node at ($(r32)+0.3*(p31)$) {$j$};
	\end{tikzpicture}}}  \, ,
\end{equation}
where the contraction of $SU(N_f)$ indices is explicitly indicated by a red line.

Let us now examine the behavior of the following two Feynman diagrams in the planar limit,
\begin{equation}
	\begin{tikzpicture}[scale=1]
		\begin{feynman}
		\coordinate (p1) at (-1,0.5);
		\coordinate (p2) at (0,0.5);
		\coordinate (p3) at (1,0.5);
		\coordinate (p4) at (1,-0.5);
		\coordinate (p5) at (0,-0.5);
		\coordinate (p6) at (-1,-0.5);
		
		\coordinate (p1a) at ($(p1)+(135:1)$);
		\coordinate (p3a) at ($(p3)+(45:1)$);
		\coordinate (p4a) at ($(p4)+(-45:1)$);		
		\coordinate (p6a) at ($(p6)+(-135:1)$);
		\diagram{
		(p1a) --[gluon] (p1);
		(p3a) --[gluon] (p3);
		(p4a) --[gluon] (p4);
		(p6a) --[gluon] (p6);
		(p1) --[fermion,thick] (p2) --[fermion,thick] (p5) --[fermion,thick] (p6) --[fermion,thick] (p1);
		(p2) --[gluon] (p3) --[gluon] (p4) --[gluon] (p5); 
		};
		\node at ($(p1a)+(135:0.3)$) {$1$};		
		\node at ($(p3a)+(45:0.3)$) {$2$};
		\node at ($(p4a)+(-45:0.3)$) {$3$};
		\node at ($(p6a)+(-135:0.3)$) {$4$};
		\end{feynman}
	\end{tikzpicture}
	\, ,\quad
	\begin{tikzpicture}[scale=1]
		\begin{feynman}
		\coordinate (p1) at (-1,0.5);
		\coordinate (p2) at (0,0.5);
		\coordinate (p3) at (1,0.5);
		\coordinate (p4) at (1,-0.5);
		\coordinate (p5) at (0,-0.5);
		\coordinate (p6) at (-1,-0.5);
		
		\coordinate (p1a) at ($(p1)+(135:1)$);
		\coordinate (p3a) at ($(p3)+(45:1)$);
		\coordinate (p4a) at ($(p4)+(-45:1)$);		
		\coordinate (p6a) at ($(p6)+(-135:1)$);
		\diagram{
		(p1a) --[gluon] (p1);
		(p3a) --[gluon] (p3);
		(p4a) --[gluon] (p4);
		(p6a) --[gluon] (p6);
		(p1) --[fermion,thick] (p2) --[fermion,thick] (p3) --[fermion,thick] (p4) --[fermion,thick] (p5) --[fermion,thick] (p6) --[fermion,thick] (p1); 
		(p2) --[gluon] (p5) ; 
		};
		\node at ($(p1a)+(135:0.3)$) {$1$};		
		\node at ($(p3a)+(45:0.3)$) {$2$};
		\node at ($(p4a)+(-45:0.3)$) {$3$};
		\node at ($(p6a)+(-135:0.3)$) {$4$};
		\end{feynman}
	\end{tikzpicture} \, .
\end{equation}
A priori, it is not obvious whether they both contribute at the same order in the planar limit. However, by employing the double-line notation, the diagrams can be rewritten as
\begin{equation}
	\vcenter{\hbox{\begin{tikzpicture}[scale=1]
		\coordinate (p1) at (-1,0.5);
		\coordinate (p2) at (0,0.5);
		\coordinate (p3) at (1,0.5);
		\coordinate (p4) at (1,-0.5);
		\coordinate (p5) at (0,-0.5);
		\coordinate (p6) at (-1,-0.5);
		
		\coordinate (p1a) at ($(p1)+(135:1)$);
		\coordinate (p3a) at ($(p3)+(45:1)$);
		\coordinate (p4a) at ($(p4)+(-45:1)$);		
		\coordinate (p6a) at ($(p6)+(-135:1)$);
		
		\coordinate (l12) at (-0.2,0.3);
		\coordinate (l13) at (-0.2,-0.3);
		\coordinate (l11) at (-0.8,0.3);
		\coordinate (l14) at (-0.8,-0.3);
		\draw[red,very thick] (l11) to (l12) to (l13) to (l14) to (l11);
		\draw[red,very thick, rarrow] (l11) to (l12);
		\draw[red,very thick, rarrow] (l12) to (l13);
		\draw[red,very thick, rarrow] (l13) to (l14);
		\draw[red,very thick, rarrow] (l14) to (l11);

		\coordinate (l21) at (0.2,0.3);
		\coordinate (l24) at (0.2,-0.3);
		\coordinate (l22) at (0.8,0.3);
		\coordinate (l23) at (0.8,-0.3);
		\draw[very thick, rarrow] (l21) to (l22);
		\draw[very thick, rarrow] (l22) to (l23);
		\draw[very thick, rarrow] (l23) to (l24);
		\draw[very thick, rarrow] (l24) to (l21);
		\draw[very thick] (l21) to (l22) to (l23) to (l24) to (l21);
		\coordinate (r21) at ($(p1a)+(45:0.2)$);
		\coordinate (r22) at ($(p1) + (45:0.2)$);
		\coordinate (r23) at (-0.2,0.64);
		\coordinate (r24) at (0.2,0.64);
		\coordinate (r25) at ($(p3)+(135:0.2)$);
		\coordinate (r26) at ($(p3a)+(135:0.2)$);
		\draw[very thick] (r21) to (r22) to (r23) to (r24) to (r25) to (r26);
		\draw[very thick,rarrow] (r23) to (r24);
		\coordinate (r41) at ($(p4a)+(-135:0.2)$);
		\coordinate (r42) at ($(p4) + (-135:0.2)$);
		\coordinate (r43) at (0.2,-0.64);
		\coordinate (r44) at (-0.2,-0.64);
		\coordinate (r45) at ($(p6)+(-45:0.2)$);
		\coordinate (r46) at ($(p6a)+(-45:0.2)$);
		\draw[very thick] (r41) to (r42) to (r43) to (r44) to (r45) to (r46);
		\draw[very thick, rarrow] (r43) to (r44);

		\coordinate (r11) at ($(p6a) + (135:0.2)$);
		\coordinate (r12) at ($(p6) + (135:0.2)$);
		\coordinate (r13) at ($(p1)+(-135:0.2)$);
		\coordinate (r14) at ($(p1a)+(-135:0.2)$);
		\draw[very thick] (r11) to (r12) to (r13) to (r14);
		\draw[very thick, rarrow] (r12) to (r13);
		\coordinate (r31) at ($(p3a) + (-45:0.2)$);
		\coordinate (r32) at ($(p3) + (-45:0.2)$);
		\coordinate (r33) at ($(p4)+(45:0.2)$);
		\coordinate (r34) at ($(p4a)+(45:0.2)$);
		\draw[very thick] (r31) to (r32) to (r33) to (r34);
		\draw[very thick, rarrow] (r32) to (r33);

		\node[red] at (-0.5,0) {\small $N_f$};
		\node at (0.5,0) {\small $N_c$};
		\node at ($(p1a)+(135:0.3)$) {$1$};		
		\node at ($(p3a)+(45:0.3)$) {$2$};
		\node at ($(p4a)+(-45:0.3)$) {$3$};
		\node at ($(p6a)+(-135:0.3)$) {$4$};

		

	\end{tikzpicture}}} \, , \quad
	\vcenter{\hbox{
	\begin{tikzpicture}[scale=1]
		\coordinate (p1) at (-1,0.5);
		\coordinate (p2) at (0,0.5);
		\coordinate (p3) at (1,0.5);
		\coordinate (p4) at (1,-0.5);
		\coordinate (p5) at (0,-0.5);
		\coordinate (p6) at (-1,-0.5);
		
		\coordinate (p1a) at ($(p1)+(135:1)$);
		\coordinate (p3a) at ($(p3)+(45:1)$);
		\coordinate (p4a) at ($(p4)+(-45:1)$);		
		\coordinate (p6a) at ($(p6)+(-135:1)$);
		
		\coordinate (l12) at (-0.2,0.3);
		\coordinate (l13) at (-0.2,-0.3);
		\coordinate (l11) at (-0.8,0.3);
		\coordinate (l14) at (-0.8,-0.3);
		\draw[red,very thick, rarrow] (l11) to (l12);
		\draw[very thick, rarrow] (l12) to (l13);
		\draw[red,very thick, rarrow] (l13) to (l14);
		\draw[red,very thick, rarrow] (l14) to (l11);

		\coordinate (l21) at (0.2,0.3);
		\coordinate (l24) at (0.2,-0.3);
		\coordinate (l22) at (0.8,0.3);
		\coordinate (l23) at (0.8,-0.3);
		\draw[red,very thick, rarrow] (l21) to (l22);
		\draw[red,very thick, rarrow] (l22) to (l23);
		\draw[red,very thick, rarrow] (l23) to (l24);
		\draw[very thick, rarrow] (l24) to (l21);
		
		\draw[very thick,red] (l11) to (l12) to (l21) to (l22) to (l23) to (l24) to (l13) to (l14) to (l11);
		\coordinate (r21) at ($(p1a)+(45:0.2)$);
		\coordinate (r22) at ($(p1) + (45:0.2)$);
		\coordinate (r23) at (-0.2,0.64);
		\coordinate (r24) at (0.2,0.64);
		\coordinate (r25) at ($(p3)+(135:0.2)$);
		\coordinate (r26) at ($(p3a)+(135:0.2)$);
		\coordinate (r41) at ($(p4a)+(-135:0.2)$);
		\coordinate (r42) at ($(p4) + (-135:0.2)$);
		\coordinate (r43) at (0.2,-0.64);
		\coordinate (r44) at (-0.2,-0.64);
		\coordinate (r45) at ($(p6)+(-45:0.2)$);
		\coordinate (r46) at ($(p6a)+(-45:0.2)$);
		\draw[very thick] (r41) to (r42) to (r43) to (r24) to (r25) to (r26);
		\draw[very thick] (r21) to (r22) to (r23) to (r44) to (r45) to (r46);

		\coordinate (r11) at ($(p6a) + (135:0.2)$);
		\coordinate (r12) at ($(p6) + (135:0.2)$);
		\coordinate (r13) at ($(p1)+(-135:0.2)$);
		\coordinate (r14) at ($(p1a)+(-135:0.2)$);
		\draw[very thick] (r11) to (r12) to (r13) to (r14);
		\draw[very thick, rarrow] (r12) to (r13);
		\coordinate (r31) at ($(p3a) + (-45:0.2)$);
		\coordinate (r32) at ($(p3) + (-45:0.2)$);
		\coordinate (r33) at ($(p4)+(45:0.2)$);
		\coordinate (r34) at ($(p4a)+(45:0.2)$);
		\draw[very thick] (r31) to (r32) to (r33) to (r34);
		\draw[very thick, rarrow] (r32) to (r33);

		\node[red] at (0.5,0) {\small $N_f$};
		\node at ($(p1a)+(135:0.3)$) {$1$};		
		\node at ($(p3a)+(45:0.3)$) {$2$};
		\node at ($(p4a)+(-45:0.3)$) {$3$};
		\node at ($(p6a)+(-135:0.3)$) {$4$};
		

	\end{tikzpicture}
	}}\, ,
\end{equation}
and it becomes manifest that the first diagram contributes at leading order in the planar limit, while the second is subleading. This distinction is a crucial difference between QCD and $\mathcal{N}=4$ sYM, where fermions live in the adjoint representation. From the perspective of leading singularities, which correspond precisely to the maximal cuts of the diagrams shown above, we therefore conclude that in the planar limit only double-box  cuts containing a small quark loop contribute. Their contribution at order $\mathcal{O}(N_f N_c)$ is given by the sum of the following four terms,

\begin{equation}
\begin{aligned}
	\vcenter{\hbox{\begin{tikzpicture}[scale=0.7]
			
			\coordinate (d45) at (45:1);
			\coordinate (dm45) at (-45:1);
			\coordinate (d135) at (135:1);
			\coordinate (dm135) at (-135:1);
			
	  		\coordinate (p1) at (-1.2,0.6);
	  		\coordinate (p2) at (0,0.6);
	  		\coordinate (p3) at (1.2,0.6);
	  		\coordinate (p4) at (1.2,-0.6);
	  		\coordinate (p5) at (0,-0.6);
	  		\coordinate (p6) at (-1.2,-0.6);
	  		
	  		\coordinate (p1aux1) at ($(p1)+(-1,0)$);
	  		\coordinate (p1aux2) at ($(p1)+(d135)$);
	  		\coordinate (p1aux3) at ($(p1)+(0,1)$);
	  		\coordinate (p3aux3) at ($(p3)+(1,0)$);
	  		\coordinate (p3aux2) at ($(p3)+(d45)$);
	  		\coordinate (p3aux1) at ($(p3)+(0,1)$);
	  		\coordinate (p5aux2) at ($(p5)+(0,-1)$);
	  		\coordinate (p5aux1) at ($(p5)+(-60:1)$);
	  		\coordinate (p5aux3) at ($(p5)+(-120:1)$);
	  		
	  		\coordinate (p2aux) at ($(p2)+(0,1)$);
	  		\coordinate (p4aux) at ($(p4)+(dm45)$);
	  		\coordinate (p6aux) at ($(p6)+(dm135)$);
	  		
	  		\draw[very thick, blue, larrow] (p1) to (p2);
	  		\draw[very thick,  rarrow] (p2) to (p3);
	  		\draw[very thick, rarrow] (p3) to (p4);
	  		\draw[very thick, rarrow] (p4) to (p5);
	  		\draw[very thick, blue,larrow] (p5) to (p6);
	  		\draw[very thick, blue,larrow] (p6) to (p1);
	  		\draw[very thick, blue,larrow] (p2) to (p5);
	  		
        		\draw[very thick] (p1aux1) to (p1);
        		\draw[very thick,rarrow] (p1aux3) to (p1);
        		
        		\draw[very thick,rarrow] (p3aux1) to (p3);
        		\draw[very thick] (p3aux3) to (p3);

        		\draw[very thick] (p5aux1) to (p5);
        		\draw[very thick] (p5aux3) to (p5);
        		
        		\draw[very thick,larrow ] (p4aux) to (p4);
        		\draw[very thick,larrow] (p6aux) to (p6);

        		\draw[fill, thick] (p1) circle (0.2);
        		\draw[fill=white,thick] (p2) circle (0.2);
            \draw[fill, thick] (p3) circle (0.2);
 		    \draw[fill=white, thick] (p4) circle (0.2);
        		\draw[fill, thick] (p5) circle (0.2);
        		\draw[fill=white, thick] (p6) circle (0.2);

        		
        		\node at ($(p1aux3)+(90:0.35)$) {$1$};
        		\node at ($(p3aux1)+(90:0.35)$) {$2$};
%
%
        		\node at ($(p4aux)+(-45:0.35)$) {$i$};
        		\node at ($(p6aux)+(-135:0.35)$) {$j$};
        		
        		\node[rotate=45] at ($(p1)!0.5!(p1aux2)$) {$\dots$};
        		\node[rotate=-45] at ($(p3)!0.5!(p3aux2)$) {$\dots$};
        		\node[] at ($(p5)!0.8!(p5aux2)$) {...};
        
    \end{tikzpicture}}} + 
    \vcenter{\hbox{\begin{tikzpicture}[scale=0.7]
			
			\coordinate (d45) at (45:1);
			\coordinate (dm45) at (-45:1);
			\coordinate (d135) at (135:1);
			\coordinate (dm135) at (-135:1);
			
	  		\coordinate (p1) at (-1.2,0.6);
	  		\coordinate (p2) at (0,0.6);
	  		\coordinate (p3) at (1.2,0.6);
	  		\coordinate (p4) at (1.2,-0.6);
	  		\coordinate (p5) at (0,-0.6);
	  		\coordinate (p6) at (-1.2,-0.6);
	  		
	  		\coordinate (p1aux1) at ($(p1)+(-1,0)$);
	  		\coordinate (p1aux2) at ($(p1)+(d135)$);
	  		\coordinate (p1aux3) at ($(p1)+(0,1)$);
	  		\coordinate (p3aux3) at ($(p3)+(1,0)$);
	  		\coordinate (p3aux2) at ($(p3)+(d45)$);
	  		\coordinate (p3aux1) at ($(p3)+(0,1)$);
	  		\coordinate (p5aux2) at ($(p5)+(0,-1)$);
	  		\coordinate (p5aux1) at ($(p5)+(-60:1)$);
	  		\coordinate (p5aux3) at ($(p5)+(-120:1)$);
	  		
	  		\coordinate (p2aux) at ($(p2)+(0,1)$);
	  		\coordinate (p4aux) at ($(p4)+(dm45)$);
	  		\coordinate (p6aux) at ($(p6)+(dm135)$);
	  		
	  		\draw[very thick, larrow] (p1) to (p2);
	  		\draw[very thick,  blue,rarrow] (p2) to (p3);
	  		\draw[very thick, blue,rarrow] (p3) to (p4);
	  		\draw[very thick, blue,rarrow] (p4) to (p5);
	  		\draw[very thick, larrow] (p5) to (p6);
	  		\draw[very thick, larrow] (p6) to (p1);
	  		\draw[very thick, blue,larrow] (p2) to (p5);
	  		
        		\draw[very thick] (p1aux1) to (p1);
        		\draw[very thick,rarrow] (p1aux3) to (p1);
        		
        		\draw[very thick,rarrow] (p3aux1) to (p3);
        		\draw[very thick] (p3aux3) to (p3);

        		\draw[very thick] (p5aux1) to (p5);
        		\draw[very thick] (p5aux3) to (p5);
        		
        		\draw[very thick,larrow ] (p4aux) to (p4);
        		\draw[very thick,larrow] (p6aux) to (p6);

        		\draw[fill, thick] (p1) circle (0.2);
        		\draw[fill=white,thick] (p2) circle (0.2);
            \draw[fill, thick] (p3) circle (0.2);
 		    \draw[fill=white, thick] (p4) circle (0.2);
        		\draw[fill, thick] (p5) circle (0.2);
        		\draw[fill=white, thick] (p6) circle (0.2);

        		
        		\node at ($(p1aux3)+(90:0.35)$) {$1$};
        		\node at ($(p3aux1)+(90:0.35)$) {$2$};
%
%
        		\node at ($(p4aux)+(-45:0.35)$) {$i$};
        		\node at ($(p6aux)+(-135:0.35)$) {$j$};
        		
        		\node[rotate=45] at ($(p1)!0.5!(p1aux2)$) {$\dots$};
        		\node[rotate=-45] at ($(p3)!0.5!(p3aux2)$) {$\dots$};
        		\node[] at ($(p5)!0.8!(p5aux2)$) {...};
        
    \end{tikzpicture}}}  \\
    \vcenter{\hbox{\begin{tikzpicture}[scale=0.7]
			
			\coordinate (d45) at (45:1);
			\coordinate (dm45) at (-45:1);
			\coordinate (d135) at (135:1);
			\coordinate (dm135) at (-135:1);
			
	  		\coordinate (p1) at (-1.2,0.6);
	  		\coordinate (p2) at (0,0.6);
	  		\coordinate (p3) at (1.2,0.6);
	  		\coordinate (p4) at (1.2,-0.6);
	  		\coordinate (p5) at (0,-0.6);
	  		\coordinate (p6) at (-1.2,-0.6);
	  		
	  		\coordinate (p1aux1) at ($(p1)+(-1,0)$);
	  		\coordinate (p1aux2) at ($(p1)+(d135)$);
	  		\coordinate (p1aux3) at ($(p1)+(0,1)$);
	  		\coordinate (p3aux3) at ($(p3)+(1,0)$);
	  		\coordinate (p3aux2) at ($(p3)+(d45)$);
	  		\coordinate (p3aux1) at ($(p3)+(0,1)$);
	  		\coordinate (p5aux2) at ($(p5)+(0,-1)$);
	  		\coordinate (p5aux1) at ($(p5)+(-60:1)$);
	  		\coordinate (p5aux3) at ($(p5)+(-120:1)$);
	  		
	  		\coordinate (p2aux) at ($(p2)+(0,1)$);
	  		\coordinate (p4aux) at ($(p4)+(dm45)$);
	  		\coordinate (p6aux) at ($(p6)+(dm135)$);
	  		
	  		\draw[very thick, larrow] (p1) to (p2);
	  		\draw[very thick,  blue,larrow] (p2) to (p3);
	  		\draw[very thick, blue,larrow] (p3) to (p4);
	  		\draw[very thick, blue,larrow] (p4) to (p5);
	  		\draw[very thick, larrow] (p5) to (p6);
	  		\draw[very thick, larrow] (p6) to (p1);
	  		\draw[very thick, blue,rarrow] (p2) to (p5);
	  		
        		\draw[very thick] (p1aux1) to (p1);
        		\draw[very thick,rarrow] (p1aux3) to (p1);
        		
        		\draw[very thick,rarrow] (p3aux1) to (p3);
        		\draw[very thick] (p3aux3) to (p3);

        		\draw[very thick] (p5aux1) to (p5);
        		\draw[very thick] (p5aux3) to (p5);
        		
        		\draw[very thick,larrow ] (p4aux) to (p4);
        		\draw[very thick,larrow] (p6aux) to (p6);

        		\draw[fill, thick] (p1) circle (0.2);
        		\draw[fill=white,thick] (p2) circle (0.2);
            \draw[fill, thick] (p3) circle (0.2);
 		    \draw[fill=white, thick] (p4) circle (0.2);
        		\draw[fill, thick] (p5) circle (0.2);
        		\draw[fill=white, thick] (p6) circle (0.2);

        		
        		\node at ($(p1aux3)+(90:0.35)$) {$1$};
        		\node at ($(p3aux1)+(90:0.35)$) {$2$};
%
%
        		\node at ($(p4aux)+(-45:0.35)$) {$i$};
        		\node at ($(p6aux)+(-135:0.35)$) {$j$};
        		
        		\node[rotate=45] at ($(p1)!0.5!(p1aux2)$) {$\dots$};
        		\node[rotate=-45] at ($(p3)!0.5!(p3aux2)$) {$\dots$};
        		\node[] at ($(p5)!0.8!(p5aux2)$) {...};
        
    \end{tikzpicture}}}  +
    \vcenter{\hbox{\begin{tikzpicture}[scale=0.7]
			
			\coordinate (d45) at (45:1);
			\coordinate (dm45) at (-45:1);
			\coordinate (d135) at (135:1);
			\coordinate (dm135) at (-135:1);
			
	  		\coordinate (p1) at (-1.2,0.6);
	  		\coordinate (p2) at (0,0.6);
	  		\coordinate (p3) at (1.2,0.6);
	  		\coordinate (p4) at (1.2,-0.6);
	  		\coordinate (p5) at (0,-0.6);
	  		\coordinate (p6) at (-1.2,-0.6);
	  		
	  		\coordinate (p1aux1) at ($(p1)+(-1,0)$);
	  		\coordinate (p1aux2) at ($(p1)+(d135)$);
	  		\coordinate (p1aux3) at ($(p1)+(0,1)$);
	  		\coordinate (p3aux3) at ($(p3)+(1,0)$);
	  		\coordinate (p3aux2) at ($(p3)+(d45)$);
	  		\coordinate (p3aux1) at ($(p3)+(0,1)$);
	  		\coordinate (p5aux2) at ($(p5)+(0,-1)$);
	  		\coordinate (p5aux1) at ($(p5)+(-60:1)$);
	  		\coordinate (p5aux3) at ($(p5)+(-120:1)$);
	  		
	  		\coordinate (p2aux) at ($(p2)+(0,1)$);
	  		\coordinate (p4aux) at ($(p4)+(dm45)$);
	  		\coordinate (p6aux) at ($(p6)+(dm135)$);
	  		
	  		\draw[very thick, blue, rarrow] (p1) to (p2);
	  		\draw[very thick,  rarrow] (p2) to (p3);
	  		\draw[very thick, rarrow] (p3) to (p4);
	  		\draw[very thick, rarrow] (p4) to (p5);
	  		\draw[very thick, blue,rarrow] (p5) to (p6);
	  		\draw[very thick, blue,rarrow] (p6) to (p1);
	  		\draw[very thick, blue,rarrow] (p2) to (p5);
	  		
        		\draw[very thick] (p1aux1) to (p1);
        		\draw[very thick,rarrow] (p1aux3) to (p1);
        		
        		\draw[very thick,rarrow] (p3aux1) to (p3);
        		\draw[very thick] (p3aux3) to (p3);

        		\draw[very thick] (p5aux1) to (p5);
        		\draw[very thick] (p5aux3) to (p5);
        		
        		\draw[very thick,larrow ] (p4aux) to (p4);
        		\draw[very thick,larrow] (p6aux) to (p6);

        		\draw[fill, thick] (p1) circle (0.2);
        		\draw[fill=white,thick] (p2) circle (0.2);
            \draw[fill, thick] (p3) circle (0.2);
 		    \draw[fill=white, thick] (p4) circle (0.2);
        		\draw[fill, thick] (p5) circle (0.2);
        		\draw[fill=white, thick] (p6) circle (0.2);

        		
        		\node at ($(p1aux3)+(90:0.35)$) {$1$};
        		\node at ($(p3aux1)+(90:0.35)$) {$2$};
%
%
        		\node at ($(p4aux)+(-45:0.35)$) {$i$};
        		\node at ($(p6aux)+(-135:0.35)$) {$j$};
        		
        		\node[rotate=45] at ($(p1)!0.5!(p1aux2)$) {$\dots$};
        		\node[rotate=-45] at ($(p3)!0.5!(p3aux2)$) {$\dots$};
        		\node[] at ($(p5)!0.8!(p5aux2)$) {...};
        
    \end{tikzpicture}}}.  \end{aligned} \label{eq:dboxquark}
\end{equation}
The reason is that all other on-shell diagrams featuring a ``large" quark cycle running along the perimeter are subleading, scaling as $\mathcal{O}(N_f)$, compared to the four on-shell diagrams above, which scale as $\mathcal{O}(N_cN_f)$. Summing these four on-shell diagrams, we reproduce the two-loop leading singularities for the $N_f$-contributions reported in \cite{Carrolo:2025agz},
\begin{equation}\label{eq:Sij}
    S_{12,ij1,n}^{(2),\text{db}}=\text{PT}_{n,12}\times\left[4-24\frac{\langle1i\rangle\langle 2j\rangle}{\langle12\rangle\langle ij\rangle}+42 \left(\frac{\langle1i\rangle\langle 2j\rangle}{\langle12\rangle\langle ij\rangle}\right)^2-22\left(\frac{\langle1i\rangle\langle 2j\rangle}{\langle12\rangle\langle ij\rangle}\right)^3\right].
\end{equation}

\begin{table}[t]
    \centering
\begin{tabular}{|c|c|c|c|}
\hline
 & $({-}{-}{+}{+}{+}{+})$ & $({-}{+}{-}{+}{+}{+})$ & $({-}{+}{+}{-}{+}{+})$ \\ \hline
$L{=}1$, $m{=}1$ & 0 & 3 & 4 \\ \hline
$L{=}2$, $m{=}1$ & 6 & 12 & 15 \\ \hline
$L{=}2$, $m{=}2$ & 0 & 0 & 1 \\ \hline
\end{tabular}
\caption{Counting of independent prefactors for the maximally-transcendental part of the six-point $L$-loop gluonic QCD hard functions at $N_f^m$-order in the MHV helicity sectors.}\label{tab:3}
\end{table}

In this work, we focus on six-gluon amplitudes in QCD. We summarize the counting of independent prefactors in Table~\ref{tab:3} for the $N_f$ and $N_f^2$ contributions of three helicity sectors at both one and two loops. For the $N_f$ contribution, compared with Table~\ref{tab:1}, in each helicity sector the counting always decreases by one, since the tree-level PT factor does not contribute at $N_f^1$ order. For the $N_f^2$ contribution, the maximally-transcendental contributions should have two separate fermionic loops in a diagram, and can only originate from the kissing-box cut. Therefore, only the MHV sector $({-}{+}{+}{-}{+}{+})$ is nonvanishing, having exactly one independent prefactor. The explicit expressions for all prefactors are provided in the ancillary files. We leave the discussion of on-shell diagrams with external quark legs and the corresponding QCD amplitudes to future work.

\newpage

\section{Prescriptive unitarity integrals}
\label{sec: Prescriptive unitarity integral}

Having computed the prefactors $R_k$ in \eqref{eq:main} as on-shell diagrams representing maximal residues along the contours shown in Fig.~\ref{fig:2Lcuts}, we now turn to determining the transcendental functions that accompany them. Concretely, we perform loop integrations of the four-dimensional loop integrands $\mathcal{I}_k$, contributing to the IR-finite part of the amplitude. To separate the IR-finite contributions, we rely on the prescriptive unitarity construction of \cite{Bourjaily:2019gqu}. 

The central idea is that, given a set of independent maximal-cut contours, one can construct integrands that isolate individual contours by having unit residue on a chosen contour while vanishing on all others. Suppose we consider a set of independent $\ell$-loop chiral contours $\{\alpha_{j,k}=0\}_{j=1,\cdots,4\ell}$, which evaluate the leading singularities $R_k^{(\ell)}$. For each such chiral contour, one can construct the corresponding $4\ell$-fold ${\rm d}\log$-form
\begin{equation}\label{eq:general}
	\mathcal{I}_k=\bigwedge_{j=1}^{4\ell}{\rm d}\log(\alpha_{j,k}) \, ,
\end{equation}
which has unit residue on the $\ell$-loop cut $\alpha_{j,k}=0$ and vanishes on all other contours. Once the set of maximal cuts is complete, the four-dimensional amplitude integrand is obtained as a sum of products $\sum_kR_k\times\mathcal{I}_k$. This establishes a one-to-one correspondence between the loop integrands $\mathcal{I}_k$ and the leading singularities $R_k^{(2)}$ in \eqref{eq:main}. As shown in \cite{Bourjaily:2019gqu}, from the perspective of four-dimensional amplitude integrands, once a complete basis of prescriptive integrals has been constructed, all nontrivial physical information about a gauge-theory amplitude is encoded in the prefactors. In this framework, maximal cut contours contributing to the IR-finite part correspond to prescriptive integrals that are finite in four dimensions, whereas contours contributing to the IR-divergent part give rise to prescriptive integrals containing IR-divergences as well. This encapsulates the core idea of prescriptive unitarity.

Having expressed the maximally-transcendental IR-finite contributions to the amplitude in terms of prescriptive integrals, one still needs to rewrite them as conventional Feynman integrals and perform the loop integrations in order to make progress in computing the amplitude. In this section, we recall the explicit expressions for the one- and two-loop prescriptive unitarity integrals, with particular emphasis on those contributing to the IR-finite part of the decomposition \eqref{eq:main}. Furthermore, by examining lower-loop and lower-point expressions for QCD amplitudes available in the literature, we provide additional evidence that prescriptive unitarity integrals capture the maximally-transcendental parts of QCD amplitudes. Although prescriptive unitarity integrals alone are not sufficient to express the two-loop QCD hard functions, they nevertheless prove to be a necessary ingredient in our bootstrap.

\subsection{One-loop prescriptive unitarity integrals and amplitudes}

At one loop, the non-trivial IR-finite maximal cut contours for MHV amplitudes are the two-mass-easy boxes \cite{Arkani-Hamed:2010pyv,Bourjaily:2013mma} shown in Fig.~\ref{fig:1Lcut}. The corresponding prescriptive integrals are obtained by introducing a chiral numerator. This choice ensures a unit residue on one of the two maximal cuts, while yielding vanishing residues on all remaining maximal cuts, including those associated with IR-divergent contours in the soft region of loop momentum. Diagrammatically, we denote such a numerator by a wavy line and write the corresponding integral as
\begin{equation}\label{eq:chiralbox}
    \raisebox{-4em}{\begin{tikzpicture}
        \draw[line width=1pt,black] (0,0)--(1.5,0)--(1.5,1.5)--(0,1.5)--(0,0);
          \draw[line width=1pt,black] (0,0)--(-0.5,-0.5);
        \draw[line width=1pt,black] (0,2)--(0,1.5)--(-0.5,1.5);
         \draw[line width=1pt,black] (1.5,-0.5)--(1.5,0)--(2,0);
          \draw[line width=1pt,black] (1.5,1.5)--(2,2);
           \draw[decorate, decoration=snake, segment length=12pt, segment amplitude=2pt, black,line width=1pt](0,0)--(1.5,1.5);
        \path[fill=gray] (0,0) circle[radius=0.11];
        \path[fill=black] (1.5,0) circle[radius=0.11];
        \path[fill=gray] (1.5,1.5) circle[radius=0.11];
        \path[fill=black] (0,1.5) circle[radius=0.11];
        \path[fill=gray] (-0.3,1.6) circle[radius=0.03];
        \path[fill=gray] (-0.2,1.7) circle[radius=0.03];
        \path[fill=gray] (-0.1,1.8) circle[radius=0.03];
         \path[fill=gray] (1.6,-0.3) circle[radius=0.03];
        \path[fill=gray] (1.7,-0.2) circle[radius=0.03];
        \path[fill=gray] (1.8,-0.1) circle[radius=0.03];
        \node[anchor=north east] at (-0.5,-0.5) {\small{$k$}};
         \node[anchor=south west] at (2,2) {\small{$j$}};
         \node[anchor=east] at (-0.5,1.5) {\small{$(k{+}1)$}};
          \node[anchor=south] at (0,2) {\small{$(j{-}1)$}};
        \node[anchor=west] at (2,0) {\small{$(j{+}1)$}};
          \node[anchor=north] at (1.5,-0.5) {\small{$(k{-}1)$}};
        \node[anchor=north] at (0.75,0) {\small{$\ell_d$}};
    \node[anchor=west] at (1.5,0.75) {\small{$\ell_a$}};
    \node[anchor=south] at (0.75,1.5) {\small{$\ell_b$}};
   \node[anchor=east] at (0,0.75) {\small{$\ell_c$}};
    \end{tikzpicture}},\ I_{n,jk}=\int\frac{{\rm d}^4\ell}{(2\pi)^2}\frac{\left[\![j,b,c,k\right]\!]}{\ell_a^2\ell_b^2\ell_c^2\ell_d^2},
\end{equation}
where 
\begin{equation}
    \left[\![a_1,\cdots,a_n\right]\!]:=\epsilon_{\alpha_2\dot\beta_1}\cdots \epsilon_{\alpha_n\dot\beta_{n{-}1}}\epsilon_{\alpha_1\dot\beta_n}(a_1)^{\alpha_1\dot\beta_1}(a_2)^{\alpha_2\dot\beta_2}\cdots (a_n)^{\alpha_n\dot\beta_n}\,
\end{equation}
and we abbreviate $p_k\to k$ and $\ell_a\to a$ for simplicity. This box integral is a prescriptive integral for the MHV branch of the maximal cut, i.e. $\ell_a^{(1)}$ in \eqref{eq:twobranchsol}, since its numerator vanishes on the $\overline{\text{MHV}}$ branch as a consequence of the following relation
\begin{equation}
    \left[\![j,b,c,k\right]\!]=s_{j,k}(\ell_a-\ell_a^{(2)})^2 \, 
\end{equation}
while the factor $s_{jk}$ ensures a unit residue on $\ell= \ell_{a}^{(1)}$.

In fact, one can verify from the one-loop MHV pure YM amplitudes available in the literature that the IR-finite functions associated with the one-loop leading singularities $R_{n,ij}^{(1)}$ \eqref{eq:L1LS} are given by the chiral two-mass-easy boxes, as expected from the decomposition \eqref{eq:main}. Since the cut shown in Fig.~\ref{fig:1Lcut} exists only for $|j-k|>1$, the maximally-transcendental parts of the six-point one-loop MHV hard functions are nonvanishing only in the following two helicity configurations (and their cyclic permutations) and take the form below, in agreement with \cite{Dunbar:2009uk},
\begin{equation}\label{eq:H10}
\begin{aligned}
    &H_{{-}{+}{-}{+}{+}{+}}^{(1)}=\text{PT}_{6,13}f_{0,13}+R_{13,24,6}^{(1)}I_{6,24}+R_{13,25,6}^{(1)}I_{6,25}+R_{13,26,6}^{(1)}I_{6,26},\\
    &H_{{-}{+}{+}{-}{+}{+}}^{(1)}=\text{PT}_{6,14}f_{0,14}+R_{14,25,6}^{(1)}I_{6,25}+R_{14,26,6}^{(1)}I_{6,26}+R_{14,35,6}^{(1)}I_{6,35}+R_{14,36,6}^{(1)}I_{6,36},
\end{aligned}
\end{equation}
where $f_{0,1i}$ are affected by the IR-subtraction.

Similarly, the maximally-transcendental part of the one-loop MHV hard function at arbitrary multiplicity is given by 
\begin{equation}    
H^{(1)}_{n,1i}=\text{PT}_{n,1i}f_{0,1i}\ \ +\sum_{1<j<i<k\leq n}R_{1i,jk,n}^{(1)}I_{n,jk} \, ,
\end{equation}
where $1$ and $i$ label the negative-helicity gluons, $\text{PT}_{n,1i}$ is the Parke-Taylor factor \eqref{eq:PT}, and $I_{n,jk}$ are the chiral-boxes given in \eqref{eq:chiralbox}. 

Finally, it is also worth mentioning that an analogous structure holds for the $N_f$-term of the one-loop hard functions. As discussed in section~\ref{sec:QCDls} and summarized in Table~\ref{tab:3}, the split-helicity MHV sector gives no contribution, $H_{{-}{-}{+}{+}{+}{+}}^{[\frac12],(1)}=0$, while for the remaining two helicity sectors we find
\begin{equation}\label{eq:H11}
\begin{aligned}
    &H_{{-}{+}{-}{+}{+}{+}}^{[\frac12],(1)}=S_{13,24,6}^{(1)}I_{6,24}+S_{13,25,6}^{(1)}I_{6,25}+S_{13,26,6}^{(1)}I_{6,26},\\
    &H_{{-}{+}{+}{-}{+}{+}}^{[\frac12],(1)}=S_{14,25,6}^{(1)}I_{6,25}+S_{14,26,6}^{(1)}I_{6,26}+S_{14,35,6}^{(1)}I_{6,35}+S_{14,36,6}^{(1)}I_{6,36},
\end{aligned}
\end{equation}
where the one-loop prefactors of the two-mass-easy box integrals are given in eq.~\eqref{eq:1LSjk}. The maximally-transcendental part of $N_f^1$-contribution to the one-loop MHV gluonic QCD hard function at arbitrary multiplicity is given by 
\begin{equation}    
H^{[\frac12],(1)}_{n,1i}=\sum_{1<j<i<k\leq n}S_{1i,jk,n}^{(1)}I_{n,jk} \, .
\end{equation}

\subsection{Two-loop prescriptive unitarity integrals}

As discussed in section \ref{sec: maximal weight projection}, there are five distinct types of two-loop contours. The corresponding prescriptive unitarity integrals were introduced in \cite{Bourjaily:2019gqu}. Their six-point instances can be found in Ref.~\cite{Bourjaily:2019iqr}, specifically in equations $(\mathcal{I}.1\text{-}4)$, $(\mathcal{I}.22)$, $(\mathcal{I}.23)$, $(\mathcal{I}.26)$ and $(\mathcal{I}.27)$ on pages 26--28, which cover the kissing-box, penta-box, double-pentagon, and hexa-box cuts. In addition, since the double-box cut is nonvanishing for pure YM amplitudes, the corresponding double-box prescriptive integrals from Table II of Ref.~\cite{Bourjaily:2019gqu} are also required. This contrasts with $\mathcal{N}{=}4$ sYM amplitudes, for which these prescriptive integrals do not contribute.

We now list these two-loop prescriptive unitarity integrals using a diagrammatic notation in which wavy lines denote the numerators ${\bf n}_{\rm topo}$ where ${\rm topo} = {\rm kb}, {\rm pb}, {\rm db}, {\rm dp}, {\rm hb}$.
\begin{enumerate}
    \item Kissing-box integrals 
    \begin{equation}\label{eq:Ikb}
         I_{n,jklm}^{\text{kb}}:=\raisebox{-4em}{\begin{tikzpicture}[scale=0.55]
     \draw[line width=1pt,black] (0,0)--(1.5,1.5)--(3,0)--(1.5,-1.5)--(0,0); 
        \draw[line width=1pt,black] (3,0)--(4.5,1.5)--(6,0)--(4.5,-1.5)--(3,0); 
        \draw[line width=1pt,black] (-0.5,-0.5)--(0,0)--(-0.5,0.5);
         \draw[line width=1pt,black] (6.5,-0.5)--(6,0)--(6.5,0.5);
         \draw[line width=1pt,black] (2.7,0.5)--(3,0)--(3.3,0.5);
         \draw[line width=1pt,black] (2.7,-0.5)--(3,0)--(3.3,-0.5);
         \draw[decorate, decoration=snake, segment length=12pt, segment amplitude=2pt, black,line width=1pt](1.5,1.5)--(1.5,-1.5);
    \draw[decorate, decoration=snake, segment length=12pt, segment amplitude=2pt, black,line width=1pt](4.5,1.5)--(4.5,-1.5);
     \draw[line width=1pt,black] (1.5,1.5)--(1.5,2);
     \draw[line width=1pt,black] (1.5,-1.5)--(1.5,-2);
     \draw[line width=1pt,black] (4.5,1.5)--(4.5,2);
     \draw[line width=1pt,black] (4.5,-1.5)--(4.5,-2);
    \path[fill=gray] (1.5,1.5) circle[radius=0.15];
    \path[fill=gray] (1.5,-1.5)circle[radius=0.15];
    \path[fill=gray] (4.5,1.5) circle[radius=0.15];
    \path[fill=gray] (4.5,-1.5)circle[radius=0.15];
     \path[fill=black] (0,0) circle[radius=0.15];
    \path[fill=black] (3,0)circle[radius=0.15];
    \path[fill=black] (6,0) circle[radius=0.15];
    \path[fill=gray] (-0.3,0) circle[radius=0.03];
    \path[fill=gray] (-0.3,0.15) circle[radius=0.03];
    \path[fill=gray] (-0.3,-0.15)circle[radius=0.03];
        \path[fill=gray] (6.3,0) circle[radius=0.03];
        \path[fill=gray] (6.3,0.15) circle[radius=0.03];
        \path[fill=gray] (6.3,-0.15)circle[radius=0.03];
    \path[fill=gray] (3,0.5) circle[radius=0.03];
   \path[fill=gray] (2.9,0.5) circle[radius=0.03];
   \path[fill=gray] (3.1,0.5) circle[radius=0.03];
       \path[fill=gray] (3,-0.5) circle[radius=0.03];
   \path[fill=gray] (2.9,-0.5) circle[radius=0.03];
   \path[fill=gray] (3.1,-0.5) circle[radius=0.03];
\node[anchor=north] at (1.5,-2) {\small{$m$}};
\node[anchor=north] at (0.35,-0.7) {\small{$\ell_b$}};
\node[anchor=south] at (0.35,0.7) {\small{$\ell_c$}};
\node[anchor=north] at (5.65,-0.7) {\small{$\ell_g$}};
\node[anchor=south] at (5.65,0.7) {\small{$\ell_f$}};
\node[anchor=north] at (4.5,-2) {\small{$l$}};
\node[anchor=south] at (1.5,2) {\small{$j$}};
\node[anchor=south] at (4.5,2) {\small{$k$}};
\end{tikzpicture}},\ \ \ {\bf n}_{\rm kb}=\left[\![m,b,c,j\right]\!]\left[\![k,f,g,l\right]\!]
    \end{equation}
\item Penta-box integrals
\begin{equation}\label{eq:Ipb}
    I_{n,jkl}^{\text{pb}}:=\raisebox{-4.5em}{\begin{tikzpicture}[scale=0.7]
    \draw[decorate, decoration=snake, segment length=12pt, segment amplitude=2pt, black,line width=1pt](-1.2,-0.525)--(-1.2,2.025);
    \draw[decorate, decoration=snake, segment length=12pt, segment amplitude=2pt, black,line width=1pt](-1.2,-0.525)--(1.5,1.5);
    \draw[line width=1pt,black] (0,0)--(1.5,0)--(1.5,1.5)--(0,1.5)--(0,0);
    \draw[line width=1pt,black] (-0.3,2)--(0,1.5)--(0.3,2);
    \draw[line width=1pt,black] (2,0)--(1.5,0)--(1.5,-0.5);
    \draw[line width=1pt,black] (-2.5,1)--(-2,0.75)--(-2.5,0.5);
     \draw[line width=1pt,black] (-1.2,-0.525)--(-1.2,-1);
      \draw[line width=1pt,black] (-1.2,2.025)--(-1.2,2.55);
\draw[line width=1pt,black] (1.5,1.5)--(2,2);
    \draw[line width=1pt,black] (0,0)--(-1.2,-0.525)--(-2,0.75)--(-1.2,2.025)--(0,1.5);
    \path[fill=gray] (-1.2,2.025) circle[radius=0.11];
     \path[fill=gray] (-1.2,-0.525) circle[radius=0.11];
    \path[fill=black] (0,1.5) circle[radius=0.11];
    \path[fill=black] (1.5,0) circle[radius=0.11];
    \path[fill=black] (-2,0.75) circle[radius=0.11];
    \path[fill=gray] (1.5,1.5) circle[radius=0.11];
    \path[fill=gray] (0,0) circle[radius=0.11];
    \node[anchor=south west] at (2,2) {\small{$k$}};
    \node[anchor=south] at (-1.2,2.55){\small{$l$}};
     \node[anchor=north] at (-1.2,-1){\small{$j$}};
    \node[anchor=north] at (-1.8,0.2){\small{$\ell_b$}};
    \node[anchor=south] at (-1.8,1.3){\small{$\ell_c$}};
    \node[anchor=west] at (1.5,0.75){\small{$\ell_f$}};
     \node[anchor=east] at (0,0.85){\small{$\ell_h$}};
    \node[anchor=north] at (0.75,0){\small{$\ell_g$}};
    \node[anchor=south] at (0.75,1.5){\small{$\ell_e$}};
    \node[anchor=north] at (-0.45,-0.5){\small{$\ell_a$}};
    \node[anchor=south] at (-0.45,1.8){\small{$\ell_d$}};
      \path[fill=gray] (0,2) circle[radius=0.03];
   \path[fill=gray] (-0.1,2) circle[radius=0.03];
   \path[fill=gray] (0.1,2) circle[radius=0.03];
   \path[fill=gray] (1.6,-0.3) circle[radius=0.03];
        \path[fill=gray] (1.7,-0.2) circle[radius=0.03];
        \path[fill=gray] (1.8,-0.1) circle[radius=0.03];
    \path[fill=gray] (-2.4,0.65) circle[radius=0.03];
    \path[fill=gray] (-2.4,0.75) circle[radius=0.03];
    \path[fill=gray] (-2.4,0.85) circle[radius=0.03];
\end{tikzpicture}},\ \ {\bf n}_{\rm pb}=-\left[\![j,b,c,l\right]\!]\left[\![k,f,g,a\right]\!]{+}\frac12\left[\![j,b,c,l,k,f,g,a\right]\!]
\end{equation}
\item Double-box integrals
\begin{equation}\label{eq:Idb}
    I_{n,jkl}^{\text{db}}:=\raisebox{-3.5em}{\begin{tikzpicture}[scale=0.75]
        \draw[line width=1pt,black] (0,0)--(1.5,0)--(1.5,1.5)--(0,1.5)--(0,0);
         \draw[line width=1pt,black] (1.5,0)--(3,0)--(3,1.5)--(1.5,1.5)--(1.5,0);
          \draw[line width=1pt,black] (0,0)--(-0.5,-0.5);
            \draw[line width=1pt,black] (3,0)--(3.5,-0.5);
        \draw[line width=1pt,black] (0,2)--(0,1.5)--(-0.5,1.5);
        \draw[line width=1pt,black] (1,-0.5)--(1.5,0)--(2,-0.5);
         \draw[line width=1pt,black] (3,2)--(3,1.5)--(3.5,1.5);
         \draw[decorate, decoration=snake, segment length=12pt, segment amplitude=2pt, black,line width=1pt](0,0)--(1.5,1.5);
         \draw[decorate, decoration=snake, segment length=12pt, segment amplitude=2pt, black,line width=1pt](3,0)--(1.5,1.5);
        \path[fill=gray] (3,0) circle[radius=0.11];
          \path[fill=black] (3,1.5) circle[radius=0.11];
        \path[fill=gray] (0,0) circle[radius=0.11];
        \path[fill=black] (1.5,0) circle[radius=0.11];
        \path[fill=gray] (1.5,1.5) circle[radius=0.11];
        \path[fill=black] (0,1.5) circle[radius=0.11];
        \path[fill=gray] (-0.3,1.6) circle[radius=0.03];
        \path[fill=gray] (-0.2,1.7) circle[radius=0.03];
        \path[fill=gray] (-0.1,1.8) circle[radius=0.03];
\path[fill=gray] (3.3,1.6) circle[radius=0.03];
        \path[fill=gray] (3.2,1.7) circle[radius=0.03];
        \path[fill=gray] (3.1,1.8) circle[radius=0.03];
\path[fill=gray] (1.35,-0.3) circle[radius=0.03];
        \path[fill=gray] (1.5,-0.3) circle[radius=0.03];
        \path[fill=gray] (1.65,-0.3) circle[radius=0.03];
        \node[anchor=north east] at (-0.5,-0.5) {\small{$k$}};
        \node[anchor=east] at (0,0.75) {\small{$\ell_b$}};
        \node[anchor=south] at (0.75,1.5) {\small{$\ell_c$}};
        \node[anchor=south] at (2.25,1.5) {\small{$\ell_d$}};
        \node[anchor=west] at (3,0.75) {\small{$\ell_e$}};
        \node[anchor=south] at (0,2) {\small{$l$}};
 \node[anchor=south] at (3,2) {\small{$l{+}1$}};
\node[anchor=north west] at (3.5,-0.5) {\small{$j$}};
\end{tikzpicture}},\ \ {\bf n}_{\rm db}:=\frac12\left[\![k,b,c,d,e,j\right]\!]
\end{equation}
\item Double-pentagon integrals. These are the special case $C=\emptyset$ of the general ``double-pentagon A" integrals in Tab.~IV of Ref.~\cite{Bourjaily:2019gqu}. In this special case, the integrals have a double-box topology, but we nevertheless refer to them and denote them as double-pentagons. 
\begin{equation}\label{eq:Idp}
     I_{n,jk}^{\text{dp}}:=\raisebox{-3.5em}{\begin{tikzpicture}[scale=0.7]
        \draw[line width=1pt,black] (0,0)--(1.5,0)--(1.5,1.5)--(0,1.5)--(0,0);
         \draw[line width=1pt,black] (1.5,0)--(3,0)--(3,1.5)--(1.5,1.5)--(1.5,0);
          \draw[line width=1pt,black] (0,0)--(-0.5,-0.5);
            \draw[line width=1pt,black] (3,-0.5)--(3,0)--(3.5,0);
        \draw[line width=1pt,black] (0,2)--(0,1.5)--(-0.5,1.5);
         \draw[line width=1pt,black] (3.5,2)--(3,1.5);
        \path[fill=black] (3,0) circle[radius=0.11];
          \path[fill=gray] (3,1.5) circle[radius=0.11];
        \path[fill=gray] (0,0) circle[radius=0.11];
        \path[fill=black] (0,1.5) circle[radius=0.11];
        \draw[decorate, decoration=snake, segment length=12pt, segment amplitude=2pt, black,line width=1pt](0,0)--(3,1.5);
        \node[anchor=north east] at (-0.5,-0.5) {\small{$k$}};
          \node[anchor=south] at (0,2) {\small{$j{-}1$}};
        \node[anchor=east] at (0,0.75) {\small{$\ell_b$}};
        \node[anchor=south] at (0.75,1.5) {\small{$\ell_c$}};
        \node[anchor=north] at (2.25,0) {\small{$\ell_f$}};
        \node[anchor=west] at (3,0.75) {\small{$\ell_e$}};
             \node[anchor=east] at (-0.5,1.5) {\small{$k{+}1$}};
 \node[anchor=south west] at (3.5,2) {\small{$j$}};
\node[anchor=north] at (3,-0.5) {\small{$k{-}1$}};
\node[anchor=west] at (3.5,0) {\small{$j{+}1$}};
\end{tikzpicture}},\ \ \ {\bf n}_{\rm dp}:=-\left[\![k,b,c,f,e,j\right]\!]
\end{equation}
\item Hexa-box integrals. These are the special case $C=\emptyset$ of the ``general hexa-box B" integrals in Tab. III of Ref.~\cite{Bourjaily:2019gqu}. In this special case, the integrals have a penta-triangle topology, but we nevertheless refer to them and denote them as hexa-boxes. It is worth noting that this integral vanishes when $k{-}1=j{+}1$.
\begin{equation}\label{eq:Ihb}
     I_{n,jk}^{\text{hb}}:=\raisebox{-3.5em}{\begin{tikzpicture}[scale=0.6]
    \draw[line width=1pt,black] (0,0)--(1.5,0.675)--(0,1.5)--(0,0);
    \draw[line width=1pt,black] (2,0.175)--(1.5,0.675);
    \draw[line width=1pt,black] (-2.5,1.25)--(-2,0.75)--(-2.5,0.25);
     \draw[line width=1pt,black] (-1.2,-0.525)--(-1.2,-1);
      \draw[line width=1pt,black] (-1.2,2.025)--(-1.2,2.55);   
\draw[line width=1pt,black] (1.5,0.675)--(2,1.175);
    \draw[line width=1pt,black] (0,0)--(-1.2,-0.525)--(-2,0.75)--(-1.2,2.025)--(0,1.5);
    \path[fill=gray] (-1.2,2.025) circle[radius=0.11];
     \path[fill=gray] (-1.2,-0.525) circle[radius=0.11];
    \path[fill=black] (1.5,0.675) circle[radius=0.11];
    \path[fill=black] (-2,0.75) circle[radius=0.11];
    \node[anchor=south west] at (2,1.125) {\small{$j{+}1$}};
    \node[anchor=south] at (-1.2,2.55){\small{$j$}};
     \node[anchor=north] at (-1.2,-1){\small{$k$}};
    \node[anchor=north] at (-1.8,0.2){\small{$\ell_b$}};
    \node[anchor=south] at (-1.8,1.3){\small{$\ell_c$}};
    \node[anchor=north west] at (2,0.175){\small{$k{-}1$}};
        \node[anchor=east] at (-2.5,1.25){\small{$j{-}1$}};
            \node[anchor=east] at (-2.5,0.25){\small{$k{+}1$}};
                    \draw[decorate, decoration=snake, segment length=12pt, segment amplitude=2pt, black,line width=1pt] (-1.2,-0.525)--(-1.2,2.025);
\end{tikzpicture}},\ \ \ {\bf n}_{\rm hb}:=s_{j{+}1,\cdots,k{-}1}\left[\![k,b,c,j\right]\!]
\end{equation}
\end{enumerate}

\begin{figure}
    \centering
    \begin{tikzpicture}[scale=0.5]
     \draw[line width=1pt,black] (0,0)--(1.5,1.5)--(3,0)--(1.5,-1.5)--(0,0); 
     \draw[decorate, decoration=snake, segment length=12pt, segment amplitude=2pt, black,line width=1pt](1.5,1.5)--(1.5,-1.5);
    \draw[decorate, decoration=snake, segment length=12pt, segment amplitude=2pt, black,line width=1pt](4.5,1.5)--(4.5,-1.5);
        \draw[line width=1pt,black] (3,0)--(4.5,1.5)--(6,0)--(4.5,-1.5)--(3,0); 
        \draw[line width=1pt,black] (-0.5,0)--(0,0);
         \draw[line width=1pt,black] (6.5,0)--(6,0);
     \draw[line width=1pt,black] (1.5,1.5)--(1.5,2);
     \draw[line width=1pt,black] (1.5,-1.5)--(1.5,-2);
     \draw[line width=1pt,black] (4.5,1.5)--(4.5,2);
     \draw[line width=1pt,black] (4.5,-1.5)--(4.5,-2);
    \path[fill=gray] (1.5,1.5) circle[radius=0.15];
    \path[fill=gray] (1.5,-1.5)circle[radius=0.15];
    \path[fill=gray] (4.5,1.5) circle[radius=0.15];
    \path[fill=gray] (4.5,-1.5)circle[radius=0.15];
     \path[fill=black] (0,0) circle[radius=0.15];
    \path[fill=black] (3,0)circle[radius=0.15];
    \path[fill=black] (6,0) circle[radius=0.15];
\node[anchor=north] at (1.5,-2) {\small{$6$}};
\node[anchor=north] at (4.5,-2) {\small{$5$}};
\node[anchor=south] at (1.5,2) {\small{$2$}};
\node[anchor=south] at (4.5,2) {\small{$3$}};
\node[anchor=east] at (-0.5,0) {\small{$1$}};
\node[anchor=west] at (6.5,0) {\small{$4$}};
\end{tikzpicture}

 \begin{tikzpicture}[scale=0.7]
    \draw[line width=1pt,black] (0,0)--(1.5,0)--(1.5,1.5)--(0,1.5)--(0,0);
    \draw[decorate, decoration=snake, segment length=12pt, segment amplitude=2pt, black,line width=1pt](-1.2,-0.525)--(-1.2,2.025);
    \draw[decorate, decoration=snake, segment length=12pt, segment amplitude=2pt, black,line width=1pt](-1.2,-0.525)--(1.5,1.5);
    \draw[line width=1pt,black] (2,-0.5)--(1.5,0);
    \draw[line width=1pt,black] (-2.5,1.25)--(-2,0.75)--(-2.5,0.25);
     \draw[line width=1pt,black] (-1.2,-0.525)--(-1.2,-1);
      \draw[line width=1pt,black] (-1.2,2.025)--(-1.2,2.55);
\draw[line width=1pt,black] (1.5,1.5)--(2,2);
    \draw[line width=1pt,black] (0,0)--(-1.2,-0.525)--(-2,0.75)--(-1.2,2.025)--(0,1.5);
    \path[fill=gray] (-1.2,2.025) circle[radius=0.11];
     \path[fill=gray] (-1.2,-0.525) circle[radius=0.11];
    \path[fill=black] (0,1.5) circle[radius=0.11];
    \path[fill=black] (1.5,0) circle[radius=0.11];
    \path[fill=black] (-2,0.75) circle[radius=0.11];
    \path[fill=gray] (1.5,1.5) circle[radius=0.11];
    \path[fill=gray] (0,0) circle[radius=0.11];
    \node[anchor=south west] at (2,2) {\small{$4$}};
    \node[anchor=south] at (-1.2,2.55){\small{$5$}};
     \node[anchor=north] at (-1.2,-1){\small{$2$}};
    \node[anchor=north west] at (2,-0.5){\small{$3$}};
        \node[anchor=east] at (-2.5,0.25){\small{$1$}};
    \node[anchor=east] at (-2.5,1.25){\small{$6$}};
        \end{tikzpicture}
   \begin{tikzpicture}[scale=0.7]
    \draw[line width=1pt,black] (0,0)--(1.5,0)--(1.5,1.5)--(0,1.5)--(0,0);
    \draw[decorate, decoration=snake, segment length=12pt, segment amplitude=2pt, black,line width=1pt](-1.2,-0.525)--(-1.2,2.025);
    \draw[decorate, decoration=snake, segment length=12pt, segment amplitude=2pt, black,line width=1pt](-1.2,-0.525)--(1.5,1.5);
    \draw[line width=1pt,black] (0,2)--(0,1.5);
    \draw[line width=1pt,black] (2,-0.5)--(1.5,0);
    \draw[line width=1pt,black] (-2.5,0.75)--(-2,0.75);
     \draw[line width=1pt,black] (-1.2,-0.525)--(-1.2,-1);
      \draw[line width=1pt,black] (-1.2,2.025)--(-1.2,2.55);
\draw[line width=1pt,black] (1.5,1.5)--(2,2);
    \draw[line width=1pt,black] (0,0)--(-1.2,-0.525)--(-2,0.75)--(-1.2,2.025)--(0,1.5);
    \path[fill=gray] (-1.2,2.025) circle[radius=0.11];
     \path[fill=gray] (-1.2,-0.525) circle[radius=0.11];
    \path[fill=black] (0,1.5) circle[radius=0.11];
    \path[fill=black] (1.5,0) circle[radius=0.11];
    \path[fill=black] (-2,0.75) circle[radius=0.11];
    \path[fill=gray] (1.5,1.5) circle[radius=0.11];
    \path[fill=gray] (0,0) circle[radius=0.11];
    \node[anchor=south west] at (2,2) {\small{$4$}};
    \node[anchor=south] at (-1.2,2.55){\small{$6$}};
     \node[anchor=north] at (-1.2,-1){\small{$2$}};
    \node[anchor=north west] at (2,-0.5){\small{$3$}};
        \node[anchor=east] at (-2.5,0.75){\small{$1$}};
    \node[anchor=south] at (0,2){\small{$5$}};
        \end{tikzpicture} 
       \begin{tikzpicture}[scale=0.7]
    \draw[line width=1pt,black] (0,0)--(1.5,0)--(1.5,1.5)--(0,1.5)--(0,0);
    \draw[decorate, decoration=snake, segment length=12pt, segment amplitude=2pt, black,line width=1pt](-1.2,-0.525)--(-1.2,2.025);
    \draw[decorate, decoration=snake, segment length=12pt, segment amplitude=2pt, black,line width=1pt](-1.2,-0.525)--(1.5,1.5);
    \draw[line width=1pt,black] (1.5,-0.5)--(1.5,0)--(2,0);
    \draw[line width=1pt,black] (-2.5,0.75)--(-2,0.75);
     \draw[line width=1pt,black] (-1.2,-0.525)--(-1.2,-1);
      \draw[line width=1pt,black] (-1.2,2.025)--(-1.2,2.55);
\draw[line width=1pt,black] (1.5,1.5)--(2,2);
    \draw[line width=1pt,black] (0,0)--(-1.2,-0.525)--(-2,0.75)--(-1.2,2.025)--(0,1.5);
    \path[fill=gray] (-1.2,2.025) circle[radius=0.11];
     \path[fill=gray] (-1.2,-0.525) circle[radius=0.11];
    \path[fill=black] (0,1.5) circle[radius=0.11];
    \path[fill=black] (1.5,0) circle[radius=0.11];
    \path[fill=black] (-2,0.75) circle[radius=0.11];
    \path[fill=gray] (1.5,1.5) circle[radius=0.11];
    \path[fill=gray] (0,0) circle[radius=0.11];
    \node[anchor=south west] at (2,2) {\small{$5$}};
    \node[anchor=south] at (-1.2,2.55){\small{$6$}};
     \node[anchor=north] at (-1.2,-1){\small{$2$}};
    \node[anchor=north] at (1.5,-0.5){\small{$3$}};
    \node[anchor=west] at (2,0){\small{$4$}};
    \node[anchor=east] at (-2.5,0.75){\small{$1$}};
        \end{tikzpicture}    

        \begin{tikzpicture}[scale=0.8]
        \draw[line width=1pt,black] (0,0)--(1.5,0)--(1.5,1.5)--(0,1.5)--(0,0);
        \draw[decorate, decoration=snake, segment length=12pt, segment amplitude=2pt, black,line width=1pt](0,0)--(1.5,1.5);
        \draw[decorate, decoration=snake, segment length=12pt, segment amplitude=2pt, black,line width=1pt](3,0)--(1.5,1.5);
         \draw[line width=1pt,black] (1.5,0)--(3,0)--(3,1.5)--(1.5,1.5)--(1.5,0);
          \draw[line width=1pt,black] (0,0)--(-0.5,-0.5);
            \draw[line width=1pt,black] (3,0)--(3.5,-0.5);
        \draw[line width=1pt,black] (0,2)--(0,1.5)--(-0.5,1.5);
          \draw[line width=1pt,black] (0,1.5)--(-0.5,2);
         \draw[line width=1pt,black] (3.5,2)--(3,1.5);
        \path[fill=gray] (3,0) circle[radius=0.11];
          \path[fill=black] (3,1.5) circle[radius=0.11];
        \path[fill=gray] (0,0) circle[radius=0.11];
        \path[fill=black] (1.5,0) circle[radius=0.11];
        \path[fill=gray] (1.5,1.5) circle[radius=0.11];
        \path[fill=black] (0,1.5) circle[radius=0.11];
        \node[anchor=north east] at (-0.5,-0.5) {\small{$4$}};
          \node[anchor=south] at (0,2) {\small{$1$}};
             \node[anchor=east] at (-0.5,1.5) {\small{$5$}};
 \node[anchor=south west] at (3.5,2) {\small{$2$}};
\node[anchor=north west] at (3.5,-0.5) {\small{$3$}};
\node[anchor=south east] at (-0.5,2) {\small{$6$}};
\end{tikzpicture}
\begin{tikzpicture}[scale=0.8]
        \draw[line width=1pt,black] (0,0)--(1.5,0)--(1.5,1.5)--(0,1.5)--(0,0);
        \draw[decorate, decoration=snake, segment length=12pt, segment amplitude=2pt, black,line width=1pt](0,0)--(1.5,1.5);
        \draw[decorate, decoration=snake, segment length=12pt, segment amplitude=2pt, black,line width=1pt](3,0)--(1.5,1.5);
         \draw[line width=1pt,black] (1.5,0)--(3,0)--(3,1.5)--(1.5,1.5)--(1.5,0);
          \draw[line width=1pt,black] (0,0)--(-0.5,-0.5);
          \draw[line width=1pt,black] (1.5,0)--(1.5,-0.5);
            \draw[line width=1pt,black] (3,0)--(3.5,-0.5);
        \draw[line width=1pt,black] (0,2)--(0,1.5)--(-0.5,1.5);
         \draw[line width=1pt,black] (3.5,2)--(3,1.5);
        \path[fill=gray] (3,0) circle[radius=0.11];
          \path[fill=black] (3,1.5) circle[radius=0.11];
        \path[fill=gray] (0,0) circle[radius=0.11];
        \path[fill=black] (1.5,0) circle[radius=0.11];
        \path[fill=gray] (1.5,1.5) circle[radius=0.11];
        \path[fill=black] (0,1.5) circle[radius=0.11];
        \node[anchor=north east] at (-0.5,-0.5) {\small{$5$}};
            \node[anchor=north] at (1.5,-0.5) {\small{$4$}};
          \node[anchor=south] at (0,2) {\small{$1$}};
             \node[anchor=east] at (-0.5,1.5) {\small{$6$}};
 \node[anchor=south west] at (3.5,2) {\small{$2$}};
\node[anchor=north west] at (3.5,-0.5) {\small{$3$}};
\end{tikzpicture}
\begin{tikzpicture}[scale=0.8]
        \draw[line width=1pt,black] (0,0)--(1.5,0)--(1.5,1.5)--(0,1.5)--(0,0);
        \draw[decorate, decoration=snake, segment length=12pt, segment amplitude=2pt, black,line width=1pt](0,0)--(1.5,1.5);
        \draw[decorate, decoration=snake, segment length=12pt, segment amplitude=2pt, black,line width=1pt](3,0)--(1.5,1.5);
         \draw[line width=1pt,black] (1.5,0)--(3,0)--(3,1.5)--(1.5,1.5)--(1.5,0);
          \draw[line width=1pt,black] (0,0)--(-0.5,-0.5);
          \draw[line width=1pt,black] (2,-0.5)--(1.5,0)--(1,-0.5);
            \draw[line width=1pt,black] (3,0)--(3.5,-0.5);
        \draw[line width=1pt,black] (-0.5,2)--(0,1.5);
         \draw[line width=1pt,black] (3.5,2)--(3,1.5);
        \path[fill=gray] (3,0) circle[radius=0.11];
          \path[fill=black] (3,1.5) circle[radius=0.11];
        \path[fill=gray] (0,0) circle[radius=0.11];
        \path[fill=black] (1.5,0) circle[radius=0.11];
        \path[fill=gray] (1.5,1.5) circle[radius=0.11];
        \path[fill=black] (0,1.5) circle[radius=0.11];
        \node[anchor=north east] at (-0.5,-0.5) {\small{$6$}};
            \node[anchor=north] at (2,-0.5) {\small{$4$}};
        \node[anchor=north] at (1,-0.5) {\small{$5$}};
          \node[anchor=south east] at (-0.5,2) {\small{$1$}};
 \node[anchor=south west] at (3.5,2) {\small{$2$}};
\node[anchor=north west] at (3.5,-0.5) {\small{$3$}};
\end{tikzpicture}

\begin{tikzpicture}[scale=0.8]
        \draw[line width=1pt,black] (0,0)--(1.5,0)--(1.5,1.5)--(0,1.5)--(0,0);
        \draw[decorate, decoration=snake, segment length=12pt, segment amplitude=2pt, black,line width=1pt](0,0)--(1.5,1.5);
        \draw[decorate, decoration=snake, segment length=12pt, segment amplitude=2pt, black,line width=1pt](3,0)--(1.5,1.5);
         \draw[line width=1pt,black] (1.5,0)--(3,0)--(3,1.5)--(1.5,1.5)--(1.5,0);
          \draw[line width=1pt,black] (0,0)--(-0.5,-0.5);
            \draw[line width=1pt,black] (3,0)--(3.5,-0.5);
        \draw[line width=1pt,black] (0,2)--(0,1.5)--(-0.5,1.5);
         \draw[line width=1pt,black] (3,2)--(3,1.5)--(3.5,1.5);
        \path[fill=gray] (3,0) circle[radius=0.11];
          \path[fill=black] (3,1.5) circle[radius=0.11];
        \path[fill=gray] (0,0) circle[radius=0.11];
        \path[fill=black] (1.5,0) circle[radius=0.11];
        \path[fill=gray] (1.5,1.5) circle[radius=0.11];
        \path[fill=black] (0,1.5) circle[radius=0.11];
        \node[anchor=north east] at (-0.5,-0.5) {\small{$5$}};
          \node[anchor=south] at (0,2) {\small{$1$}};
        \node[anchor=east] at (-0.5,1.5) {\small{$6$}};
 \node[anchor=south] at (3,2) {\small{$2$}};
  \node[anchor=west] at (3.5,1.5) {\small{$3$}};
\node[anchor=north west] at (3.5,-0.5) {\small{$4$}};
\end{tikzpicture}

    \caption{All six-point two-loop prescriptive unitarity integrals for the IR-finite part of MHV amplitudes, up to dihedral permutations of the external legs.}
    \label{fig:six-points}
\end{figure}
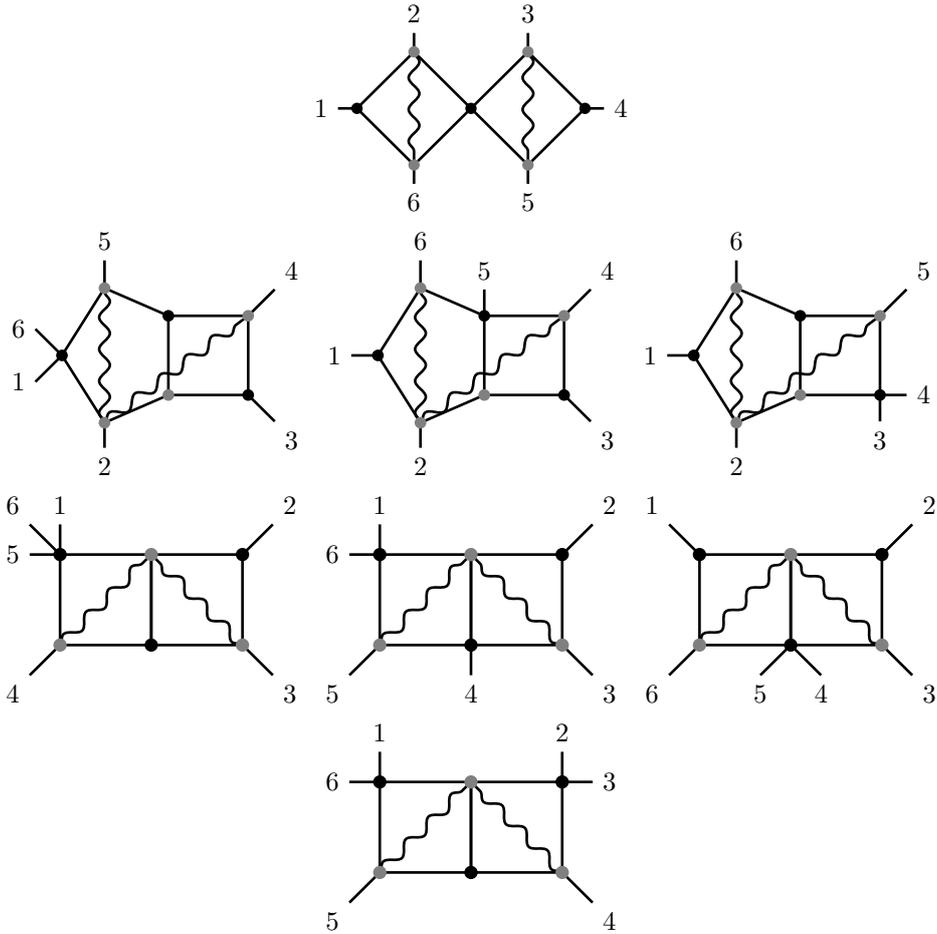

The two-loop leading singularities $R_k^{(2)}$ appearing in the IR-finite part of the decomposition \eqref{eq:main} arise from the first three types of maximal cuts, {\it i.e.} the kissing-box, penta-box and double-box cuts. Consequently, the corresponding IR-finite prescriptive integrals furnish all transcendental functions $\int\mathcal{I}_k$ required to determine the IR-finite part of \eqref{eq:main}. We now specialize the two-loop prescriptive integrals to the six-point case. In the six-point case, the loop integrations of these prescriptive integrals can be carried out using IBP reductions, taking advantage of the known basis of six-point two-loop master integrals. Further details of the IBP reductions employed in this calculation are presented in Appendix~\ref{sec: IBP reduction}. Up to dihedral permutations of the external legs, there is one kissing-box, three penta-boxes and four double-box integrals, which are shown in Fig.~\ref{fig:six-points}. We note that the first four integrals in Fig.~\ref{fig:six-points} coincide with   $(\mathcal{I}.1\text{-}4)$ of Ref.~\cite{Bourjaily:2019iqr}.

We also remark that, as discussed in \cite{Bourjaily:2019iqr,Bourjaily:2019gqu}, the two-loop cuts shown in Figs.~\ref{fig:doublepentagoncut} and~\ref{fig:hexaboxcut} coincide with one-loop two-mass-easy box cuts and therefore they do not compute genuinely two-loop leading singularities. Therefore, they contribute to the IR-subtracted part of the two-loop pure YM hard functions in the decomposition \eqref{eq:main}.  In section \ref{sec: Bootstrapping six-gluon QCD amplitudes}, they are required to compute the $N_f$-contributions to gluonic QCD amplitudes.

\subsection{More examples at two loops}

Finally, we provide further evidence for the two-loop prescriptive unitarity decomposition of the QCD amplitudes by examining the maximally-transcendental parts of known two-loop results. The examples presented in this part include the four- and five-point hard functions at two loops, for which we refer to results in \cite{Jin:2019nya} and \cite{Abreu:2019odu}, and the recently bootstrapped six-point hard function \cite{Carrolo:2025agz} in pure YM theory.

{\it \underline{1.  Four-point amplitudes}}:

At four points, the two-loop cuts are the double-boxes. There are two distinct MHV helicity sectors, $(--++)$ and $(-+-+)$, and the corresponding hard functions are
\begin{equation}
    \begin{aligned}
        &H_{--++}^{(2)}=\text{PT}_{4,12}\ g_{4,12}^{(2)}+ R_{12,341,4}^{(2),\text{db}}I^{\text{db}}_{4,341},\\
        &H_{-+-+}^{(2)}=\text{PT}_{4,13}\ g_{4,13}^{(2)}+ R_{13,24,4}^{(1)}g_{4,24}^{(2)}.
    \end{aligned}
\end{equation}

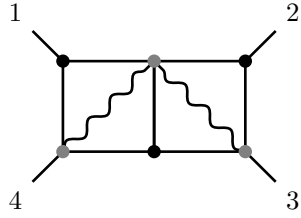
\begin{figure}[t]
    \centering
        \begin{tikzpicture}[scale=0.8]
        \draw[line width=1pt,black] (0,0)--(1.5,0)--(1.5,1.5)--(0,1.5)--(0,0);
        \draw[decorate, decoration=snake, segment length=12pt, segment amplitude=2pt, black,line width=1pt](0,0)--(1.5,1.5);
         \draw[decorate, decoration=snake, segment length=12pt, segment amplitude=2pt, black,line width=1pt](3,0)--(1.5,1.5);
         \draw[line width=1pt,black] (1.5,0)--(3,0)--(3,1.5)--(1.5,1.5)--(1.5,0);
          \draw[line width=1pt,black] (0,0)--(-0.5,-0.5);
            \draw[line width=1pt,black] (3,0)--(3.5,-0.5);
        \draw[line width=1pt,black] (-0.5,2)--(0,1.5);
         \draw[line width=1pt,black] (3.5,2)--(3,1.5);
        \path[fill=gray] (3,0) circle[radius=0.11];
          \path[fill=black] (3,1.5) circle[radius=0.11];
        \path[fill=gray] (0,0) circle[radius=0.11];
        \path[fill=black] (1.5,0) circle[radius=0.11];
        \path[fill=gray] (1.5,1.5) circle[radius=0.11];
        \path[fill=black] (0,1.5) circle[radius=0.11];
        \node[anchor=north east] at (-0.5,-0.5) {\small{$4$}};
          \node[anchor=south east] at (-0.5,2) {\small{$1$}};
 \node[anchor=south west] at (3.5,2) {\small{$2$}};
\node[anchor=north west] at (3.5,-0.5) {\small{$3$}};
\end{tikzpicture}
\caption{The four-point prescriptive integral contributing to the helicity sector $({-}{-}{+}{+})$.}\label{fig:4ptdb}
\end{figure}

The double-box prescriptive integral shown in Fig.~\ref{fig:4ptdb} contributes in the helicity sector $(--++)$, whereas no IR-finite double-box integral exists in the helicity sector $(-+-+)$. Indeed, as noted above, the double-box cut contributes only when the indices $j$ and $k$ in \eqref{eq:dbonshell} satisfy $i<j<k\leq n<1$. The pure functions $g_{4,1i}^{(2)}$ and $g_{4,24}^{(2)}$ receive contributions from IR subtraction. We do not attempt to trace their origin to explicit Feynman integrals.

\begin{figure}[t]
\centering
    \begin{tikzpicture}[scale=0.8]
        \draw[line width=1pt,black] (0,0)--(1.5,0)--(1.5,1.5)--(0,1.5)--(0,0);
        \draw[decorate, decoration=snake, segment length=12pt, segment amplitude=2pt, black,line width=1pt](0,0)--(1.5,1.5);
         \draw[decorate, decoration=snake, segment length=12pt, segment amplitude=2pt, black,line width=1pt](3,0)--(1.5,1.5);
         \draw[line width=1pt,black] (1.5,0)--(3,0)--(3,1.5)--(1.5,1.5)--(1.5,0);
          \draw[line width=1pt,black] (0,0)--(-0.5,-0.5);
            \draw[line width=1pt,black] (3,0)--(3.5,-0.5);
        \draw[line width=1pt,black] (0,2)--(0,1.5)--(-0.5,1.5);
         \draw[line width=1pt,black] (3.5,2)--(3,1.5);
        \path[fill=gray] (3,0) circle[radius=0.11];
          \path[fill=black] (3,1.5) circle[radius=0.11];
        \path[fill=gray] (0,0) circle[radius=0.11];
        \path[fill=black] (1.5,0) circle[radius=0.11];
        \path[fill=gray] (1.5,1.5) circle[radius=0.11];
        \path[fill=black] (0,1.5) circle[radius=0.11];
        \node[anchor=north east] at (-0.5,-0.5) {\small{$4$}};
          \node[anchor=south] at (0,2) {\small{$1$}};
             \node[anchor=east] at (-0.5,1.5) {\small{$5$}};
 \node[anchor=south west] at (3.5,2) {\small{$2$}};
\node[anchor=north west] at (3.5,-0.5) {\small{$3$}};
\end{tikzpicture}
\begin{tikzpicture}[scale=0.8]
        \draw[line width=1pt,black] (0,0)--(1.5,0)--(1.5,1.5)--(0,1.5)--(0,0);
        \draw[decorate, decoration=snake, segment length=12pt, segment amplitude=2pt, black,line width=1pt](0,0)--(1.5,1.5);
         \draw[decorate, decoration=snake, segment length=12pt, segment amplitude=2pt, black,line width=1pt](3,0)--(1.5,1.5);
         \draw[line width=1pt,black] (1.5,0)--(3,0)--(3,1.5)--(1.5,1.5)--(1.5,0);
          \draw[line width=1pt,black] (0,0)--(-0.5,-0.5);
            \draw[line width=1pt,black] (3,0)--(3.5,-0.5);
        \draw[line width=1pt,black] (-0.5,2)--(0,1.5);
         \draw[line width=1pt,black] (3,2)--(3,1.5)--(3.5,1.5);
        \path[fill=gray] (3,0) circle[radius=0.11];
          \path[fill=black] (3,1.5) circle[radius=0.11];
        \path[fill=gray] (0,0) circle[radius=0.11];
        \path[fill=black] (1.5,0) circle[radius=0.11];
        \path[fill=gray] (1.5,1.5) circle[radius=0.11];
        \path[fill=black] (0,1.5) circle[radius=0.11];
        \node[anchor=north east] at (-0.5,-0.5) {\small{$5$}};
          \node[anchor=south east] at (-0.5,2) {\small{$1$}};
 \node[anchor=south] at (3,2) {\small{$2$}};
  \node[anchor=west] at (3.5,1.5) {\small{$3$}};
\node[anchor=north west] at (3.5,-0.5) {\small{$4$}};
\end{tikzpicture}
\begin{tikzpicture}[scale=0.8]
        \draw[line width=1pt,black] (0,0)--(1.5,0)--(1.5,1.5)--(0,1.5)--(0,0);
        \draw[decorate, decoration=snake, segment length=12pt, segment amplitude=2pt, black,line width=1pt](0,0)--(1.5,1.5);
         \draw[decorate, decoration=snake, segment length=12pt, segment amplitude=2pt, black,line width=1pt](3,0)--(1.5,1.5);
         \draw[line width=1pt,black] (1.5,0)--(3,0)--(3,1.5)--(1.5,1.5)--(1.5,0);
          \draw[line width=1pt,black] (0,0)--(-0.5,-0.5);
          \draw[line width=1pt,black] (1.5,0)--(1.5,-0.5);
            \draw[line width=1pt,black] (3,0)--(3.5,-0.5);
        \draw[line width=1pt,black] (-0.5,2)--(0,1.5);
         \draw[line width=1pt,black] (3.5,2)--(3,1.5);
        \path[fill=gray] (3,0) circle[radius=0.11];
          \path[fill=black] (3,1.5) circle[radius=0.11];
        \path[fill=gray] (0,0) circle[radius=0.11];
        \path[fill=black] (1.5,0) circle[radius=0.11];
        \path[fill=gray] (1.5,1.5) circle[radius=0.11];
        \path[fill=black] (0,1.5) circle[radius=0.11];
        \node[anchor=north east] at (-0.5,-0.5) {\small{$5$}};
            \node[anchor=north] at (1.5,-0.5) {\small{$4$}};
          \node[anchor=south east] at (-0.5,2) {\small{$1$}};
 \node[anchor=south west] at (3.5,2) {\small{$2$}};
\node[anchor=north west] at (3.5,-0.5) {\small{$3$}};
\end{tikzpicture}
\caption{Three five-point prescriptive integrals contributing to the helicity sector $({-}{-}{+}{+}{+})$.}\label{fig:5ptdb}
\end{figure}
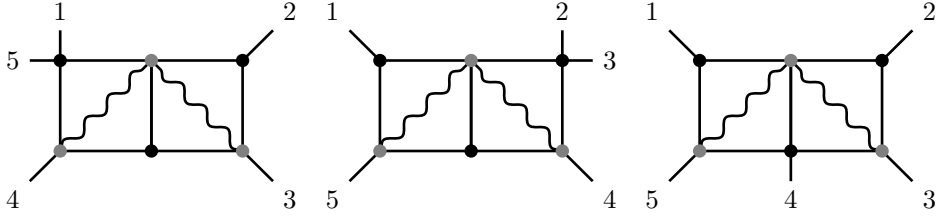

\begin{figure}[t]
    \centering
    \begin{tikzpicture}[scale=0.8]
        \draw[line width=1pt,black] (0,0)--(1.5,0)--(1.5,1.5)--(0,1.5)--(0,0);
        \draw[decorate, decoration=snake, segment length=12pt, segment amplitude=2pt, black,line width=1pt](0,0)--(1.5,1.5);
         \draw[decorate, decoration=snake, segment length=12pt, segment amplitude=2pt, black,line width=1pt](3,0)--(1.5,1.5);
         \draw[line width=1pt,black] (1.5,0)--(3,0)--(3,1.5)--(1.5,1.5)--(1.5,0);
          \draw[line width=1pt,black] (0,0)--(-0.5,-0.5);
            \draw[line width=1pt,black] (3,0)--(3.5,-0.5);
        \draw[line width=1pt,black] (0,2)--(0,1.5)--(-0.5,1.5);
         \draw[line width=1pt,black] (3.5,2)--(3,1.5);
        \path[fill=gray] (3,0) circle[radius=0.11];
          \path[fill=black] (3,1.5) circle[radius=0.11];
        \path[fill=gray] (0,0) circle[radius=0.11];
        \path[fill=black] (1.5,0) circle[radius=0.11];
        \path[fill=gray] (1.5,1.5) circle[radius=0.11];
        \path[fill=black] (0,1.5) circle[radius=0.11];
        \node[anchor=north east] at (-0.5,-0.5) {\small{$5$}};
          \node[anchor=south] at (0,2) {\small{$2$}};
             \node[anchor=east] at (-0.5,1.5) {\small{$1$}};
 \node[anchor=south west] at (3.5,2) {\small{$3$}};
\node[anchor=north west] at (3.5,-0.5) {\small{$4$}};
\end{tikzpicture}
\begin{tikzpicture}[scale=0.8]
        \draw[line width=1pt,black] (0,0)--(1.5,0)--(1.5,1.5)--(0,1.5)--(0,0);
         \draw[line width=1pt,black] (1.5,0)--(3,0)--(3,1.5)--(1.5,1.5)--(1.5,0);
         \draw[decorate, decoration=snake, segment length=12pt, segment amplitude=2pt, black,line width=1pt](0,0)--(1.5,1.5);
         \draw[decorate, decoration=snake, segment length=12pt, segment amplitude=2pt, black,line width=1pt](3,0)--(1.5,1.5);
          \draw[line width=1pt,black] (0,0)--(-0.5,-0.5);
            \draw[line width=1pt,black] (3,0)--(3.5,-0.5);
        \draw[line width=1pt,black] (-0.5,2)--(0,1.5);
         \draw[line width=1pt,black] (3,2)--(3,1.5)--(3.5,1.5);
        \path[fill=gray] (3,0) circle[radius=0.11];
          \path[fill=black] (3,1.5) circle[radius=0.11];
        \path[fill=gray] (0,0) circle[radius=0.11];
        \path[fill=black] (1.5,0) circle[radius=0.11];
        \path[fill=gray] (1.5,1.5) circle[radius=0.11];
        \path[fill=black] (0,1.5) circle[radius=0.11];
        \node[anchor=north east] at (-0.5,-0.5) {\small{$5$}};
          \node[anchor=south east] at (-0.5,2) {\small{$1$}};
 \node[anchor=south] at (3,2) {\small{$2$}};
  \node[anchor=west] at (3.5,1.5) {\small{$3$}};
\node[anchor=north west] at (3.5,-0.5) {\small{$4$}};
\end{tikzpicture}\\
\raisebox{-4.5em}{\begin{tikzpicture}[scale=0.75]
\draw[decorate, decoration=snake, segment length=12pt, segment amplitude=2pt, black,line width=1pt](-1.2,-0.525)--(-1.2,2.025);
    \draw[decorate, decoration=snake, segment length=12pt, segment amplitude=2pt, black,line width=1pt](-1.2,-0.525)--(1.5,1.5);
    \draw[line width=1pt,black] (0,0)--(1.5,0)--(1.5,1.5)--(0,1.5)--(0,0);
    \draw[line width=1pt,black] (2,-0.5)--(1.5,0);
    \draw[line width=1pt,black] (-2.5,0.75)--(-2,0.75);
     \draw[line width=1pt,black] (-1.2,-0.525)--(-1.2,-1);
      \draw[line width=1pt,black] (-1.2,2.025)--(-1.2,2.55);
\draw[line width=1pt,black] (1.5,1.5)--(2,2);
    \draw[line width=1pt,black] (0,0)--(-1.2,-0.525)--(-2,0.75)--(-1.2,2.025)--(0,1.5);
    \path[fill=gray] (-1.2,2.025) circle[radius=0.11];
     \path[fill=gray] (-1.2,-0.525) circle[radius=0.11];
    \path[fill=black] (0,1.5) circle[radius=0.11];
    \path[fill=black] (1.5,0) circle[radius=0.11];
    \path[fill=black] (-2,0.75) circle[radius=0.11];
    \path[fill=gray] (1.5,1.5) circle[radius=0.11];
    \path[fill=gray] (0,0) circle[radius=0.11];
    \node[anchor=south west] at (2,2) {\small{$4$}};
    \node[anchor=south] at (-1.2,2.55){\small{$5$}};
     \node[anchor=north] at (-1.2,-1){\small{$2$}};
    \node[anchor=north west] at (2,-0.5){\small{$3$}};
        \node[anchor=east] at (-2.5,0.75){\small{$1$}};
\end{tikzpicture}}\ +\ \raisebox{-4.5em}{\begin{tikzpicture}[scale=0.75]
\draw[decorate, decoration=snake, segment length=12pt, segment amplitude=2pt, black,line width=1pt](-1.2,-0.525)--(-1.2,2.025);
    \draw[decorate, decoration=snake, segment length=12pt, segment amplitude=2pt, black,line width=1pt](-1.2,-0.525)--(1.5,1.5);
    \draw[line width=1pt,black] (0,0)--(1.5,0)--(1.5,1.5)--(0,1.5)--(0,0);
    \draw[line width=1pt,black] (2,-0.5)--(1.5,0);
    \draw[line width=1pt,black] (-2.5,0.75)--(-2,0.75);
     \draw[line width=1pt,black] (-1.2,-0.525)--(-1.2,-1);
      \draw[line width=1pt,black] (-1.2,2.025)--(-1.2,2.55);
\draw[line width=1pt,black] (1.5,1.5)--(2,2);
    \draw[line width=1pt,black] (0,0)--(-1.2,-0.525)--(-2,0.75)--(-1.2,2.025)--(0,1.5);
    \path[fill=gray] (-1.2,2.025) circle[radius=0.11];
     \path[fill=gray] (-1.2,-0.525) circle[radius=0.11];
    \path[fill=black] (0,1.5) circle[radius=0.11];
    \path[fill=black] (1.5,0) circle[radius=0.11];
    \path[fill=black] (-2,0.75) circle[radius=0.11];
    \path[fill=gray] (1.5,1.5) circle[radius=0.11];
    \path[fill=gray] (0,0) circle[radius=0.11];
    \node[anchor=south west] at (2,2) {\small{$5$}};
    \node[anchor=south] at (-1.2,2.55){\small{$4$}};
     \node[anchor=north] at (-1.2,-1){\small{$2$}};
    \node[anchor=north west] at (2,-0.5){\small{$1$}};
        \node[anchor=east] at (-2.5,0.75){\small{$3$}};
\end{tikzpicture}}
    \caption{Three combinations of prescriptive integrals contributing to the helicity sector $(-+-++)$.
    }
    \label{fig:5ptdbpb}
\end{figure}
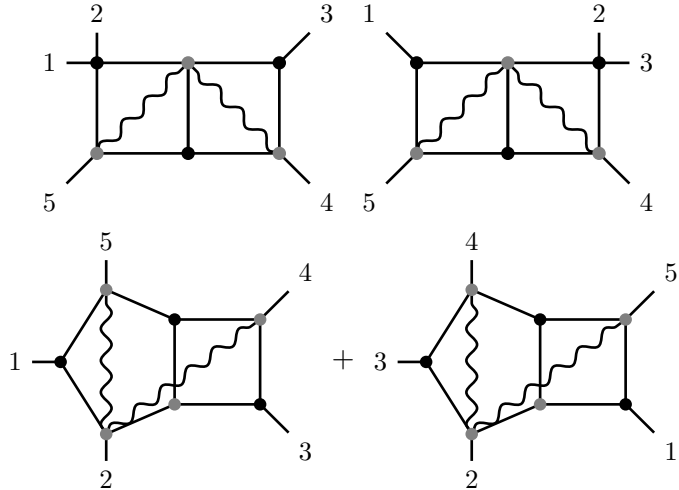

{\it \underline{2. Five-point amplitudes}}:

At five points, there are two inequivalent MHV helicity sectors. By examining the maximally-transcendental part of the two-loop hard function in the helicity sector $(--+++)$, one finds three independent prefactors in addition to the tree-level PT factor. By contrast, the one-loop hard function in this helicity sector involves only the PT factor. Consequently, the transcendental functions accompanying the three genuinely two-loop prefactors are determined by IR-finite prescriptive integrals. Furthermore, as discussed in section \ref{sec: Leading singularities for MHV QCD amplitudes}, for the helicity sector $(--+\cdots)$ we only need to consider double-box cuts. Accordingly the hard function takes the form,
\begin{equation}
    H_{--+++}^{(2)}=\text{PT}_{5,12}\ g_{5,12}^{(2)}+ R_{12,341,5}^{(2),\text{db}}I^{\text{db}}_{5,341}+ R_{12,451,5}^{(2),\text{db}}I^{\text{db}}_{5,451}+ R_{12,351,5}^{(2),\text{db}}I^{\text{db}}_{5,351},
\end{equation}
where $g_{5,12}^{(2)}$ is a pure function arising from IR subtraction, and the three double-box integrals are depicted in Fig.~\ref{fig:5ptdb}.

Similarly, one finds five independent prefactors, in addition to the PT factor, in the two-loop hard function for the helicity sector $(-+-++)$. Two of these prefactors are inherited from one loop and are given by \eqref{eq:L1LS}, while the remaining three are genuinely two-loop. In this helicity sector, both double-box and penta-box cuts contribute. The expression for the two-loop hard function can be written in the following form, where the three pure functions accompanying the two-loop prefactors are given by three combinations of prescriptive integrals shown in Fig.~\ref{fig:5ptdbpb}, 
\begin{equation}
\begin{aligned}
    H_{-+-++}^{(2)}=&\text{PT}_{5,13}g_{5,13}^{(2)}+R_{13,24,5}^{(1)}g_{5,24}^{(2)}+R_{13,25,5}^{(1)}g_{5,25}^{(2)}\\
    &+R_{13,451,5}^{(2),\text{db}}I^{\text{db}}_{5,451}+R_{13,452,5}^{(2),\text{db}}I^{\text{db}}_{5,452}+R_{13,245,5}^{(2),\text{pb}}(I_{5,245}^{\text{pb}}{+}I_{5,254}^{\text{pb}}).
\end{aligned} \label{eq:H5ptMHVpu}
\end{equation}
Specifically, the two penta-box integrals combine, since the penta-box prefactor \eqref{eq:Rpb} is determined by two distinct penta-box on-shell diagrams, see \eqref{eq:Rpbsym}. These double-box prefactors and the ones for split helicity sectors at four and five points can be found in the appendix \ref{sec:dbpref}.

{\it \underline{3.   Six-point amplitude}}:

Our final example is the six-point MHV hard function in the helicity sector $({-}{-}{+}{+}{+}{+})$, which has the following form \cite{Carrolo:2025agz},
\begin{equation}
    H_{{-}{-}{+}{+}{+}{+}}^{(2)}=\text{PT}_{6,12}f_0+\sum_{2<i<j\leq6}R_{12,ij1,6}^{(2),\text{db}} I_{6,ij1}^{\text{db}},
\end{equation}
where the pure function $f_0$ is affected by IR subtraction, and the prescriptive integrals accompanying the prefactors $R^{\text{db}}$ are six double-boxes (four of these are shown in Fig.~\ref{fig:six-points}, while the remaining two are obtained by reflection $p_i\leftrightarrow p_{3-i}$). These are all possible double-box integrals $I_{n,jkl}^{\text{db}}$ with $l=1$ and $2<j<k\leq 6$. We conjecture that the IR-finite part of the two-loop hard function at arbitrary multiplicity $n$ in the MHV sector $(--+\cdots)$ is obtained by replacing $6$ with $n$ in the summation range of the previous equation.

\newpage

\section{Bootstrapping six-gluon QCD amplitudes}
\label{sec: Bootstrapping six-gluon QCD amplitudes}

Having discussed the IR decomposition \eqref{eq:main} of the maximal weight projection of QCD amplitude integrands, reviewed the basis of prescriptive unitarity integrals, and classified the associated prefactors, we are now in a position to complete the computation of the QCD hard functions. We adopt a hybrid strategy that combines the symbol bootstrap with IBP reductions. 

As explained in section \ref{sec: Prescriptive unitarity integral}, the prescriptive integrals completely determine the IR-finite part of the decomposition \eqref{eq:main}, and hence the corresponding contributions to the hard functions. These prescriptive integrals are calculated in Appendix~\ref{sec: IBP reduction} by projecting them onto a basis of six-point two-loop master integrals using IBP reductions. This task is significantly simpler than a full amplitude calculation via IBP, since the prescriptive integrals are finite in four dimensions, have unit leading singularities, evaluate to pure functions, and possess numerators of relatively low rank. 

For the IR-subtracted part of the hard function, which involves tree-level PT factors and one-loop prefactors \eqref{eq:L1LS}, we remain agnostic about the explicit loop integrand and instead employ a symbol ansatz, thereby avoiding direct loop integrations. The availability of explicit symbol representations for the prescriptive integrals drastically reduces the number of free parameters in the bootstrap ansatz compared to that used in the Letter~\cite{Carrolo:2025agz} for the split-helicity MHV configuration. We find that this reduction is essential for uniquely fixing all remaining free parameters in the bootstrap for the other MHV helicity sectors. 

We begin by bootstrapping the $N_f^0$ contributions and construct the corresponding ansatz using the prefactors enumerated in Table~\ref{tab:1}.
We then turn to the quark contributions, which are $N_f^k$ terms of the hard function with $k=1,2$. Guided by the known five-point two-loop results, we find that these contributions are determined by a dozen of functions and prescriptive integrals.  Calculations in this section can be generalized to QCD amplitudes with quark external legs in a similar manner. 

Finally, we examine the properties of the resulting hard functions, including their common symbol alphabet and the adjacency conditions. In particular, by uncovering a connection between our bootstrapped pure YM results and the two-loop six-point MHV remainder function in ${\cal N}=4$ sYM theory, we shed light on
the principle of maximal transcendentality from the perspective of prescriptive unitarity. This connection is discussed at the end of this section.

\subsection{Bootstrapping pure YM amplitudes}

Up to dihedral permutations, there are three inequivalent MHV helicity configurations of six external gluons. The sector $({-}{-}{+}{+}{+}{+})$ was previously studied in the Letter~\cite{Carrolo:2025agz}. We first revisit this case using prescriptive unitarity, and then complete the calculation for the remaining two MHV sectors, $({-}{+}{-}{+}{+}{+})$ and $({-}{+}{+}{-}{+}{+})$\footnote{It is worth noting that, in our initial approach to this problem, we adopted the bootstrap strategy of \cite{Carrolo:2025agz}, which does not rely on the prescriptive unitarity and therefore requires a more general symbol ansatz. While the available physical constraints were sufficient to uniquely fix the sector $({-}{+}{+}{-}{+}{+})$, the sector $({-}{+}{-}{+}{+}{+})$ still contained approximately a dozen unfixed parameters. Nevertheless, these partially constrained results already exhibited a clear correspondence with prescriptive unitarity integrals, which in turn motivated the more focused and efficient bootstrap strategy presented in this section.}.

We begin by outlining the symbol spaces and physical constraints employed in our bootstrap. Not all of the 712 independent two-loop  weight-four symbols discussed in section~\ref{sec:basis-symbols} are required in our analysis. Since hard functions depend only on ratios of Mandelstam variables, the symbol entries can be chosen to be dimensionless. This choice reduces the number of independent weight-four symbols to {430}.
Furthermore, for each helicity sector we impose the corresponding dihedral symmetry relations (see \eqref{eq:Hsym}) and construct a symbol ansatz that manifestly respects these symmetries.

To fix all free parameters in the symbol ansatz, we impose several necessary physical constraints, namely spurious pole cancellation and consistency with the collinear and triple-collinear limits. First, the leading singularities $R_{i,j}^{(1)}$ and $R^{{(2),\text{topo}}}$, defined in~\eqref{eq:L1LS} and \eqref{eq:LSdb}, contain spurious poles at $\langle{ij}\rangle =0$ for non-adjacent $i$ and $j$.
These unphysical singularities must be eliminated by requiring the accompanying weight-four symbols to vanish on these loci. Apart from the symmetry constraints, we start by imposing these conditions purely from convenience as they are homogeneous conditions that do not rely on lower point data. Nevertheless, they are rarely strong enough to fix the result, and thus other physical properties must be taken into account.

Second, we require the symbol ansatz to reproduce the known five-point planar hard functions of Ref.~\cite{Abreu:2019odu} in the limit where a pair of adjacent particles becomes collinear. In practice, we implement the collinear limits using the momentum-twistor parametrization of the kinematics. For definiteness, we take the limit $p_5||p_6$ and obtain the remaining cases by cyclic permutations of the ansatz. In momentum-twistor variables, we take 
\begin{equation}
    Z_6\to Z_5+\epsilon (a_1 Z_1+a_2 Z_4)+\epsilon^2 a_3 Z_2 \,, 
\end{equation}
where the parameters $a_i$ are arbitrary and must cancel from the hard function in the limit $\epsilon\to0$.

Third, we impose the triple-collinear limit of the symbol ansatz. When three adjacent external legs become collinear, the ansatz should reproduce the known four-point hard functions. Implementing this limit in momentum twistors and considering $p_4||p_5||p_6$, we take
\begin{equation}
    Z_5\to Z_4+\epsilon (a_1 Z_3+a_2 Z_1)+\epsilon^2 a_3 Z_2,\qquad   Z_6\to Z_4+\epsilon (b_1 Z_3+b_2 Z_1)+\epsilon^2 b_3 Z_2\,,
\end{equation}
where $a_i$ and $b_i$ are arbitrary parameters that must cancel from the hard function in this limit. A detailed review of multi-collinear limits of QCD scattering amplitudes is provided in section~\ref{sec: behaviors}. In these limits, the hard functions obey the universal behavior specified in eqs.~\eqref{eq:HAcoll} and \eqref{eq:HAtripcol} whenever they develop singularities in the corresponding kinematic regimes. In particular, such singular behavior is required for tree-level amplitudes.

For all MHV helicity sectors, all free parameters in the symbol ans\"{a}tze are uniquely fixed once the above conditions are imposed simultaneously.

Amplitudes and hard functions are known to exhibit universal behavior in the soft and multi-soft limits. These limits are not required as physical constraints in our bootstrap procedure, since a unique solution is obtained without invoking them. Nevertheless, we use them as nontrivial consistency checks of the bootstrapped results. We implement the soft kinematics using a momentum-twistor parametrization,
\begin{equation}
    Z_6\to Z_5{+}a_1 Z_1{+}\epsilon (a_2 Z_4{+}a_3 Z_2)\,,
\end{equation}
where $p_6$ becomes soft, $p_6 \to 0$, as $\ep \to 0$. The double-soft limit can be implemented analogously by applying a similar parametrization to a pair of momentum twistors. Further details on soft and multi-soft limits are collected in section~\ref{sec: behaviors}. In these limits, the hard functions behave as specified in eqs.~\eqref{eq:HAsoft}  and \eqref{eq:HAdbsoft}.

Finally, we also calculated the multi-Regge asymptotics of the bootstrapped hard functions, which are presented in Appendix.~\ref{sec: Regge}.

Having outlined the general bootstrap strategy, we proceed to construct ans\"{a}tze for each MHV helicity sector using the two-loop six-point planar symbol basis and the prescriptive unitarity framework. In the section, since $n=6$ always holds, we drop the subscript $n$ of all leading singularities \eqref{eq:L1LS}, \eqref{eq:Rkb}, \eqref{eq:Rpb}, and \eqref{eq:LSdb}  in the expressions.

{\it \underline{1.  Sector $({-}{-}{+}{+}{+}{+})$}}:

The hard function in this sector was bootstrapped in \cite{Carrolo:2025agz}.
Here, building on the decomposition \eqref{eq:main}, we instead employ a refined ansatz,
\begin{tcolorbox}    
\begin{equation}\label{eq:ansatz0}
    H_{{-}{-}{+}{+}{+}{+}}^{(2)}=\text{PT}_{6,12}\, f_0+\sum_{2<i<j\leq6}R_{12,ij1}^{(2),\text{db}} I_{6,ij1}^{\text{db}},
\end{equation}
\end{tcolorbox}
\noindent where $f_0$ is a pure function for which we employ a symbol ansatz, the prefactors $R_{12,ij1}^{(2),\text{db}}$ are defined in eq.~\eqref{eq:LSdb}, and $I_{6,ij1}^{\text{db}}$ are the explicitly known double-box prescriptive integrals \eqref{eq:Idb}. Imposing the dihedral symmetry $p_i\leftrightarrow p_{3-i}$ of the hard function (see \eqref{eq:Hsym}) reduces the number of free parameters in the ansatz for $f_0$ from 430 to ${ 238}$. These constitute the remaining unknowns in our ansatz, which are to be fixed by imposing physical constraints.

As shown in \cite{Carrolo:2025agz}, the collinear and triple-collinear limits are sufficient to uniquely fix all remaining unknowns in this ansatz. In total, there are five independent collinear limits and four independent triple-collinear limits we can consider in this sector. The first column of Table~\ref{tab:2} summarizes the number of independent constraints arising from each condition.

{\it \underline{2. Sector $({-}{+}{-}{+}{+}{+})$}}:

For this sector, the genuinely two-loop prefactors are determined by the on-shell diagrams shown in Fig.~\ref{fig:onshell22}. These prefactors are accompanied by explicitly known prescriptive integrals. Accordingly, we adopt the following ansatz with 13 prefactors,
\begin{tcolorbox}
\begin{equation}\label{eq:ansatz2}
\begin{aligned}
    H_{{-}{+}{-}{+}{+}{+}}^{(2)}&=\text{PT}_{6,13}g_0+R_{13,24}^{(1)}g_{2,4}+R_{13,25}^{(1)}g_{2,5}+R_{13,26}^{(1)}g_{2,6}\\    
&+\sum_{\substack{l=1,2\\3<j<k\leq6}}R_{13,jkl}^{(2),\text{db}}I_{6,jkl}^{\text{db}}+\sum_{\{j,k,l\}\in\sigma_1}R_{13,jkl}^{(2),\text{pb}}(I^{\text{pb}}_{6,jkl}+I^{\text{pb}}_{6,jlk})\,,
\end{aligned}
\end{equation}
\end{tcolorbox}
\noindent where $\sigma_1=\{\{2,4,5\},\{2,4,6\},\{2,5,6\}\}$, and $g_0$, $g_{2,4}$, $g_{2,5}$, and $g_{2,6}$ are pure functions to be determined by the bootstrap. Under the reflection symmetry $p_i\leftrightarrow p_{4-i}$ of the hard function (see \eqref{eq:Hsym}), the functions $g_0$ and $g_{2,5}$ are invariant, while $g_{2,4}$ and $g_{2,6}$ are related by this symmetry. Taking this reflection invariance into account in the symbol ansatz, we find $2\times 221+ 430={ 872}$ unknowns in total.

We now proceed to fix the unknowns in the ansatz, starting with spurious pole cancellation. The one-loop prefactor $R_{13,ij}^{(1)}$ becomes singular as $\langle ij\rangle\to0$, which constrains $g_{2,4}$ and $g_{2,5}$ on the loci $\langle24\rangle=0$ and $\langle25\rangle=0$, respectively. It is worth noting that the two-loop prefactors $R^{(2),\text{topo}}$ involve the same spurious poles, so their cancellation occurs among several terms in the ansatz \eqref{eq:ansatz2}. More precisely, we find that the nonvanishing residues at these spurious poles are all proportional,
\begin{equation}
\begin{aligned}
    &\underset{\langle24\rangle=0}{\text{Res}}R_{13,24}^{(1)}=\underset{{\langle24\rangle=0}}{\text{Res}}R_{13,245}^{(2),\text{pb}}=\underset{\langle24\rangle=0}{\text{Res}}R_{13,246}^{(2),\text{pb}}\neq0,\\ &\underset{\langle25\rangle=0}{\text{Res}}R_{13,25}^{(1)}=\underset{\langle25\rangle=0}{\text{Res}}R_{13,452}^{(2),\text{db}}=-\underset{\langle25\rangle=0}{\text{Res}}R_{13,561}^{(2),\text{db}}=\underset{\langle25\rangle=0}{\text{Res}}R_{13,245}^{(2),\text{pb}}=\underset{\langle25\rangle=0}{\text{Res}}R_{13,256}^{(2),\text{pb}}\neq0 \,,
    \end{aligned}
\end{equation}
which leads to constraints on the associated pure functions that are imposed at the level of symbols,
\begin{equation}
\begin{aligned}
    &\left(g_{2,4}+I_{245}^{\text{pb}}+I_{254}^{\text{pb}}+I_{246}^{\text{pb}}+I_{264}^{\text{pb}}\right)\biggr|_{\langle24\rangle\to0}=0,\\ &\left(g_{2,5}+I_{452}^{\text{db}}-I_{561}^{\text{db}}+I_{245}^{\text{pb}}+I_{254}^{\text{pb}}+I_{256}^{\text{pb}}+I_{265}^{\text{pb}}\right)\biggr|_{\langle25\rangle\to0}=0 \,.
    \end{aligned} 
\end{equation}
Owing to the reflection symmetry of the ansatz, cancellation of the spurious pole at $\langle26\rangle = 0$ follows directly from that at $\langle24\rangle = 0$.

Finally, we consider the multi-collinear limits. In this helicity sector, the hard function exhibits universal behavior in six collinear limits and five triple-collinear limits. The triple collinear limit $p_1||p_2||p_3$, however, does not provide useful physical information, since the amplitude is nonsingular in this regime owing to the vanishing of the tree-level $\text{PT}_{6,13}$. To fully exploit the multi-collinear constraints, one must also account for the corresponding gluon splitting functions. At two loops,  the collinear splitting functions are well-known, whereas the triple gluon splitting functions are not known apriori. Using the already bootstrapped helicity sector $({-}{-}{+}{+}{+}{+})$, we extract the following MHV triple gluon splitting functions \cite{Carrolo:2025agz},
\begin{align} \label{eq:MHVsplittings}
{+}\to{+}{+}{+} \,,\qquad
{-}\to{-}{+}{+} \,,\qquad
{-}\to{+}{+}{-} \,.
\end{align}
The remaining triple splitting, ${-}\to{+}{-}{+}$, cannot be accessed within this helicity configuration. We therefore impose the physical constraints arising from the three triple-collinear limits associated with the splittings in eq.~\eqref{eq:MHVsplittings}. 

Altogether, these constraints are sufficient to fix all unknowns in the ansatz and to determine a unique result. The counting of constraints arising from each condition is summarized in the second column of Table~\ref{tab:2}.

{\it \underline{3.   Sector $({-}{+}{+}{-}{+}{+})$}}:

Finally, for this sector we construct an ansatz involving 16 prefactors, based on the on-shell diagrams shown in Fig.~\ref{fig:onshell33},
\begin{tcolorbox}
\begin{equation}\label{eq:ansatz3}
\begin{aligned}
    H_{{-}{+}{+}{-}{+}{+}}^{(2)}&=\text{PT}_{6,14}h_0+R_{14,25}^{(1)}h_{2,5}+R_{14,26}^{(1)}h_{2,6}+R_{14,35}^{(1)}h_{3,5}+R_{14,36}^{(1)}h_{3,6}\\    
&+R_{14,2356}^{(2),\text{kb}}I_{6,2356}^{\text{kb}}+\sum_{\substack{1\leq l\leq3}}R_{14,56l}^{(2),\text{db}}I_{6,56l}^{\text{db}}+\sum_{\substack{4\leq l\leq6}}R_{41,23l}^{(2),\text{db}}I_{6,23l}^{\text{db}}\\&+\sum_{\{j,k,l\}\in\sigma_2}R_{14,jkl}^{(2),\text{pb}}(I^{\text{pb}}_{6,jkl}+I^{\text{pb}}_{6,jlk})\,,
\end{aligned}
\end{equation}
\end{tcolorbox}
\noindent where $\sigma_2=\{\{6,2,3\},\{5,2,3\},\{2,5,6\},\{3,5,6\}\}$, and $h_0$ and $h_{i,j}$ are pure functions to be determined by the bootstrap. Owing to the dihedral symmetries of the hard function (see \eqref{eq:Hsym}), we require the ansatz to be invariant under the reflections $p_i\leftrightarrow p_{5{-}i}$ and $p_i\leftrightarrow p_{8{-}i}$, as well as under the cyclic shift $p_i\to p_{i{+}3}$. The function $h_0$ is invariant under all these permutations. Under the first reflection, the functions $h_{2,5}$ and $h_{3,6}$ are mapped into one another, as are $h_{2,6}$ and $h_{3,5}$. The function $h_{2,5}$ is invariant under the cyclic shift, while is invariant $h_{3,5}$ under the second reflection. After imposing these discrete symmetry constraints, the resulting symbol ansatz contains $130+231+221={ 582}$ unknowns.

We find the following linear relations among the nonvanishing residues of the prefactors at their spurious poles,
\begin{equation}
\begin{aligned}
    &\underset{\langle25\rangle=0}{\text{Res}}R_{14,25}^{(1)}=\underset{\langle25\rangle=0}{\text{Res}}R_{41,234}^{(2),\text{db}}=-\underset{\langle25\rangle=0}{\text{Res}}R_{41,235}^{(2),\text{db}}=\underset{\langle25\rangle=0}{\text{Res}}R_{41,523}^{(2),\text{pb}}=\underset{\langle25\rangle=0}{\text{Res}}R_{14,256}^{(2),\text{pb}}\neq0,\\    &\underset{\langle35\rangle=0}{\text{Res}}R_{14,35}^{(1)}=\underset{\langle35\rangle=0}{\text{Res}}R_{41,523}^{(2),\text{pb}}=\underset{\langle35\rangle=0}{\text{Res}}R_{14,356}^{(2),\text{pb}}\neq0,\ \ \underset{\langle35\rangle=0}{\text{Res}}R_{14,2356}^{(2),\text{kb}}\neq0 \,,
\end{aligned}
\end{equation}
and the requirement of spurious pole cancellation then imposes corresponding constraints on the associated symbols,
\begin{equation}
\begin{aligned}
&\left(h_{2,5}+I_{234}^{\text{db}}-I_{561}^{\text{db}}+I_{523}^{\text{pb}}+I_{532}^{\text{pb}}+I_{256}^{\text{pb}}+I_{265}^{\text{pb}}\right)\biggr|_{\langle25\rangle\to0}=0,\ \\
&\left(h_{3,5}+I_{523}^{\text{pb}}+I_{532}^{\text{pb}}+I_{356}^{\text{pb}}+I_{365}^{\text{pb}}\right)\biggr|_{\langle35\rangle\to0}=0.
\end{aligned}
\end{equation}
Finally, the hard function exhibits universal behavior in six collinear and six triple-collinear limits. Analogously to the previous sector, we impose only four triple collinear limits associated with the splittings in eq.~\eqref{eq:MHVsplittings} and disregard the remaining two, which involve the splitting ${-}\to{+}{-}{+}$. Taking all these constraints into account, we fix a unique solution to the bootstrap ansatz. In the third column of Table~\ref{tab:2}, we summarize the counting of independent constraints arising from each condition. Note that for all sectors the constraints from collinear limits are significantly more constraining than those of spurious poles. This highlights the importance of lower-point data in fixing the physical results at higher multiplicity.

\begin{table}[t]
\centering 
    \begin{tabular}{lccc}
    \hline\hline
Sectors & $({-}{-}{+}{+}{+}{+})$ & $({-}{+}{-}{+}{+}{+})$ & $({-}{+}{+}{-}{+}{+})$\\  
unknowns  & 238 & 872 & 582\\ \hline
{\bf Conditions } &\\ \hline
no spurious poles & \\ 
\hspace{1em} $  
\langle24\rangle=0$ & - & 403 &- \\
\hspace{1em} $ 
\langle25\rangle=0$ & - & 208 & 216 \\
\hspace{1em} $ 
\langle35\rangle=0$ & - &- & 205\\\hline

collinear limits
& \\
\hspace{1em} $p_6||p_5$ & 216 & 698 & 366\\
\hspace{1em} $p_5||p_4$ & 167 & 698 & 529\\
\hspace{1em} $p_4||p_3$ & 216 & 690 & 529\\
\hspace{1em} $p_3||p_2$ & 216 & 200 & 366\\
\hspace{1em} $p_2||p_1$ & -   & 200 & 529\\ 
\hspace{1em} $p_1||p_6$ & 216 & 690 & 529\\ \hline
triple collinear limits 
&  
\\ \hspace{1em} 
$p_6||p_5||p_4$ & 205 & 283 & 129\\ 
\hspace{1em} 
$p_5||p_4||p_3$ & 205 & 476 & - \\ \hspace{1em} 
$p_4||p_3||p_2$ & 205 & -& 129\\ \hspace{1em} 
$p_3||p_2||p_1$ & -  &  - & 129\\ \hspace{1em} 
$p_2||p_1||p_6$ & - & - & -\\ \hspace{1em} 
$p_1||p_6||p_5$ & 205 & 476& 129\\   \hline 
{\bf Total} & 238 &872 &582\\
\hline\hline
\end{tabular}

\caption{Counting of independent constraints arising from each physical condition in each MHV helicity sector.}
    \label{tab:2}
\vspace{-2ex}
\end{table}

\subsection{From pure YM to QCD hard functions}

The pure YM two-loop hard functions bootstrapped in the previous subsection correspond to the $N_f^0$-terms, $H_{n}^{[1],(2)}$, in the $N_f$-decomposition \eqref{eq:hardde} of gluonic QCD hard functions in the planar limit. We now turn to calculating the remaining two-loop $N_f^k$-terms with $k=1,2$ in eq.~\eqref{eq:hardde}.
Their prefactors are computed from on-shell diagrams containing quark loops. In light of the IR-decomposition \eqref{eq:main}, in which the IR-finite part is completely determined by the known two-loop prescriptive unitarity integrals, we need to introduce bootstrap ans\"atze only for the pure functions accompanying the one-loop prefactors \eqref{eq:1LSjk}.

At order $N_f^1$ in the helicity sector $({-}{-}{+}{+}{+}{+})$, the one-loop contribution vanishes, $H_{{-}{-}{+}{+}{+}{+}}^{[\frac12],(1)}=0$. As a result, all prefactors contributing at two loops are genuinely two-loop quantities and are computed from the double-box on-shell diagrams in eq.~\eqref{eq:dboxquark}. The six prefactors $S_{12,ij1}^{(2),\text{db}}$ (see \eqref{eq:Sij}) are accompanied by exactly the same double-box prescriptive integrals as in the $N_f^0$-term in \eqref{eq:ansatz0}. Consequently, this helicity sector is completely fixed by prescriptive unitarity,
\begin{equation}
    H_{{-}{-}{+}{+}{+}{+}}^{[\frac12]}=\sum_{2<i<j\leq6}S_{12,ij1}^{(2),\text{db}} I_{6,ij1}^{\text{db}}\,.
\end{equation}

In the helicity sectors $({-}{+}{-}{+}{+}{+})$ and $({-}{+}{+}{-}{+}{+})$ at order $N_f^1$, there are $12$ and $15$ prefactors, respectively, of which three and four already appear at one loop, see Tab.~\ref{tab:3}. These one-loop prefactors are precisely those whose accompanying pure functions need to be bootstrapped. A bootstrap procedure  employing ans\"atze analogous to the $N_f^0$-expressions \eqref{eq:ansatz2} and \eqref{eq:ansatz3} can be carried out, and unique symbol results can be obtained after imposing the physical constraints. However, as discussed below, we compute both sectors in a much simpler way, without resorting to the bootstrap.

To uncover relations between the $N_f^0$ and $N_f^1$ terms of the six-point hard functions, we first examine the five-point MHV sector $({-}{+}{-}{+}{+})$, using the two-loop results of Ref.~\cite{Abreu:2019odu}. The $N_f^0$-contribution, $H_{-+-++}^{[1]}$, is given in eq.~\eqref{eq:H5ptMHVpu}, and an analogous representation holds for the $N_f^1$-contribution, $H_{-+-++}^{[\frac12]}$. Each involves three genuinely two-loop prefactors, corresponding to the three two-loop maximal cut contours of the integrals in Fig.~\ref{fig:5ptdbpb}. In both cases, the accompanying pure functions are given by the same set of prescriptive integrals. In contrast to the $N_f^0$-contribution \eqref{eq:H5ptMHVpu}, the $N_f^1$-term does not involve the tree-level prefactor $\text{PT}_{5,13}$. Its remaining two prefactors are one-loop quantities, $S_{13,24}^{(1)}$ and $S_{13,25}^{(1)}$, defined in \eqref{eq:1LSjk}. At one loop, they are computed from one-mass box on-shell diagrams. We verify that the following simple and elegant relation holds among the associated transcendental functions,
\begin{equation}\label{eq:observation}
    f_{2,4}^{[1]}-f_{2,4}^{[\frac12]}=I_{24}^{\text{dp}}+\tilde{I}_{24}^{\text{dp}}+I_{42}^{\text{hb}}
\end{equation}
where the functions $f_{2,4}^{[1]}$ and $f_{2,4}^{[\frac12]}$ accompany the one-loop prefactors $R_{13,24}^{(1)}$ and $S_{13,24}^{(1)}$, respectively, contributing to $H_{-+-++}^{[1]}$ and $H_{-+-++}^{[\frac12]}$. The double-pentagon and hexa-box prescriptive integrals appearing in the rhs of \eqref{eq:observation} are defined in eqs.~\eqref{eq:Idp} and \eqref{eq:Ihb}, respectively, and $\tilde{I}_{24}^{\text{dp}}$ is obtained from $I_{24}^{\text{dp}}$ by the reflection $p_i{\to} p_{6-i}$. Note that the subscript of the hexa-box integral is 42 rather than 24, because according to the definition in \eqref{eq:Ihb}, this corresponds to a configuration where the two external legs $p_5$ and $p_1$ are on the triangle side of the diagram. Diagrammatically, the combination on the rhs of \eqref{eq:observation} can be represented as
\begin{align}
    \raisebox{-7ex}{\begin{tikzpicture}[scale=0.6]
        \draw[line width=1pt,black] (0,0)--(1.5,0)--(1.5,1.5)--(0,1.5)--(0,0);
         \draw[line width=1pt,black] (1.5,0)--(3,0)--(3,1.5)--(1.5,1.5)--(1.5,0);
          \draw[line width=1pt,black] (0,0)--(-0.5,-0.5);
            \draw[line width=1pt,black] (3,0)--(3.5,-0.5);
        \draw[line width=1pt,black] (0,2)--(0,1.5)--(-0.5,1.5);
         \draw[line width=1pt,black] (3.5,2)--(3,1.5);
        \draw[decorate, decoration=snake, segment length=12pt, segment amplitude=2pt, black,line width=1pt](0,0)--(3,1.5);
        \path[fill=black] (3,0) circle[radius=0.11];
          \path[fill=gray] (3,1.5) circle[radius=0.11];
        \path[fill=gray] (0,0) circle[radius=0.11];
        \path[fill=black] (0,1.5) circle[radius=0.11];
        \node[anchor=north east] at (-0.5,-0.5) {\small{$4^+$}};
          \node[anchor=south] at (0,2) {\small{$1^-$}};
             \node[anchor=east] at (-0.5,1.5) {\small{$5^+$}};
 \node[anchor=south west] at (3.5,2) {\small{$2^+$}};
\node[anchor=north west] at (3.5,-0.5) {\small{$3^-$}};
\end{tikzpicture}}
+\raisebox{-7ex}{\begin{tikzpicture}[scale=0.6]
        \draw[line width=1pt,black] (0,0)--(1.5,0)--(1.5,1.5)--(0,1.5)--(0,0);
         \draw[line width=1pt,black] (1.5,0)--(3,0)--(3,1.5)--(1.5,1.5)--(1.5,0);
          \draw[line width=1pt,black] (0,0)--(-0.5,-0.5);
            \draw[line width=1pt,black] (3,0)--(3.5,-0.5);
        \draw[line width=1pt,black] (0,2)--(0,1.5)--(-0.5,1.5);
         \draw[line width=1pt,black] (3.5,2)--(3,1.5);
         \draw[decorate, decoration=snake, segment length=12pt, segment amplitude=2pt, black,line width=1pt](0,0)--(3,1.5);
        \path[fill=black] (3,0) circle[radius=0.11];
          \path[fill=gray] (3,1.5) circle[radius=0.11];
        \path[fill=gray] (0,0) circle[radius=0.11];
        \path[fill=black] (0,1.5) circle[radius=0.11];
        \node[anchor=north east] at (-0.5,-0.5) {\small{$2^+$}};
          \node[anchor=south] at (0,2) {\small{$5^+$}};
             \node[anchor=east] at (-0.5,1.5) {\small{$1^-$}};
 \node[anchor=south west] at (3.5,2) {\small{$4^+$}};
\node[anchor=north west] at (3.5,-0.5) {\small{$3^-$}};
\end{tikzpicture}}+\raisebox{-8ex}{\begin{tikzpicture}[scale=0.6]
    \draw[line width=1pt,black] (0,0)--(1.5,0.675)--(0,1.5)--(0,0);
    \draw[line width=1pt,black] (2,-0.125)--(1.5,0.675);
    \draw[line width=1pt,black] (-2.5,0.75)--(-2,0.75);
     \draw[line width=1pt,black] (-1.2,-0.525)--(-1.2,-1);
      \draw[line width=1pt,black] (-1.2,2.025)--(-1.2,2.55);   
\draw[line width=1pt,black] (1.5,0.675)--(2,1.325);
    \draw[line width=1pt,black] (0,0)--(-1.2,-0.525)--(-2,0.75)--(-1.2,2.025)--(0,1.5);
        \draw[decorate, decoration=snake, segment length=12pt, segment amplitude=2pt, black,line width=1pt] (-1.2,-0.525)--(-1.2,2.025);
    \path[fill=gray] (-1.2,2.025) circle[radius=0.11];
     \path[fill=gray] (-1.2,-0.525) circle[radius=0.11];
    \path[fill=black] (1.5,0.675) circle[radius=0.11];
    \path[fill=black] (-2,0.75) circle[radius=0.11];
    \node[anchor=south west] at (2,1.325) {\small{$5^+$}};
    \node[anchor=south] at (-1.2,2.55){\small{$4^+$}};
     \node[anchor=north] at (-1.2,-1){\small{$2^+$}};
    \node[anchor=north west] at (2,-0.125){\small{$1^-$}};
        \node[anchor=east] at (-2.5,0.75){\small{$3^-$}};
\end{tikzpicture}}
\label{eq:dpdphb}
\end{align}
An analogous relation holds for the pure functions accompanying the remaining one-loop prefactors $R_{13,25}^{(1)}$ and $S_{13,25}^{(1)}$. The maximal-cut contours associated with the two-loop prescriptive integrals in \eqref{eq:dpdphb} yield the same residues, $R_{1i,jk}^{(1)}$ and $S_{1i,jk}^{(1)}$, as the one-mass box cut, see the discussion in subsection~\ref{sec:IRfinTerms}. Eq.~\eqref{eq:observation} therefore shows that the prescriptive double-pentagon and hexa-box integrals contribute to the $N_f^0$-term, but not to the $N_f^1$-term.

This discrepancy between $f_{j,k}^{[1]}$ and $f_{j,k}^{[\frac12]}$ can be understood from the large-$N_c$ scaling of the double-pentagon and hexa-box on-shell diagrams in the planar limit, following reasoning similar to that discussed at the end of subsection~\ref{sec:QCDls}. For instance, the double-pentagon on-shell diagrams with quarks circulating in one of the loops, which evaluate to the one-loop $N_f^1$-order prefactors $S_{1i,jk}^{(1)}$, are shown below

\begin{equation}
	\vcenter{\hbox{\begin{tikzpicture}[scale=0.7]
			
			\coordinate (d45) at (45:1);
			\coordinate (dm45) at (-45:1);
			\coordinate (d135) at (135:1);
			\coordinate (dm135) at (-135:1);
			
	  		\coordinate (p1) at (-1.2,0.6);
	  		\coordinate (p2) at (0,0.6);
	  		\coordinate (p3) at (1.2,0.6);
	  		\coordinate (p4) at (1.2,-0.6);
	  		\coordinate (p5) at (0,-0.6);
	  		\coordinate (p6) at (-1.2,-0.6);
	  		
	  		\coordinate (p1aux1) at ($(p1)+(-1,0)$);
	  		\coordinate (p1aux2) at ($(p1)+(d135)$);
	  		\coordinate (p1aux3) at ($(p1)+(0,1)$);
	  		\coordinate (p3aux3) at ($(p3)+(1,0)$);
	  		\coordinate (p3aux2) at ($(p3)+(d45)$);
	  		\coordinate (p3aux1) at ($(p3)+(0,1)$);
	  		\coordinate (p5aux2) at ($(p5)+(0,-1)$);
	  		\coordinate (p5aux1) at ($(p5)+(-60:1)$);
	  		\coordinate (p5aux3) at ($(p5)+(-120:1)$);
	  		\coordinate (p4aux1) at ($(p4) + (0:1)$);	  		
	  		\coordinate (p4aux3) at ($(p4)+(-90:1)$);

	  		\coordinate (p2aux) at ($(p2)+(0,1)$);
	  		\coordinate (p4aux) at ($(p4)+(dm45)$);
	  		\coordinate (p6aux) at ($(p6)+(dm135)$);
	  		
	  		\draw[very thick, ] (p1) to (p2);
	  		\draw[very thick,  ] (p2) to (p3);
	  		\draw[very thick, ] (p3) to (p4);
	  		\draw[very thick,] (p4) to (p5);
	  		\draw[very thick, ] (p5) to (p6);
	  		\draw[very thick, ] (p6) to (p1);
	  		\draw[very thick, dashed] (p2) to (p5);
	  		
        		\draw[very thick] (p1aux1) to (p1);
        		\draw[very thick] (p1aux3) to (p1);
        		
        		\draw[very thick,larrow] (p3aux2) to (p3);

        		
        		\draw[very thick,larrow] (p6aux) to (p6);
        		
        		\draw[very thick] (p4aux1) to (p4);
        		\draw[very thick] (p4aux3) to (p4);

        		\draw[fill, thick] (p1) circle (0.2);
        		\draw[thick] (p2) circle (0.2);
            \draw[fill=white, thick] (p3) circle (0.2);
 		    \draw[fill, thick] (p4) circle (0.2);
        		\draw[ thick] (p5) circle (0.2);
        		\draw[fill=white, thick] (p6) circle (0.2);

        		
        		\node at ($(p1aux1)+(-0.6,0)$) {\small $k+1$};
        		\node at ($(p1aux2)$) {$1^-$};
        		\node at ($(p1aux3)+(90:0.3)$) {\small$j-1$};
%
        		\node at ($(p3aux2)+(45:0.35)$) {$j$};
%
%
        		\node at ($(p4aux)$) {$i^-$};
        		\node at ($(p6aux)+(-135:0.35)$) {$k$};
        		
        		\node at ($(p4aux1) + (0:0.6)$) {\small $j+1$};        		
        		\node at ($(p4aux3) + (-90:0.3)$) {\small $k-1$};

        		\node[rotate=45] at ($(p1)!0.5!(p1aux2)$) {$\dots$};  
        		\node[rotate=45] at ($(p4)!0.5!(p4aux)$) {$\dots$};

        
    \end{tikzpicture}}},
    \vcenter{\hbox{\begin{tikzpicture}[scale=0.7]
			
			\coordinate (d45) at (45:1);
			\coordinate (dm45) at (-45:1);
			\coordinate (d135) at (135:1);
			\coordinate (dm135) at (-135:1);
			
	  		\coordinate (p1) at (-1.2,0.6);
	  		\coordinate (p2) at (0,0.6);
	  		\coordinate (p3) at (1.2,0.6);
	  		\coordinate (p4) at (1.2,-0.6);
	  		\coordinate (p5) at (0,-0.6);
	  		\coordinate (p6) at (-1.2,-0.6);
	  		
	  		\coordinate (p1aux1) at ($(p1)+(-1,0)$);
	  		\coordinate (p1aux2) at ($(p1)+(d135)$);
	  		\coordinate (p1aux3) at ($(p1)+(0,1)$);
	  		\coordinate (p3aux3) at ($(p3)+(1,0)$);
	  		\coordinate (p3aux2) at ($(p3)+(d45)$);
	  		\coordinate (p3aux1) at ($(p3)+(0,1)$);
	  		\coordinate (p5aux2) at ($(p5)+(0,-1)$);
	  		\coordinate (p5aux1) at ($(p5)+(-60:1)$);
	  		\coordinate (p5aux3) at ($(p5)+(-120:1)$);
	  		\coordinate (p4aux1) at ($(p4) + (0:1)$);	  		
	  		\coordinate (p4aux3) at ($(p4)+(-90:1)$);

	  		\coordinate (p2aux) at ($(p2)+(0,1)$);
	  		\coordinate (p4aux) at ($(p4)+(dm45)$);
	  		\coordinate (p6aux) at ($(p6)+(dm135)$);
	  		
	  		\draw[very thick, blue] (p1) to (p2);
	  		\draw[very thick,  blue] (p2) to (p3);
	  		\draw[very thick, blue] (p3) to (p4);
	  		\draw[very thick, blue] (p4) to (p5);
	  		\draw[very thick, blue] (p5) to (p6);
	  		\draw[very thick, blue] (p6) to (p1);
	  		\draw[very thick, dashed] (p2) to (p5);
	  		
        		\draw[very thick] (p1aux1) to (p1);
        		\draw[very thick] (p1aux3) to (p1);
        		
        		\draw[very thick,larrow] (p3aux2) to (p3);

        		
        		\draw[very thick,larrow] (p6aux) to (p6);
        		
        		\draw[very thick] (p4aux1) to (p4);
        		\draw[very thick] (p4aux3) to (p4);

        		\draw[fill, thick] (p1) circle (0.2);
        		\draw[thick] (p2) circle (0.2);
            \draw[fill=white, thick] (p3) circle (0.2);
 		    \draw[fill, thick] (p4) circle (0.2);
        		\draw[ thick] (p5) circle (0.2);
        		\draw[fill=white, thick] (p6) circle (0.2);

        		
        		\node at ($(p1aux1)+(-0.6,0)$) {\small $k+1$};
        		\node at ($(p1aux2)$) {$1^-$};
        		\node at ($(p1aux3)+(90:0.3)$) {\small$j-1$};
%
        		\node at ($(p3aux2)+(45:0.35)$) {$j$};
%
%
        		\node at ($(p4aux)$) {$i^-$};
        		\node at ($(p6aux)+(-135:0.35)$) {$k$};
        		
        		\node at ($(p4aux1) + (0:0.6)$) {\small $j+1$};        		
        		\node at ($(p4aux3) + (-90:0.3)$) {\small $k-1$};

        		\node[rotate=45] at ($(p1)!0.5!(p1aux2)$) {$\dots$};  
        		\node[rotate=45] at ($(p4)!0.5!(p4aux)$) {$\dots$};

        
    \end{tikzpicture}}},
\end{equation}
where two orientations of the quark loop are possible, together with their reflections. The second diagram, which contains a ``large" quark loop and scales as $\mathcal{O}(N_f)$, is suppressed in the planar limit compared to the first diagram, which scales as $\mathcal{O}(N_cN_f)$. An analogous scaling analysis applies to the hexa-box on-shell diagrams. By contrast, in the pure YM case all two-loop planar on-shell diagrams scale uniformly as ${\cal O}(N_c^2)$. Consequently, unlike the $N_f^1$ case, the $N_f^0$ terms receive additional contributions from purely gluonic on-shell diagrams associated with the double-pentagon and hexa-box contours shown in Fig.~\ref{fig:2Lcuts}, and the corresponding prescriptive unitarity integrals contribute. This explains the discrepancy observed in eq.~\eqref{eq:observation}.

Based on this observation regarding the large-$N_c$ scaling of on-shell diagrams with quark loops, we can immediately write down the $N_f^1$-contributions for the remaining two MHV helicity six-point sectors! For the sector $({-}{+}{-}{+}{+}{+})$, we omit the PT term from the pure YM expression \eqref{eq:ansatz2} and replace the one-loop $R^{(1)}$ and two-loop prefactors $R^{(2)}$ with their $N_f^1$ counterparts $S^{(1)}$ (see \eqref{eq:1LSjk}) and $S^{(2)}$, respectively. These prefactors are computed from on-shell diagrams containing a single quark loop. Moreover, from the pure functions $g_{2,4}$, $g_{2,5}$, and $g_{2,6}$ appearing in \eqref{eq:ansatz2}, we subtract the sum of double-pentagon and hexa-box prescriptive integrals, as discussed previously. In this way, we obtain
\begin{equation}\label{eq:ansatz4}
\begin{aligned}
    H_{{-}{+}{-}{+}{+}{+}}^{[\frac12]}&=\sum_{i=4,5,6}S_{13,2i}^{(1)}(g_{2,i}-I_{2i}^{\text{dp}}-\tilde{I}_{2i}^{\text{dp}}-I_{2i}^{\text{hb}}-I_{i2}^{\text{hb}})\\ 
&+\sum_{\substack{l=1,2\\3<j<k\leq6}}S_{13,jkl}^{(2),\text{db}}I_{6,jkl}^{\text{db}}+\sum_{\{j,k,l\}\in\sigma_1}S_{13,jkl}^{(2),\text{pb}}(I^{\text{pb}}_{6,jkl}+I^{\text{pb}}_{6,jlk})
\end{aligned}
\end{equation}
where $\tilde{I}_{2i}^{\text{dp}}$ denotes the reflection $p_k{\to} p_{2{+}i-k}$ of ${I}_{2i}^{\text{dp}}$, and where $I_{24}^{\text{hb}}=I_{62}^{\text{hb}}=0$. Remarkably, the resulting expression \eqref{eq:ansatz4} satisfies all physical constraints that would arise in the bootstrap analysis, including the dihedral symmetry, spurious pole cancellation, and the correct behavior in the multi-collinear and soft limits, {\it etc.}, thereby providing a strong consistency check.

An analogous procedure also applies to the $({-}{+}{+}{-}{+}{+})$ sector and allows us to obtain the $N_f^1$ contribution from the corresponding $N_f^0$ expression \eqref{eq:ansatz3},
\begin{equation}\label{eq:ansatz5}
\begin{aligned}
    H_{{-}{+}{+}{-}{+}{+}}^{[\frac12]}&=\sum_{j=2,3,\ i=5,6}S_{14,ij}^{(1)}(h_{i,j}-I_{ij}^{\text{dp}}-\tilde{I}_{ij}^{\text{dp}}-I_{ij}^{\text{hb}}-I_{ij}^{\text{hb}})+S_{14,2356}^{(2),\text{kb}}I_{6,2356}^{\text{kb}}\\ 
&+\sum_{\substack{1\leq l\leq3}}S_{14,45l}^{(2),\text{db}}I_{6,45l}^{\text{db}}+\sum_{\substack{4\leq l\leq6}}S_{41,23l}^{(2),\text{db}}I_{6,23l}^{\text{db}}+\sum_{\{j,k,l\}\in\sigma_2}S_{14,jkl}^{(2),\text{pb}}(I^{\text{pb}}_{6,jkl}+I^{\text{pb}}_{6,jlk})
\end{aligned}
\end{equation}
where the pure functions $h_{i,j}$ appear in \eqref{eq:ansatz3}, and $\tilde{I}_{ij}^{\text{dp}}$ denotes the reflection $p_k{\to} p_{i{+}j-k}$ of ${I}_{ij}^{\text{dp}}$.

Finally, as discussed at the end of subsection~\ref{sec:QCDls}, only the MHV sector $({-}{+}{+}{-}{+}{+})$ receives an $N_f^2$-contribution at maximal transcendentality, originating from the kissing-box cut. The corresponding pure functions are given by the kissing-box prescriptive integrals,
\begin{equation}\label{eq:ansatz6}
    H_{{-}{+}{+}{-}{+}{+}}^{[\frac12,\frac12]}=T_{14,2356}^{(2),\text{kb}}I_{6,2356}^{\text{kb}}
\end{equation}
where the prefactors $T_{14,2356}^{(2),\text{kb}}$ are computed from the kissing-box on-shell diagrams with quarks propagating in both loops. 

We have therefore determined all maximally-transcendental contributions to the two-loop planar QCD hard functions for six-gluon MHV amplitudes. The corresponding symbol expressions are collected in the ancillary files.

\subsection{Six-point two-loop MHV symbol alphabet}

By considering dihedral permutations of the bootstrapped results, we identify the complete set of symbol letters appearing in the maximally-transcendental part of the MHV gluonic QCD hard functions. Our main finding is that this symbol alphabet coincides with the set of 137 letters that have appeared in previous six-point two-loop gauge-theory calculations \cite{Carrolo:2025pue,Chicherin:2025cua,Carrolo:2025agz}. There $87$ parity-even letters, consisting of $39$ one-loop letters \cite{Henn:2022ydo}   
\begin{align}
   \{W_1,\cdots, W_9,W_{16},\cdots,W_{33},W_{46},\cdots,W_{51},W_{88},\cdots,W_{93}\} \,.
\end{align} 
and $48$ genuine two-loop letters
\begin{align}
    \{W_{10},\cdots,W_{15}, W_{34},\cdots,W_{45},W_{58},\cdots,W_{69},W_{76},\cdots,W_{87},W_{100},\cdots,W_{105}\} \,. \label{eq:oddfrom}
\end{align}
Furthermore, there are $50$ parity-odd letters, which is a combination of $36$ odd letters with $\eps(i,j,k,l)$
\begin{align}
    \{W_{182},\cdots,W_{190},W_{194},\cdots, W_{211},W_{218},\cdots,W_{220},W_{242},\cdots,W_{247} \}\,.
\end{align}
and $14$ letters involve three-mass-triangle roots
\begin{align} \label{eq:nonrational}
        \{W_{157},\cdots,W_{166},W_{275},\cdots,W_{278}\} \,.
            \end{align}
This reduces the alphabet from 167 symbol letters present in the finite parts of six-point two-loop Feynman integrals. They are contributed by the five integral sectors, as listed in \cite{Carrolo:2025pue}.

We can group the 30 symbol letters that drop out from the full set of 167 into three types. First, the individual three-mass-triangle roots or parity-odd Lorentz invariants $\varepsilon(i,j,k,l)$ cannot appear as individual letters. This includes
\begin{equation}
    \{W_{118},W_{119},W_{123},\dots,W_{128},W_{138}\}
\end{equation}
Second, the odd letters from the roots of the two-mass-easy double triangle 
\begin{equation}
r_3=\sqrt{s_{12}^2+s_{45}^2+s_{36}^2-2s_{12}s_{45}-2s_{12}s_{36}-2s_{45}s_{36}}\, , 
\end{equation} 
and its two cyclic images $r_4,r_5$. These include
\begin{equation}
    \{W_{167},\dots,W_{175}\}
\end{equation}
Finally, there is one more dihedral orbit 
\begin{equation}
    \{W_{52},\dots,W_{57},W_{70},\dots,W_{75}\}
\end{equation}
coming from the leading singularity of the box-triangle integral with two-mass-easy kinematics. The second and third groups of letters are genuine two-loop letters. For illustration purposes, we represent the two cuts from the double-triangle and box-triangle topologies in Fig.~\ref{fig:missedsing}, whose leading Landau singularities give $r_3$ and $W_{52}$, respectively. Each of the cuts is evaluated in $\ell_i\to\infty$, and thus both of them should be viewed as composite leading cuts \cite{Arkani-Hamed:2010pyv}. It is easy to see that the double-triangle cut has NMHV helicity counting, hence cannot contribute to our results. The box-triangle cut, on the other hand, has MHV helicity counting and thus it should naively contribute. From this cut and its corresponding on-shell diagrams, we find prefactors present in the final result. Therefore, one cannot exclude the possibility that the orbit from $W_{52}$ yields a contribution to some MHV sectors on the basis of a cut analysis.

\begin{figure}[t]
    \centering
\begin{tikzpicture}[scale=0.8]
        \draw[line width=1pt,black] (0,0.75)--(1.5,0)--(1.5,1.5)--(0,0.75);
         \draw[line width=1pt,black] (1.5,0)--(3,0.75)--(1.5,1.5)--(1.5,0);
          \draw[line width=1pt,black] (-0.5,0.5)--(0,0.75)--(-0.5,1);
        \draw[line width=1pt,black] (3.5,0.5)--(3,0.75)--(3.5,1);
        \draw[line width=1pt,black] (1.5,1.5)--(1.5,2);
                \draw[line width=1pt,black] (1.5,0)--(1.5,-0.5);
          \path[fill=black] (3,0.75) circle[radius=0.15];
        \path[fill=black] (0,0.75) circle[radius=0.15];
        \path[fill=black] (1.5,0) circle[radius=0.15];
        \path[fill=black] (1.5,1.5) circle[radius=0.15];
        \node[anchor=north] at (1.5,-0.5) {\small{$6$}};
          \node[anchor=south east] at (-0.5,1) {\small{$2$}};
             \node[anchor=north east] at (-0.5,0.5) {\small{$1$}};
 \node[anchor=south west] at (3.5,1) {\small{$4$}};
\node[anchor=north west] at (3.5,0.5) {\small{$5$}};
\node[anchor=south] at (1.5,2) {\small{$3$}};
\node[anchor=north] at (2,1) {\small{$\infty$}};
\node[anchor=north] at (1,1) {\small{$\infty$}};
\end{tikzpicture}
\quad
    \begin{tikzpicture}[scale=0.8]
        \draw[line width=1pt,black] (0,0.75)--(1.5,0)--(1.5,1.5)--(0,0.75);
         \draw[line width=1pt,black] (1.5,0)--(3,0)--(3,1.5)--(1.5,1.5)--(1.5,0);
          \draw[line width=1pt,black] (-0.5,0.5)--(0,0.75)--(-0.5,1);
            \draw[line width=1pt,black] (3.5,0)--(3,0)--(3,-0.5);
        \draw[line width=1pt,black] (1.5,1.5)--(1.5,2);
         \draw[line width=1pt,black] (3.5,2)--(3,1.5);
        \path[fill=black] (3,0) circle[radius=0.15];
          \path[fill=gray] (3,1.5) circle[radius=0.15];
        \path[fill=black] (0,0.75) circle[radius=0.15];
        \path[fill=gray] (1.5,0) circle[radius=0.15];
        \path[fill=black] (1.5,1.5) circle[radius=0.15];
        \node[anchor=west] at (3.5,0) {\small{$5$}};
          \node[anchor=south east] at (-0.5,1) {\small{$2$}};
             \node[anchor=north east] at (-0.5,0.5) {\small{$1$}};
 \node[anchor=south west] at (3.5,2) {\small{$4$}};
  \node[anchor=north] at (1,1) {\small{$\infty$}};
\node[anchor=north] at (3,-0.5) {\small{$6$}};
\node[anchor=south] at (1.5,2) {\small{$3$}};
\end{tikzpicture}
    \caption{Two cuts giving the missed two-loop letters}
    \label{fig:missedsing}
\end{figure}

However, as discussed in Section~\ref{sec: Prescriptive unitarity integral}, the transcendental functions contributing to the hard functions have a clear origin in prescriptive unitarity integrals. This observation suggests that hard functions for quark scattering should likewise be closely related to prescriptive integrals, albeit with different prefactors. These results strongly indicate that, for any six-parton two-loop massless MHV scattering process in QCD, the underlying symbol alphabet is still universally given by these 137 letters. We expect that a more systematic study of the six-parton scattering amplitudes, including contributions of non-maximal transcendentality, will further confirm this conjecture.

\subsection{From pure YM to $\mathcal{N}{=}4$ sYM hard functions}
\label{sec:MT}
We now discuss an interesting relation between the six-point hard functions in pure YM and in $\mathcal{N}{=}4$ sYM.

Based on our previous discussion of the prescriptive unitarity basis and the maximal weight projection of loop integrands, the four-dimensional amplitude integrands of pure YM and $\mathcal{N}{=}4$ sYM can be expanded in the same basis of prescriptive integrals, while the prefactors encode the theory dependence. In $\mathcal{N}{=}4$ sYM, all prefactors of MHV amplitudes are given by the supersymmetric PT factor, which we simply set to unity in this subsection. Consequently, by replacing the nontrivial prefactors appearing in the maximal weight projection of the pure YM integrand \eqref{eq:main} with rational numbers, one recovers the corresponding four-dimensional integrand of the ${\cal N} =4$ sYM amplitude.

Remarkably, our bootstrapped results show that this relation persists at the level of integrated quantities. In the discussion of the finite part of the six-point MHV amplitude in $\mathcal{N}{=}4$ sYM, one usually introduces the remainder function \cite{Drummond:2008aq,Bern:2008ap}, defined as the dual conformal invariant ratio of the amplitude to the BDS ansatz \cite{Bern:2005iz}. In order to draw a direct parallel between pure YM theory and ${\cal N}=4$ sYM, however, we instead adopt an identical IR subtraction scheme for both gauge theories (see section~\ref{sec:IR}). It is worth noting that the hard function in ${\cal N}=4$ sYM differs from the remainder function by the exponentiation of the one-loop hard function. With this choice of IR subtraction, the hard function of $\mathcal{N}{=}4$ sYM can be obtained from the maximally-transcendental part of the corresponding six-point MHV hard function in pure YM theory by replacing all of its prefactors with rational numbers. 

At one loop, consider the hard function for an arbitrary MHV helicity configuration $(h_1,\ldots,h_6)$. Upon replacing all prefactors $\text{PT}_{n,1i}$ and $R_{1i,jk}^{(1)}$ by $1$, one recovers the one-loop hard function of ${\cal N} =4$ sYM,
\begin{align}
 H^{(1)}_{h_1 \ldots h_6} =\text{PT}_{n,1i}f_0^{(1)}{+} \sum_{i} R_{1i,jk}^{(1)} I_{j,k} \xrightarrow{\text{PT}_{n,1i},\ \ R_{1i,jk}^{(1)} \to 1 }\  \left( H^{(1)}_6 \right)^{\text{${\cal N}=4$ sYM}} \,.
\end{align}
In particular, in the split-helicity sector $({-}{-}{+}{+}{+}{+})$, the maximally-transcendental part of the one-loop hard function in pure YM theory coincides exactly with the hard function in $\mathcal{N}{=}4$ sYM theory.

At two loops, this relation continues to hold in an analogous manner. Since amplitude integrands in $\mathcal{N}{=}4$ sYM have no poles at infinity, double-box prescriptive integrals do not contribute to sYM amplitudes. For the bootstrapped result \eqref{eq:ansatz0} in the split-helicity sector $({-}{-}{+}{+}{+}{+})$, we verify that 
\begin{align}
f_0=\  \left( H^{(2)}_6 \right)^{\text{${\cal N}=4$ sYM}} \,.
\end{align}
In other words, the two-loop hard functions in the two theories coincide upon setting  $R_{12,ij1}^{(2),\text{db}}\to 0$ in \eqref{eq:ansatz0}. For the remaining two MHV helicity sectors, \eqref{eq:ansatz2} and \eqref{eq:ansatz3}, double-pentagon and kissing-box prescriptive integrals also contribute. In these cases, the supersymmetric hard function is obtained from the pure YM hard functions by discarding the double-box contributions, $R^{\text{db}}_{1i,jkl}\to0$, while replacing all remaining prefactors $R\to1$. In a word, for any helicity sector $(h_1,h_2,\cdots,h_6)$ at two loops, we have
\begin{tcolorbox}
\begin{equation}
\begin{aligned}
    H_{h_1\cdots h_6}^{(2)}\xrightarrow[R^{(2),\text{db}}\to0]{\text{PT},\ \ R^{(1)},\ \ R^{(2),\text{pb}},\ \ R^{(2),\text{kb}}\to 1}\left( H^{(2)}_6 \right)^{\text{${\cal N}=4$ sYM}}.\\
\end{aligned}
\end{equation}
\end{tcolorbox}

This identification can be viewed as a manifestation of the principle of maximal transcendentality previously observed in various QCD calculations, as mentioned in the Introduction. We expect that, in $D=4{-}2\epsilon$ dimensions, there also exists a basis of prescriptive unitarity integrals. These Feynman integrals are uniformly transcendental in dimensional regularization and therefore contribute to the maximally-transcendental part of gauge-theory amplitudes.  All theory-specific information is encoded in the corresponding prefactors, which are computed as $D$-dimensional leading singularities. Upon restricting to four dimensions, this putative basis for the MHV helicity sectors reduces to the two-loop prescriptive unitarity integrals of Refs.~\cite{Bourjaily:2019gqu,Bourjaily:2019iqr}. We leave a detailed investigation of this perspective to future work.

\newpage

\section{Multi-collinear and multi-soft limits and splitting functions}
\label{sec: behaviors}

In this section we review the behavior of gluonic QCD scattering amplitudes and their associated hard functions in (multi-)collinear and (multi-)soft limits. In these kinematic regimes, amplitudes factorize in a universal manner governed by splitting and soft functions. This universality makes these limits powerful tools both as physical constraints and as consistency checks in the amplitude bootstrap. Moreover, information extracted from low-multiplicity amplitudes in such limits can be systematically repurposed to constrain higher-multiplicity amplitudes in the same singular kinematic regimes. In section~\ref{sec: Bootstrapping six-gluon QCD amplitudes}, we employed the collinear limit as a physical constraint in the symbol bootstrap of the six-point two-loop MHV hard functions in QCD. In this context, the collinear limit provides inhomogeneous bootstrap input by relating the sought-after six-point hard functions to the previously known five-point ones. Conversely, by analyzing the singular limits of the bootstrapped hard functions, we extract new results for the gluonic two-loop triple-collinear splitting and double-soft functions. 

Throughout this work, we restrict to the large-$N_c$ limit. All expressions are understood to refer to their maximally-transcendental contributions and are considered at the level of symbols. In particular, all transcendental constants are systematically omitted.

\subsection{Review of collinear limit}

For completeness, we briefly review the collinear factorization of QCD amplitudes and their corresponding hard functions. We consider the collinear limit in which a pair of adjacent light-like momenta in an $(n+1)$-point amplitude, say $p_n$ and $p_{n+1}$, satisfy
\begin{align}
p_n \to z p_{\bar{n}} \,,\qquad
p_{n+1} \to (1-z) p_{\bar{n}}
\end{align}
where $0<z<1$ is the splitting fraction. In this limit, $s_{n\,n+1} \to 0$, and the combined momentum becomes light-like, $p^2_{\bar n} = 0$, with $p_n+p_{n+1} \to p_{\bar{n}}$.

When a gauge theory amplitude develops a collinear singularity, its leading contribution in this kinematic limit is controlled by long-distance dynamics and exhibits a universal factorized structure. For color-ordered planar amplitudes, loop corrections obey the well-known collinear factorization formula~\cite{Kosower:1999xi},
\begin{align}
A^{(L)}_{n+1} (p_1^{h_1},\ldots,p_n^{h_n},p_{n+1}^{h_{n+1}})  \to \sum_{\ell =0}^{L} \sum_{h=\pm} {\rm Split}^{(\ell)}_{-h} (z,p_n^{h_n},p_{n+1}^{h_{n+1}}) \, A^{(L-\ell)}_{n} (p_1^{h_1},\ldots,p_{n-1}^{h_{n-1}}, p_{\bar{n}}^{h}).\label{eq:Acoll}
\end{align}
The associated splitting functions carry nontrivial helicity dependence and decompose as follows
\begin{align} \label{eq:split1->2}
{\rm Split}^{(\ell)}_{-h} (z,p_n^{h_n},p_{n+1}^{h_{n+1}}) = {\rm Split}^{(0)}_{-h} (z,p_n^{h_n},p_{n+1}^{h_{n+1}}) \, r^{(\ell)\, h_n \, h_{n+1}}_{-h} (z,s_{n\,n+1},\ep) 
\end{align}
whenever the tree-level splitting function, ${\rm Split}^{(0)}$, does not vanish. The tree-level splitting function contains a simple pole at $s_{n\,n+1} \to 0$, while its loop corrections $r^{(\ell)}$ give rise to logarithmic singularities in the collinear limit. 

The tree-level splitting functions for the helicity configurations ${-} \to {+}{+}$ and ${+}\to {-}{-}$ vanish, and the corresponding loop corrections $r^{(\ell)}$ are of subleading transcendentality. Likewise, in the collinear limits of MHV amplitudes, the splittings ${-} \to {-}{-}$, ${+} \to {+}{-}$, and ${+} \to {-}{+}$ can be neglected, as they appear in the factorization formula \eqref{eq:Acoll} multiplied by single-minus amplitudes. These amplitudes also vanish at tree-level, and their loop corrections are again of subleading transcendendentality. Consequently, for the purposes of this work, we discard these splittings and retain only the remaining helicity configurations,
\begin{align} 
& h \to h_{n} h_{n+1} \,: \quad {+} \to  {+} {+} \,, \quad
{-} \to  {+}{-} \,, \quad
{-} \to  {-}{+} \,.  \label{eq:collsplitconf}
\end{align} 
In particular, only a single intermediate helicity state $h= \pm1$ contributes to the sum on the rhs of eq.~\eqref{eq:Acoll}. For example, at tree level the collinear factorization reduces to
\begin{align}
A^{(0)}_{n+1} \to {\rm Split}^{(0)\, h_n h_{n+1}}_{-h} \, A^{(0)}_{n} \,,
\end{align}
where the allowed helicity configurations are those listed in \eqref{eq:collsplitconf}.

The planar loop corrections $r^{(\ell)}$ for splitting in QCD depend on the helicity configurations in \eqref{eq:collsplitconf}. However, these discrepancies come in the form of subleading transcendentality contributions \cite{Bern:2004cz}. Their maximally-transcendental part is given by the splitting function of ${\cal N}=4$ sYM amplitudes. At one loop, according to~\cite{Kosower:1999rx,Bern:1994zx},  
\begin{align}
r^{(1)}(z,s,\ep) = -\frac{1}{\ep^2} \left( \frac{(-s) z (1-z)}{\mu^2} \right)^{-\ep} + 2 \log(z) \log(1-z)+ O(\ep)
\end{align}
and the two-loop splitting function is given by the ABDK/BDS exponentiation~\cite{Anastasiou:2003kj,Bern:2005iz},
\begin{align}
r^{(2)} = \frac{1}{2} \left( r^{(1)} \right)^2 +O(\ep) \label{eq:splitexp}
\end{align}
where we systematically ignore all transcendental constants. The quark flavor contribution also come in as subleading transcendentalities, so we can ignore $N_f$-dependence of the splitting functions. 

In our symbol bootstrap, the collinear limits of hard functions are employed, which entails subtracting IR divergences from the collinear factorization of amplitudes in \eqref{eq:Acoll}. For maximally-transcendental parts of one- and two-loop MHV hard functions, normalized by their respective tree-level amplitudes, the collinear factorization takes the following simple form,
\begin{align}
& \left( H^{(1)}_{n+1}/A^{(0)}_{n+1} \right) \to  \left( H^{(1)}_{n} / A^{(0)}_{n} \right) + {\cal C}^{(1)}  \,, \notag \\
& \left( H^{(2)}_{n+1} / 
A^{(0)}_{n+1} \right) \to \left(  H^{(2)}_{n} / A^{(0)}_{n}  \right) + {\cal C}^{(1)} \left( H^{(1)}_{n} / A^{(0)}_{n}  \right) + \frac{1}{2} \left( {\cal C}^{(1)} \right)^2 \,, \label{eq:HAcoll}
\end{align}
which involves only four-dimensional quantities and is independent of the helicity splitting configurations listed in \eqref{eq:collsplitconf}. The collinear poles cancel explicitly in the ratios entering \eqref{eq:HAcoll}, while  logarithmically divergent and  finite terms are retained. The one-loop IR-subtracted splitting function is defined in terms of the IR subtraction function \eqref{eq:I61} as~\cite{Bern:2004cz},
\begin{align}
{\cal C}^{(1)} := I^{(1)}_n + r^{(1)} - \underset{p_n || p_{n+1}}{\lim}\, I^{(1)}_{n+1}  \,,
\end{align}
which is IR finite and independent of the multiplicity. Neglecting  ${\cal O}(\ep)$ terms, it takes the explicit form,
\begin{align}
{\cal C}^{(1)}= \log(z) \log(1-z) + \log(z) \log\left(\frac{p_{n-1}\cdot p_{\bar{n}}}{p_{n} \cdot p_{n+1}} \right) + \log(1-z) \log\left(\frac{p_{1}\cdot p_{\bar{n}}}{p_{n} \cdot p_{n+1}} \right).
\label{eq:C1}
\end{align}
Let us note that ${\cal C}^{(1)}$ diverges logarithmically in the collinear limit. The exponentiation of loop corrections of the IR-subtracted splitting function in \eqref{eq:HAcoll} is inherited from that in eqs.~\eqref{eq:I62} and \eqref{eq:splitexp}.

We emphasize that only hard functions which are singular in the collinear regime admit the universal collinear factorization in eq.~\eqref{eq:HAcoll}. For instance, the
MHV hard function $({+}{+}\ldots {+}{-}{-})$ is finite and vanishes at tree level in the limit $p_n || p_{n+1}$.

Finally, although the maximally-transcendental parts of the splitting functions are helicity independent, the hard functions themselves retain a non-trivial dependence on the helicity configuration. The collinear limits imposed in the symbol bootstrap are summarized in Tab.~\ref{tab:2}. We have also verified that the bootstrapped symbols of both pure YM and QCD hard functions satisfy all allowed collinear limits.


\subsection{Review of triple-collinear limits}

We now turn to the triple-collinear limit~\cite{Catani:2003vu}. For six-point or higher-multiplicity scattering amplitudes, the kinematics of this limit is non-degenerate. In what follows, we therefore specialize to six-particle scattering. In the triple collinear limit $p_4 || p_5 || p_6$, the momenta of three external particles simultaneously approach a common light-like direction $p_{\bar{4}}$ at the same rate,
\begin{align}
p_4 \to z_4 p_{\bar{4}} \,,\qquad
p_5 \to z_5 p_{\bar{4}} \,,\qquad
p_6 \to z_6 p_{\bar{4}} \,,\qquad
\end{align}
such that $p_4 + p_5 + p_6 \to p_{\bar{4}} $. The momentum fractions $z_4, z_5, z_6$ parameterize the splitting, 
\begin{align}
z_4 + z_5 + z_6 = 1 \,,\qquad 0<z_i <1 \,.
\end{align}
The triple collinear limit is controlled by a parameter $\delta \to 0$, with Mandelstam variables involving only the collinear momenta scaling homogeneously,
\begin{align}
s_{45} \sim s_{56} \sim s_{46} \sim s_{456} \sim \delta^2\,.
\end{align}

At tree level, scattering amplitudes develop a triple-collinear pole $1/\delta^2$, which receives logarithmic corrections at loop level. Provided that the color-ordered loop amplitude is singular in this limit, its most singular contribution assumes a universal form,
\begin{align}
A^{(L)} _{6} (p_1^{h_1},\ldots,p_6^{h_6})  \to \sum_{\ell =0}^{L} \sum_{h=\pm} {\rm Split}^{(\ell)}_{-h} (p_4^{h_5},p_{5}^{h_{5}},p_{6}^{h_{6}}) \, A^{(L-\ell)}_{4} (p_1^{h_1},p_2^{h_2},p_3^{h_3},\bar{p}_{4}^{h}) \label{eq:AmplTriple}
\end{align}
where less singular non-universal terms have been omitted on the rhs. In this limit, the kinematics factorizes. The reduced four-point amplitude $A_4$ depends on $s_{12}$ and $s_{23}$, while the triple splitting functions depend on $s_{45}/s_{456},s_{56}/s_{456},z_4,z_6$. Since we restrict to the maximally-transcendental parts of six-point MHV helicity amplitudes, and since four-point all-plus and single-minus helicity amplitudes contribute at subleading transcendentality, only the following triple-gluonic splitting channels are relevant,
\begin{equation}\label{eq:split1->3}
h\to h_4h_5h_6 \,:  \quad    {+}\to {+}{+}{+},\ \  {-}\to{-}{+}{+},\ {-}\to{+}{+}{-},\ \ {-}\to {+}{-}{+} \,.
\end{equation}
In this case, a single intermediate helicity state $h=\pm 1$ contributes to the sum in \eqref{eq:AmplTriple}. The second and third splitting configurations in \eqref{eq:split1->3} are related by reflection symmetry,
\begin{align}
{\rm Split}^{(\ell)}_{-h} (p_4^{h_5},p_{5}^{h_{5}},p_{6}^{h_{6}}) = {\rm Split}^{(\ell)}_{-h} (p_{6}^{h_{6}},p_{5}^{h_{5}},p_4^{h_5}) \,. \label{eq:tripReflSym}
\end{align}

The triple-collinear splitting functions carry a helicity dependence. We focus on their one- and two-loop corrections, which we find convenient to represent in the following form, since the tree-level splitting is nonvanishing for the helicity configurations listed in \eqref{eq:split1->3}, 
\begin{align}
& {\rm Split}^{(1)}_{-h} (p_4^{h_5},p_{5}^{h_{5}},p_{6}^{h_{6}}) = {\rm Split}^{(0)}_{-h} (p_4^{h_5},p_{5}^{h_{5}},p_{6}^{h_{6}}) \left( V_0 + V^{(1) -h }_{h_4 h_5 h_6} \right) ,\label{eq:trip1L}\\
& {\rm Split}^{(2)}_{-h} (p_4^{h_5},p_{5}^{h_{5}},p_{6}^{h_{6}}) = {\rm Split}^{(0)}_{-h} (p_4^{h_5},p_{5}^{h_{5}},p_{6}^{h_{6}}) \left( \frac{1}{2} \left( V_0 + V^{(1) -h}_{ h_4 h_5 h_6} \right)^2 + V^{(2)-h}_{ h_4 h_5 h_6}  \right) \label{eq:trip2L} .
\end{align}
In contrast to the two-particle splittings in \eqref{eq:split1->2}, three-particle splitting functions exhibit a significantly more intricate kinematic dependence, and comparatively little is known about them beyond one loop. Even at leading color and maximal transcendentality, the QCD triple-collinear splitting functions depend on the helicity configuration $h\to h_4 h_5 h_6$ as well as on the number of quark flavors $N_f$. Moreover, they do not obey a simple exponentiation of the one-loop result as in \eqref{eq:splitexp}. In particular, the function $V^{(2)}$ in \eqref{eq:trip2L} is genuinely nontrivial. As an illustration, the six-gluon two-loop  amplitude in ${\cal N}=4$ sYM involves a non-vanishing remainder function, which corrects the one-loop BDS exponentiation of the triple-collinear splitting and can be identified with $V^{(2)}$ in this theory \cite{Bern:2008ap}.

We now discuss the one- and two-loop functions $V_0$, $V^{(1)}$, and $V^{(2)}$, which are free of spinor-helicity variables and appear in the triple-collinear splitting functions \eqref{eq:trip1L} and \eqref{eq:trip2L}. The function $V_0$ is helicity independent and captures all IR divergences of the splitting function \cite{Catani:2003vu},
\begin{align}
V_0 = -\frac{1}{\ep^2} \left( \left( \frac{-s_{45}}{\mu^2} \right)^{-\ep} + \left( \frac{-s_{56}}{\mu^2} \right)^{-\ep} + \left( \frac{-s_{456}}{\mu^2} \right)^{-\ep} \left( z_4^{-\ep} + z_6^{-\ep} - 2\right)\right) . \label{eq:V0}
\end{align} 
The IR-finite part of the one-loop splitting in \eqref{eq:trip1L}, $V^{(1) -h}_{h_4 h_5 h_6}$, depends on the helicity configuration. Its maximally-transcendental part for the first three helicity configurations listed in \eqref{eq:split1->3} is given by the following pure function~\cite{Badger:2015cxa},
\begin{align}
& V^{(1) {-}}_{{+}{+}{+}} = V^{(1) {+}}_{{-}{+}{+}} = V^{(1) {+}}_{{+}{+}{-}} = \frac{1}{2} \left[ \log^2(z_4) + \log^2(z_6) \right] -\log\left( \frac{s_{45}}{s_{456}} \right) \log\left( \frac{s_{56}}{s_{456}} \right) \notag\\
& +\log\left( \frac{1-z_6}{z_4}\right) \log\left( \frac{s_{45}}{s_{456}} \right) 
+\log\left( \frac{1-z_4}{z_6}\right) \log\left( \frac{s_{56}}{s_{456}} \right) + {\rm Li}_2 \left( - \frac{z_5}{z_4} \right) + {\rm Li}_2 \left( - \frac{z_5}{z_6} \right) \notag \\
& + {\rm Li}_2 \left( - \frac{z_4}{1-z_4} \right) + {\rm Li}_2 \left( - \frac{z_6}{1-z_6} \right) - {\rm Li}_2 \left( 1 - \frac{s_{45}}{(1-z_6) s_{456}} \right) - {\rm Li}_2 \left( 1 - \frac{s_{56}}{(1-z_4) s_{456}} \right) ,
\label{eq:tcSplit1L+++}
\end{align}   
where all transcendental constants has been omitted. The remaining helicity configuration in \eqref{eq:split1->3} instead receives a non-pure contribution~\cite{Badger:2015cxa},
\begin{align}
& V^{(1){+}}_{{+}{-}{+}} = V^{(1) {-}}_{{+}{+}{+}} +  \left( r^{[1]-}_{{+}{-}{+}} +\frac{N_f}{N_c} r^{[\frac12]-}_{{+}{-}{+}} \right) \left( \frac12 \log^2\left( \frac{s_{45}}{s_{56}}  \right) + {\rm Li}_2 \left( 1 - \frac{s_{456}}{s_{45}} \right)
+ {\rm Li}_2 \left( 1 - \frac{s_{456}}{s_{56}} \right) \right) ,\label{eq:tcSplit1L+-+} 
\end{align}
associated with the following prefactors, which can be obtained as triple-collinear limits of the six-point one-loop prefactors \eqref{eq:L1LS} and \eqref{eq:1LSjk},
\begin{align}
r^{[1]-}_{{+}{-}{+}} := -1 + \lim_{p_4 || p_5 || p_6}  \frac{R^{(1)}_{i5,46,6}}{\text{PT}_{6,i5}} \,,\qquad
r^{[\frac12]-}_{{+}{-}{+}} := \lim_{p_4 || p_5 || p_6}  \frac{S^{(1)}_{i5,46,6}}{\text{PT}_{6,i5}} \label{eq:r+-+def}
\end{align}
where $i=1,2,3$, and the result is independent of the choice of $i$. They multiply by the pure function in \eqref{eq:tcSplit1L+-+} given by the finite one-mass box $I_{6,46}$ \eqref{eq:chiralbox}.

The IR-finite contribution $V^{(2)-h}_{ h_4 h_5 h_6}$ to the two-loop splitting in \eqref{eq:trip2L} encodes the deviation from one-loop exponentiation. In the following subsection, we determine $V^{(2)}$ from the bootstrapped hard functions. The reflection symmetry of the triple-collinear splitting functions in \eqref{eq:tripReflSym}, together with that of $V_0$, implies the corresponding reflection symmetry for their IR-finite parts,
\begin{align}
V^{(\ell)-h}_{ h_4 h_5 h_6}\left(\frac{s_{45}}{s_{456}},\frac{s_{56}}{s_{456}},z_4,z_6\right) = V^{(\ell)-h}_{ h_6 h_5 h_4} \left(\frac{s_{56}}{s_{456}},\frac{s_{45}}{s_{456}},z_6,z_4\right) .   \label{eq:tripleflip} 
\end{align}

Turning to hard functions, the triple-collinear limit is obtained by subtracting IR divergences from \eqref{eq:AmplTriple},
\begin{align}
& \left( H^{(1)}_6 / A^{(0)}_6 \right) \to \left( H^{(1)}_4 / A^{(0)}_4 \right) + {\cal C}^{(1) -h}_{h_4 h_5 h_6} , \label{eq:tcH1L} \\
& \left( H^{(2)}_6 / A^{(0)}_6 \right)  \to \left( H^{(2)}_4 / A^{(0)}_4 \right) + {\cal C}^{(1) -h}_{h_4 h_5 h_6} \left( H^{(1)}_4 / A^{(0)}_4 \right)  +  \frac{1}{2} \left( {\cal C}^{(1) -h}_{h_4 h_5 h_6} \right)^2 + V^{(2) -h}_{ h_4 h_5 h_6} \,.\label{eq:HAtripcol}
\end{align}
For six-point MHV scattering at maximal transcendentality, a single helicity configuration $h \to h_4 h_5 h_6$, drawn from the list \eqref{eq:split1->3}, contributes in this limit. The IR divergences of the triple-collinear splitting functions arise exclusively from $V_0$ \eqref{eq:V0}. Subtracting these divergences using the dipole function \eqref{eq:I61}, we define
\begin{align}
{\cal C}_0 := I^{(1)}_4 + V_0 - \underset{p_4 || p_{5} || p_6}{\lim}\, I^{(1)}_{6}  \,,
\end{align}
and decompose the one-loop IR-subtracted triple-collinear splitting function accordingly, 
\begin{align}
{\cal C}^{(1) -h}_{h_4 h_5 h_6} := {\cal C}_0 + V^{(1) -h}_{ h_4 h_5 h_6} \,, 
\end{align}
where ${\cal C}_0$ takes the following simple form, with all transcendental constants omitted,   
\begin{align}
{\cal C}_0 = \log(z_4) \log\left( \frac{s_{12}}{s_{456}} \right) + \log(z_6) \log\left( \frac{s_{23}}{s_{456}} \right).
\label{eq:TC1}
\end{align}
In the triple-collinear limit $\delta \to 0$, ${\cal C}_0$ exhibits logarithmic divergences, while $V^{(1)}$ and $V^{(2)}$ remain finite.

\subsection{Two-loop triple-collinear splitting functions}

We now turn to the two-loop triple-collinear gluonic splitting functions. More precisely, we focus on the function $V^{(2)}$ in \eqref{eq:trip2L}, which encodes a deviation from the one-loop exponentiation ansatz for the IR finite part of the triple-collinear splitting functions. This function controls the triple-collinear asymptotics of the two-loop hard functions, according to \eqref{eq:HAtripcol}. As we discussed above, $V^{(2)}$ is IR finite and defined in four dimensions. It is free of helicity-spinor variables, respects the reflection symmetry \eqref{eq:tripleflip}, and depends only on the dimensionless variables $s_{45}/s_{456},s_{56}/s_{456},z_4,z_6$. Consequently, $V^{(2)}$ remains finite in the triple-collinear limit $\delta \to0$. 

As discussed in the previous section, we employ partial information from the triple-collinear limit as input in the symbol bootstrap of the hard functions. Consistency with triple-collinear kinematic factorization is sufficient to completely fix the six-point two-loop MHV ${-}{-}{+}{+}{+}{+}$ hard function in \eqref{eq:ansatz0}, without requiring an explicit expression for $V^{(2)}$, see also~\cite{Carrolo:2025agz}. Subsequently, cyclically permuting the hard function, evaluating its triple-collinear limit, and matching the results to \eqref{eq:HAtripcol}, we extract the maximally-transcendental part of $V^{(2)}$ for the helicity configurations listed in \eqref{eq:MHVsplittings}. 

For the ${+} \to {+}{+}{+}$ splitting, $V^{(2)}$ is reflection invariant and coincides with its counterpart in ${\cal N}=4$ sYM,
\begin{equation}\label{eq:v2+++}
    V^{(2) {-}}_{{+}{+}{+}} =  V^{(2)}_{{\cal N}=4\; \text{sYM}} \,,
\end{equation}
where it is given by the two-loop remainder function 
after expressing the dual-conformal cross-ratios $u_1,u_2,u_3$, 
\begin{equation} \label{eq:DCCR}
    u_1{=}\frac{s_{12}s_{45}}{s_{123}s_{345}} \,,\quad  u_2{=}\frac{s_{23}s_{56}}{s_{234}s_{123}}\,,\quad  u_3{=}\frac{s_{34}s_{16}}{s_{345}s_{234}} \,
\end{equation}
in the triple-collinear limit    \cite{Bern:2008ap},
\begin{equation}
    u_1\to\frac{s_{45}}{s_{456}(1{-}z_6)},\quad  u_2\to\frac{s_{56}}{s_{456}(1{-}z_4)},\quad  u_3\to\frac{z_4 z_6}{(1{-}z_4)(1{-}z_6)} \,.
\end{equation}

For the splitting ${-}\to{-}{+}{+}$, $V^{(2)}$ receives a correction relative to the supersymmetric expression, 
\begin{align}
V^{(2)+}_{ -++} = V^{(2)-}_{+++} + \left(r^{[1]-}_{-++}+\frac{N_f}{N_c}r_{-++}^{[\frac12]-}\right) \, \widehat{w}_1\,,
\label{eq:v2-++}    
\end{align}
which is given by a weight-four finite pure function $\widehat{w}_1$. The accompanying prefactors $r^{[1]-}_{-++}$ and $r^{[\frac12]-}_{-++}$ are obtained as the triple-collinear limit of the double-box prefactors in \eqref{eq:Rdb12} and \eqref{eq:Sij},
\begin{align}
r^{[1]-}_{-++} := \lim_{p_4 || p_5 || p_6} \frac{R_{34,563}^{(2),\text{db}}}{\text{PT}_{6,34}}, \,\ \ r^{[\frac12]-}_{-++} := \lim_{p_4 || p_5 || p_6} \frac{S_{34,563}^{(2),\text{db}}}{\text{PT}_{6,34}}.
\end{align}
The result for the remaining helicity configuration in \eqref{eq:MHVsplittings} then follows by applying the reflection symmetry in \eqref{eq:tripleflip}.

The triple-splitting functions discussed above are employed as input in the symbol bootstrap of the two remaining MHV hard functions in eqs.~\eqref{eq:ansatz2} and \eqref{eq:ansatz3}. The corresponding triple-collinear limits are summarized in Tab.~\ref{tab:2}. We do not invoke the remaining MHV splitting ${-}\to{+}{-}{+}$ in \eqref{eq:split1->3} as input for the bootstrap. Instead, we extract the associated splitting function from the bootstrapped MHV hard functions,
\begin{align}
V^{(2)+}_{ +-+} = & V^{(2)-}_{+++} +  r^{[1]-}_{{+}{-}{+}} \, \widehat{w}_2\,\, +\frac{N_f}{N_c} r^{[\frac12]-}_{{+}{-}{+}} \widehat{w}_3 \notag \\ 
& -\frac12 \left( r^{[1]-}_{{+}{-}{+}} +\frac{N_f}{N_c} r^{[\frac12]-}_{{+}{-}{+}} \right)^2 \left[\frac12 \log^2\left( \frac{s_{45}}{s_{56}}  \right) + {\rm Li}_2 \left( 1 - \frac{s_{456}}{s_{45}} \right)
+ {\rm Li}_2 \left( 1 - \frac{s_{456}}{s_{56}} \right) \right]^2 \, 
\label{eq:v2+-+}   
\end{align}
where $\widehat{w}_2$ and $\widehat{w}_3$ are  reflection-invariant, weight-four finite pure functions, and prefactors $r^{[1]-}_{+-+}$, $r^{[\frac12]-}_{+-+}$ are defined in \eqref{eq:r+-+def}.

Finally, we have verified that all possible triple-collinear limits of the bootstrapped MHV hard functions are consistent with the extracted functions $V^{(2)}$. We provide triple collinear splitting functions \eqref{eq:v2+++}, \eqref{eq:v2-++} and \eqref{eq:v2+-+} in the attached file.


\subsection{Soft and double-soft limits}

In the limit where the momenta of one or more external particles become soft, gauge-theory amplitudes exhibit a universal behavior governed by soft theorems. Provided the amplitude is singular in this limit, its leading singular behavior is captured by the eikonal approximation and encoded in universal soft functions, which are closely related to soft currents. The soft function can be represented as a matrix element describing the emission of soft particles from a gauge-invariant product of light-like Wilson lines oriented along the directions of the hard external momenta. In particular, this representation immediately implies that soft functions are invariant under rescalings of the hard particle momenta. For six-point kinematics, the soft and double-soft limits are nontrivial, and we review them in detail below. This discussion parallels that of the (multi-)collinear limits considered above. Restricting to the maximally-transcendental part of MHV gluon amplitudes, only soft limits of positive-helicity gluons are relevant. 

We begin with the single-soft limit. In the limit $p_{n+1} \to 0$, a color-ordered $(n+1)$-point amplitude factorizes universally as~\cite{Bern:1998sc},
\begin{align}
A^{(L)}_{n+1} (p_1^{h_1},\ldots,p_n^{h_n},p_{n+1}^{h_{n+1}})  \to \sum_{\ell =0}^{L} \sum_{h=\pm} {\rm Soft}^{(\ell)} (p_n,p_{n+1}^{h_{n+1}}, p_{1}) \, A^{(L-\ell)}_{n} (p_1^{h_1},\ldots,p_{n}^{h_{n}})\,.
\end{align}
The singular soft behavior is governed by the soft function
\begin{align}
{\rm Soft}^{(\ell)} (p_n,p_{n+1}^{h_{n+1}}, p_{1}) = {\rm Soft}^{(0)}(p_n,p_{n+1}^{h_{n+1}}, p_{1}) \, e^{(\ell)}\left(p_n,p_{n+1},p_1 ; \ep \right).
\end{align}
At tree level, the soft function exhibits a simple soft pole, while helicity-independent loop corrections $e^{(\ell)}$ generate logarithmic enhancements of the soft singularity. The soft function is invariant under independent rescalings of the hard momenta, $p_i \to \la_i \, p_i $ with $i=1,n$. At one loop, the soft function takes the form~\cite{Bern:1998sc},
\begin{align}
e^{(1)} = - \frac{1}{\ep^2} \left( -\frac{\mu^2 s_{1 n}}{s_{1 \, n+1} s_{n\, n+1}}\right)^{\ep} + O(\ep)
\end{align}
where all transcendental constants are omitted. At two loops, the maximally-transcendental part of the soft function is independent of the number of quark flavors $N_f$ and is given by the exponentiation of the one-loop result \cite{Badger:2004uk},
\begin{align}
e^{(2)} = \frac{1}{2} \left(e^{(1)}\right)^2 + O(\ep) \,.
\end{align}

In close analogy with the collinear case, we subtract the IR divergences of the soft function using the dipole IR subtraction function \eqref{eq:I61} and subsequently remove the dimensional regulator,
\begin{align}
{\cal S}^{(1)} := e^{(1)} + I^{(1)}_n  - \underset{p_{n+1} \to 0}{\lim}\, I^{(1)}_{n+1}  \,,
\end{align}
which evaluates to the following expression and exhibits a logarithmic diverges in the soft limit,
\begin{align}
{\cal S}^{(1)} = - \log\left( \frac{s_{1 \, n+1}}{s_{1 n}} \right)\log\left( \frac{s_{n\,n+1}}{s_{1 n}} \right) .
\label{eq:S1}
\end{align}
The maximally-transcendental part of the MHV hard functions in the single-soft limit $p_{n+1} \to 0$ of a positive-helicity gluon, $h_{n+1}=1$, then obeys
\begin{align}
& \left( H^{(1)}_{n+1}/A^{(0)}_{n+1} \right) \to  \left( H^{(1)}_{n} / A^{(0)}_{n} \right) + {\cal S}^{(1)}  \,, \notag \\
& \left( H^{(2)}_{n+1} / 
A^{(0)}_{n+1} \right) \to \left(  H^{(2)}_{n} / A^{(0)}_{n}  \right) + {\cal S}^{(1)} \left( H^{(1)}_{n} / A^{(0)}_{n}  \right) + \frac{1}{2} \left( {\cal S}^{(1)} \right)^2 \,. \label{eq:HAsoft}
\end{align}
We have explicitly verified that the bootstrapped six-point hard functions satisfy these single-soft limits.

We turn now to the double-soft limit and focus on six-point kinematics. When a pair of external momenta vanish simultaneously, $p_5,p_6\to0$, the color-ordered amplitude factorizes universally as~\cite{Kosower:2003bh}, 
\begin{align}
A^{(L)} _{6} (p_1^{h_1},\ldots,p_6^{h_6})  \to \sum_{\ell =0}^{L} \sum_{h=\pm} {\rm Soft}^{(\ell)} (p_4,p_{5}^{h_{5}},p_{6}^{h_{6},p_1}) \, A^{(L-\ell)}_{4} (p_1^{h_1},p_2^{h_2},p_3^{h_3},p_{4}^{h_4})
\end{align}
where singular behavior in the limit is controlled by a universal double-soft function,
\begin{align}
{\rm Soft}^{(\ell)} (p_4,p_{5}^{h_{5}},p_{6}^{h_{6}},p_1)= {\rm Soft}^{(0)} (p_4,p_{5}^{h_{5}},p_{6}^{h_{6}},p_1) \, J^{(\ell)}_{h_5h_6}(p_4,p_5,p_6,p_1;\ep) \,.
\end{align}
For the maximally-transcendental part of MHV amplitudes, only soft limits for positive-helicity gluons are relevant, $h_5h_6={+}{+}$. The corresponding one-loop double-soft function has been computed in~\cite{Zhu:2020ftr,Czakon:2022dwk},
\begin{align}
    J^{(1)}_{++}{=}\left({-}\frac{s_{56}}{\mu^2}\right)^{-\ep}\left\{-\frac2{\ep^2}{-}\frac1\ep\log\left(\frac{s_{56}s_{14}}{s_{45}s_{16}}\right)
    {-}\text{Li}_2\left[1{-}\frac{s_{56}s_{14}}{(s_{45}{+}s_{46})(s_{15}{+}s_{16})}\right]{-}\text{Li}_2\left[1{-}\frac{s_{16}}{s_{15}{+}s_{16}}\right]\right. \notag\\
    \left.{-}\text{Li}_2\left[1{-}\frac{s_{45}}{s_{45}{+}s_{46}}\right]{-}\frac12\left[\log^2\frac{s_{56}s_{14}}{(s_{45}{+}s_{46})(s_{15}{+}s_{16})}{+}\log^2\frac{s_{16}}{(s_{15}{+}s_{16})}{+\log^2\frac{s_{45}}{(s_{45}{+}s_{46})}}\right]\right\}
\end{align}
where we retain only terms of maximal transcendentality. In this approximation, the dependence on the number of quark flavors drops out. Moreover, one readily verifies that $J^{(1)}_{++}$ is invariant under independent rescalings of the hard momenta, $p_i \to \la_i \, p_i $ with $i=1,4$.

Implementing the IR subtraction of the double-soft function, 
\begin{equation}
    {\cal S}_{++}^{(1)} :=J^{(1)}_{++} + I^{(1)}_4-\lim_{p_5,p_6\to0}I^{(1)}_6
\end{equation}
we obtain the double-soft limit of the maximally-transcendental part of the MHV hard functions, 
\begin{align}
& \left(H_6^{(1)}/A_6^{(0)}\right) \to \left( H_4^{(1)}/A_4^{(0)} \right) + {\cal S}_{++}^{(1)},\notag \\
    &\left( H_6^{(2)}/A_6^{(0)} \right) \to \left( H_4^{(2)}/A_4^{(0)}\right) + {\cal S}_{++}^{(1)} \left( H_4^{(1)}/A_4^{(0)} \right) + \frac12\left({\cal S}_{++}^{(1)}\right)^2+ {\cal S}_{++}^{(2)} \,.\label{eq:HAdbsoft}
\end{align}
The function ${\cal S}_{++}^{(2)}$ encodes the deviation of the two-loop double-soft function from the simple exponentiation of the one-loop result. It is finite in the double-soft limit and depends on three independent combinations $y_1$, $y_2$, and $y_3$, which are invariant under rescalings of the hard momenta, 
\begin{equation}\label{2}
\begin{aligned} 
    y_1 = \frac{s_{45}}{s_{46}},\ \   
        y_2 = \frac{s_{16}}{s_{15}},\ \   
              y_3 = \frac{s_{45}s_{16}}{ s_{56}s_{14}}  \,.
\end{aligned}
\end{equation}
We find that the maximally-transcendental part of the double-soft function in QCD coincides with that in ${\cal N}=4$ sYM~\cite{Carrolo:2025agz},
\begin{equation}
    \mathcal{S}_{++}^{(2)}=\mathcal{S}^{(2)}_{\mathcal{N}{=}4\ \text{sYM}} \,.
\end{equation}
Indeed, the dual-conformal cross-ratios (see \eqref{eq:DCCR}) of the ${\cal N} = 4$ sYM remainder function are finite in the double-soft limit, 
\begin{equation}
    u_1 \to\frac{s_{45}}{s_{46}{+}s_{45}},\quad u_2 \to\frac{s_{14}s_{56}}{(s_{45}{+}s_{46})(s_{15}{+}s_{16})},\quad u_3  \to\frac{s_{16}}{s_{15}{+}s_{16}}
\end{equation}
and can be mapped to the variables in \eqref{2} through a birational transformation.

We have explicitly verified that the bootstrapped six-point hard functions agree with the double-soft limits.


\newpage

\section{Discussion and outlook}
\label{sec: Discussion}
In this work, we systematically investigated the analytic structure and carried out a bootstrap computation of the maximally-transcendental part of planar scattering amplitudes in QCD. Starting from the maximal weight projection at the integrand level, we argued that the relevant amplitude prefactors are determined by two-loop contours satisfying triangle power-counting. Focusing on the MHV helicity sector and on IR subtracted hard functions, we showed that the prefactors appearing beyond the Parke-Taylor factor in the maximally-transcendental part are fully described by five types of two-loop contours and their corresponding maximal residues. By exploiting on-shell diagrams and on-shell functions, we derived these prefactors for the maximally-transcendental part of MHV pure YM amplitudes at arbitrary multiplicity. We further extended this analysis to QCD amplitudes, including $N_f$ dependence arising from quark-loop contributions in gluonic amplitudes, as well as amplitudes with external quarks.

We then applied the framework of prescriptive unitarity to six-point two-loop MHV QCD amplitudes and argued that the transcendental functions appearing in the IR-finite parts of these amplitudes can be computed from three types of prescriptive unitarity integrals. We computed these integrals using IBP reduction and expressed the results in a function basis constructed via the canonical differential equation method. Furthermore, using this function basis, we constructed a symbol ansatz for the amplitude contributions affected by IR-subtraction. Employing a symbol-bootstrap approach, we computed the hard functions for three inequivalent MHV helicity sectors with purely gluonic external states. We observed that in all cases the corresponding symbol alphabets form subsets of the previously identified 137-letter alphabet. Finally, we discussed the implications of these bootstrap results for triple-collinear spitting and the double-soft functions.

Beyond providing new results and insights, our work also raises several questions for future investigation.

{\it \underline{1. QCD leading singularities and  Grassmanian geometry.}}
On-shell diagrams and on-shell functions enjoy deep connections to mathematical structures such as Grassmannian geometry. Prior to this work, these tools have been applied primarily within sYM theories, both with maximal or reduced supersymmetry. Studies of on-shell diagrams involving fermions in QCD, along with their associated leading singularities and geometric interpretations, therefore remain scarce (see the recent work \cite{Bourjaily:2025tpr} for further developments). Moreover, when extending the analysis to higher multiplicities and non-MHV helicity sectors in QCD, we expect a substantially richer set of contours and prefactors to contribute. This renders the classification of structures analogous to the R-invariants of $ \mathcal{N} = 4$ sYM an important direction for future research in QCD. Emerging phenomena, such as three-mass-triangle coefficients in pure YM NMHV amplitudes \cite{Dunbar:2009uk}, indicate that this classification will exhibit a far richer structure than in sYM theories. Based on the classification results of leading singularities, we can consider bootstrap computations for helicity amplitudes in other QCD channels, including cases with external quark legs, as well as amplitudes in the NMHV helicity sector. 

\newpage

{\it \underline{2. Effective $\mathcal{N}{=}1$ supersymmetry at maximal transcendentality.}}
In the main text, we analyzed all quark-flavor $N_f^k$ terms in planar gluonic QCD amplitudes. At one loop, an effective $\mathcal{N}{=}1$ supersymmetric structure emerges in the IR-finite part. Prefactors ~\eqref{eq:L1LS} and \eqref{eq:1LSjk} from the same two-mass-easy box can be interpreted as arising from an $\mathcal{N}{=}1$ vector supermultiplet circulating in the loop. At two loops, however, the situation becomes more intricate.
Owing to the different large-$N_f$ and large-$N_c$ scalings of the on-shell diagrams, additional structure emerges that warrants deeper investigation, both in the gluonic sector and in channels involving external quarks. This raises several natural questions. Can systematic connections and distinctions between QCD amplitudes and $\mathcal{N}{=}1$ sYM amplitudes \cite{Elvang:2011fx} be identified through amplitudes with different arrangements of external gluons and fermions? Moreover, can on-shell functions and results from $\mathcal{N}{=}1$ sYM amplitudes facilitate computations in fermionic sectors of QCD, and conversely? We leave these questions for future work.

{\it \underline{3.    Full classification of QCD amplitude prefactors.}}
In this work, we focused on the maximally-transcendental contributions. A complete calculation of QCD scattering amplitudes, however, also requires determining the subleading-transcendental contributions and rational terms. This therefore reduces once again to the problem of computing and classifying the rational prefactors. The study of two-loop all-plus Yang–Mills amplitudes \cite{Dunbar:2017nfy,Dunbar:2019fcq} indicated that, for the prefactors of the non-maximally-transcendental parts, on-shell diagrams with one-loop all-plus Yang–Mills amplitudes inserted as vertices can likewise be used to classify all of their leading-transcendentality prefactors. Besides, at one loop, generalized unitarity \cite{Forde:2007mi} has shown that amplitude prefactors are closely tied to specific non-maximal cuts and their associated phase-space integrals. Furthermore, rational terms in the amplitudes come from $D=4{-}2\epsilon$ cuts of the integrand \cite{Bern:1995db}.  We therefore seek a more profound understanding of lower-transcendentality prefactors and of the rational terms. Frontier approaches to this problem also include prescriptive unitarity constructions beyond the maximally transcendental part \cite{Bourjaily:2021ujs}, new surface integral formulation for pure YM amplitudes \cite{Arkani-Hamed:2024tzl}, and so on. We expect that such insights will not only facilitate extensions of our computational framework to higher multiplicities and loop orders, but also deepen our understanding of fundamental principles of QCD from a perturbative perspective.

\paragraph{Note added} After submission, the results for planar Higgs plus two-jet production amplitudes in the heavy top quark mass limit have been calculated \cite{DeLaurentis:2026brm}. We have extracted the relevant three gluonic splitting functions from triple collinear limits, namely $+\to+++$, $-\to-++$ and $-\to+-+$, and have found complete agreement with our results in section \ref{sec: behaviors}.

\section*{Acknowledgments}
We thank Simon Badger, Jacob Bourjaily, Subramanya Hegde, Xuhang Jiang, Jiahao Liu, Yichao Tang, Congkao Wen, Yu Wu, Huaimin Yu and Yu Jiao Zhu for valuable discussions. 
This work was supported by the European Union (ERC, UNIVERSE PLUS, 101118787). Views and opinions expressed are however those of the authors only and do not necessarily reflect those of the European Union or the European Research Council Executive Agency. Neither the European Union nor the granting authority can be held responsible for them. D.C. is supported by ANR-24-CE31-7996.   YZ is supported by
NSFC through Grant No. 12575078 and 12247103.

\newpage

\appendix
\section{IBP reduction calculation for the prescriptive unitarity integrals}
\label{sec: IBP reduction}

To obtain the symbol expressions for prescriptive unitarity integrals, our strategy is to reduce them, via IBP identities, to the known two-loop six-point UT integral basis in~\cite{Henn:2025xrc}.

The first step is to rewrite the numerators of prescriptive unitarity integrals to the local integrand form in terms of the trace of gamma matrices.  With the definition of~\cite{Bourjaily:2019iqr}, those integrals have the numerator form,
\begin{equation}
\left[\![a_1, a_2, b_1, b_2, \cdots, c_1, c_2]\!\right] \equiv \left[(a_1 \cdot a_2)^\alpha{}_\beta (b_1 \cdot b_2)^\beta{}_\gamma \cdots (c_1 \cdot c_2)^\delta{}_\alpha\right]\,,
\end{equation}
where for instance $a_1$ stands for the contraction with Pauli's matrices, $a_1^{\alpha\dot \alpha} =l_{1,\mu} \sigma^{\mu,\alpha \dot \alpha}$, where $l_1$ is either an external momentum or a {\it four-dimensional} loop momentum. $(a_1\cdot a_2)^\alpha{}_\beta\equiv a_1^{\alpha \dot{\alpha}}\,\epsilon_{\dot{\alpha}\dot{\gamma}}\,a_2^{\dot{\gamma}\gamma}\,\epsilon_{\gamma\beta}\,
$. It is not easy to reduce such expressions in standard IBP reduction programs, so following the discussion in \cite{Bourjaily:2019iqr}, this numerator expression is reformulated as the ``Dirac trace", 
\begin{equation}
\left[\![ \ldots]\!\right]=\rm{tr}_+\big[\ldots \big] =\frac{1}{2}\rm{tr}\big[(1+\gamma^5)\ldots\big]\,.
\end{equation}

The second step is to expand the Dirac trace to the standard input form for IBP reduction programs. The standard procedure in~\cite{Badger:2016ozq} for the local integrand expansion is used here for the expansion. For the Dirac trace with even number of gamma matrices and free of $\gamma_5$, the recursive relation,
\begin{equation}
{\rm tr}\bigl[\gamma^{\nu_1}\dots\gamma^{\nu_{2n}}\bigr]
=2\sum_{i=2}^{2n}(-1)^i\,\eta^{\nu_1 \nu_i}\,
{\rm tr}\bigl[\gamma^{\nu_2}\dots\hat{\gamma}^{\nu_i}\dots\gamma^{\nu_{2n}}\bigr],
\end{equation}
reduces the expression to scalar products and $\rm{tr}[\mathbb 1]$. The Dirac trace with the odd number of gamma matrices and free of $\gamma……5$ vanishes. We use the following formula to simplify a long trace with a $\gamma^5$, $n\geq 4$,
\begin{gather}
{\rm tr}\bigl[\gamma^5 \gamma^{\nu_1} \dots \gamma^{\nu_n}\bigr]
= \sum_{1\leq i<j<k<l\leq n} (-1)^{i+j+k+l}
\frac{{\rm tr}\bigl[\gamma^5 \gamma^{\nu_i} \gamma^{\nu_j} \gamma^{\nu_k} \gamma^{\nu_l}\bigr]}
{{\rm tr} \bigl[\mathbb 1 \bigr]} \nonumber \\
\times
{\rm tr}\bigl[\gamma^{\nu_1} \dots \hat{\gamma}^{\nu_i} \dots \hat{\gamma}^{\nu_j} \dots \hat{\gamma}^{\nu_k} \dots \hat{\gamma}^{\nu_l} \dots \gamma^{\nu_n}\bigr],
\end{gather}
For the trace with four $\gamma$ matrices and a $\gamma^5$, 
\begin{equation}
{\rm tr}[\gamma^5 \gamma^{\nu_1} \gamma^{\nu_2}\gamma^{\nu_3} \gamma^{\nu_4}] l_{1,\nu_1} l_{2,\nu_2 }l_{3,\nu_3}l_{4,\nu_4} =-\frac{G\left(
\begin{array}{cccc}
l_1 & l_2 & l_3 & l_4\\
p_1 & p_2 & p_3 & p_4
\end{array}
\right)}{\epsilon(1,2,3,4)}\,,
\end{equation}
where $G$ denotes the Gram determinant, and $\epsilon(1,2,3,4)=4i\varepsilon_{\mu\nu\rho\sigma}p_1^\mu p_2^\nu p_3^\rho p_4^\sigma$. Traces with fewer $\gamma$ matrices and a $\gamma^5$ vanish. After this step, all prescriptive unitary numerators are converted to the combination of scalar products, the constants $\epsilon(1,2,3,4)$ and ${\rm tr}[\mathbb 1]$. We set ${\rm tr}[\mathbb 1]=4$.

In the third step, the prescriptive unitarity integrals are reduced in standard IBP programs like {\sc FIRE7}~\cite{Smirnov:2025prc} and {\sc KIRA3}~\cite{Lange:2025fba}. The two-loop six-point UT basis in~\cite{Henn:2025xrc} is reduced in the meanwhile. Afterwards, the finite-field tool {\sc FiniteFlow}~\cite{Peraro:2016wsq,Peraro:2019svx} is called to convert the reduction results of prescriptive unitarity integrals on the UT basis. In  practice, we find that, with the prescriptions in the first two steps, in all reduction tests, prescriptive unitarity integral in~\cite{Bourjaily:2019iqr} are reduced to linear combinations of UT integrals with {\it constant} coefficients. Therefore, the IBP reduction for prescriptive unitarity integrals can be done with numeric kinematics. 

The typical running time for all three steps in total is less than one hour on a workstation, with $20$ cores and $128$GB RAM. For example, the following integral with the integrand,
\begin{equation}
I_{6,361}^{\text{db}}:=\raisebox{-3.5em}{\begin{tikzpicture}[scale=0.8]
        \draw[line width=1pt,black] (0,0)--(1.5,0)--(1.5,1.5)--(0,1.5)--(0,0);
        \draw[decorate, decoration=snake, segment length=12pt, segment amplitude=2pt, black,line width=1pt](0,0)--(1.5,1.5);
        \draw[decorate, decoration=snake, segment length=12pt, segment amplitude=2pt, black,line width=1pt](3,0)--(1.5,1.5);
         \draw[line width=1pt,black] (1.5,0)--(3,0)--(3,1.5)--(1.5,1.5)--(1.5,0);
          \draw[line width=1pt,black] (0,0)--(-0.5,-0.5);
          \draw[line width=1pt,black] (2,-0.5)--(1.5,0)--(1,-0.5);
            \draw[line width=1pt,black] (3,0)--(3.5,-0.5);
        \draw[line width=1pt,black] (-0.5,2)--(0,1.5);
         \draw[line width=1pt,black] (3.5,2)--(3,1.5);
        \path[fill=gray] (3,0) circle[radius=0.11];
          \path[fill=black] (3,1.5) circle[radius=0.11];
        \path[fill=gray] (0,0) circle[radius=0.11];
        \path[fill=black] (1.5,0) circle[radius=0.11];
        \path[fill=gray] (1.5,1.5) circle[radius=0.11];
        \path[fill=black] (0,1.5) circle[radius=0.11];
        \node[anchor=north east] at (-0.5,-0.5) {\small{$6$}};
            \node[anchor=north] at (2,-0.5) {\small{$4$}};
        \node[anchor=north] at (1,-0.5) {\small{$5$}};
          \node[anchor=south east] at (-0.5,2) {\small{$1$}};
 \node[anchor=south west] at (3.5,2) {\small{$2$}};
\node[anchor=north west] at (3.5,-0.5) {\small{$3$}};
 \node[anchor=east] at (0,0.75) {\small{$\ell_b$}};
        \node[anchor=south] at (0.75,1.5) {\small{$\ell_c$}};
        \node[anchor=south] at (2.25,1.5) {\small{$\ell_d$}};
        \node[anchor=west] at (3,0.75) {\small{$\ell_e$}};
\end{tikzpicture}},
\, \quad {\bf n}_{\rm db}:=\frac12\left[\![6,b,c,d,e,3\right]\!]\,,
\end{equation}
is reduced to, 
\begin{gather}
-\frac{1}{8}\mathcal I_{27} - \frac{3}{8}\mathcal I_{28} + \frac{1}{8}\mathcal I_{29} + \frac{1}{8}\mathcal I_{56} + \frac{1}{4}\mathcal I_{78} + \frac{1}{4}\mathcal I_{106} + \frac{1}{8}\mathcal I_{122} 
- \frac{1}{4}\mathcal I_{131}- \frac{1}{4}\mathcal I_{138} - \frac{1}{8}\mathcal I_{154} \nonumber\\ - \frac{1}{8}\mathcal I_{156} + \frac{1}{16}\mathcal I_{170} + \frac{1}{8}\mathcal I_{171} 
- \frac{1}{8}\mathcal I_{173} + \frac{1}{8}\mathcal I_{174}  - \frac{1}{16}\mathcal I_{189} + \frac{3}{16}\mathcal I_{202} - \frac{1}{4}\mathcal I_{216} + \frac{1}{2}\mathcal I_{226} 
+ \frac{1}{4}\mathcal I_{227} \nonumber\\ - \mathcal I_{233} + \frac{3}{16}\mathcal I_{243} + \frac{1}{8}\mathcal I_{248} - \frac{1}{8}\mathcal I_{258} - \frac{3}{8}\mathcal I_{259} 
+ \frac{1}{4}\mathcal I_{260} + \frac{3}{16}\mathcal I_{263} + \frac{3}{16}\mathcal I_{264} - \frac{1}{4}\mathcal I_{266}\,,
\label{eq:prescriptive_IBP_reduction}
\end{gather}
where $\mathcal I_i$ is the $i$-th UT integral for the double pentagon family defined in~\cite{Henn:2025xrc}, with the permutation of external legs $p_i\to p_{i{+}4}$. 
From the expression of the UT integrals~\cite{Henn:2025xrc}, \eqref{eq:prescriptive_IBP_reduction} is finite and 
has the uniform weight four.

We remark that the prescriptive unitarity integrals' numerator essentially consist of {\it four-dimensional} loop momenta for the spinor helicity formalism~\cite{Bourjaily:2019iqr}, but the IBP reduction and the UT integral basis are calculated with dimensional regularization. There is no unique way to upgrade four-dimensional loop momenta to $D$-dimensional loop momenta, for the IBP reduction. Furthermore, the Dirac trace calculation depends on the chosen scheme. However, since the prescriptive integrals are finite, the different definitions of the $(-2\epsilon)$-dimensional component of loop momenta, or different Dirac trace schemes, are expected to affect only the vanishing parts, $O(\epsilon)$, of those integrals.

\section{A review of on-shell diagrams and on-shell functions}
\label{sec: review of on-shell diagrams}

The computation of on-shell diagrams has been crucial in understanding the underlying combinatorial structure of scattering amplitudes in gauge theories with some degree of supersymmetry. Nonetheless, this construction is perfectly well defined, albeit with fewer simplifications, in theories without supersymmetry, such as Yang-Mills theory or QCD.
The purpose of this Appendix is thus to provide a practical guide to computing these functions, starting from the elementary building blocks of any perturbative theory---the three-particle amplitudes. To illustrate this, we work in pure YM, however, everything in our discussion remains valid for general gauge theories. 

\subsection{Three-particle amplitudes and gluing}
As it is, by now, well known, three-particle amplitudes in four dimensions are completely fixed by requiring little group covariance and locality \cite{Benincasa:2007xk}. The usual story goes as follows: The momentum conservation condition for three massless particles is equivalent to
\begin{equation}
    \la_1^\alpha \tl_1^{\dot{\alpha}} + \la_2^\alpha \tl_2^{\dot{\alpha}} + \la_3^\alpha \tl_3^{\dot{\alpha}} =0 \, , \forall\alpha, \dot{\alpha} \in \{1,2\} .
\end{equation}
This, in turn, produces a degenerate geometric configuration, either in $\la$-space or in $\tl$-space. To see this, one can organize the kinematic data into two three-dimensional vectors - $(\la_1^\alpha,\la_2^\alpha,\la_3^\alpha)$, and equivalently for $\tl$. For generic momenta, both the $\la$ and the $\tl$ vectors span two 2-planes in 3 dimensions. What momentum conservation says is that these 2-planes must be orthogonal. In 3 dimensions, this can only be achieved provided one of the 2-planes degenerates into a line. Thus, there are two (parity related) solutions to three-particle momentum conservation condition
\begin{equation} \label{eq:3ptkin}
    \begin{cases}
        \la_1 \sim \la_2 \sim \la_3 \, , \quad \tl_i \text{ generic} \\
        \tl_1 \sim \tl_2 \sim \tl_3 \, , \quad \la_i \text{ generic} \, .
    \end{cases}
\end{equation}
In what follows, we will denote the first with a white vertex and the second with a black vertex. Graphically, we have that the three gluon amplitudes can be represented as
\begin{equation}
    \vcenter{\hbox{\begin{tikzpicture}[decoration={,markings,mark=at position 0.7 with {\arrow{latex}}}]
        \draw[fill, thick] (0,0) circle (0.2);
        \coordinate (p1) at ($({cos(0)},{sin(0)})$);
        \coordinate (p2) at ($({cos(120)},{sin(120)})$);
        \coordinate (p3) at ($({cos(-120)},{sin(-120)})$);
        \coordinate (p1a) at ($0.2*({cos(0)},{sin(0)})$);
        \coordinate (p2a) at ($0.2*({cos(120)},{sin(120)})$);
        \coordinate (p3a) at ($0.2*({cos(-120)},{sin(-120)})$);
        \node (p1t) at ($1.3*(p1)$){$1^-$};
        \node (p2t) at ($1.3*(p2)$){$2^-$};
        \node (p3t) at ($1.3*(p3)$){$3^+$};
        \draw[very thick, postaction={decorate}] (p1.center) to (p1a.center);
        \draw[very thick, postaction={decorate}] (p2.center) to (p2a.center);
        \draw[very thick, postaction={decorate}] (p3a.center) to (p3.center);
    \end{tikzpicture}}} = \frac{\ab{12}^4 }{\ab{12}\ab{23}\ab{31}} \delta^{2\times2}(\la_i \tl_i) \, , \quad
    \vcenter{\hbox{\begin{tikzpicture}[decoration={,markings,mark=at position 0.7 with {\arrow{latex}}}]
        \draw[thick] (0,0) circle (0.2);
        \coordinate (p1) at ($({cos(0)},{sin(0)})$);
        \coordinate (p2) at ($({cos(120)},{sin(120)})$);
        \coordinate (p3) at ($({cos(-120)},{sin(-120)})$);
        \coordinate (p1a) at ($0.2*({cos(0)},{sin(0)})$);
        \coordinate (p2a) at ($0.2*({cos(120)},{sin(120)})$);
        \coordinate (p3a) at ($0.2*({cos(-120)},{sin(-120)})$);
        \node (p1t) at ($1.3*(p1)$){$1^+$};
        \node (p2t) at ($1.3*(p2)$){$2^+$};
        \node (p3t) at ($1.3*(p3)$){$3^-$};
        \draw[very thick, postaction={decorate}] (p1a.center) to (p1.center);
        \draw[very thick, postaction={decorate}] (p2a.center) to (p2.center);
        \draw[very thick, postaction={decorate}] (p3.center) to (p3a.center);
    \end{tikzpicture}}} =\frac{\Sb{12}^4 }{\Sb{12}\Sb{23}\Sb{31}} \delta^{2\times2}(\la_i \tl_i) \, ,
\end{equation}
where the momentum conserving delta functions $\delta^{2\times2}$ are, in each case separately, supposed to impose the two different parallel conditions of eq.~\eqref{eq:3ptkin}. Moreover, note how each line is decorated with the corresponding helicity of the respective gluon---arrows going in denote negative helicity, while arrows going out denote positive helicity gluons.

In QCD, we also need to consider three-particle amplitudes involving fermions. We represent these as follows
\begin{equation}
    \vcenter{\hbox{\begin{tikzpicture}[decoration={,markings,mark=at position 0.7 with {\arrow{latex}}}]
        \draw[fill, thick] (0,0) circle (0.2);
        \coordinate (p1) at ($({cos(0)},{sin(0)})$);
        \coordinate (p2) at ($({cos(120)},{sin(120)})$);
        \coordinate (p3) at ($({cos(-120)},{sin(-120)})$);
        \coordinate (p1a) at ($0.2*({cos(0)},{sin(0)})$);
        \coordinate (p2a) at ($0.2*({cos(120)},{sin(120)})$);
        \coordinate (p3a) at ($0.2*({cos(-120)},{sin(-120)})$);
        \node (p1t) at ($1.3*(p1)$){$1_g^-$};
        \node (p2t) at ($1.3*(p2)$){$2^{-}_q$};
        \node (p3t) at ($1.3*(p3)$){$3^{+}_q$};
        \draw[very thick, postaction={decorate}] (p1.center) to (p1a.center);
        \draw[blue, very thick, postaction={decorate}] (p2.center) to (p2a.center);
        \draw[blue,very thick, postaction={decorate}] (p3a.center) to (p3.center);
    \end{tikzpicture}}} = \frac{\ab{12}^2 }{\ab{23}} \delta^{2\times2}(\la_i \tl_i) \, , \quad
    \vcenter{\hbox{\begin{tikzpicture}[decoration={,markings,mark=at position 0.7 with {\arrow{latex}}}]
        \draw[thick] (0,0) circle (0.2);
        \coordinate (p1) at ($({cos(0)},{sin(0)})$);
        \coordinate (p2) at ($({cos(120)},{sin(120)})$);
        \coordinate (p3) at ($({cos(-120)},{sin(-120)})$);
        \coordinate (p1a) at ($0.2*({cos(0)},{sin(0)})$);
        \coordinate (p2a) at ($0.2*({cos(120)},{sin(120)})$);
        \coordinate (p3a) at ($0.2*({cos(-120)},{sin(-120)})$);
        \node (p1t) at ($1.3*(p1)$){$1_g^+$};
        \node (p2t) at ($1.3*(p2)$){$2^{-}_q$};
        \node (p3t) at ($1.3*(p3)$){$3^{+}_q$};
        \draw[very thick, postaction={decorate}] (p1a.center) to (p1.center);
        \draw[blue, very thick, postaction={decorate}] (p2.center) to (p2a.center);
        \draw[blue,very thick, postaction={decorate}] (p3a.center) to (p3.center);
    \end{tikzpicture}}} =\frac{\Sb{31}^2 }{\Sb{23}} \delta^{2\times2}(\la_i \tl_i) \, ,
    \label{eq:Ferm3ptamps}
\end{equation}
where fermions with positive helicity are represented by outgoing blue arrows, while negative helicity are represented by incoming blue arrows. Now that we have established the expressions for three-particle amplitudes, we can now discuss how to glue them to obtain more complicated objects.

Given any set of three-particle amplitudes, we can glue them by multiplying the corresponding expressions and integrating over the internal lines' Lorentz phase spaces. For instance, after gluing just two three-particle amplitudes, we obtain a factorization channel.
\begin{equation}
\begin{aligned}
	\vcenter{\hbox{\begin{tikzpicture}[scale=0.8,decoration={,markings,mark=at position 0.7 with {\arrow{latex}}}]
    		\coordinate (shiftv) at (0.6,0);
 
        		\coordinate (p1) at ($({cos(0)},{sin(0)})$);
       		\coordinate (p2) at ($({cos(120)},{sin(120)})$);
        		\coordinate (p3) at ($({cos(-120)},{sin(-120)})$);
        		\coordinate (p1a) at ($0.2*({cos(0)},{sin(0)})$);
        		\coordinate (p2a) at ($0.2*({cos(120)},{sin(120)})$);
        		\coordinate (p3a) at ($0.2*({cos(-120)},{sin(-120)})$);

        		\draw[ thick] (shiftv) circle (0.2);
        		\draw[very thick, postaction={decorate}] ($(180:1)+ (shiftv)$) to ($(180:0.2) + (shiftv)$);
        		\draw[very thick, postaction={decorate}] ($(60:0.2)+(shiftv)$) to ($(60:1)+(shiftv)$);
        		\draw[very thick, postaction={decorate}] ($(-60:0.2)+(shiftv)$) to ($(-60:1)+(shiftv)$);
        		\draw[fill,thick] ($-1*(shiftv)$) circle (0.2);
        		\draw[very thick, postaction={decorate}] ($(120:1)- (shiftv)$) to ($(120:0.2) - (shiftv)$);
        		\draw[very thick, postaction={decorate}] ($(-120:1)-(shiftv)$) to ($(-120:0.2)-(shiftv)$);
        		
        		\node (p1t) at ($(120:1.3)-(shiftv)$){$1^-$};
        		\node (p2t) at ($(-120:1.3)-(shiftv)$){$2^{-}$};
        		\node (p3t) at ($(60:1.3)+(shiftv)$){$3^{+}$};
        		\node (p4t) at ($(-60:1.3)+(shiftv)$){$4^{+}$};
        		\node (p4t) at ($(0,0.3)$){$I$};

    \end{tikzpicture}}} &= \int d^3\text{LIPS}(I) \frac{\ab{12}^4 }{\ab{12}\ab{2I}\ab{I1}} \frac{\Sb{34}^4}{\Sb{34}\Sb{4I}\Sb{I3}} \delta^{2\times2}(p_1 + p_2 + p_I)\delta^{2\times2}(p_3+p_4-p_I) \\
    &=\delta^4(\text{P}_{\text{tot}}) \int d^4p_I \delta(p_I^2) \delta^{2\times 2}(p_3+p_4-p_I) \frac{\ab{12}^3}{\ab{2I}\ab{I1}} \frac{\Sb{34}^3}{\Sb{4I}\Sb{I3}}  \\
    & = \delta^4(\text{P}_{\text{tot}}) \delta(\ab{34}) \frac{\ab{12}^4}{\ab{12}\ab{23}\ab{41}}.
    \label{eq:4ptfactchan}
\end{aligned}
\end{equation}

This gluing over the Lorentz invariant phase space of the internal particle can be used to glue \textit{any} on-shell object. In particular, we can glue higher-point MHV trees with three-point anti-MHV. For example, gluing a five-point MHV tree with the white vertex, we obtain a factorization channel
\begin{equation}
    \vcenter{\hbox{\begin{tikzpicture}[scale=0.9]
        \coordinate (c1) at (-0.6,0);
        \coordinate (c2) at (0.6,0);
        \coordinate (l2) at ($(c1)+(180:1.2)$);
        \coordinate (l3) at ($(c1)+(120:1.2)$);
        \coordinate (l1) at ($(c1)+(-120:1.2)$);
        
        \coordinate (l4) at ($(c2)+(60:1.2)$);
        \coordinate (l5) at ($(c2)+(-60:1.2)$);
        \draw[very thick,rarrow] (l1) to (c1);
        \draw[very thick,larrow] (l2) to (c1);
        \draw[very thick,larrow] (l3) to (c1);
        \draw[very thick,rarrow] (l4) to (c2);
        \draw[very thick,larrow] (l5) to (c2);
        \draw[very thick,larrow] (c1) to (c2);
        \node at ($(l2)+(180:0.3)$) {$2$};
        \node at ($(l1)+(-120:0.3)$) {$1$};
        \node at ($(l3)+(120:0.3)$) {$3$};
        \node at ($(l4)+(60:0.3)$) {$4$};
        \node at ($(l5)+(-60:0.3)$) {$5$};
        \node at (0,0.35) {$I$};

        \draw[fill,very thick] (c1) circle (0.2);
        \draw[fill=white,very thick] (c2) circle (0.2);
        
    \end{tikzpicture}}} = \delta^4(P_\text{tot})\frac{\ab{14}^4}{\ab{12}\ab{23}\ab{34}\ab{51}}\delta(\ab{45}) \, .
\end{equation}
In the main text, since we want to connect to the usual unitarity cut analysis, we directly use $n$-point MHV tree amplitudes as ingredients for producing on-shell diagrams. This is slightly different from the on-shell diagrams of \cite{Arkani-Hamed:2010zjl}, where on-shell diagrams are exclusively made from cubic vertices. However, these can be connected by recognizing that the higher-point trees can be ``blown up" into diagrams with only cubic vertices. In the example above, we could have blown up the four-point MHV into 
\begin{equation}
    \vcenter{\hbox{\begin{tikzpicture}[scale=0.9]
        \coordinate (c1) at (-0.6,0);
        \coordinate (c2) at (0.6,0);
        \coordinate (l2) at ($(c1)+(180:1.2)$);
        \coordinate (l3) at ($(c1)+(120:1.2)$);
        \coordinate (l1) at ($(c1)+(-120:1.2)$);
        
        \coordinate (l4) at ($(c2)+(60:1.2)$);
        \coordinate (l5) at ($(c2)+(-60:1.2)$);
        \draw[very thick,rarrow] (l1) to (c1);
        \draw[very thick,larrow] (l2) to (c1);
        \draw[very thick,larrow] (l3) to (c1);
        \draw[very thick,rarrow] (l4) to (c2);
        \draw[very thick,larrow] (l5) to (c2);
        \draw[very thick,larrow] (c1) to (c2);
        \node at ($(l2)+(180:0.3)$) {$2$};
        \node at ($(l1)+(-120:0.3)$) {$1$};
        \node at ($(l3)+(120:0.3)$) {$3$};
        \node at ($(l4)+(60:0.3)$) {$4$};
        \node at ($(l5)+(-60:0.3)$) {$5$};
        \node at (0,0.35) {$I$};

        \draw[fill,very thick] (c1) circle (0.2);
        \draw[fill=white,very thick] (c2) circle (0.2);
        
    \end{tikzpicture}}} \xrightarrow{\text{blow up}}
    \vcenter{\hbox{\begin{tikzpicture}[scale=0.8]
        \coordinate (v4) at (-0.6,0);
        \coordinate (v5) at (0.6,0);
        \coordinate (v1) at ($(v4)+(-135:1.2)$);        
        \coordinate (v3) at ($(v4)+(135:1.2)$);
        \coordinate (v2) at ($(v3)+(-135:1.2)$);

        \coordinate (p1) at ($(v1)+(-90:1.2)$);
        \coordinate (p2) at ($(v2)+(180:1.2)$);
        \coordinate (p3) at ($(v3)+(90:1.2)$);
        \coordinate (p4) at ($(v5)+(45:1.2)$);
        \coordinate (p5) at ($(v5)+(-45:1.2)$);
        \node (l1) at ($(p1)+(-90:0.3)$) {$1$};
        \node (l2) at ($(p2)+(180:0.3)$){$2$};
        \node (l3) at ($(p3)+(90:0.3)$){$3$};
        \node (l4) at ($(p4)+(45:0.3)$){$4$};
        \node (l5) at ($(p5)+(-45:0.3)$){$5$};
        \draw[very thick, rarrow] (p1) to (v1);
        \draw[very thick, larrow] (p2) to (v2);
        \draw[very thick, larrow] (p3) to (v3);
        \draw[very thick, rarrow] (p4) to (v5);
        \draw[very thick, larrow] (p5) to (v5);
        
        \draw[very thick, rarrow] (v5) to (v4);
        \draw[very thick, rarrow] (v1) to (v2);
        \draw[very thick, rarrow] (v2) to (v3);
        \draw[very thick, larrow] (v3) to (v4);
        \draw[very thick, rarrow] (v4) to (v1);
        \draw[very thick, fill] (v1) circle (0.2); 
        \draw[very thick, fill=white] (v2) circle (0.2); 
        \draw[very thick, fill] (v3) circle (0.2); 
        \draw[very thick, fill=white] (v4) circle (0.2); 
        \draw[very thick, fill=white] (v5) circle (0.2); 

    \end{tikzpicture}}} \, .
\end{equation}
Whereas, for other higher point trees, a similar procedure can be done. The box sub-diagram from this blow-up procedure never contains oriented helicity distributions. As a consequence, it is always equivalent to directly sewing a higher-point PT factor.

\subsection{BCFW-bridges and Inverse-soft factor}

In order to construct arbitrarily complicated on-shell diagrams, we need two basic ingredients: BCFW bridges, which increase the "loop-order" and inverse soft-factors, which increase the number of external legs. A pure gluon BCFW bridge consists of gluing the structure of eq.~\eqref{eq:4ptfactchan} to an arbitrary diagram. This can be computed as follows
\begin{equation}
\begin{aligned}
\vcenter{\hbox{\begin{tikzpicture}[rotate=-90,decoration={,markings,mark=at position 0.7 with {\arrow{latex}}}]
    		\coordinate (shiftv) at (0.8,0);
 
        		\coordinate (p1) at ($({cos(0)},{sin(0)})$);
       		\coordinate (p2) at ($({cos(120)},{sin(120)})$);
        		\coordinate (p3) at ($({cos(-120)},{sin(-120)})$);
        		\coordinate (p1a) at ($0.2*({cos(0)},{sin(0)})$);
        		\coordinate (p2a) at ($0.2*({cos(120)},{sin(120)})$);
        		\coordinate (p3a) at ($0.2*({cos(-120)},{sin(-120)})$);

        		\draw[ thick] (shiftv) circle (0.2);
        		\draw[very thick,postaction={decorate}] ($(0.2,0)-(shiftv)$) to ($(-0.2,0)+(shiftv)$);
        		\draw[very thick,postaction={decorate}] ($(60:0.2)+(shiftv)$) to ($(60:1)+(shiftv)$);
        		\draw[very thick,postaction={decorate}] ($(-90:0.2)+(shiftv)$) to ($(-90:1)+(shiftv)$);
        		\draw[fill,thick] ($-1*(shiftv)$) circle (0.2);
        		\draw[very thick,postaction={decorate}] ($(120:1)- (shiftv)$) to ($(120:0.2) - (shiftv)$);
        		\draw[very thick,postaction={decorate}] ($(-90:1)-(shiftv)$) to ($(-90:0.2)-(shiftv)$);
        		
        		\node (p1t) at ($(120:0.5)-(shiftv)-(0.3,0)$){$a$};
        		\node (p2t) at ($(-90:0.6)-(shiftv)-(0.3,0)$){$\hat{a}$};
        		\node (p3t) at ($(60:0.5)+(shiftv)+(0.3,0)$){$b$};
        		\node (p4t) at ($(-90:0.6)+(shiftv)+(0.3,0)$){$\hat{b}$};
        		\node (p4t) at ($(0,0.3)$){$I$};
		\draw[very thick] (0,-1.6) ellipse (1.5 and 0.7);
		\node at (0,-1.65) {\textbf{$\mathcal{F}$}};
		\draw[very thick] ($(-145:1)+ (0,-1.65)$) to ($(-145:1)+ (0,-1.65) + (-120:0.8)$);
		\draw[very thick] ($(-35:1)+ (0,-1.65)$) to ($(-35:1)+ (0,-1.65) + (-60:0.8)$);
		\node[rotate=90] at (0,-2.65) {$\dots$};
    \end{tikzpicture}}} &= \int_{\hat{a},\hat{b},I} \underbrace{\mathcal{A}_L (a,I,\hat{a}) \mathcal{A}_R(b,\hat{b},I)\mathcal{F}(\hat{a},\hat{b},\dots)}_{f(a,b,\hat{a},\hat{b},I)} \delta^4_R \delta^4_L \delta^4_{\mathcal{F}} \\
    &= \delta^4(\text{P}_{\text{tot}}) \int \frac{dz}{z} \frac{1}{s_{ab}} f(p_{\hat{a}} = \tl_a(\la_a - z \la_b), p_{\hat{b}} = \la_b(\tl_b + z \tl_a) ) \\
    &= \delta^4(\text{P}_{\text{tot}}) \int \frac{dz}{z} z^{\theta_h} \mathcal{F}(\la_{\hat{a}} = \la_a - z \la_b , \tl_{\hat{b}} = \tl_b + z \tl_a) \, ,
\end{aligned}
\end{equation}
where $\theta_h$ is an exponent which depends on the helicity of legs $a$ and $\hat{a}$. For the sake of conciseness, we graphically represent the possible BCFW bridges and their corresponding $z$-integrands. We can write them explicitly as follows
\begin{equation}
\begin{aligned}
    \vcenter{\hbox{\begin{tikzpicture}[scale=0.7,rotate=-90,decoration={,markings,mark=at position 0.7 with {\arrow{latex}}}]
    		\coordinate (shiftv) at (0.8,0);
 
        		\coordinate (p1) at ($({cos(0)},{sin(0)})$);
       		\coordinate (p2) at ($({cos(120)},{sin(120)})$);
        		\coordinate (p3) at ($({cos(-120)},{sin(-120)})$);
        		\coordinate (p1a) at ($0.2*({cos(0)},{sin(0)})$);
        		\coordinate (p2a) at ($0.2*({cos(120)},{sin(120)})$);
        		\coordinate (p3a) at ($0.2*({cos(-120)},{sin(-120)})$);

        		\draw[ thick] (shiftv) circle (0.2);
        		\draw[very thick,postaction={decorate}] ($(0.2,0)-(shiftv)$) to ($(-0.2,0)+(shiftv)$);
        		\draw[very thick,postaction={decorate}] ($(60:0.2)+(shiftv)$) to ($(60:1)+(shiftv)$);
        		\draw[very thick,postaction={decorate}] ($(-90:0.2)+(shiftv)$) to ($(-90:1)+(shiftv)$);
        		\draw[fill,thick] ($-1*(shiftv)$) circle (0.2);
        		\draw[very thick,postaction={decorate}] ($(120:1)- (shiftv)$) to ($(120:0.2) - (shiftv)$);
        		\draw[very thick,postaction={decorate}] ($(-90:1)-(shiftv)$) to ($(-90:0.2)-(shiftv)$);
        		
        		\node (p1t) at ($(120:0.5)-(shiftv)-(0.3,0)$){$a$};
        		\node (p2t) at ($(-90:0.6)-(shiftv)-(0.3,0)$){$\hat{a}$};
        		\node (p3t) at ($(60:0.5)+(shiftv)+(0.3,0)$){$b$};
        		\node (p4t) at ($(-90:0.6)+(shiftv)+(0.3,0)$){$\hat{b}$};
        		\node (p4t) at ($(0,0.3)$){$I$};
		\draw[very thick] (0,-1.6) ellipse (1.5 and 0.7);
		\node at (0,-1.65) {\textbf{$\mathcal{F}$}};
		\draw[very thick] ($(-145:1)+ (0,-1.65)$) to ($(-145:1)+ (0,-1.65) + (-120:0.8)$);
		\draw[very thick] ($(-35:1)+ (0,-1.65)$) to ($(-35:1)+ (0,-1.65) + (-60:0.8)$);
		\node[rotate=90] at (0,-2.65) {$\dots$};
    \end{tikzpicture}}} &= \delta^{4}(\text{P}_{tot}) \int \frac{dz}{z}z^4 \mathcal{F}(z) \, , \quad
    \vcenter{\hbox{\begin{tikzpicture}[scale=0.7,rotate=-90,decoration={,markings,mark=at position 0.7 with {\arrow{latex}}}]
    		\coordinate (shiftv) at (0.8,0);
 
        		\coordinate (p1) at ($({cos(0)},{sin(0)})$);
       		\coordinate (p2) at ($({cos(120)},{sin(120)})$);
        		\coordinate (p3) at ($({cos(-120)},{sin(-120)})$);
        		\coordinate (p1a) at ($0.2*({cos(0)},{sin(0)})$);
        		\coordinate (p2a) at ($0.2*({cos(120)},{sin(120)})$);
        		\coordinate (p3a) at ($0.2*({cos(-120)},{sin(-120)})$);

        		\draw[ thick] (shiftv) circle (0.2);
        		\draw[very thick,postaction={decorate}] ($(-0.2,0)+(shiftv)$) to ($(0.2,0)-(shiftv)$) ;
        		\draw[very thick] ($(60:0.2)+(shiftv)$) to ($(60:1)+(shiftv)$);
        		\draw[very thick] ($(-90:0.2)+(shiftv)$) to ($(-90:1)+(shiftv)$);
        		\draw[fill,thick] ($-1*(shiftv)$) circle (0.2);
        		\draw[very thick] ($(120:1)- (shiftv)$) to ($(120:0.2) - (shiftv)$);
        		\draw[very thick] ($(-90:1)-(shiftv)$) to ($(-90:0.2)-(shiftv)$);
        		
        		\node (p1t) at ($(120:0.5)-(shiftv)-(0.3,0)$){$a$};
        		\node (p2t) at ($(-90:0.6)-(shiftv)-(0.3,0)$){$\hat{a}$};
        		\node (p3t) at ($(60:0.5)+(shiftv)+(0.3,0)$){$b$};
        		\node (p4t) at ($(-90:0.6)+(shiftv)+(0.3,0)$){$\hat{b}$};
        		\node (p4t) at ($(0,0.3)$){$I$};
		\draw[very thick] (0,-1.6) ellipse (1.5 and 0.7);
		\node at (0,-1.65) {\textbf{$\mathcal{F}$}};
		\draw[very thick] ($(-145:1)+ (0,-1.65)$) to ($(-145:1)+ (0,-1.65) + (-120:0.8)$);
		\draw[very thick] ($(-35:1)+ (0,-1.65)$) to ($(-35:1)+ (0,-1.65) + (-60:0.8)$);
		\node[rotate=90] at (0,-2.65) {$\dots$};
    \end{tikzpicture}}} &= \delta^{4}(\text{P}_{tot}) \int \frac{dz}{z} \mathcal{F}(z) \, ,
	\end{aligned}
\end{equation}
Note how in the second bridge, we only assigned helicity to the intermediate leg, because the formula applies regardless of the helicity assignment on legs $a$ and $b$.

To illustrate the usefulness of the construction above, we can apply it to the simplest case discussed so far---the factorization channel of eq.~\eqref{eq:4ptfactchan}. Attaching a non-trivial BCFW bridge, we find
\begin{equation}
	\begin{aligned}
	\vcenter{\hbox{\begin{tikzpicture}[decoration={,markings,mark=at position 0.7 with {\arrow{latex}}}]
			
        		\draw[thick] (0.6,0.6) circle (0.2);
        		\draw[fill, thick] (-0.6,0.6) circle (0.2);
        		\draw[fill, thick] (0.6,-0.6) circle (0.2);
        		\draw[ thick] (-0.6,-0.6) circle (0.2);
        		
        		\draw[very thick, postaction={decorate}] (-0.4,0.6) to (0.4,0.6);
			\draw[very thick, postaction={decorate}] (0.6,0.4) to (0.6,-0.4);
			\draw[very thick, postaction={decorate}] (0.4,-0.6) to (-0.4,-0.6);
			\draw[very thick, postaction={decorate}] (-0.6,-0.4) to (-0.6,0.4);
			
			\draw[very thick, postaction={decorate}] ($(135:1)+(-0.6,0.6)$) to ($(135:0.2)+(-0.6,0.6)$);
			\draw[very thick, postaction={decorate}]  ($(45:0.2)+(0.6,0.6)$) to ($(45:1)+(0.6,0.6)$);
			\draw[very thick, postaction={decorate}] ($(-45:1)+(0.6,-0.6)$) to ($(-45:0.2)+(0.6,-0.6)$);
			\draw[very thick, postaction={decorate}] ($(-135:0.2)+(-0.6,-0.6)$) to ($(-135:1)+(-0.6,-0.6)$);

        		\node at ($(135:0.5)+(-1,0.6)$) {$1$};
        		\node at ($(45:0.5)+(1,0.6)$) {$2$};
        		\node at ($(-45:0.5)+(1,-0.6)$) {$3$};
        		\node at ($(-135:0.5)+(-1,-0.6)$) {$4$};
        
    \end{tikzpicture}}} &= \delta^4(\text{P}_{\text{tot}}) \int \frac{dz}{z} \left(\frac{\ab{1\hat{3}}^4 }{\ab{12}\ab{23}\ab{41}} \delta(\ab{\hat{3}4}) \right) \\
    &= \frac{\ab{13}^4}{\ab{12}\ab{23}\ab{41}} \int \frac{dz}{z} (\ab{13}-z\ab{12})^4\delta(\ab{34}-z\ab{24}) \\
    &= \frac{\ab{13}^4}{\ab{12}\ab{23}\ab{34}\ab{41}} \left(\frac{\ab{14}\ab{32}}{\ab{13}\ab{24}}\right)^4 \, ,
    \label{eq:4pteg}
\end{aligned}
\end{equation}
which is an example of a non-Parke-Taylor leading singularity already shown in section~\ref{sec: Leading singularities for MHV QCD amplitudes}.

The other useful tool in computing on-shell diagrams are the inverse soft-factors. These allow us to increase the number of legs by one, at the cost of introducing a delta function. To any on-shell function $\mathcal{F}$, we can attach a three-particle vertex in the following way

\begin{equation}
    \begin{aligned}
	\vcenter{\hbox{\begin{tikzpicture}[decoration={,markings,mark=at position 0.7 with {\arrow{latex}}}]
    		\coordinate (shiftv) at (1,0);
 
        		\coordinate (p1) at ($({cos(0)},{sin(0)})$);
       		\coordinate (p2) at ($({cos(120)},{sin(120)})$);
        		\coordinate (p3) at ($({cos(-120)},{sin(-120)})$);
        		\coordinate (p1a) at ($0.2*({cos(0)},{sin(0)})$);
        		\coordinate (p2a) at ($0.2*({cos(120)},{sin(120)})$);
        		\coordinate (p3a) at ($0.2*({cos(-120)},{sin(-120)})$);

        		\draw[ thick] (shiftv) circle (0.2);
        		
        		\draw[very thick,postaction={decorate}]  ($(0,0)$) to ($(-0.2,0)+(shiftv)$) ;
        		\draw[very thick,postaction={decorate}] ($(60:0.2)+(shiftv)$) to ($(60:1)+(shiftv)$) ;
        		\draw[very thick,postaction={decorate}] ($(-60:0.2)+(shiftv)$) to ($(-60:1)+(shiftv)$);

        		\node (p3t) at ($(60:0.5)+(shiftv)+(0.3,0)$){$n$};
        		\node (p4t) at ($(-60:0.6)+(shiftv)+(0.6,0)$){$n+1$};
        		\node (p4t) at ($(0.5,0.3)$){$\bar{n}$};		
        		\draw[very thick] (-1,0) ellipse (1 and 0.7);
		\node at (-1,0) {\textbf{$\mathcal{F}$}};
		\draw[very thick] ($(-1.9,0.3)$) to ($(-1.9,0.3) + (145:0.8)$);
		\draw[very thick] ($(-1.9,-0.3)$) to ($(-1.9,-0.3) + (-145:0.8)$);

		\node at ($(-1.9,0.3) + (145:0.8) + (-0.6,0)$) {$n-1$};
		\node at ($(-1.9,-0.3) + (-145:0.8) + (-0.3,0)$) {$1$};

		\node[rotate=90] at (-2.5,0) {$\dots$};
    \end{tikzpicture}}} = \int d^4 p_{\bar{n}} \delta(p_{\bar{n}}^2) \delta^4_{\mathcal{F}}\delta^{4}(p_n + p_{n+1}-p_{\bar{n}}) \mathcal{F}(p_1,\dots,p_{\bar{n}})  \mathcal{A}_3\\
    = \delta(\ab{n,n+1}) \mathcal{F}\left(\lambda_{\bar{n}} = \lambda_n , \, \tl_{\bar{n}} = \tl_n + \frac{\ab{r,n+1}}{\ab{r,n}}\tl_{n+1}\right) \frac{\ab{rn}}{\ab{r,n+1}} \delta^4(\text{P}_{\text{tot}})\, ,
    \end{aligned}
    \label{eq:softfact1}
\end{equation}
where $\lambda_r$ is an arbitrary reference spinor introduced when solving the momentum-conservation conditions. It may be chosen freely to simplify the calculations, provided it is distinct from $\la_n$ and $\la_{n+1}$. Furthermore, as in the case of the BCFW bridge, different helicity configurations of legs $n$ and $n+1$ lead to different factors of $\frac{\ab{r,n}}{\ab{r,n+1}}$. The remaining two soft-factors are
\begin{equation}
	\begin{aligned}
		\vcenter{\hbox{\begin{tikzpicture}[decoration={,markings,mark=at position 0.7 with {\arrow{latex}}}]
    		\coordinate (shiftv) at (1,0);
 
        		\coordinate (p1) at ($({cos(0)},{sin(0)})$);
       		\coordinate (p2) at ($({cos(120)},{sin(120)})$);
        		\coordinate (p3) at ($({cos(-120)},{sin(-120)})$);
        		\coordinate (p1a) at ($0.2*({cos(0)},{sin(0)})$);
        		\coordinate (p2a) at ($0.2*({cos(120)},{sin(120)})$);
        		\coordinate (p3a) at ($0.2*({cos(-120)},{sin(-120)})$);

        		\draw[ thick] (shiftv) circle (0.2);
        		
        		\draw[very thick,postaction={decorate}]  ($(-0.2,0)+(shiftv)$) to($(0,0)$)  ;
        		\draw[very thick,postaction={decorate}]  ($(60:1)+(shiftv)$) to ($(60:0.2)+(shiftv)$);
        		\draw[very thick,postaction={decorate}] ($(-60:0.2)+(shiftv)$) to ($(-60:1)+(shiftv)$);

        		\node (p3t) at ($(60:0.5)+(shiftv)+(0.3,0)$){$n$};
        		\node (p4t) at ($(-60:0.6)+(shiftv)+(0.6,0)$){$n+1$};
        		\node (p4t) at ($(0.5,0.3)$){$\bar{n}$};		
        		\draw[very thick] (-1,0) ellipse (1 and 0.7);
		\node at (-1,0) {\textbf{$\mathcal{F}$}};
		\draw[very thick] ($(-1.9,0.3)$) to ($(-1.9,0.3) + (145:0.8)$);
		\draw[very thick] ($(-1.9,-0.3)$) to ($(-1.9,-0.3) + (-145:0.8)$);

		\node at ($(-1.9,0.3) + (145:0.8) + (-0.6,0)$) {$n-1$};
		\node at ($(-1.9,-0.3) + (-145:0.8) + (-0.3,0)$) {$1$};

		\node[rotate=90] at (-2.5,0) {$\dots$};
    \end{tikzpicture}}} \, &= \frac{\ab{rn}}{\ab{rn+1}}\bar{\mathcal{F}}\delta(\ab{n,n+1}) \, ,\\ 
    \vcenter{\hbox{\begin{tikzpicture}[decoration={,markings,mark=at position 0.7 with {\arrow{latex}}}]
    		\coordinate (shiftv) at (1,0);
 
        		\coordinate (p1) at ($({cos(0)},{sin(0)})$);
       		\coordinate (p2) at ($({cos(120)},{sin(120)})$);
        		\coordinate (p3) at ($({cos(-120)},{sin(-120)})$);
        		\coordinate (p1a) at ($0.2*({cos(0)},{sin(0)})$);
        		\coordinate (p2a) at ($0.2*({cos(120)},{sin(120)})$);
        		\coordinate (p3a) at ($0.2*({cos(-120)},{sin(-120)})$);

        		\draw[ thick] (shiftv) circle (0.2);
        		
        		\draw[very thick,postaction={decorate}]  ($(-0.2,0)+(shiftv)$) to($(0,0)$)  ;
        		\draw[very thick,postaction={decorate}] ($(60:0.2)+(shiftv)$) to ($(60:1)+(shiftv)$) ;
        		\draw[very thick,postaction={decorate}]  ($(-60:1)+(shiftv)$) to ($(-60:0.2)+(shiftv)$);

        		\node (p3t) at ($(60:0.5)+(shiftv)+(0.3,0)$){$n$};
        		\node (p4t) at ($(-60:0.6)+(shiftv)+(0.6,0)$){$n+1$};
        		\node (p4t) at ($(0.5,0.3)$){$\bar{n}$};		
        		\draw[very thick] (-1,0) ellipse (1 and 0.7);
		\node at (-1,0) {\textbf{$\mathcal{F}$}};
		\draw[very thick] ($(-1.9,0.3)$) to ($(-1.9,0.3) + (145:0.8)$);
		\draw[very thick] ($(-1.9,-0.3)$) to ($(-1.9,-0.3) + (-145:0.8)$);

		\node at ($(-1.9,0.3) + (145:0.8) + (-0.6,0)$) {$n-1$};
		\node at ($(-1.9,-0.3) + (-145:0.8) + (-0.3,0)$) {$1$};

		\node[rotate=90] at (-2.5,0) {$\dots$};
    \end{tikzpicture}}} &= \left(\frac{\ab{rn+1}}{\ab{rn}}\right)^3\bar{\mathcal{F}}\delta(\ab{n,n+1})\, ,
	\end{aligned}
\end{equation}
where $\bar{\mathcal{F}}$ is $\mathcal{F}$ evaluated on the kinematic locus of eq.~\eqref{eq:softfact1}.

To put both of these methods to use, we end this discussion by computing the following five-point one-loop leading singularity
\begin{equation}
    \vcenter{\hbox{\begin{tikzpicture}[]
			
        		\coordinate (p2) at (-0.6,0.6);
        		\coordinate (p3) at (0.6,0.6);
        		\coordinate (p4) at (0.6,-0.6);
        		\coordinate (p1) at (-0.6,-0.6);
        		\coordinate (p1aux) at ($(p1) + (-135:1)$);
        		\coordinate (p3aux) at ($(p3)+(45:1)$);
        		\coordinate (p2aux1) at ($(p2) + (180:1)$);
        		\coordinate (p2aux2) at ($(p2) + (135:1)$);
        		\coordinate (p2aux3) at ($(p2) + (90:1)$);
        		
        		\coordinate (p4aux1) at ($(p4) + (0:1)$);
        		\coordinate (p4aux2) at ($(p4) + (-45:1)$);
        		\coordinate (p4aux3) at ($(p4) + (-90:1)$);
        		
        		\draw[very thick,rarrow] (p1) to (p2);
			\draw[very thick,rarrow ] (p2) to (p3);
			\draw[very thick, rarrow] (p3) to (p4);
			\draw[very thick, rarrow] (p4) to (p1);
			
			\draw[very thick, larrow] (p1aux) to (p1);
			
			\draw[very thick, larrow ]  (p2aux1) to (p2);
			\draw[very thick, rarrow]  (p2aux3) to (p2);

			\draw[very thick, larrow] (p3aux) to (p3);
			
			\draw[very thick, rarrow]  (p4aux2) to (p4);

			\draw[fill=white,thick] (p1) circle (0.2);
        		\draw[fill, thick] (p2) circle (0.2);
        		\draw[fill=white, thick] (p3) circle (0.2);
        		\draw[fill, thick] (p4) circle (0.2);
        		\node at ($(p1aux) + (-135:0.3)$) {$4$};
        		\node at ($(p2aux1) + (180:0.3)$) {$5$};        \node at ($(p2aux3) + (90:0.3)$) {$1$};
        		\node at ($(p3aux) + (45:0.3)$) {$2$};
        		\node at ($(p4aux2) + (-45:0.3)$) {$3$};
    \end{tikzpicture}}} \, .
\end{equation}
To compute it, we can break it down into two steps: first insert an inverse soft-factor and then attach a BCFW bridge. One such possibility is 
\begin{equation}
\vcenter{\hbox{\begin{tikzpicture}[]

        		\coordinate (p2) at (-0.6,0.6);
        		\coordinate (p3) at (0.6,0.6);
        		\coordinate (p4) at (0.6,-0.6);
        		\coordinate (p1) at (-0.6,-0.6);
        		\coordinate (p1aux) at ($(p1) + (-135:1)$);
        		\coordinate (p3aux) at ($(p3)+(45:1)$);
        		\coordinate (p2aux1) at ($(p2) + (180:1)$);
        		\coordinate (p2aux2) at ($(p2) + (135:1)$);
        		\coordinate (p2aux3) at ($(p2) + (90:1)$);
        		
        		\coordinate (p4aux1) at ($(p4) + (0:1)$);
        		\coordinate (p4aux2) at ($(p4) + (-45:1)$);
        		\coordinate (p4aux3) at ($(p4) + (-90:1)$);
        		
        		\draw[very thick,rarrow] (p1) to (p2);
			\draw[very thick,rarrow ] (p2) to (p3);
			
			
			\draw[very thick, larrow ]  (p2aux1) to (p2);
			\draw[very thick, rarrow]  (p2aux3) to (p2);

			

        		\draw[fill, thick] (p2) circle (0.2);
        		\node at ($(p2aux1) + (180:0.3)$) {$5$};        \node at ($(p2aux3) + (90:0.3)$) {$1$};
        		\node at ($(p3) + (0:0.3)$) {$2$};
       		\node at ($(p1) + (0,-0.35)$) {$3$};
    \end{tikzpicture}}}\rightarrow
\vcenter{\hbox{\begin{tikzpicture}[]
			\draw [dashed, very thick] (-1,-0.2) to (0,-0.2);
            \node at (-1.5,-0.2) {\small I.S.F };

        		\coordinate (p2) at (-0.6,0.6);
        		\coordinate (p3) at (0.6,0.6);
        		\coordinate (p4) at (0.6,-0.6);
        		\coordinate (p1) at (-0.6,-0.6);
        		\coordinate (p1aux) at ($(p1) + (-135:1)$);
        		\coordinate (p3aux) at ($(p3)+(45:1)$);
        		\coordinate (p2aux1) at ($(p2) + (180:1)$);
        		\coordinate (p2aux2) at ($(p2) + (135:1)$);
        		\coordinate (p2aux3) at ($(p2) + (90:1)$);
        		
        		\coordinate (p4aux1) at ($(p4) + (0:1)$);
        		\coordinate (p4aux2) at ($(p4) + (-45:1)$);
        		\coordinate (p4aux3) at ($(p4) + (-90:1)$);
        		
        		\draw[very thick,rarrow] (p1) to (p2);
			\draw[very thick,rarrow ] (p2) to (p3);
			\draw[very thick, rarrow] (p4) to (p1);
			
			\draw[very thick, larrow] (p1aux) to (p1);
			
			\draw[very thick, larrow ]  (p2aux1) to (p2);
			\draw[very thick, rarrow]  (p2aux3) to (p2);

			

			\draw[fill=white,thick] (p1) circle (0.2);
        		\draw[fill, thick] (p2) circle (0.2);
        		\node at ($(p1aux) + (-135:0.3)$) {$4$};
        		\node at ($(p2aux1) + (180:0.3)$) {$5$};        \node at ($(p2aux3) + (90:0.3)$) {$1$};
       		\node at ($(p1)!0.5!(p4) + (0,-0.35)$) {$3$};
       		\node at ($(p2)!0.5!(p3) + (0,0.35)$) {$2$};
    \end{tikzpicture}}}
    \rightarrow
    \vcenter{\hbox{\begin{tikzpicture}[]
			\draw [blue,dashed, very thick] (0.2,1) to (0.2,-1);  
            \node[blue] at (0,-1.3) {\small$[2,3\rangle$ shift};

        		\coordinate (p2) at (-0.6,0.6);
        		\coordinate (p3) at (0.6,0.6);
        		\coordinate (p4) at (0.6,-0.6);
        		\coordinate (p1) at (-0.6,-0.6);
        		\coordinate (p1aux) at ($(p1) + (-135:1)$);
        		\coordinate (p3aux) at ($(p3)+(45:1)$);
        		\coordinate (p2aux1) at ($(p2) + (180:1)$);
        		\coordinate (p2aux2) at ($(p2) + (135:1)$);
        		\coordinate (p2aux3) at ($(p2) + (90:1)$);
        		
        		\coordinate (p4aux1) at ($(p4) + (0:1)$);
        		\coordinate (p4aux2) at ($(p4) + (-45:1)$);
        		\coordinate (p4aux3) at ($(p4) + (-90:1)$);
        		
        		\draw[very thick,rarrow] (p1) to (p2);
			\draw[very thick,rarrow ] (p2) to (p3);
			\draw[very thick, rarrow] (p3) to (p4);
			\draw[very thick, rarrow] (p4) to (p1);
			
			\draw[very thick, larrow] (p1aux) to (p1);
			
			\draw[very thick, larrow ]  (p2aux1) to (p2);
			\draw[very thick, rarrow]  (p2aux3) to (p2);

			\draw[very thick, larrow] (p3aux) to (p3);
			
			\draw[very thick, rarrow]  (p4aux2) to (p4);

			\draw[fill=white,thick] (p1) circle (0.2);
        		\draw[fill, thick] (p2) circle (0.2);
        		\draw[fill=white, thick] (p3) circle (0.2);
        		\draw[fill, thick] (p4) circle (0.2);
        		\node at ($(p1aux) + (-135:0.3)$) {$4$};
        		\node at ($(p2aux1) + (180:0.3)$) {$5$};        \node at ($(p2aux3) + (90:0.3)$) {$1$};
        		\node at ($(p3aux) + (45:0.3)$) {$2$};
        		\node at ($(p4aux2) + (-45:0.3)$) {$3$};
    \end{tikzpicture}}} \, 
\end{equation}
from which we obtain
\begin{equation}
    \int \frac{dz}{z} \frac{\ab{1\hat{3}}^4}{\ab{12}\ab{23}\ab{45}\ab{51}} \delta(\ab{\hat{3}4}) = \text{PT}_{5,13}\left(\frac{\ab{14}\ab{32}}{\ab{24}\ab{13}}\right)^4\, ,
\end{equation}
which exactly matches the second term of eq.~\eqref{eq:L1LS}.

At two loops, low-multiplicity on-shell diagrams can be efficiently evaluated using the BCFW-bridge and inverse-soft constructions, rather than by sewing the tree-level vertices. In particular, this allows one to work recursively in the loop order. For instance, let us consider the four-point amplitude in the helicity configuration $(--++)$. Its two-loop leading singularities are computed from the double-box on-shell diagram. One of the helicity assignments featuring two orientated cycles is shown below,
\begin{equation}
	\begin{tikzpicture}[scale=0.75]
			
			\coordinate (d45) at (45:1);
			\coordinate (dm45) at (-45:1);
			\coordinate (d135) at (135:1);
			\coordinate (dm135) at (-135:1);
			
	  		\coordinate (p1) at (-1.2,0.6);
	  		\coordinate (p2) at (0,0.6);
	  		\coordinate (p3) at (1.2,0.6);
	  		\coordinate (p4) at (1.2,-0.6);
	  		\coordinate (p5) at (0,-0.6);
	  		\coordinate (p6) at (-1.2,-0.6);
	  		
	  		\coordinate (p1aux1) at ($(p1)+(-1,0)$);
	  		\coordinate (p1aux2) at ($(p1)+(d135)$);
	  		\coordinate (p1aux3) at ($(p1)+(0,1)$);
	  		\coordinate (p3aux3) at ($(p3)+(1,0)$);
	  		\coordinate (p3aux2) at ($(p3)+(d45)$);
	  		\coordinate (p3aux1) at ($(p3)+(0,1)$);
	  		\coordinate (p5aux2) at ($(p5)+(0,-1)$);
	  		\coordinate (p5aux1) at ($(p5)+(-60:1)$);
	  		\coordinate (p5aux3) at ($(p5)+(-120:1)$);
	  		
	  		\coordinate (p2aux) at ($(p2)+(0,1)$);
	  		\coordinate (p4aux) at ($(p4)+(dm45)$);
	  		\coordinate (p6aux) at ($(p6)+(dm135)$);
	  		
	  		\draw[very thick,larrow] (p1) to (p2);
	  		\draw[very thick,larrow] (p2) to (p5);
	  		\draw[very thick,larrow] (p5) to (p6);
	  		\draw[very thick,larrow] (p6) to (p1);
	  		
	  		\draw[very thick,rarrow] (p2) to (p3);
	  		\draw[very thick,rarrow] (p3) to (p4);
	  		\draw[very thick,rarrow] (p4) to (p5);
	  		
        		\draw[very thick,rarrow] (p1aux2) to (p1);
        		
        		\draw[very thick,rarrow] (p3aux2) to (p3);

        		
        		\draw[very thick, larrow] (p4aux) to (p4);
        		\draw[very thick,larrow] (p6aux) to (p6);

        		\draw[fill, thick] (p1) circle (0.2);
        		\draw[fill=white,thick] (p2) circle (0.2);
            \draw[fill, thick] (p3) circle (0.2);
 		    \draw[fill=white, thick] (p4) circle (0.2);
        		\draw[fill, thick] (p5) circle (0.2);
        		\draw[fill=white, thick] (p6) circle (0.2);

        		
        		\node at ($(p1aux2)+(135:0.35)$) {$1$};
%
        		\node at ($(p3aux2)+(45:0.35)$) {$2$};
%
%
        		\node at ($(p4aux)+(-45:0.35)$) {$3$};
        		\node at ($(p6aux)+(-135:0.35)$) {$4$};
        		
        
    \end{tikzpicture}
\end{equation}
This on-shell diagram is readily evaluated by applying the BCFW $\langle23]$-shift to its one-loop sub-diagram (left box of the double box),
\begin{equation}
    \frac{\langle13\rangle^4}{\langle12\rangle\langle23\rangle\langle34\rangle\langle41\rangle}\left(\frac{\langle12\rangle\langle34\rangle}{\langle13\rangle\langle24\rangle}\right)^4\xrightarrow{\langle23]-\text{shift}}{\rm d}z\  z^3\frac{\langle34\rangle^3}{\langle23\rangle\langle41\rangle}\frac{(\langle12\rangle+z\langle13\rangle)^3}{(\langle24\rangle+z\langle34\rangle)^4},
\end{equation}
that corresponds to the first term in the one-form $\Omega_{12,341,4}^{(2),\text{db}}$ \eqref{eq:LSdb}. As suggested in the main text, if we sum over the other internal lines' helicity configurations and take the residue of the resulting form at $z=\infty$, we get
\begin{equation}
    R^{\text{db}}_{12,341,4} = \text{PT}_{12,4} \left(-2 +12 \frac{\ab{13}\ab{24}}{\ab{12}\ab{34}} -30 \left(\frac{\ab{13}\ab{24}}{\ab{12}\ab{34}} \right)^2 +20\left(\frac{\ab{13}\ab{24}}{\ab{12}\ab{34}}\right)^3\right) \, .
\end{equation}
Note that this is exactly the double-box leading singularity of equation \eqref{eq:Rdb12}.

Similar simplifications also occur for penta-box on-shell diagrams, which we illustrate with the following five-point example in the helicity sector $(-+-++)$,

\begin{equation}
	\begin{tikzpicture}[scale=0.7]
			
			\coordinate (d45) at (45:1);
			\coordinate (dm45) at (-45:1);
			\coordinate (d135) at (135:1);
			\coordinate (dm135) at (-135:1);
			
	  		\coordinate (p1) at (180:1);
	  		\coordinate (p2) at (108:1);
	  		\coordinate (p3) at (36:1);
	  		\coordinate (p6) at (-36:1);
	  		\coordinate (p7) at (-108:1);
	  		
	  		\coordinate (p4) at ($(p3)+(1.2,0)$);
	  		\coordinate (p5) at ($(p6)+(1.2,0)$);
	  		
	  		\coordinate (p1aux1) at ($(p1)+(dm135)$);
	  		\coordinate (p1aux2) at ($(p1)+(-1,0)$);
	  		\coordinate (p1aux3) at ($(p1)+(d135)$);

	  		\coordinate (p3aux1) at ($(p3)+(60:1)$);
	  		\coordinate (p3aux2) at ($(p3)+(90:1)$);
	  		\coordinate (p3aux3) at ($(p3)+(120:1)$);

	  		\coordinate (p5aux1) at ($(p5)+(1,0)$);
	  		\coordinate (p5aux2) at ($(p5)+(dm45)$);
	  		\coordinate (p5aux3) at ($(p5)+(0,-1)$);

	  		\coordinate (p2aux1) at ($(p2)+(108:1)$);
	  		\coordinate (p4aux1) at ($(p4)+(45:1)$);
	  		\coordinate (p7aux1) at ($(p7)+(-108:1)$);

	  		\draw[very thick, rarrow ] (p1) to (p2);
	  		\draw[very thick, rarrow ] (p2) to (p3);
	  		\draw[very thick, rarrow ] (p3) to (p6);
	  		\draw[very thick, rarrow] (p6) to (p7);
	  		\draw[very thick, rarrow] (p7) to (p1);
	  		\draw[very thick, larrow] (p5) to (p6);
	  		\draw[very thick, larrow] (p3) to (p4);
	  		\draw[very thick, larrow] (p4) to (p5);
	  		
	  		\draw[very thick,rarrow ] (p1aux2) to (p1);
	  		
	  		
	  		\draw[very thick,rarrow ] (p5aux2) to (p5);
	  		
	  		\draw[very thick, larrow ] (p2aux1) to (p2);
	  		\draw[very thick, larrow] (p4aux1) to (p4);
	  		\draw[very thick, larrow] (p7aux1) to (p7);

        		\draw[fill, thick] (p1) circle (0.2);
        		\draw[fill=white,thick] (p2) circle (0.2);
            \draw[fill, thick] (p3) circle (0.2);
 		    \draw[fill=white, thick] (p6) circle (0.2);
        		\draw[fill=white, thick] (p7) circle (0.2);
        		\draw[fill, thick] (p5) circle (0.2);
        		\draw[fill=white, thick] (p4) circle (0.2);

        		\node at ($(p1aux2)+(180:0.2)$) {$1$};
%
%
        		\node at ($(p5aux2)+(-45:0.3)$) {$3$};
%
        		\node at ($(p4aux1)+(45:0.35)$) {$4$};
        		\node at ($(p2aux1)+(108:0.35)$) {$5$};
        		\node at ($(p7aux1)+(-108:0.35)$) {$2$};
        		
        
    \end{tikzpicture}
\end{equation}
We compute this on-shell diagram by attaching an inverse-soft factor to its one-loop box subdiagram,
\begin{equation}\label{eq:invsofteg}
    \frac{\langle35\rangle^4}{\langle23\rangle\langle34\rangle\langle45\rangle\langle52\rangle}\times\left(\frac{\langle23\rangle\langle45\rangle}{\langle24\rangle\langle35\rangle}\right)^4\xrightarrow[\text{factor}\ \lambda_1]{\text{inverse-soft}}\text{PT}_{5,13}\left(\frac{\langle12\rangle\langle23\rangle\langle45\rangle}{\langle13\rangle\langle24\rangle\langle25\rangle}\right)^4 \,.
\end{equation}
The latter expression corresponds to the first term of $R_{13,245,5}^{(2),\text{pb}}$ given in \eqref{eq:Rpb}.

\subsection{Four and Five-point prefactors at two-loops \label{sec:dbpref}}

Using the building blocks above, we can construct all the leading singularities discussed in this work, but of particular interest are the double boxes since their $\delta$-function constraints localize all but one integration. In order to get the leading singularities used in this work, one needs to take a further residue at infinity in this integration variable. Although somewhat big, the expressions for four and five points can still be written down in terms of little-group invariant cross-ratios of angle brackets. When we write an on-shell diagram without internal helicities assigned, it is intended to represent the sum over all possible helicity assignments.  

\subsubsection{$--++$ prefactors at two-loops}

For four external gluons, there is only one new prefactor at two-loops which comes from the split helicity sector $--++$. Its expression is 
\begin{equation}
	R_{12,341,4}^{\text{(db)}} = \Res_{z=\infty}
	\vcenter{\hbox{\begin{tikzpicture}[scale=0.75]
			
			\coordinate (d45) at (45:1);
			\coordinate (dm45) at (-45:1);
			\coordinate (d135) at (135:1);
			\coordinate (dm135) at (-135:1);
			
	  		\coordinate (p1) at (-1.2,0.6);
	  		\coordinate (p2) at (0,0.6);
	  		\coordinate (p3) at (1.2,0.6);
	  		\coordinate (p4) at (1.2,-0.6);
	  		\coordinate (p5) at (0,-0.6);
	  		\coordinate (p6) at (-1.2,-0.6);
	  		
	  		\coordinate (p1aux1) at ($(p1)+(-1,0)$);
	  		\coordinate (p1aux2) at ($(p1)+(d135)$);
	  		\coordinate (p1aux3) at ($(p1)+(0,1)$);
	  		\coordinate (p3aux3) at ($(p3)+(1,0)$);
	  		\coordinate (p3aux2) at ($(p3)+(d45)$);
	  		\coordinate (p3aux1) at ($(p3)+(0,1)$);
	  		\coordinate (p5aux2) at ($(p5)+(0,-1)$);
	  		\coordinate (p5aux1) at ($(p5)+(-60:1)$);
	  		\coordinate (p5aux3) at ($(p5)+(-120:1)$);
	  		
	  		\coordinate (p2aux) at ($(p2)+(0,1)$);
	  		\coordinate (p4aux) at ($(p4)+(dm45)$);
	  		\coordinate (p6aux) at ($(p6)+(dm135)$);
	  		
	  		\draw[very thick] (p1) to (p2);
	  		\draw[very thick] (p2) to (p5);
	  		\draw[very thick] (p5) to (p6);
	  		\draw[very thick] (p6) to (p1);
	  		
	  		\draw[very thick] (p2) to (p3);
	  		\draw[very thick] (p3) to (p4);
	  		\draw[very thick] (p4) to (p5);
	  		
        		\draw[very thick,rarrow] (p1aux2) to (p1);
        		
        		\draw[very thick,rarrow] (p3aux2) to (p3);

        		
        		\draw[very thick, larrow] (p4aux) to (p4);
        		\draw[very thick,larrow] (p6aux) to (p6);

        		\draw[fill, thick] (p1) circle (0.2);
        		\draw[fill=white,thick] (p2) circle (0.2);
            \draw[fill, thick] (p3) circle (0.2);
 		    \draw[fill=white, thick] (p4) circle (0.2);
        		\draw[fill, thick] (p5) circle (0.2);
        		\draw[fill=white, thick] (p6) circle (0.2);

        		
        		\node at ($(p1aux2)+(135:0.35)$) {$1$};
%
        		\node at ($(p3aux2)+(45:0.35)$) {$2$};
%
%
        		\node at ($(p4aux)+(-45:0.35)$) {$3$};
        		\node at ($(p6aux)+(-135:0.35)$) {$4$};
        		
        
    \end{tikzpicture}}} = -2 + 12 u_{34} - 30 u_{34}^2 + 20 u_{34}^3 \, ,
\end{equation}
where the cross ratio $u_{34}$ is given by
\begin{equation}
	u_{34} = \frac{\ab{13}\ab{24}}{\ab{12}\ab{34}} \, .
    \label{eq:u34}
\end{equation}

\subsubsection{$--+++$ prefactors at two-loops}

For five external gluons, we have new prefactors in both helicity configurations $(--+++)$ and $(-+-++)$. For the split helicity sector, the expressions are simply \eqref{eq:Rdb12} evaluated in the appropriate external leg configurations. For the sake of clarity, we write them in terms of $R_{12,jkl,5}$, the corresponding on-shell diagrams and the explicit expressions.
\begin{equation}
	R^{\text{db}}_{12,341,5} = \vcenter{\hbox{\begin{tikzpicture}[scale=0.75]
			
			\coordinate (d45) at (45:1);
			\coordinate (dm45) at (-45:1);
			\coordinate (d135) at (135:1);
			\coordinate (dm135) at (-135:1);
			
	  		\coordinate (p1) at (-1.2,0.6);
	  		\coordinate (p2) at (0,0.6);
	  		\coordinate (p3) at (1.2,0.6);
	  		\coordinate (p4) at (1.2,-0.6);
	  		\coordinate (p5) at (0,-0.6);
	  		\coordinate (p6) at (-1.2,-0.6);
	  		
	  		\coordinate (p1aux1) at ($(p1)+(-1,0)$);
	  		\coordinate (p1aux2) at ($(p1)+(d135)$);
	  		\coordinate (p1aux3) at ($(p1)+(0,1)$);
	  		\coordinate (p3aux1) at ($(p3)+(1,0)$);
	  		\coordinate (p3aux2) at ($(p3)+(d45)$);
	  		\coordinate (p3aux3) at ($(p3)+(0,1)$);
	  		\coordinate (p5aux2) at ($(p5)+(0,-1)$);
	  		\coordinate (p5aux1) at ($(p5)+(-60:1)$);
	  		\coordinate (p5aux3) at ($(p5)+(-120:1)$);
	  		
	  		\coordinate (p2aux) at ($(p2)+(0,1)$);
	  		\coordinate (p4aux) at ($(p4)+(dm45)$);
	  		\coordinate (p6aux) at ($(p6)+(dm135)$);
	  		
	  		\draw[very thick] (p1) to (p2);
	  		\draw[very thick] (p2) to (p5);
	  		\draw[very thick] (p5) to (p6);
	  		\draw[very thick] (p6) to (p1);
	  		
	  		\draw[very thick] (p2) to (p3);
	  		\draw[very thick] (p3) to (p4);
	  		\draw[very thick] (p4) to (p5);
	  		
        		\draw[very thick,larrow] (p1aux1) to (p1);
        		\draw[very thick,rarrow] (p1aux3) to (p1);
        		
        		\draw[very thick,rarrow] (p3aux2) to (p3);

        		
        		\draw[very thick, larrow] (p4aux) to (p4);
        		\draw[very thick,larrow] (p6aux) to (p6);

        		\draw[fill, thick] (p1) circle (0.2);
        		\draw[fill=white,thick] (p2) circle (0.2);
            \draw[fill, thick] (p3) circle (0.2);
 		    \draw[fill=white, thick] (p4) circle (0.2);
        		\draw[fill, thick] (p5) circle (0.2);
        		\draw[fill=white, thick] (p6) circle (0.2);

        		
        		\node at ($(p1aux1)+(-0.35,0)$) {$5$};
        		\node at ($(p1aux3)+(90:0.35)$) {$1$};
%
        		\node at ($(p3aux2)+(45:0.35)$) {$2$};
%
%
        		\node at ($(p4aux)+(-45:0.35)$) {$3$};
        		\node at ($(p6aux)+(-135:0.35)$) {$4$};
        		
        
    \end{tikzpicture}}} = -2 + 12 u_{34} -30 u_{34}^2 + 20u_{34}^3
\end{equation}
where $u_{34}$ is the same as in \eqref{eq:u34}. The second double-box leading singularity is given by
\begin{equation}
	R^{\text{db}}_{12,451,5} = 
	\vcenter{\hbox{\begin{tikzpicture}[scale=0.75]
			
			\coordinate (d45) at (45:1);
			\coordinate (dm45) at (-45:1);
			\coordinate (d135) at (135:1);
			\coordinate (dm135) at (-135:1);
			
	  		\coordinate (p1) at (-1.2,0.6);
	  		\coordinate (p2) at (0,0.6);
	  		\coordinate (p3) at (1.2,0.6);
	  		\coordinate (p4) at (1.2,-0.6);
	  		\coordinate (p5) at (0,-0.6);
	  		\coordinate (p6) at (-1.2,-0.6);
	  		
	  		\coordinate (p1aux1) at ($(p1)+(-1,0)$);
	  		\coordinate (p1aux2) at ($(p1)+(d135)$);
	  		\coordinate (p1aux3) at ($(p1)+(0,1)$);
	  		\coordinate (p3aux3) at ($(p3)+(1,0)$);
	  		\coordinate (p3aux2) at ($(p3)+(d45)$);
	  		\coordinate (p3aux1) at ($(p3)+(0,1)$);
	  		\coordinate (p5aux2) at ($(p5)+(0,-1)$);
	  		\coordinate (p5aux1) at ($(p5)+(-60:1)$);
	  		\coordinate (p5aux3) at ($(p5)+(-120:1)$);
	  		
	  		\coordinate (p2aux) at ($(p2)+(0,1)$);
	  		\coordinate (p4aux) at ($(p4)+(dm45)$);
	  		\coordinate (p6aux) at ($(p6)+(dm135)$);
	  		
	  		\draw[very thick] (p1) to (p2);
	  		\draw[very thick] (p2) to (p5);
	  		\draw[very thick] (p5) to (p6);
	  		\draw[very thick] (p6) to (p1);
	  		
	  		\draw[very thick] (p2) to (p3);
	  		\draw[very thick] (p3) to (p4);
	  		\draw[very thick] (p4) to (p5);
	  		
        		\draw[very thick,rarrow] (p1aux2) to (p1);
        		
        		\draw[very thick,rarrow] (p3aux1) to (p3);
        		\draw[very thick,larrow] (p3aux3) to (p3);

        		
        		\draw[very thick, larrow] (p4aux) to (p4);
        		\draw[very thick,larrow] (p6aux) to (p6);

        		\draw[fill, thick] (p1) circle (0.2);
        		\draw[fill=white,thick] (p2) circle (0.2);
            \draw[fill, thick] (p3) circle (0.2);
 		    \draw[fill=white, thick] (p4) circle (0.2);
        		\draw[fill, thick] (p5) circle (0.2);
        		\draw[fill=white, thick] (p6) circle (0.2);

        		
        		\node at ($(p1aux2)+(135:0.35)$) {$1$};
%
        		\node at ($(p3aux1)+(90:0.35)$) {$2$};
        		\node at ($(p3aux3)+(0:0.35)$) {$3$};
%
%
        		\node at ($(p4aux)+(-45:0.35)$) {$4$};
        		\node at ($(p6aux)+(-135:0.35)$) {$5$};
        		
        
    \end{tikzpicture}
}} = -2 + 12 u_{45} -30u_{45}^2 +20u_{45}^3 \, ,
\end{equation}
where $u_{45}$ is given by
\begin{equation}
	u_{45} = \frac{\ab{14}\ab{25}}{\ab{12}\ab{45}} \, .
\end{equation}
Finally, the last double-box prefactor in the split helicity sector is given by
\begin{equation}
	R^{\text{db}}_{12,351,5} = \vcenter{\hbox{
	\begin{tikzpicture}[scale=0.75]
			
			\coordinate (d45) at (45:1);
			\coordinate (dm45) at (-45:1);
			\coordinate (d135) at (135:1);
			\coordinate (dm135) at (-135:1);
			
	  		\coordinate (p1) at (-1.2,0.6);
	  		\coordinate (p2) at (0,0.6);
	  		\coordinate (p3) at (1.2,0.6);
	  		\coordinate (p4) at (1.2,-0.6);
	  		\coordinate (p5) at (0,-0.6);
	  		\coordinate (p6) at (-1.2,-0.6);
	  		
	  		\coordinate (p1aux1) at ($(p1)+(-1,0)$);
	  		\coordinate (p1aux2) at ($(p1)+(d135)$);
	  		\coordinate (p1aux3) at ($(p1)+(0,1)$);
	  		\coordinate (p3aux1) at ($(p3)+(1,0)$);
	  		\coordinate (p3aux2) at ($(p3)+(d45)$);
	  		\coordinate (p3aux3) at ($(p3)+(0,1)$);
	  		\coordinate (p5aux2) at ($(p5)+(0,-1)$);
	  		\coordinate (p5aux1) at ($(p5)+(-60:1)$);
	  		\coordinate (p5aux3) at ($(p5)+(-120:1)$);
	  		
	  		\coordinate (p2aux) at ($(p2)+(0,1)$);
	  		\coordinate (p4aux) at ($(p4)+(dm45)$);
	  		\coordinate (p6aux) at ($(p6)+(dm135)$);
	  		
	  		\draw[very thick] (p1) to (p2);
	  		\draw[very thick] (p2) to (p5);
	  		\draw[very thick] (p5) to (p6);
	  		\draw[very thick] (p6) to (p1);
	  		
	  		\draw[very thick] (p2) to (p3);
	  		\draw[very thick] (p3) to (p4);
	  		\draw[very thick] (p4) to (p5);
	  		
        		\draw[very thick,rarrow] (p1aux2) to (p1);
        		
        		\draw[very thick,rarrow] (p3aux2) to (p3);

        		\draw[very thick,larrow] (p5aux2) to (p5);
        		
        		\draw[very thick, larrow] (p4aux) to (p4);
        		\draw[very thick,larrow] (p6aux) to (p6);

        		\draw[fill, thick] (p1) circle (0.2);
        		\draw[fill=white,thick] (p2) circle (0.2);
            \draw[fill, thick] (p3) circle (0.2);
 		    \draw[fill=white, thick] (p4) circle (0.2);
        		\draw[fill, thick] (p5) circle (0.2);
        		\draw[fill=white, thick] (p6) circle (0.2);

        		
        		\node at ($(p1aux2)+(135:0.35)$) {$1$};
%
        		\node at ($(p3aux2)+(45:0.35)$) {$2$};
%
        		\node at ($(p5aux2)+(-90:0.35)$) {$4$};
%
        		\node at ($(p4aux)+(-45:0.35)$) {$3$};
        		\node at ($(p6aux)+(-135:0.35)$) {$5$};
        		

    \end{tikzpicture}
	}} = -2 + 12 u_{35} - 30u_{35}^2 + 20u_{35}^3 \, ,
\end{equation}
where $u_{35}$ is the following cross-ratio
\begin{equation}
	u_{35} = \frac{\ab{13}\ab{25}}{\ab{12}\ab{35}} \, .
\end{equation}

\subsubsection{ $-+-++$ prefactors at two-loops}

In the case of the non-split helicity sector $(-+-++)$, the double-box prefactor expressions are significantly more complicated. They are expressed as polynomials in two cross-ratios $u$ and $v$ given by
\begin{equation}
	u = \frac{\ab{12}\ab{35}}{\ab{13}\ab{25}} \, , \quad v = \frac{\ab{14}\ab{35}}{\ab{13}\ab{45}} \, .
\end{equation}
The corresponding expressions for the two double-box prefactors are
\begin{equation}
	R^{\text{(db)}}_{13,451,5} =
	\vcenter{\hbox{\begin{tikzpicture}[scale=0.75]
			
			\coordinate (d45) at (45:1);
			\coordinate (dm45) at (-45:1);
			\coordinate (d135) at (135:1);
			\coordinate (dm135) at (-135:1);
			
	  		\coordinate (p1) at (-1.2,0.6);
	  		\coordinate (p2) at (0,0.6);
	  		\coordinate (p3) at (1.2,0.6);
	  		\coordinate (p4) at (1.2,-0.6);
	  		\coordinate (p5) at (0,-0.6);
	  		\coordinate (p6) at (-1.2,-0.6);
	  		
	  		\coordinate (p1aux1) at ($(p1)+(-1,0)$);
	  		\coordinate (p1aux2) at ($(p1)+(d135)$);
	  		\coordinate (p1aux3) at ($(p1)+(0,1)$);
	  		\coordinate (p3aux3) at ($(p3)+(1,0)$);
	  		\coordinate (p3aux2) at ($(p3)+(d45)$);
	  		\coordinate (p3aux1) at ($(p3)+(0,1)$);
	  		\coordinate (p5aux2) at ($(p5)+(0,-1)$);
	  		\coordinate (p5aux1) at ($(p5)+(-60:1)$);
	  		\coordinate (p5aux3) at ($(p5)+(-120:1)$);
	  		
	  		\coordinate (p2aux) at ($(p2)+(0,1)$);
	  		\coordinate (p4aux) at ($(p4)+(dm45)$);
	  		\coordinate (p6aux) at ($(p6)+(dm135)$);
	  		
	  		\draw[very thick] (p1) to (p2);
	  		\draw[very thick] (p2) to (p5);
	  		\draw[very thick] (p5) to (p6);
	  		\draw[very thick] (p6) to (p1);
	  		
	  		\draw[very thick] (p2) to (p3);
	  		\draw[very thick] (p3) to (p4);
	  		\draw[very thick] (p4) to (p5);
	  		
        		\draw[very thick,rarrow] (p1aux2) to (p1);
        		
        		\draw[very thick,larrow] (p3aux1) to (p3);
        		\draw[very thick,rarrow] (p3aux3) to (p3);

        		
        		\draw[very thick, larrow] (p4aux) to (p4);
        		\draw[very thick,larrow] (p6aux) to (p6);

        		\draw[fill, thick] (p1) circle (0.2);
        		\draw[fill=white,thick] (p2) circle (0.2);
            \draw[fill, thick] (p3) circle (0.2);
 		    \draw[fill=white, thick] (p4) circle (0.2);
        		\draw[fill, thick] (p5) circle (0.2);
        		\draw[fill=white, thick] (p6) circle (0.2);

        		
        		\node at ($(p1aux2)+(135:0.35)$) {$1$};
%
        		\node at ($(p3aux1)+(90:0.35)$) {$2$};
        		\node at ($(p3aux3)+(0:0.35)$) {$3$};
%
%
        		\node at ($(p4aux)+(-45:0.35)$) {$4$};
        		\node at ($(p6aux)+(-135:0.35)$) {$5$};
        		
        
    \end{tikzpicture}
	}} = \begin{aligned}
		 2 u &\left[ u^3+2 u^2 (v-2)+u (v (5 v-8)+6) \right. \\
		& \left. +2 v (5 (v-2) v+6)-4\right]
	\end{aligned}
	 \, ,
\end{equation}
and
\begin{equation}
	R^{\text{(db)}}_{13,452,5} = \vcenter{\hbox{\begin{tikzpicture}[scale=0.75]
			
			\coordinate (d45) at (45:1);
			\coordinate (dm45) at (-45:1);
			\coordinate (d135) at (135:1);
			\coordinate (dm135) at (-135:1);
			
	  		\coordinate (p1) at (-1.2,0.6);
	  		\coordinate (p2) at (0,0.6);
	  		\coordinate (p3) at (1.2,0.6);
	  		\coordinate (p4) at (1.2,-0.6);
	  		\coordinate (p5) at (0,-0.6);
	  		\coordinate (p6) at (-1.2,-0.6);
	  		
	  		\coordinate (p1aux1) at ($(p1)+(-1,0)$);
	  		\coordinate (p1aux2) at ($(p1)+(d135)$);
	  		\coordinate (p1aux3) at ($(p1)+(0,1)$);
	  		\coordinate (p3aux1) at ($(p3)+(1,0)$);
	  		\coordinate (p3aux2) at ($(p3)+(d45)$);
	  		\coordinate (p3aux3) at ($(p3)+(0,1)$);
	  		\coordinate (p5aux2) at ($(p5)+(0,-1)$);
	  		\coordinate (p5aux1) at ($(p5)+(-60:1)$);
	  		\coordinate (p5aux3) at ($(p5)+(-120:1)$);
	  		
	  		\coordinate (p2aux) at ($(p2)+(0,1)$);
	  		\coordinate (p4aux) at ($(p4)+(dm45)$);
	  		\coordinate (p6aux) at ($(p6)+(dm135)$);
	  		
	  		\draw[very thick] (p1) to (p2);
	  		\draw[very thick] (p2) to (p5);
	  		\draw[very thick] (p5) to (p6);
	  		\draw[very thick] (p6) to (p1);
	  		
	  		\draw[very thick] (p2) to (p3);
	  		\draw[very thick] (p3) to (p4);
	  		\draw[very thick] (p4) to (p5);
	  		
        		\draw[very thick,rarrow] (p1aux1) to (p1);
        		\draw[very thick,larrow] (p1aux3) to (p1);
        		
        		\draw[very thick,rarrow] (p3aux2) to (p3);

        		
        		\draw[very thick, larrow] (p4aux) to (p4);
        		\draw[very thick,larrow] (p6aux) to (p6);

        		\draw[fill, thick] (p1) circle (0.2);
        		\draw[fill=white,thick] (p2) circle (0.2);
            \draw[fill, thick] (p3) circle (0.2);
 		    \draw[fill=white, thick] (p4) circle (0.2);
        		\draw[fill, thick] (p5) circle (0.2);
        		\draw[fill=white, thick] (p6) circle (0.2);

        		
        		\node at ($(p1aux1)+(-0.35,0)$) {$1$};
        		\node at ($(p1aux3)+(90:0.35)$) {$2$};
%
        		\node at ($(p3aux2)+(45:0.35)$) {$3$};
%
%
        		\node at ($(p4aux)+(-45:0.35)$) {$4$};
        		\node at ($(p6aux)+(-135:0.35)$) {$5$};
        		
        
    \end{tikzpicture}}} = 
    \begin{aligned}
    	2 (u-1) & \left[5 (u-3) v^2+2 ((u-3) u+3) v \right. \\ 
    	& \left. \quad  +(u-1)^3+10 v^3\right]
    \end{aligned}
     \, .
\end{equation}

\section{Multi-Regge limit}
\label{sec: Regge}

In subsection~\ref{sec:MT}, we discussed an interesting relation between the hard functions in pure YM theory and the remainder function in ${\cal N}=4$ sYM. Here we uncover a further interesting connection between these four-dimensional quantities in two gauge theories, now in the multi-Regge limit. Throughout this Appendix, all equations are understood at the level of symbols and for the maximally-transcendental parts.

We consider the bootstrapped pure YM hard functions in the multi-Regge kinematics \cite{DelDuca:2022skz} for the scattering channel $56 \to 1234$. In this kinematic regime, the transverse momenta are of comparable magnitude 
\begin{align}
|{\bf p}_1| \simeq |{\bf p}_2| \simeq |{\bf p}_3| \simeq |{\bf p}_4| \,,
\end{align}
while the rapidities are strongly ordered,
\begin{align}
|p^{+}_1| \gg |p^{+}_2| \gg |p^{+}_3| \gg |p^{+}_4| \,, \\
|p^{-}_1| \ll |p^{-}_2| \ll |p^{-}_3| \ll |p^{-}_4| \,,
\end{align}
where we use light-cone coordinates $p_i = (p^{+}_{i},p^{-}_{i},{\bf p}_i)$. To control the multi-Regge regime, we adopt the  following parametrization of the kinematics with a small parameter $x \to 0$,
\begin{align}
& s_{12} = \frac{s_1}{x} \,,\quad s_{23} = \frac{s_2}{x} \,,\quad s_{34} = \frac{s_3}{x} \,,\quad s_{56} = \frac{s_1 s_2 s_3}{\kappa^2 |z_1-z_2|^2 x^3}  \,,\quad s_{345} = -|z_2|^2 \kappa \notag\\
& s_{123} = \frac{s_1 s_2}{\kappa  |z_1-z_2|^2 x^2} \,,\quad s_{234} = \frac{s_2 s_3}{\kappa x^2} \,,\quad s_{16} = - |z_1|^2 \kappa  \,,\quad s_{45} = -|1-z_2|^2\kappa
\end{align}
where only the leading terms in this limit are displayed; $s_1,s_2,s_3,\kappa$ have the dimensions of Mandelstam variables, while $z_1,\bar{z}_1,z_2,\bar{z}_2$ are dimensionless. We employ the shorthand notation $|z_i|^2 = z_i \bar{z}_i$; in the Euclidean region, $z_i,\bar{z}_i$ are treated as independent variables.

At one loop, we find that all prefactors (see \eqref{eq:L1LS}) are of the same order as $x\to 0$, once normalized by the tree-level $\text{PT}_{n,1i}$ \eqref{eq:PT},
\begin{align}
\frac{R_{1i,jk}^{(1)}}{\text{PT}_{n,1i}} \to 1 \qquad \text{at} \quad x \to 0 \,.
\end{align}
The one-loop six-point hard functions for all MHV helicity configurations $(h_1 \ldots h_6)$ exhibit identical asymptotic behavior as $x \to 0$, coinciding with that of the one-loop MHV hard function in ${\cal N}=4$ sYM,
\begin{align}
H^{(1)}_{{h_1 \ldots h_6}}/ A^{(0)}  = 
 \left( H^{(1)}/ A^{(0)}  \right)^{\text{${\cal N}=4$ sYM}}  \qquad \text{at} \quad x \to 0 \,
 \label{eq:MRK1L}
\end{align}
where all divergent powers of $\log(x)$ and finite terms are retained, and $A^{(0)}$ is the tree-level partial amplitude given by the PT factor.

At two loops, some prefactors, normalized by the PT factor, become rational numbers $\{0,\pm1,\pm\frac12\}$ as $x \to 0$, while others diverge in this limit. We verify that the symbols accompanying the divergent prefactors vanish as $x\to 0$. As a result, the two-loop hard function in this limit is pure, {\it i.e.}, all prefactors reduce to numerical constants.  Moreover, as in the one-loop case, we find that the two-loop hard functions in pure YM theory and ${\cal N}=4$ sYM coincide in the multi-Regge limit for arbitrary MHV helicity configurations,
\begin{align}
 \label{eq:MRK2L}
 H^{(2)}_{h_1 \ldots h_6}/ A^{(0)}   = 
 \left( H^{(2)}/ A^{(0)}  \right)^{\text{${\cal N}=4$ sYM}} =  \frac{1}{2} \left( \left( H^{(1)}/ A^{(0)}  \right)^{\text{${\cal N}=4$ sYM}} \right)^2  \qquad \text{at} \quad x \to 0 \,
\end{align}
Let us recall that the symbol of the two-loop remainder function of ${\cal N}=4$ sYM vanishes in the multi-Regge limit, which explains the one-loop exponentiation of the hard function in the previous equation.

Finally, we note that the $N_f$-contributions to QCD hard functions vanish in the multi-Regge limit. Consequently, eqs.~\eqref{eq:MRK1L} and \eqref{eq:MRK2L} also hold for the QCD hard functions.

\bibliographystyle{JHEP}
\bibliography{refs}
\end{document}